\documentclass[preprint,12pt]{elsarticle}

\usepackage{graphicx}
\usepackage{amsmath}
\usepackage{amsfonts}
\usepackage{amssymb}
\usepackage{mathtools}
\usepackage[T1]{fontenc}
\usepackage{ulem}
\DeclarePairedDelimiter\floor{\lfloor}{\rfloor}
\usepackage[a4paper,left=3cm,right=3cm,top=2.5cm,bottom=2.5cm]{geometry}

\extrafloats{100}
\usepackage[table,xcdraw]{xcolor}
\usepackage{url}

\makeatletter
\let\sv@endpart\@endpart
\def\@endpart{\thispagestyle{empty}\sv@endpart}
\makeatother

\journal{Physics Reports}

\begin{document}

\begin{frontmatter}

\title{Multiscale characteristics of the emerging global cryptocurrency market}

\author[ifj,pk]{Marcin W\k atorek\corref{cor2}}
\cortext[cor2]{marcin.watorek@pk.edu.pl}
\author[ifj,pk]{Stanis{\l}aw Dro\.zd\.z\corref{cor1}}
\cortext[cor1]{stanislaw.drozdz@ifj.edu.pl}
\author[ifj]{Jaros{\l}aw Kwapie\'n}
\author[ifj]{Ludovico Minati}
\author[ifj,uj]{Pawe{\l} O\'swi\k{e}cimka}
\author[pk]{Marek Stanuszek}

\address[ifj]{Complex Systems Theory Department, Institute of Nuclear Physics, Polish Academy of Sciences, ul.~Radzikowskiego 152, 31-342 Krak\'ow, Poland}
\address[pk]{Faculty of Computer Science and Telecommunications, Cracow University of Technology, ul.~Warszawska 24, 31-155 Krak\'ow, Poland}

\address[uj]{Faculty of Physics, Astronomy and Applied Computer Science, Jagiellonian University, ul. {\L}ojasiewicza 11, 30-348 Krak\'ow, Poland}

\thispagestyle{empty}

\begin{abstract}

Modern financial markets are characterized by a rapid flow of information, a vast number of participants having diversified investment horizons, and multiple feedback mechanisms, which collectively lead to the emergence of complex phenomena, for example speculative bubbles or crashes. As such, they are considered as one of the most complex systems known. Numerous studies have illuminated stylized facts, also called complexity characteristics, which are observed across the vast majority of financial markets. These include the so-called ``fat tails'' of the returns distribution, volatility clustering, the ``long memory'', strong stochasticity alongside non-linear correlations, persistence, and the effects resembling fractality and even multifractality.\par
The striking development of the cryptocurrency market over the last few years -- from being entirely peripheral to capitalizing at the level of an intermediate-size stock exchange -- provides a unique opportunity to observe its evolution in a short period. The availability of high-frequency data allows conducting advanced statistical analysis of fluctuations on cryptocurrency exchanges right from their birth up to the present day. This opens a window that allows quantifying the evolutionary changes in the complexity characteristics which accompany market emergence and maturation. The purpose of the present review, then, is to examine the properties of the cryptocurrency market and the associated phenomena. The aim is to clarify to what extent, after such an impetuous development, the characteristics of the complexity of exchange rates on the cryptocurrency market have become similar to traditional and mature markets, such as stocks, bonds, commodities or currencies.\par
The review introduces the history of cryptocurrencies, offering a description of the blockchain technology behind them. Differences between cryptocurrencies and the exchanges on which they are traded have been consistently shown. The central part of the review surveys the analysis of cryptocurrency price changes on various platforms. The statistical properties of the fluctuations in the cryptocurrency market have been compared to the traditional markets. With the help of the latest statistical physics methods, namely, the multifractal cross-correlation analysis and the $q$-dependent detrended cross-correlation coefficient, the non-linear correlations and multiscale characteristics of the cryptocurrency market are analyzed. In the last part of this paper, through applying matrix and network formalisms, the co-evolution of the correlation structure among the 100 cryptocurrencies having the largest capitalization is retraced. The detailed topology of cryptocurrency network on the Binance platform from bitcoin perspective is also considered. Finally, an interesting observation on the Covid-19 pandemic impact on the cryptocurrency market is presented and discussed: recently we have witnessed a ``phase transition'' of the cryptocurrencies from being a hedge opportunity for the investors fleeing the traditional markets to become a part of the global market that is substantially coupled to the traditional financial instruments like the currencies, stocks, and commodities.\par
The main contribution is an extensive demonstration that, fuelled by the increased transaction frequency, turnover, and the number of participants, structural self-organization in the cryptocurrency markets has caused the same to attain complexity characteristics that are nearly indistinguishable from the Forex market at the level of individual time-series. However, the cross-correlations between the exchange rates on cryptocurrency platforms differ from it. The cryptocurrency market is less synchronized and the information flows more slowly, which results in more frequent arbitrage opportunities. The methodology used in the review allows the latter to be detected, and lead-lag relationships to be discovered. Hypothetically, the methods for describing correlations and hierarchical relationships between exchange rates presented in this review could be used to construct investment portfolios and reduce exposure to risk. A new investment asset class appears to be dawning, wherein the bitcoin assumes the role of the natural base currency to trade.
\thispagestyle{empty}
\end{abstract}
\thispagestyle{empty}


\end{frontmatter}

\newpage
\addtocontents{toc}{\protect\thispagestyle{empty}}
\thispagestyle {empty}
\tableofcontents 
\thispagestyle {empty}

\newpage
\setcounter{page}{1}

\section{Introduction}
\label{Intro}

Modern financial markets are characterized by a rapid flow of information. There is a huge number of transactions between market participants with different investment horizons. There are pension funds, for those time scale is years, and at the same time specialized algorithms operating at the level of seconds or even milliseconds (high-frequency trading). The market behaviour is the result of various influencing factors, ranging from economic data, results of companies, interventions of central banks, poll and referendum results, individual tweets of high-ranking people, and mutual interactions between participants. Through feedback, this leads to critical-like phenomena such as speculative bubbles or crashes. This happens often in hours or even minutes -- the so-called ``flash crashes''. These features undoubtedly fit into the characteristics of complex systems, such as a large number of elements, nonlinear interactions, structural self-organization, and emergent phenomena.

The first quantitative research on financial markets was the subject of Louis Bachelier's work~\cite{Bachelier}, in which he derived a formula for option price based on cumulative distribution function (CDF) of a stochastic process now called the Wiener process. Later Beno\^{i}t Mandelbrot's works turned out to be a groundbreaking achievement~\cite{Mandelbrot1963}. While examining cotton price fluctuations, he observed that, contrary to wide belief, their probability distribution function (PDF) is characterized by heavy, non-Gaussian tails. Mandelbrot was also the first to notice a fractal structure of stock price fluctuations.

More than a quarter century ago, statistical physicists started their serious research in financial markets, which led to an outburst of a new discipline -- econophysics~\cite{Schinckus2016,Kutner2019}. Numerous studies carried out since then allowed researchers for a better understanding of the mechanisms governing various phenomena on both the macroscopic level (like speculative bubble formation, market crashes, asset cross-correlations, nonlinear autocorrelations, portfolio evolution, etc.) and the microscopic level (order book properties, efficacy of investing strategies, price formation, etc.). Along with the understanding, new models have been proposed that better describe and predict market behaviour and practical trading algorithms have been developed and applied especially in high-frequency trading.

All types of markets have been a subject of research since the beginning of econophysics: stock markets~\cite{Lux1996,Gopi1999,Mantegna1999,Drozdz2000}, commodity markets~\cite{WESTERHOFF2005,Drozdzoil,Sieczka2009}, option and future contract markets~\cite{PERELLO2002,THOMAKOS2002,MCCAULEY2003}, bond markets~\cite{CUNIBERTI1999,BERSHADSKII2001,ZHOU2004}, real-estate markets~\cite{KAIZOJI2004,ZHOU2006,RICHMOND2007}, as well as the foreign currency market (Forex)~\cite{Ausloos2000,Drozdz2007curr,drozdzepps}. Unlike typical complex systems studied by experimental physicists, these markets offer good quality data with minimum systematic errors, which is one of the reasons why so much attention they attract. However, there is an important issue that is inherently associated with all the above mentioned markets: they have already been existing for a long time before they become a subject of research. Therefore, we cannot investigate their evolution since their origin. Nevertheless, more or less a decade ago a brand new market was established -- a cryptocurrency market, which offers precisely what the other markets lack -- a possibility of observing their structural self-organization process from the very beginning to a rather mature form in a short period of time. A principal question is to what extent, after such dynamic development, complexity characteristics of the cryptocurrency market are similar to those of traditional markets, especially Forex.

The first cryptocurrency - Bitcoin (BTC) - was proposed in 2008 by a person or group of people under a nickname Satoshi Nakamoto~\cite{Nakamoto2008}. The date of this new concept emergence does not seem to be accidental, because it is correlated with an epicenter of the global financial crisis of 2008-2010. In order to mitigate its effects, the central banks began to increase the monetary base (``printing money'') massively, which weakened trust in the traditional fiat currencies. The bitcoin protocol was based on a peer-to-peer network and the previously known public and private key cryptographic techniques together with a new consensus mechanism called ``Proof of Work''. The idea behind bitcoin was to provide, for the first time in human history, a tool that would allow people anywhere to trust each other and carry out transactions via Internet without a central management institution. Instead of the current confidence in the state/central banks, confidence in technology was proposed.

The first widely recognized platform offering BTC-to-fiat-currency exchange was Mt.~Gox founded in Feb 2011. Since then a spectacular development of the cryptocurrency market has occurred. There are already 3,600 different cryptocurrencies traded on 350 platforms with nearly 30,000 cryptocurrency pairs listed. Current capitalization of the entire market is about 350 billion USD (Oct 2020~\cite{coinmarket}). It increased rapidly in 2017 during a speculative bubble referred to as ICO-mania~\cite{Gerlach2018,Aste2019} reaching over 800 billion USD in the beginning of 2018. BTC valuation on some Korean platforms was even 20 000 USD. Currently, the market is highly decentralized with the same cryptocurrency pairs listed on many platforms. There is no single price to refer to unlike that provided by Reuters on Forex. Only for BTC there are quoted future contracts launched in Dec 2017 on the CME group~\cite{CME}. Another characteristic property is that cryptocurrency transactions are most often made through the platforms in contrast to Forex with its over-the-counter market.

Aim of this work is to review the available results on the cryptocurrency market obtained with various methods of statistical physics with the multifractal formalism and the network approach in particular. Our work sketches briefly history of the cryptocurrency introduction (Sect.~\ref{historia}) and describes the blockchain technology standing behind them (Sect.~\ref{blockchain}), but its main part presents a review of the most important properties of the cryptocurrency price fluctuations. Price return distributions, autocorrelations, and inter-transaction times of cryptocurrency exchange rates are compared with their counterparts in Forex (Sect.~\ref{Dane stat}). Non-linear auto- and cross-correlations together with multifractal-like characteristics on the cryptocurrency market are then described. Among others, it is shown that PDF/CDF form, autocorrelation functions, Hurst exponents, and multiscale properties depend on trading frequency (Sect.~\ref{nonlinear}). Cross-correlations among the cryptocurrency exchange rate pairs within one trading platform and between two platforms, as well as between the cryptocurrency platforms and Forex are also discussed (Sect.~\ref{sect::NonlinearCrossCorrel}). Triangle arbitrage effects are a subject of Sect.~\ref{sect::TriangularArbitrage}. Latest Covid-19 pandemic impact on correlations between cryptocurrencies and traditional financial markets is covered in Sect.~\ref{duka_krypto}. The last part of this review discusses results of a network approach (Sect.~\ref{Matrixcorr}). It is shown that, currently, BTC is a natural base cryptocurrency for other cryptocurrencies.

\section{The emergence and technology of cryptocurrencies}

\subsection{History: from barter through the blockchain}
\label{historia}

\begin{figure}[ht!]
\centering
\includegraphics[width=1\textwidth]{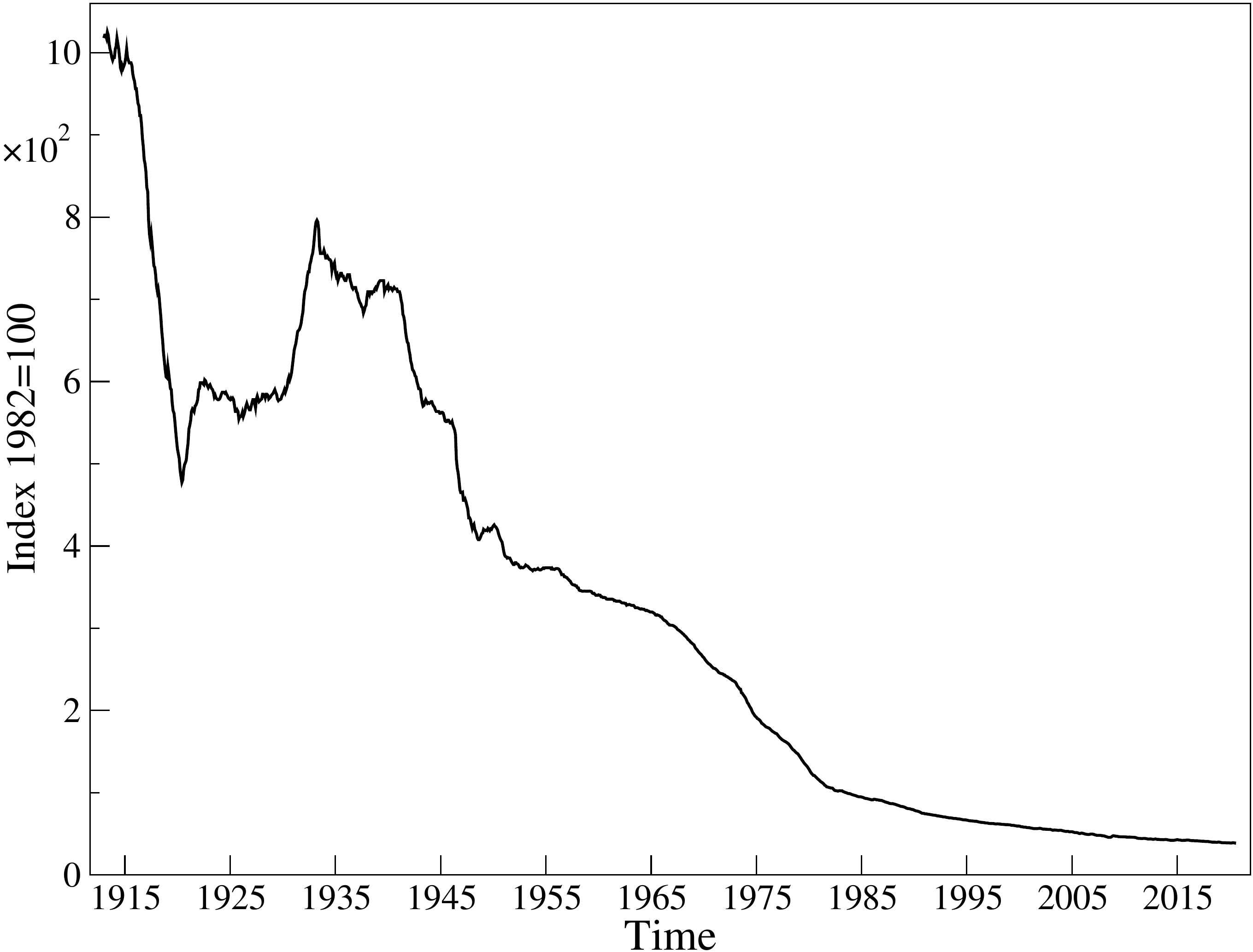}
\caption{Purchasing power index of the US dollar (USD). Source: Board of Governors of the Federal Reserve System (US)~\cite{StlouisFED}.}
\label{fig:usdpower}
\end{figure}

\begin{figure}[!ht]
\centering
\includegraphics[width=1\textwidth]{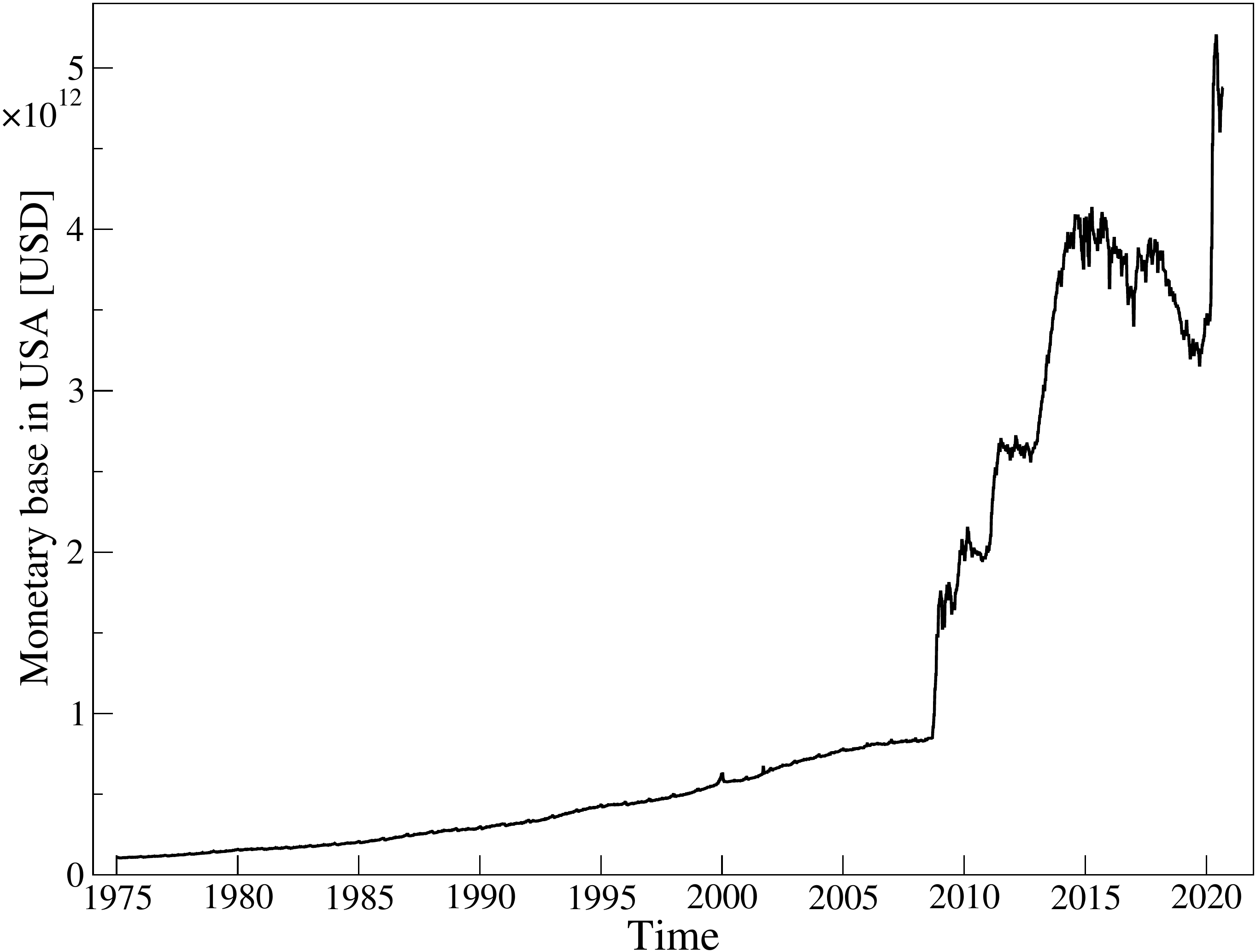}
\caption{Monetary base in USD. Source: Board of Governors of the Federal Reserve System (US) \cite{StlouisFED}.}
\label{fig:monetarybase}
\end{figure}

Human history has been paced by discrete events that shaped its course through discontinuously accelerating development. Without question, one of these events is the emergence of money in the form of standardized coins, which were introduced in Lydia around 600 years BC. This first generation of currency shaped Mediterranean culture in many ways~\cite{moneyhistory}, and spearheaded a continuous increase in trade exchanges. Merchants no longer had to resort to barter, good for good, or service for service. The second generation was paper money, which appeared in the Renaissance. It was introduced by Italian banks and subsequently taken over by the established national banks. The invention of banking and paper money disrupted the status quo of the feudal systems, by strengthening the nation-states which had the power to issue it. From this point onwards, the possibility of modulating the money supply upwards or downwards was easily available, eventually giving rise to modern capitalism~\cite{moneyhistory}. The 20th century brought the appearance of electronic transactions, which further accelerated the circulation of currency throughout the economy, boosting growth. Crucially, paper and electronic money caused a total decoupling of fiat currencies from their fundamental value. After the collapse of the Bretton Woods system in 1971, exchange rates ceased to be linked to gold and based on trust in the state. From that point onwards, fiat currencies systematically lost their value, as shown for the example of the US dollar in Fig.~\ref{fig:usdpower}. Moreover, since breaking with gold as a reference, the number of USD in circulation has kept increasing systematically, as visible in Fig. ~\ref{fig:monetarybase}.\par

Currently, central banks can easily increase the supply of currency without any physical printing. After the financial crisis of the year 2008, all the major central banks significantly expanded their monetary base through quantitative easing programs (QE), as visualized in Fig. ~\ref{fig:monetarybase}. Incidentally, around the same time, a radically new financial asset appeared. That was the first cryptocurrency, the Bitcoin~\cite{Nakamoto2008}. The underlying revolutionary idea was to combine existing technologies in a novel way, namely asymmetric cryptography, together with a distributed database having a new ``Proof of Work'' consensus mechanism into a decentralized, secure register (the distributed ledger technology, DLT) blockchain~\cite{Wattenhofer}.\par

The intention is that cryptocurrencies are not subjugated to any institution or government; rather, they are based on trust in technological infrastructure. They allow sending financial resources anywhere in the world with almost zero latency. The network users themselves provide the authentication mechanism. The cryptocurrency concept combines the advantages of cash, namely the transaction anonymity, with the speed and convenience of electronic transactions. Importantly, unlike traditional currencies, the Bitcoin by design has a built-in supply limitation mechanism, which helps prevent its loss of value.\par

Initially, the Bitcoin seemed to be only a technological curiosity, and there was no organized trade. Individual transactions for exchanging real goods via on-line discussion groups were described, such as buying two pizzas in May 2010 for 10,000 BTC~\cite{pizzaday}. However, the innovative idea soon began diffusing outside its original circle of computer geeks, extending towards the broader financial sector and, eventually, by virtue of the anonymity, also to criminal circles. The first widely recognized exchange, which enabled bitcoins to be traded for traditional currencies, Mt.Gox, was launched in July 2010. Soon after, the first on-line black market, Silk Road, was created: it allowed virtually unregulated trading of any goods, bolstered by bitcoin payment and full anonymity. Perhaps concerningly, this was the first practical application of the Bitcoin. It significantly increased demand and contributed to the first speculative bubble on bitcoin prices~\cite{Gerlach2018}, which burst after Silk Road was shut down by the FBI in October 2013 and trading by the largest cryptocurrency exchange, Mt.Gox, was halted in February 2014, plausibly after a hack which led to the disappearance of 850,000 BTC.\par

\begin{figure}[!ht]
\centering
\includegraphics[width=1\textwidth]{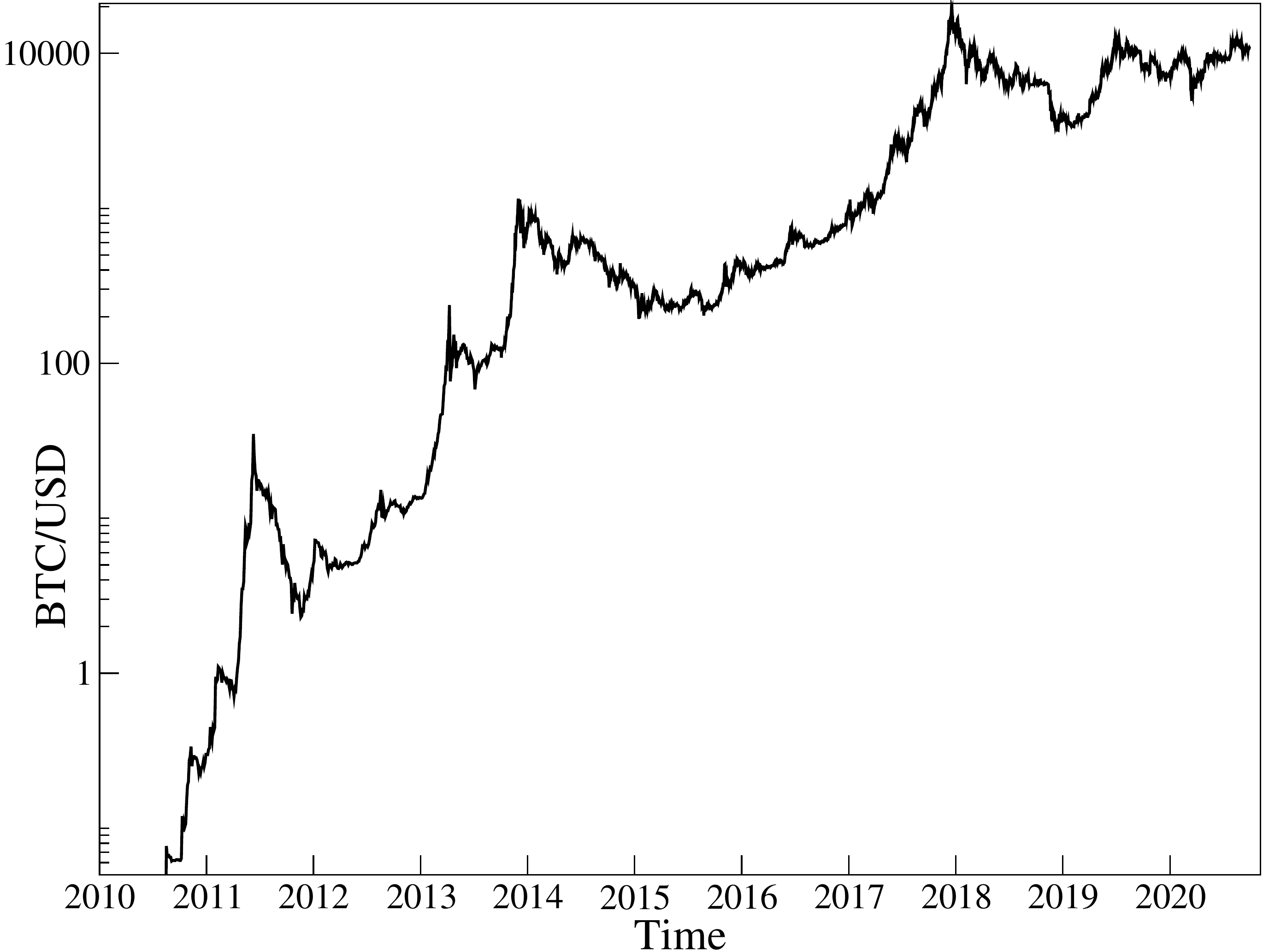}
\caption{Bitcoin price in USD.}
\label{fig:BTCUSDdaily}
\end{figure}

\begin{figure}[!ht]
\centering
\includegraphics[width=1\textwidth]{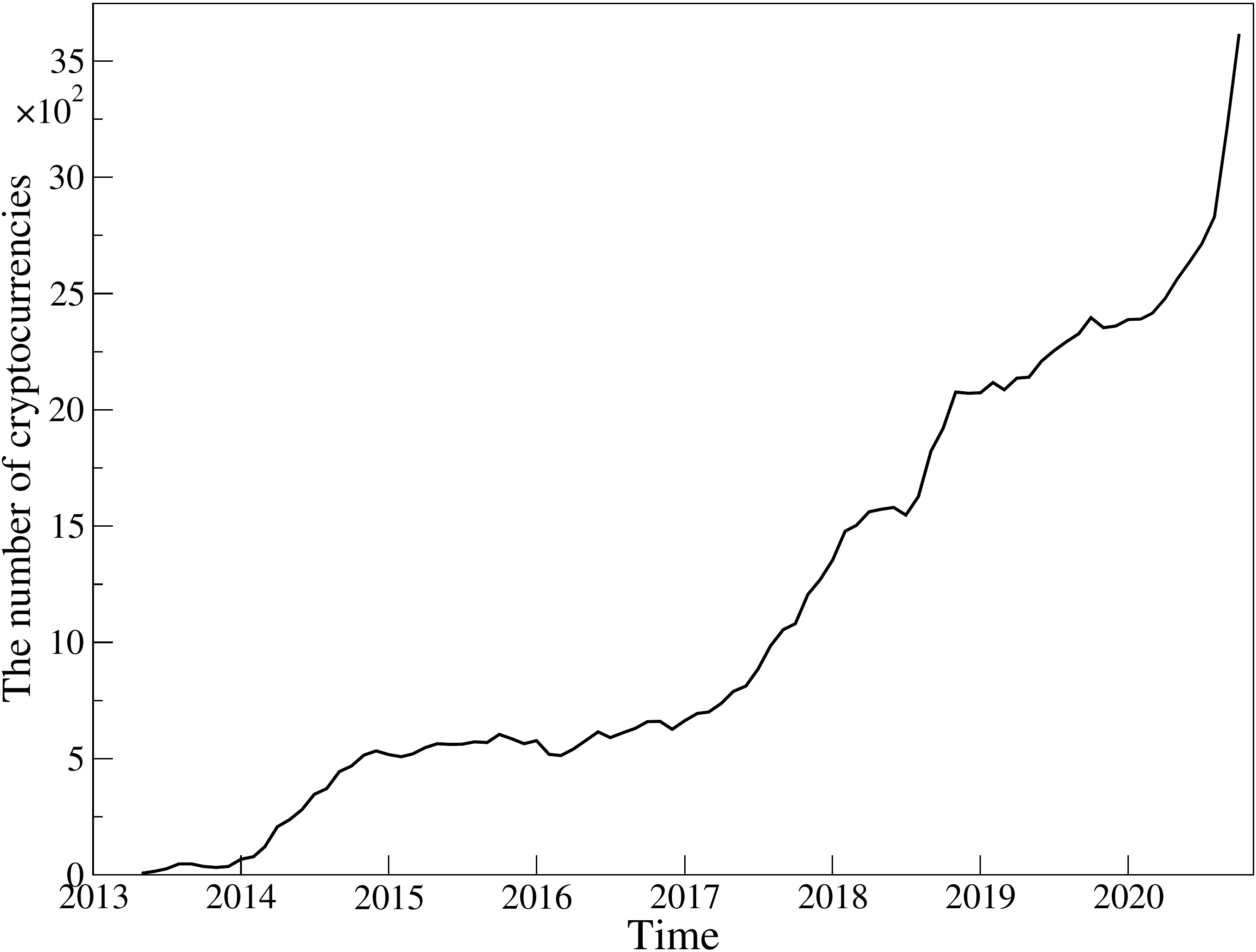}
\caption{The number of actively traded cryptocurrencies. Data from the CoinMarketCap website~\cite{coinmarket}.}
\label{fig:Lkrypto}
\end{figure}

\begin{figure}[!ht]
\centering
\includegraphics[width=1\textwidth]{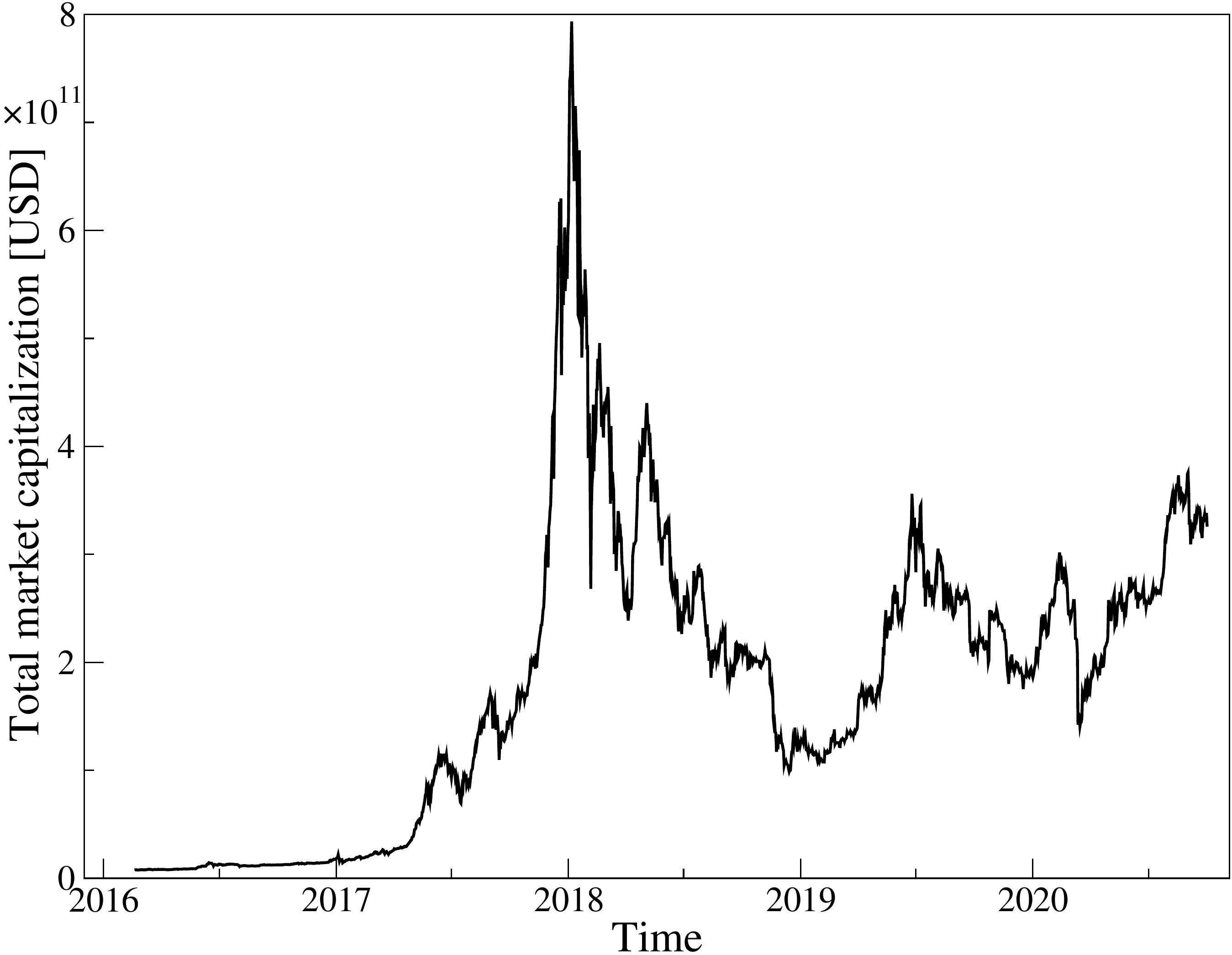}
\caption{Capitalization of the entire cryptocurrency market in USD. Data from CoinMarketCap.com~\cite{coinmarket}.}
\label{fig:Tcap}
\end{figure}

As recognition increased, the technology became increasingly widespread. Eventually, it turned out that the blockchain technology on which cryptocurrencies are based allows using for a decentralized registry not only for financial purposes but also for the execution of computer code (scripts) in a distributed form. At the end of 2013, the idea of an Ethereum distributed computing network was introduced. The project started in July 2015~\cite{ethereum}. The Ethereum platform allows anyone to create decentralized applications operating without the possibility of downtime, censorship, fraud, or tampering with their code. It also allows one to issue their own tokens via smart contracts, which consist of computer code performing a prescribed action under certain conditions on the Ethereum blockchain. This innovation quickly found application in raising capital via a simplified mechanism for various projects through the so-called Initial Coin Offer (ICO). In 2017, the ICO underwent a boom, which contributed to another speculative bubble on the cryptocurrency market, referred to as the ICO-mania~\cite{Aste2019}. At that time, the number of cryptocurrencies doubled from 700 to 1,400 by the end of 2017 (figure~\ref{fig:Lkrypto}), and the capitalization of the entire market reached USD 800 billion (figure~\ref{fig:Tcap}). Unavoidably, the bubble eventually burst in January 2018.\par

The current state of the blockchain technology can be compared to the dot-com bubble at the turn of the century. The considerable potential was already visible in the then possibilities of the Internet, but it was not known precisely which particular way the technology would develop. At that time, merely mention a company's intention to engage in Internet-related activities would cause a sharp increase in share price~\cite{Shiller}. A similar case was verified for the cryptocurrencies. With the first euphoric phase now well over, the bitcoin from top to bottom in December 2018 lost over 80 percent of its value (figure~\ref{fig:BTCUSDdaily}). Other cryptocurrencies have dropped by as much as 99 percent, with some unavoidably succumbing. Currently, the first practical applications of the blockchain are emerging; these will be discussed alongside the mechanics of cryptocurrencies in the next sections.

\subsection{Core technology}
\label{blockchain}
 
The most straightforward approach for dispatching ``electronic cash'' would be using data files. However, digital data can inherently be replicated unlimitedly, which engenders a double-spending problem. Therefore, there was a need for a technology that could form an electronic register covering all transactions: transferring funds would, then, consist of swapping entries in the register. This type of record is commonplace in the electronic banking system, but central to the notion of a cryptocurrency was the aim of abandoning their central character, rendering them completely public instead. In this way, each user could verify transactions independently, yielding another valuable property, namely the impossibility of corrupting the registry history. This challenge was resolved through the use of peer-to-peer (P2P) networks, cryptographic techniques, and connecting transactions in blocks, and after that in a chain of blocks, the blockchain. In this section, without losing generality, the underlying technology will be presented, taking as an example the first cryptocurrency, the Bitcoin.

\subsubsection{The network}
\label{BTC_blockchain}

\begin{figure}[ht!]
\centering
\includegraphics[width=1\textwidth]{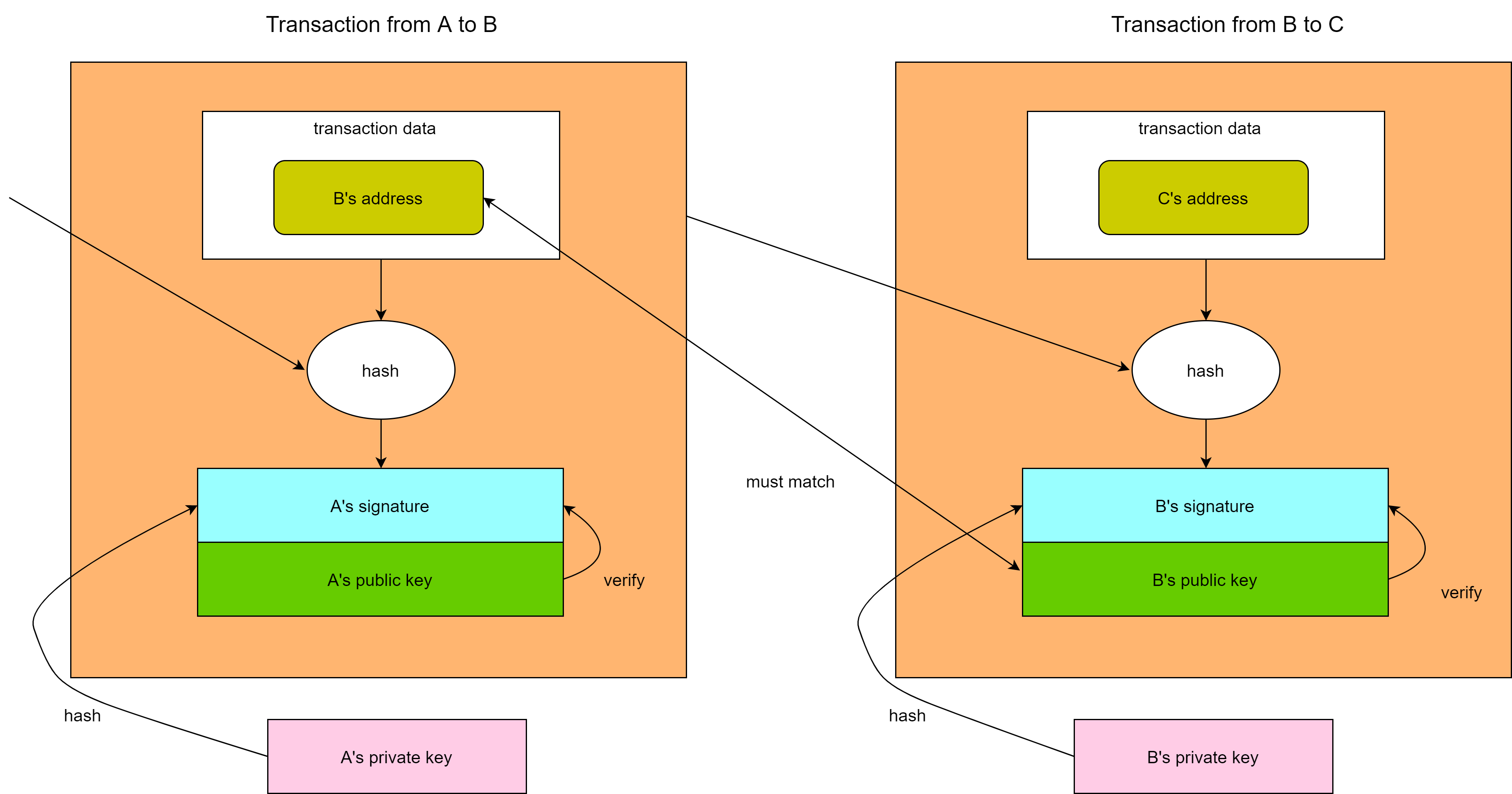}
\caption{Transaction scheme in Bitcoin network.}
\label{fig:transakcja}
\end{figure}

The Bitcoin is a virtual monetary entity and, as such, is devoid of any physical form. The minimum unit is 0.00000001 BTC (1 satoshi). Bitcoin blockchains are data files that contain information about past transactions and about the creation of new Bitcoins, providing a means of reward for closing a block with transactions. This is known as the Bitcoin network register. The blockchain consists of a sequence of blocks, wherein each subsequent block is built upon the previous one and contains information about new transactions in the network. The first block was created in 2009. Currently, as of October 2020 the chain counts nearly 650,000 blocks. As said, the Bitcoin blockchain is public: everyone can access all information about how many BTCs belong to a given address in the network. Further, the registry is not centralized: all network participants have their own blockchain copies and can modify them. There is no supervisory unit; however, there is a set of rules that each user must follow. As a consequence, the participants reciprocally control each other. To enter the Bitcoin network, a prospective user must use a network client or an external wallet. Enacting a transferring implies sending to the network information indicating that the recipient address is now the owner of the specified BTC amount. This information is distributed in a capillary form until all nodes have been informed about the transfer. Throughout this process, transaction integrity is provided by asymmetric cryptography, wherein the private key encrypts the transaction by the sender, and the other network users can read the message via the public key. In other words, this means that each transaction can be ascribed to a precise actor because no other has access to their private key. When information circulates through the network, everyone can read it but not re-encode an altered transaction. The scheme for sending BTC in the network is represented in figure ~\ref{fig:transakcja}. Each transaction is coded via the hash function, which for Bitcoin is dSHA256~\cite{bitcointech}.

For a virtual currency, the fundamental substrate to functioning is a means of determining how many units are circulating at any time-point and how many will be created. There must exist a consensus mechanism ensuring that all users agree to the ownership of the virtual currency. In the Bitcoin network, wherein users remain anonymous and do not have to entrust each other, proper operation requires that all nodes maintain consensus regarding the current state of the system among themselves. Most participants in the distributed network must agree and perform the same action: the Bitcoin's key innovation is a mechanism that allows large-scale compliance of users on the network, solving the so-called Byzantine generals problem~\cite{Byzantine}.

\begin{figure}[ht!]
\centering
\includegraphics[width=1\textwidth]{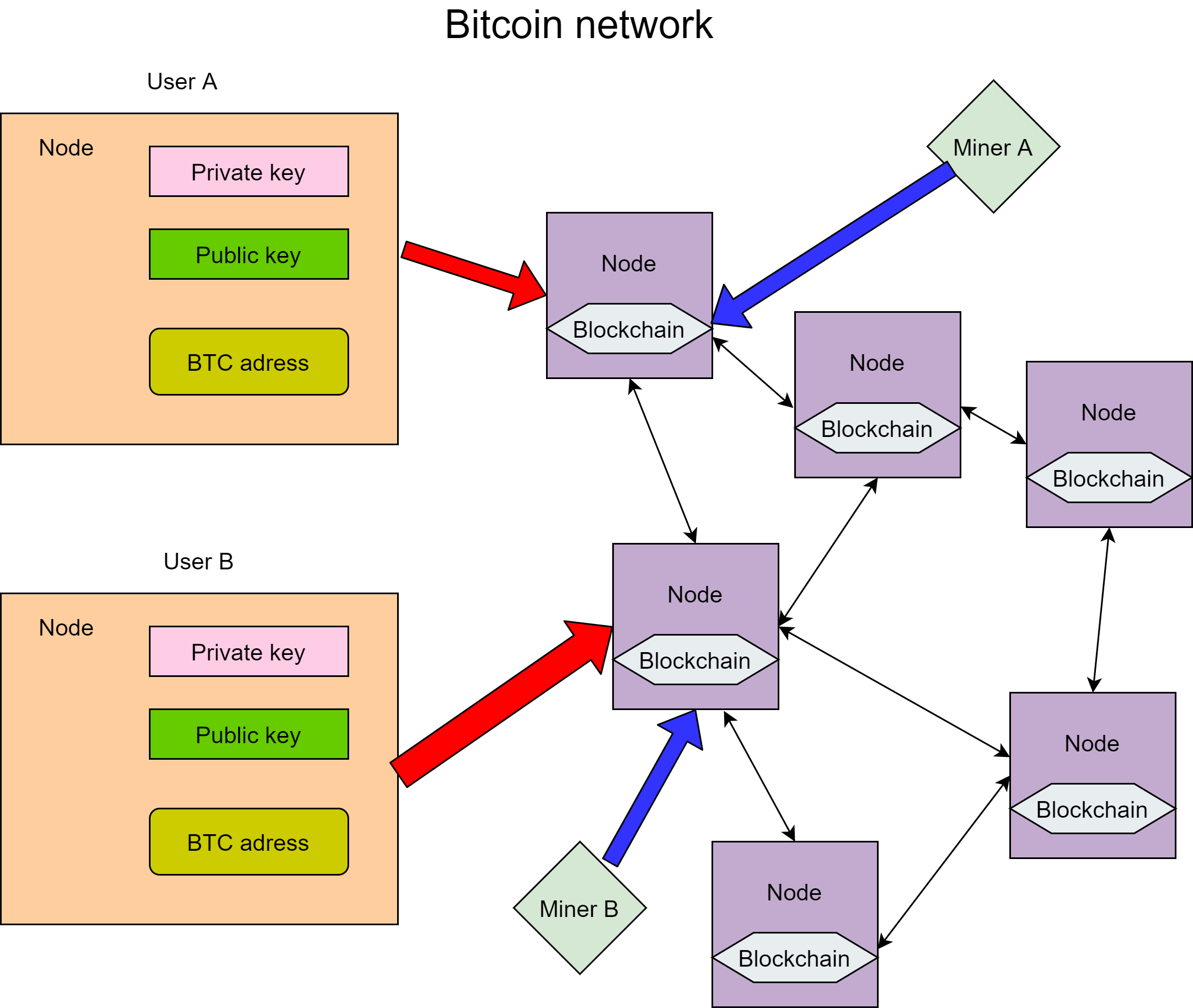}
\caption{Diagram of the Bitcoin network.}
\label{fig:siecBTC}
\end{figure}

For the network to function efficiently, there is a need for entities that collect transactions, ascertain correctness, and combine into potential blocks: these are known as miners. As a reward for their activities, they receive an amount in newly-created Bitcoins alongside a usually small transaction fee, debited after block approval and inclusion in the primary chain. This happens when most network participants have reached a consensus to add a block candidate in their copy of the blockchain. Anyone can become a miner. The only requirement is the client software and the latest copy of the network registry. However, due to the sheer computational cost of the validating block process, miners arrange themselves into so-called ``Mining pools'' and make use of tailored high-performance computing solutions based on graphics processing units (GPU) or application-specific integrated circuits (ASICs)~\cite{btcminingtech}. To be accepted, a new block must meet specific criteria. The result of the hash function, namely dSHA256 for the case of the Bitcoin, when applied to the data contained in the block must have a unique property: it must begin with a predetermined number of zeros. The number of starting bits resulting from hashing function which must be zero reflects the difficulty of confirming blocks. Miners collect transactions from the network (Fig.~\ref{fig:siecBTC}) and search for a number called ``nonce'', given which the hash function will return the result with the specified number of zeros at the beginning. If they succeed, they immediately send their block to the network, and the other users can easily verify its correctness. The consensus mechanism among miners is that any miner who receives a new block from the corresponding result of the hash function (hash) includes it in the blockchain. From game theory, a strategy wherein all miners add correct blocks to their copy of the blockchain is Nash's equilibrium state~\cite{Nash}. In other words, insofar as a miner believes that everyone else is acting fairly, their best option is to add a block to their copy. Operations on a blockchain version that is are not universally accepted are economically unprofitable because there is no reward for simply creating new blocks in it. This arrangement results in a consensus regarding the state of ownership of Bitcoins throughout the network, despite the absence of a centralized manager.

\begin{figure}[ht!]
\centering
\includegraphics[width=1\textwidth]{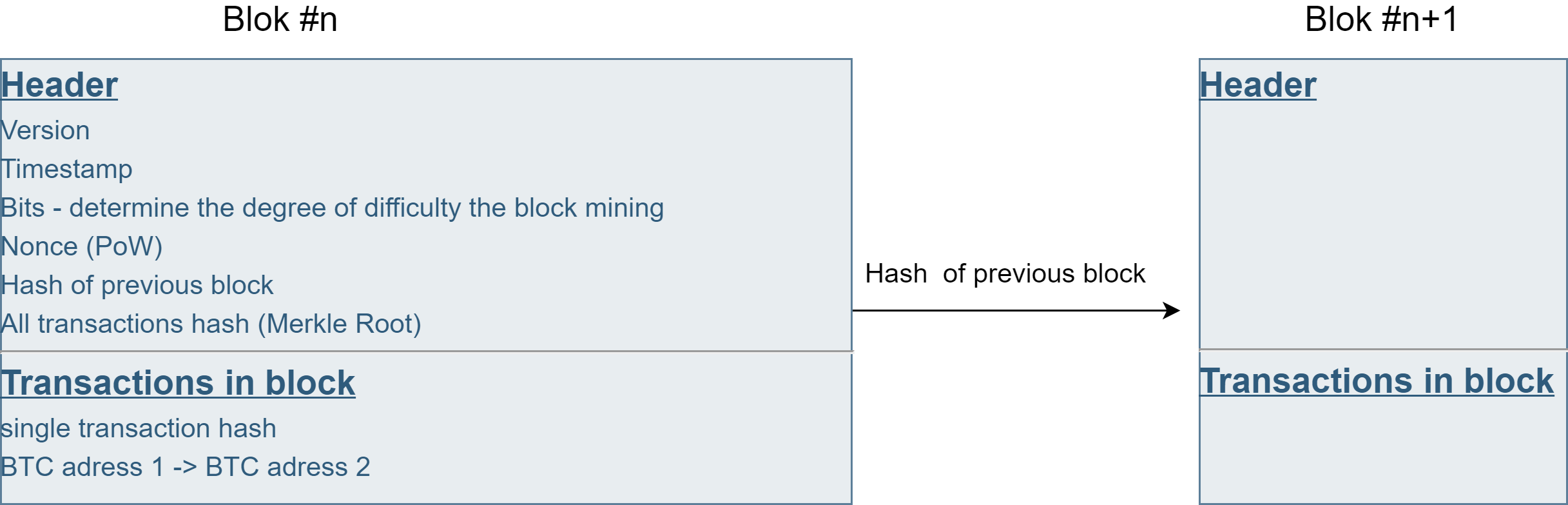}
\caption{Blockchain scheme in Bitcoin network.}
\label{fig:blok}
\end{figure}

When confirming subsequent blocks, commonly known as ``mining Bitcoins'', financial costs are incurred in the form of electricity consumption and the necessity to have specialized equipment. Nonce, the number required for closing a block, can be obtained only through an iterative process, i.e., a closed-form equation is prohibited by the formulation. For this reason, the consensus mechanism is referred to as a proof-of-work (PoW). Finding the correct result of a hash function with a specific number of zeros at the beginning requires, on average, a huge number of calculations~\cite{btcminingtech}. Adding incorrect information (e.g., fictitious transactions) to the candidate for a block would cause his rejection, simply dissipating the energy used for the calculations. This is the reason finding the correct result from hashing is proof that a miner helps maintain the Bitcoin network.

Because each subsequent block contains the header from the previous one, it is impossible to alter past transactions without rebuilding the entire chain. Due to the time, difficulty, and cost of finding the right result of the hash function, it is economically unviable. The blockchain diagram is shown in Fig.~\ref{fig:blok}. The Bitcoin protocol was created in such a way that the algorithm adjusts the difficulty in finding the right result of the hash function to the network computing power, so that a new block is created on average every 10 minutes~\cite{bitcointech}. The maximum block size is 1 MByte, determining how many transactions it can contain. As said, for the inclusion of a new block into the blockchain, miners receive a prize in the form of newly-created Bitcoins. The reward in halved every 210,000 blocks, up to the limit of 21 million Bitcoins~\cite{bitcointech}. It is possible that two miners would simultaneously and independently produce a new block and include it in their copy of the blockchain, causing multiple versions to appear. Such occurrences are dealt with applying the longest chain rule, i.e., the one for computing which the larger amount of power was invested is retained. This is a fundamental preservation mechanism. The pool rejects the shorter one during confirmation, and the miner receives no reward. A consequence is that blockchain technology is susceptible to so-called 51\% attacks. When a miner has control over more than half of the computing power in the given network, they can confirm blocks faster than anyone else, thus manipulating transactions in the nearest attached blocks. The cost of carrying out such an attack on the Bitcoin network and other PoW-based networks was obtained in Ref.~\cite{51procent}.

A limitation of the Bitcoin protocol is its low efficiency and the high costs of proof-of-work mechanism, which ensures network security. The primary costs are the non-recurring investment in computing power and the ongoing cost of electricity, which is currently at the level of that consumed by entire countries, such as Ireland and Denmark~\cite{elektr}. Another issue is the low network capacity. Bitcoin can handle, on average, only five transactions per second (TPS)~\cite{btcTPS}. By comparison, the Visa circuits handle on average about 1700 TPS~\cite{Visa}. In other words, one of the foundational advantages of the blockchain technology, namely the inability to make arbitrary changes to the blockchain, also results in a core rigidity. If one performs a transaction, this needs to be propagated to the entire network, and the previous version of the chain needs to be discarded, leading to the so-called ``hard forks''. It should be noted that the Bitcoin protocol is not static, as it is subject to constant modifications. To date, a mechanism reducing the transaction size has been introduced, known as the ``SegWit'' (segregated witness), which packing more transactions into each lock. Work is currently underway on the introduction of the ``Lightning Network'', which would allow micropayments to take place outside the main chain and thus which would increase network bandwidth, albeit at the price of reduced security.

\subsubsection{Methods for attaining network-wide consensus}
\label{konsensus}

The near totality of cryptocurrencies are predicated, like the Bitcoin, on the initially-introduced proof-of-work (PoW) consensus algorithm. It protects against external attacks, however, it is inefficient. For this reason, faster algorithms have been developed and are being trialled. The second most popular algorithm after proof-of-work is the proof-of-stake (PoS) alongside its various modifications. Unlike with PoW, no miners are competing for calculating as soon as possible the right hash function value closing a block. Instead, the next block's validator is a pseudo-random node from the network, the precise choice being based upon a combination of factors potentially including randomness and how many cryptocurrency units a given node has.\par

According to the PoS algorithm, blocks are ``forged'' and not mined as in the case with the PoW mechanism. Blockchain networks based on proof-of-stake also make use of transaction fees as rewards for ``forgers''. However, whereas in the case of networks based on proof-of-work the reward for miners reward is a newly created cryptocurrency unit, in proof-of-stake networks the users who want to take part in the forging process must block for some time the number of cryptocurrency units in the network as a ``stake''. The assumed rate's size determines the chances of choosing a given node as the next validator, which will check the next block. The higher the stake, the higher their chances are. For the selection process not to favor only the richest, additional selection methods are added to the network. The two most commonly used are randomized block selection and coin age selection.\par

Every cryptocurrency using the proof-of-stake approach has its own set of rules determining who will become the validator of the next block. Security in the PoS algorithm is implemented via the built-in possibility of losing a blocked stake and the right to participate in future transactions as a punishment of passing a fraudulent transaction. As long as the stake remains higher than the reward, the validator will lose more cryptocurrency units than they would have gained when attempting to commit fraud. As for the PoW protocol, the PoS algorithm is vulnerable to 51\% attack~\cite{Postech}. Such an attack can be perpetrated when a network node possesses more than half of the supply of a given cryptocurrency. In that case, they can then approve fraudulent transactions. The main advantage of the PoS algorithm is its efficiency. As the need for computationally-demanding calculations is relieved, no specialized user groups are needed to confirm blocks, enabling it be done faster than as compared to PoW.\par

A variation of the PoS algorithm is known as the delegated proof-of-stake (DPoS)~\cite{DPostech}. It is based on the notion that the owners of a given cryptocurrency outsource block processing work to a third party. The latter are known as the so-called delegates, for whom the network users can cast votes, and who secure the network if they win the election. Their task is attaining consensus during the generation and approval of new blocks in the network. The weight with which each network user decides about its future during voting is proportional to the number of the units of a given cryptocurrency that they possess. The voting system varies depending on the project, but usually, each delegate presents an individual proposal related to their plan for the development of the network and then asks for votes. As a rule, rewards received by delegates for creating and confirming blocks are proportionally shared with their voters. Therefore, the DPoS algorithm is based on a voting system directly dependent on the reputation of delegates. If a selected node (delegate) does not work efficiently or breaks the rules, it is removed, and another one is selected in its place. In terms of performance, networks using DPoS are more scalable. As the number of users increases, they can process more transactions per second compared to networks using PoW and PoS algorithms. The next section surveys the Bitcoin protocol modifications and other possible blockchain technology applications in further detail.\par


\subsection{Cryptoassets and diversified applications of the blockchain technology}
\label{altcoiny}

A core aspect of the Bitcoin protocol is its open-source form: consequently, anyone can review, analyze, and copy the specifications, resulting in a rapid spread of the new technology. At first, mere Bitcoin clones were created, sharing the same network structure with minor modifications such as block size, hash function, mining limit, and creation time for new blocks. However, as the blockchain technology developed, novel projects expanding its use cases rapidly emerged. For instance, they allowed not only saving the transaction register but also executing code computer code in the form of a script. Throughout the impetuous development of the cryptocurrency market and all possible uses of blockchain technology, the original concept of a cryptocurrency began blurring. A more general concept was introduced: cryptoassets, which could be classified in three categories as a function of the applications: cryptocurrencies, cryptocommodities, and tokens~\cite{cryptoassets}.

Cryptocurrencies denote the first use of the blockchain technology and are intended to transfer funds. Modifications of the Bitcoin protocol are the most common. The first to maintain a significant position today is Litecoin (LTC), created in 2011. It is a clone embedding two changes. The first consists of reduced time between block extractions by a factor of four times and an increased maximum limit to 84 million, allowing a substantial increase in the speed of transaction confirmation~\cite{Visa}. The second is the use of another hash function, known as scrypt, which is more appropriate for implementing on standard processors for confirming blocks in the Litecoin network instead of requiring ASIC as in the case of Bitcoin.\par

The first cryptocurrency not based on the Bitcoin protocol was the Ripple~\cite{ripple}, created in 2012. It does not implement the PoW consensus mechanism, implying that there are no miners in the network. Instead, it relies on a partially-centralized system called trusted nodes, which are responsible for confirming transactions. The idea behind Ripple is to provide connections between banks and stock exchanges to send money in real-time, which would entail using Ripple instead of fiat currencies for transfers outside national borders. Ripple tokens (XRP) were released by Ripple Labs. It currently owns most of this cryptocurrency. Currently\footnote{The data presented in this section are from October 2020.}, XRP is the fourth cryptocurrency in terms of capitalization. Its main competitor is Stellar (XLM), which also supports transactions between financial institutions. Unlike Ripple, the protocol is open source type and based on an own token. Both XRP and XLM tokens do not have a fixed supply limit and are, therefore, subject to inflation.\par

Since the Bitcoin blockchain is public and one can trace the transaction history of each mined coin, it is not entirely anonymous. The answer to this ``drawback'' was the emergence of cryptocurrencies ensuring complete undetectability to their users, the so-called ``private coins''. This group of such cryptocurrencies includes:\ Dash, Monero, and Zcash. Dash (DASH) has a hybrid architecture based on two network layers: the first involves miners using the PoW mechanism, as in the case of Bitcoin, the second comprises the so-called ``masternodes'' using the PoS technique. Monero (XMR) ensures anonymity thanks to a RingCT, by which the addresses (public keys) of those who make transactions are hidden in the blockchain. It is considered the cryptocurrency providing the most significant anonymity and, as such, is regrettably in widespread use by criminals, for example, for demand ransom payments~\cite{okupm}. Zcash (ZEC) is based on a ``zero-knowledge'' protocol called ZK-Snarks. It is a cryptographic solution that allows confirming whether the information is correct without having to disclose it, thus ensuring the anonymity of both the sender and recipient of a transaction to be preserved, while concealing its entity. Anonymous addresses are called shielded addresses and are compatible with public addresses, such that one can perform a transaction from a public wallet to a protected one and vice versa. Zero-knowledge is also used by cryptocurrencies such as ZClassic (ZCL), Bitcoin Private (BTCP), PIVX, and Komodo (KMD). In the cryptocurrency group protecting anonymity, Dash and Zcash have a built-in maximum supply mechanism, whereas Monero does not.\par 

The so-called cryptocurrency group also includes several Bitcoin hard forks. The Bitcoin Cash (BCH) represents the first Bitcoin division, wherein part of the community decided to disconnect and develop the project on a new blockchain. It entails increasing the block size from 1 to 8 MBytes, and then to 32MB. In the next division, Bitcoin Gold, the hash function was changed to Zhash, allowing efficient block confirmation on non-specialized hardware. Another variation, Bitcoin SV, stems from Bitcoin Cash and features a block size of 128 MBytes.

The second category of cryptoassets, cryptocommodities, serve as the basis for applications of the blockchain technology. They are ``material'' enabling the creation of decentralized applications and smart contracts. These are automatically executed computer codes that perform specific actions when certain conditions are met. Cryptocommodities operate as ``fuel'', allowing one to pay for using a decentralized computing network.\par

Ethereum~\cite{ethereum} was the first realization of such an idea. It was proposed in 2013 and launched in 2015 as an open-source computing platform based on the blockchain. It has been equipped with its own programming language, known as Solidity, through which one can program smart contracts and decentralized applications. It has its cryptocurrency called ether or ethereum (ETH), which serves as a payment unit for carrying out computational operations. Their price is expressed in the so-called units of ``Gas'' and depends on the computational complexity necessary to complete an operation. Ethereum, like Bitcoin, is based on blockchain technology and the PoW consensus mechanism. However, it uses another hash function, Ethash, that supports the efficient use of GPU hardware in the mining process. It does not have a fixed block size. Instead, each block requires a specific number of Gas units, which determines the computing power needed to complete the transactions it contains. The average time between blocks is about 15 seconds, and the maximum number of transactions per second is around 25. Unlike for Bitcoin, there is no upper limit on mining. In the future, the developers predict a transition to a PoS consensus mechanism, which would increase network bandwidth.\par

The Ethereum concept gained considerable popularity among the cryptocurrency community from the outset. In 2014, 18.5 million USD were collected from pre-sale, hallmarking the highest result of crowdfunding at that time. This result was beaten by a venture capital fund, the DAO, which aimed to raise capital for the development of blockchain startups and non-profit organizations. The fund took the form of a decentralized, autonomous organization built on the Ethereum blockchain in an intelligent contract. In 2016, it collected 11.5 million ethereum worth nearly 168 million USD. On this occasion, one of the pros/cons of blockchain was revealed, namely code unchangeability. It turned out that the DAO smart contract code had gaps that allow unauthorized funds to be transferred. A hacker attack took place in June 2016, during which 3.6 million ethereum were stolen. To reverse the effects and change the DAO code, the Ethereum blockchain had to be split. On the new, the withdrawal of funds was canceled, whereas on the old, now called Ethereum Classic (ETC), everything remained unchanged. Currently, the ethereum remains the second cryptocurrency in terms of capitalization.\par

\begin{figure}[!ht]
\centering
\includegraphics[width=1\textwidth]{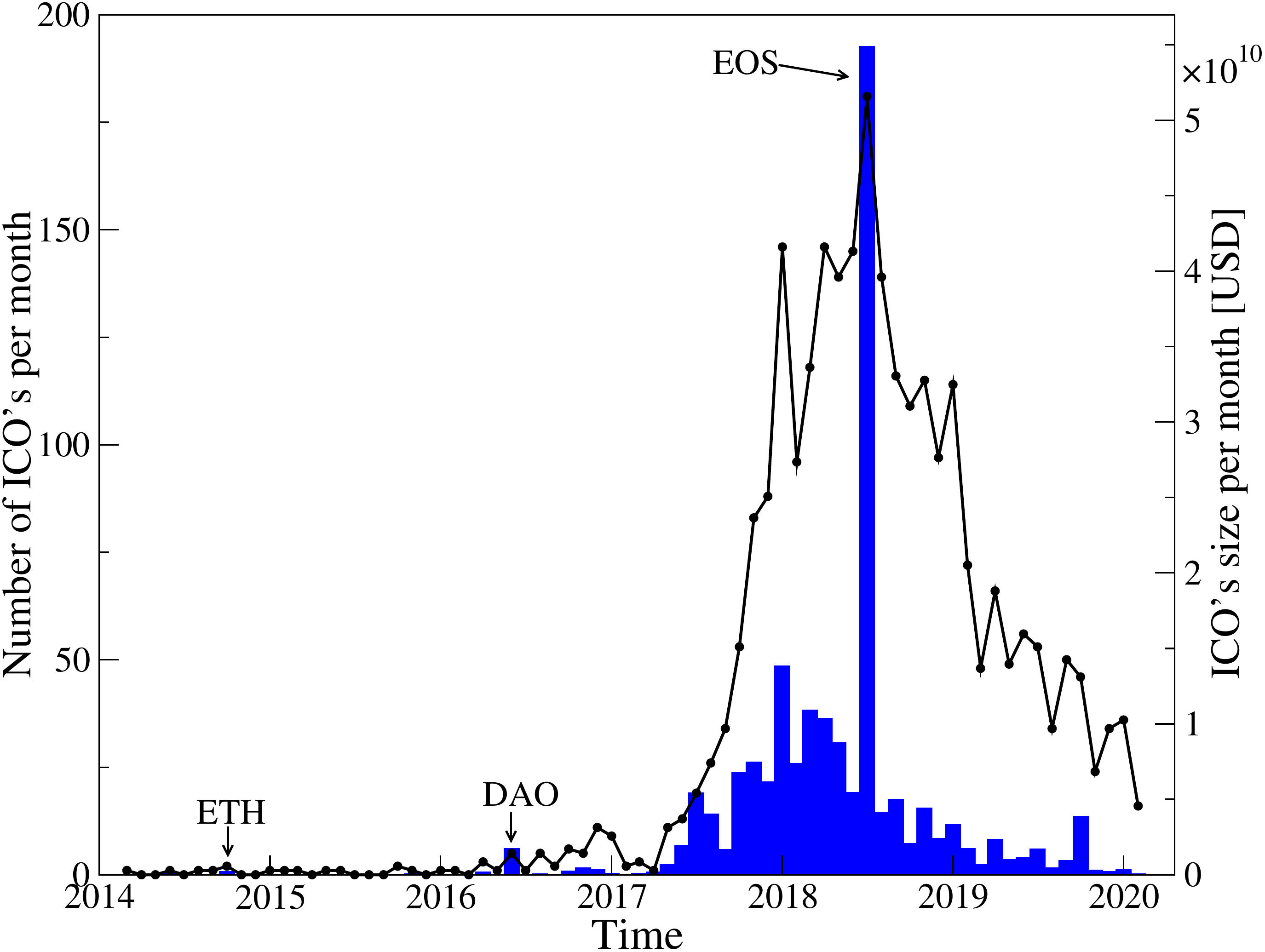}
\caption{Number and size of ICO's per month. Data from ICOBench.com~\cite{ICOBench}. The largest ICO offers are marked with arrows.}
\label{fig:liczbaICO}
\end{figure}

The considerable success of the concept of smart contract, in particular the possibility of collecting funds under the Initial Coin Offer (see Fig.~\ref{fig:liczbaICO}) on the Ethereum platform, spearheaded competition also offering the possibility of creating applications in a decentralized computing network. Major projects of this type include EOS and Cardano. The EOS~\cite{EOS} platform is based on the delegated proof-of-stake algorithm (DPoS). As there is no mining process, unlike with Bitcoin or Ethereum, the network can process up to 15,000 transactions per second. However, this is attained at the expense of security. Only 21 selected block producers are responsible for verifying transactions, who receive awards in the form of newly created EOS units for their activities. The maximum number of EOS tokens is set to one billion units. The programming language used on the platform is known as WebAssembly. The leading EOS network was launched in June 2018, hallmarking the most substantial ICO offer in history. In year-long offer, ended in June 2018, it raised 4.2 billion USD eclipsing the world's biggest initial public offerings of the year and more than doubling the next biggest offering of that type (see Fig.~\ref{fig:liczbaICO}).\par

On the other hand, Cardano~\cite{cardano} uses the PoS algorithm, known as Ouroboros. There are two independent layers of the network: billing and computational. The accounting layer is built upon the blockchain network. The Haskell programming language is used to create intelligent contracts. The maximum supply of ADA cryptocurrency is 45 billion units. Other platforms offering similar services as Ethereum are the Tron -- TRX, Lisk -- LSK, ICON -- ICX. They are similarly based on the DPoS algorithm. PoS-based platforms are Tezos, NEO (called the Chinese equivalent of Ethereum), NEM (based on Proof of Importance - a variety of PoS), and Waves. As one can see, it is typical for cryptocommodities to move from the original PoW consensus mechanism towards PoS, since this allows faster processing of network operations, which is required for additional than transaction-register functions.\par

The third group are tokens, which are the youngest cryptoassets and can be referred to as direct applications of blockchain technology. They are most often used as means of payment in decentralized applications (dApps) built on cryptocommodities such as Ethereum or are issued in the ICO offers for the development of a blockchain venture. Typically, they don't have their own blockchain. Blockchain technology, thanks to eliminating a central intermediary who earn on commissions can be used wherever there is possibility to connect sellers and buyers directly (for example Uber and Airbnb). 

The combination of token and cryptocurrency categories denotes the so-called ``stable coins''. Their exchange rate is usually related to fiat currency. The most popular is tether~\cite{tether} (USDT), which linked to the US dollar in a one-to-one relationship. It relies on Bitcoin blockchain via the Omni Layer platform and is issued by the private company Tether Limited, which declares that its supply is fully covered in US dollars. Other notable stable coins are the DAI, PAX, TUSD, and USDC.

The latest applications of blockchain technology are so called decentralized finance (DeFi). This term refers to an ecosystem of financial applications that are built on blockchain networks. Most DeFi platforms take the form of decentralized apps (dApps). They use smart contracts to automate financial transactions. The idea behind DeFi is to create decentralized cryptocurrency borrowing and lending system and to provide monetary banking services~\cite{DeFi2020}. The largest projects of this type are: Chainlink, Wrapped Bitcoin, Uniswap, Maker and Compound.

\subsection{Cryptocurrency trading}
\label{krypto_gieldy}

Cryptocurrencies, as a new concept, would not have much value in the early stages of their existence if they could not be exchanged for traditional currencies. One of the first exchanges that offered such opportunities was the Mt.Gox, as mentioned earlier, but the present cryptocurrency market is profoundly divided and decentralized. Nowadays\footnote{The data presented in this section are from October 2020.} there are 350 cryptocurrency exchanges around the world. Most of them are platforms trading only cryptocurrencies with each other, with transactions performed around the clock. This is a key feature distinguishing the cryptocurrency market from the Forex, wherein trading occurs only from Monday to Friday. The second fundamental difference lies in the form and participants of the trade: in the case of cryptocurrencies, it takes place mostly between individual investors~\cite{BTCusers}, whereas on the Forex, trading takes place on the OTC (Over-The-Counter) market and its participants are mainly banks. The third difference is the lack of a reference price, which in the case of the Forex market is provided by Reuters. Only in the case of the bitcoin, there is a futures contract quoted on the CME Group~\cite{CME}.\par

The decentralization of the cryptocurrency market means that the same cryptocurrency pairs are traded on different exchanges. Historically, significant valuation differences could be observed between exchanges~\cite{arbitrage}. For the stocks, this issue has been investigated and is called ``dual-listed companies''~\cite{Froot1999,Jong2009}. However, in the case of cryptocurrencies, which can be sent almost immediately between anywhere in the world, it represents a different kind of issue. The topic of trade comparison between cryptocurrency exchanges and arbitrage opportunities is covered in Sections~\ref{AITT} and~\ref{sect::TriangularArbitrage}. A common feature between the Forex and the cryptocurrency market is the possibility of triangular arbitrage~\cite{Fenn2009,gebarowski2019}. This will be discussed in detail in Section~\ref{sect::TriangularArbitrage}. A separate category not found in traditional financial markets is the so-called decentralized exchanges (DEX), where cryptocurrencies, through smart contracts, can be traded automatically, without any reliance on a central exchange server~\cite{flashboys}.\par

The statistical properties of the exchange rates of major cryptocurrencies will be analyzed in this review, considering the Binance, Bitstamp, and Kraken exchange rates. Binance~\cite{Binance} is currently the largest cryptocurrency exchange in terms of volume value. It was founded in July 2017 in China. After the regulation regarding trading in cryptocurrencies was tightened, the head office was moved to Malta in March 2018. At its inception, only cryptocurrency trade was conducted on it. Binance offers the ability to make transactions using approximately 800 different cryptocurrency pairs and additionally owns a cryptocurrency called the binance coin, which is used for paying commissions on the exchange. The exchange also offers the opportunity of carrying out ICO offers on it. Furthermore, daughter companies allowing exchanging cryptocurrencies for fiat currency. The Kraken exchange~\cite{Kraken} was launched in September 2013, with headquarters in San Francisco and branches in Canada and Europe. It is the largest exchange in terms of volume value on cryptocurrency pairs expressed in Euro and currently offers 185 cryptocurrency pairs. The Bitstamp exchange~\cite{Bitstamp} is one of the longest-running cryptocurrency exchanges. It was established in August 2011, and its head office of the exchange is in Luxembourg. The platform offers trading only on the most liquid cryptocurrencies, currently supporting 38 exchange rates.\par
\begin{figure}[!ht]
\centering
\includegraphics[width=1\textwidth]{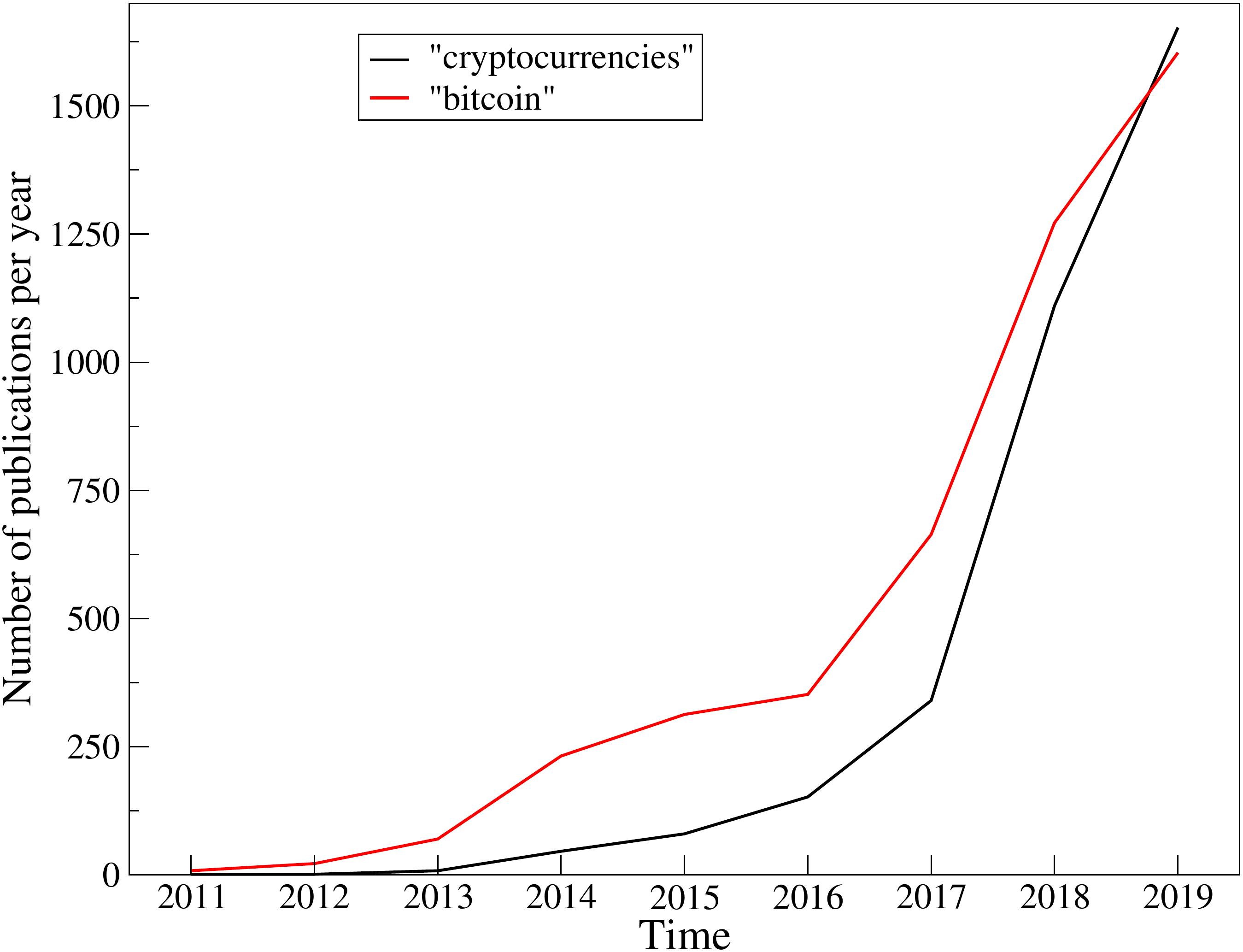}
\caption{The number of scientific publications containing the word ``bitcoin'' (in red) or ``cryptocurrencies'' (in black) in the title or abstract that was published during the year. Data from the website dimensions.ai~\cite{dimensions}.}
\label{fig:dimensions}
\end{figure}
There was a sharp increase in the number of actively traded cryptocurrencies in 2017 (Fig.~\ref{fig:Lkrypto}). Since then, the number of ICO offers has kept increased systematically (Fig. ~\ref{fig:liczbaICO}). The market capitalization also increased spectacularly, heralding the speculative bubble phase~\cite{Gerlach2018}. At the time of writing, approximately 3600 cryptocurrencies and nearly 30,000 exchange rates are present on the market~\cite{coinmarket}. The capitalization of the entire cryptocurrency market is on the order of 350 billion USD, and therefore comparable to middle-size stock exchange and largest American companies~\cite{marketsize}. The spectacular development of the cryptocurrency market has attracted worldwide interest in the new asset, including the scientific community. The first works on the Bitcoin were published already in the period 2013-2015~\cite{Kristoufek2013,Kristoufek2015}, but the real boom has only occurred since 2017 (figure~\ref{fig:dimensions}). At first, scholars mainly examined the characteristics of the first cryptocurrency, the bitcoin~\cite{Bariviera2017,DrozdzBTC2018,Garnier2019}. Next,
the community developed an interest in comparing it to other cryptocurrencies~\cite{Wu2018,futinternet2019,Kristoufek2019}, which was followed by studies revealing correlations within the cryptocurrency market~\cite{stosic2018,zunino2018,bouri2019,zieba2019,chaos2020,FERREIRA2020,PAPADIMITRIOU2020,polovnikov2020} and its relationships to mature markets~\cite{Corbet2018,corelli2018,ji2018,kristj2019,entropy2020}. In particular, the possible uses of the bitcoin as a means of diversifying an investment portfolio and as a hedging instrument for the currency market~\cite{urquhart2019}, gold and commodities~\cite{shahzad2019} and stock markets~\cite{wang2019,Conlon2020,shahzad2019a} were considered. A more detailed review of existing literature devoted to research on cryptocurrencies in the context of financial markets is available in Refs. \cite{Corbet2019,fang2020cryptocurrency}.


\section{Statistical properties of fluctuations}
\label{Dane stat}

\subsection{Data}
\label{Data}

The presentation of results begins with the statistical properties of cryptocurrency price fluctuations from a brief description of the empirical data sets. Three issues are of interest here: (1) the cryptocurrency prices expressed in regular (fiat) currencies: US dollar and euro, with particular focus on bitcoin (BTC) as the most liquid and the most frequently traded cryptocurrency, (2) the cryptocurrency-cryptocurrency exchange rates, and (3) result comparison for two independent trading platforms.\par

The data were collected from the following sources. For BTC, the oldest cryptocurrency, the longest series of historical data was obtained from the Bitstamp exchange platform~\cite{Bitstamp}, offering the BTC price tick-by-tick quotes covering the interval 2012-2018 and expressed in USD. While the cryptocurrency market was expanding, more and more exchanges were created over time. They allowed trading not only BTC with USD, but also other cryptocurrencies with regular currencies, and even cryptocurrencies among themselves. A comparison between the properties of bitcoin and ethereum (ETH) was possible based on tick-by-tick quotes obtained from the Kraken platform~\cite{Kraken}. The quotes covered an interval from mid-2016 to the end of 2018. A data set with quotes representing some less liquid cryptocurrencies, like litecoin (LTC) and ripple (XRP), unavoidably has to be smaller and covers only the entire year 2018, during which their trading frequency was sufficient to construct time-series without excessive constant-value intervals. To compare the properties of data from the different trading platforms, there is also consider a parallel set of the cryptocurrency quotes from Binance, the largest platform of this type. However, the sampling frequency of this set is 1 min, which provides us with 463,000 quotes for each exchange rate.\par

As a benchmark representing the regular currencies from the Forex, the EUR/USD exchange rate was considered, with quotes obtained from a broker, Dukascopy~\cite{Dukascopy}. However, there is a significant difference between this dataset and the cryptocurrency datasets: the former excludes weekends and holidays, whereas the latter are recorded continuously.\par

In the following, three principal characteristics are reported: the average inter-transaction time (when available), which is denote by $\bar{\delta_t}$, the probability distribution function (PDF), and the autocorrelation function for the returns, defined as logarithmic exchange rate change over a fixed time interval $\Delta t$: $R_{\Delta t}(t)=\log P(t+\Delta t)-\log P(t)$, where $P(t)$ is an exchange rate of two distinct currencies or cryptocurrencies at time $t$. The autocorrelations of both the signed returns and absolute returns (volatility) are of interest.\par

\subsection{Average inter-transaction time}
\label{AITT}

\begin{figure}[!ht]
\centering
\includegraphics[width=1\textwidth]{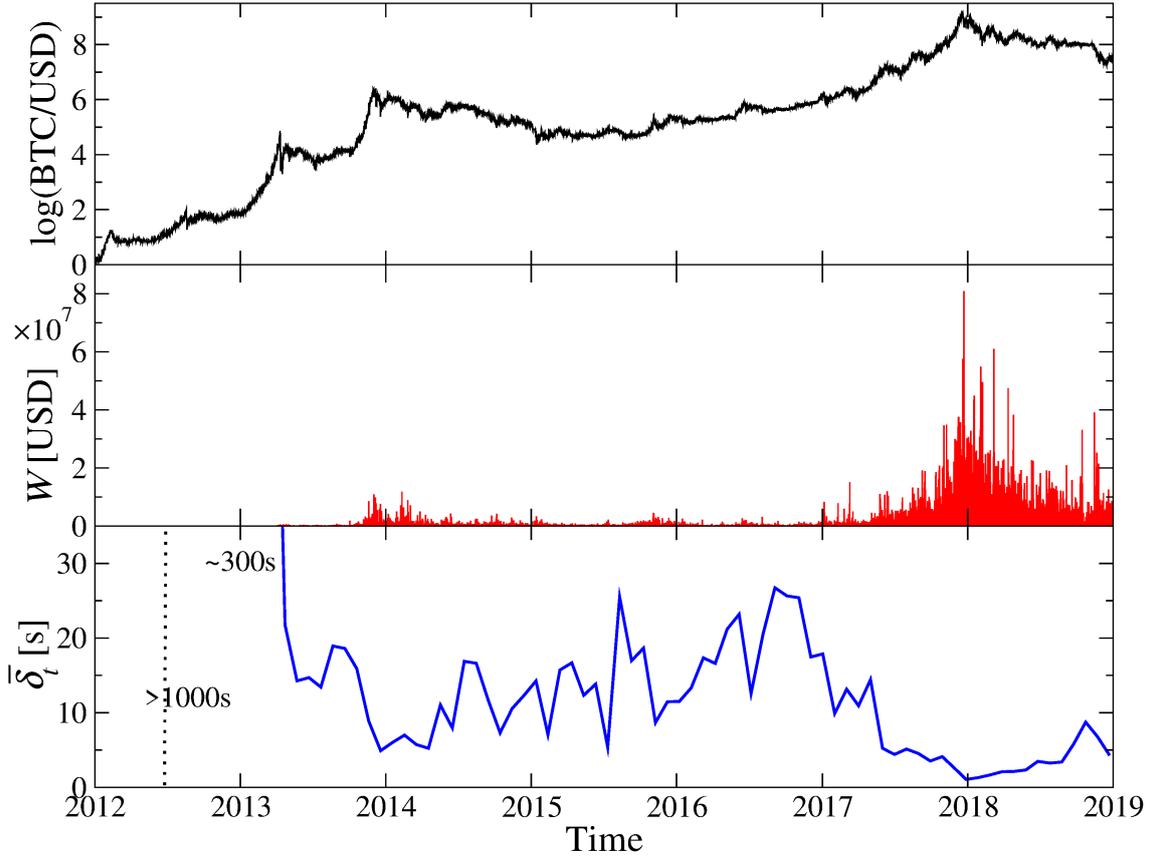}
\caption{BTC/USD exchange rate logarithm (Bitstamp exchange platform) from January 2012 to December 2018 (top panel), volume in USD (middle panel), and average inter-transaction time in monthly windows (bottom panel).}
\label{fig:BTCbitstamp}
\end{figure}

\begin{figure}[!ht]
\centering
\includegraphics[width=1\textwidth]{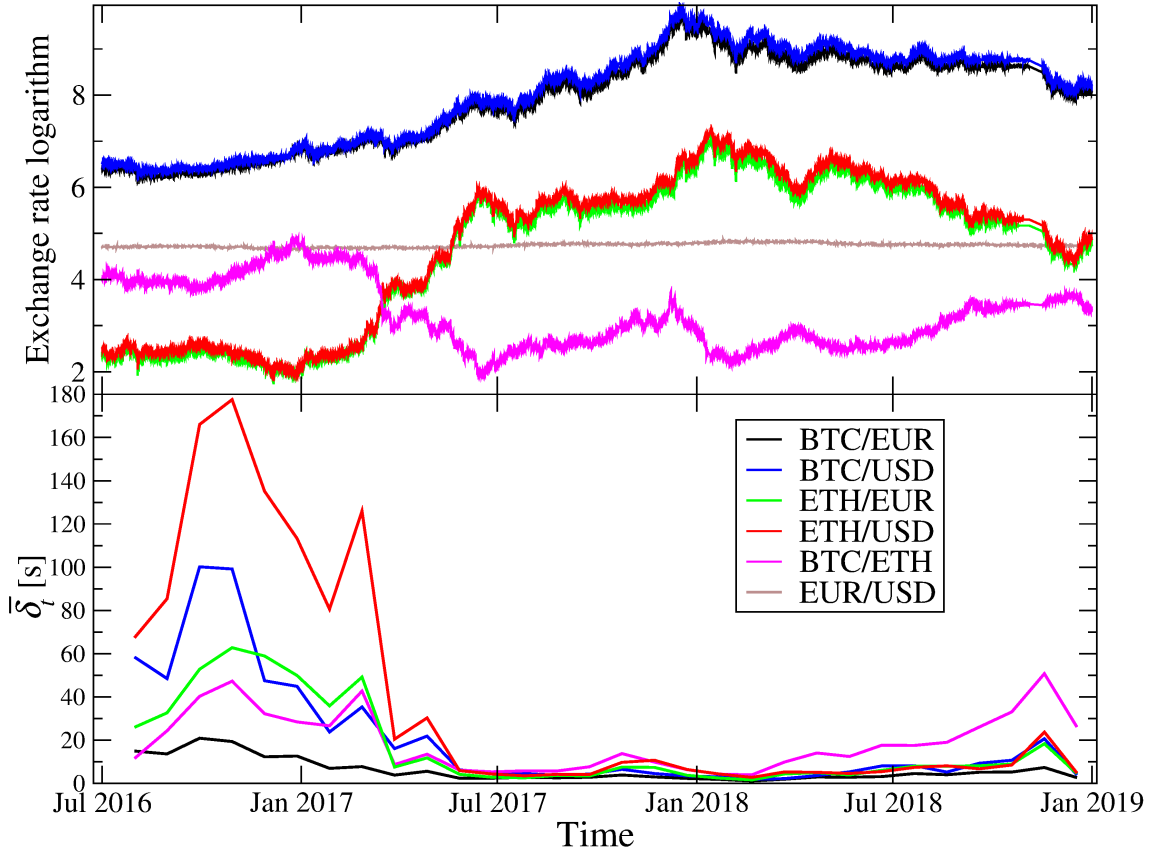}
\caption{Logarithm of the exchange rates BTC/EUR, BTC/USD, ETH/EUR, ETH/USD and BTC/ETH on the Kraken exchange and EUR/USD, multiplied by a factor 100 for better visibility, from July 2016 to December 2018 alongside the average inter-transaction times in monthly window.}
\label{fig:Kraken_serie}
\end{figure}

The analysis starts with the tick-by-tick quotes of the BTC/USD exchange rate as it is the most liquid and frequently traded cryptocurrency asset (Bitstamp quotes). Figure~\ref{fig:BTCbitstamp} shows the BTC/USD exchange rate, the BTC volume (in USD) traded during hour-long intervals, and average inter-transaction time calculated over monthly windows. There is an upward trend of BTC/USD throughout the years until the turn of 2017 and 2018. The trend was accompanied by a volume increase and a reduction in the average inter-transaction time. In 2012 and the first half of 2013, the trade volume was negligible. Subsequently, the peak of the first bull phase, at the turn of 2013 and 2014, was accompanied by an increase in volume and a reduction in the inter-transaction times. A side trend dominated the BTC/USD rate over the next three years and was associated with a volume decline and less frequent trading. A resume of the bull market towards the end of 2016, with spectacular growth in 2017 in particular, brought a further substantial increase in volume accompanied by a decrease in the average inter-transaction times. At the peak of the speculative bubble, at the turn of 2017 and 2018, the time between transactions became as short as around a second. A subsequent bear market caused a decrease in volume, which, however, remains above the pre-crash level observed in 2017. The average inter-transaction time $\bar{\delta_t}$ was also affected but remains below 10 seconds.\par

The mutual exchange rates (logarithms, top panel) and the inter-transaction intervals averaged over monthly windows (bottom panel) for the BTC, ETH, USD, and EUR based on the Kraken and Dukascopy data are shown in Figure~\ref{fig:Kraken_serie}. In parallel with Figure~\ref{fig:BTCbitstamp}, one observes a gradual shortening of $\bar{\delta_t}$ for the cryptocurrencies down to below 10 seconds in the bull-market phase (between mid-2017 and a trend reversal in early 2018). On the contrary, in the bear-market phase of 2018, a decrease in the BTC and ETH valuation in regular currencies was accompanied by an increase in $\bar{\delta_t}$. Only for BTC/EUR one still observed $\bar{\delta_t}<10$ s. By looking at the exchange rate fluctuations over time, one can notice the considerably lower volatility of EUR/USD, almost resembling a flat line in the scale of Figure~\ref{fig:Kraken_serie}, as compared to the cryptocurrency-involving rates. 

\begin{figure}[!ht]
\centering
\includegraphics[width=1\textwidth]{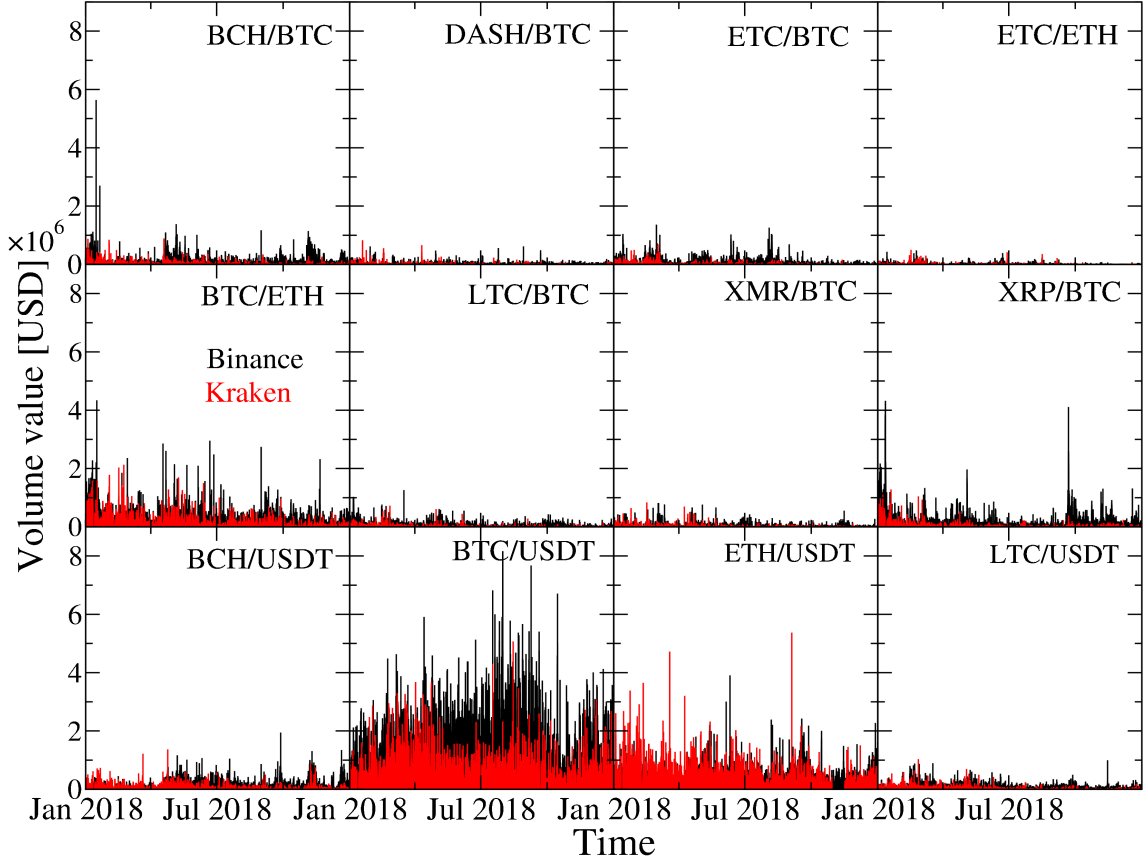}
\caption{Volume value in USD for the exchange rates listed simultaneously on Binance and Kraken (USD instead of USDT), per minute.}
\label{fig:Kraken_Binance_WartV}
\end{figure}

No strictly analogous result could be shown for the Binance dataset since it does not contain tick-by-tick quotes. Instead, the trading frequency can be related to the average length of intervals without trade (zero-return sequences) and the number of periods without trading. In Figure~\ref{fig:Kraken_Binance_WartV}, the volume time-series from both exchanges were compared for 12 exchange rates involving both the most liquid cryptocurrencies (BTC, ETH) and the least liquid ones (DASH, XMR, LTC, XRP, BCH). Since on the Binance platform, the cryptocurrency value may only be expressed in terms of other cryptocurrencies, to facilitate the inter-platform comparison, the USD has to be replaced by a proxy -- Tether (USDT), whose USD-based exchange rate fluctuates closely around 1. Therefore, the ETH/USD on Kraken corresponds to ETH/USDT on Binance. Figure~\ref{fig:Kraken_Binance_WartV} shows that a higher 1-min volume is seen on Binance for all exchange rates under consideration, which implies that the average inter-transaction time on Binance is shorter than on Kraken (see Table~\ref{tab:rozkladBiKR} in Sect.~\ref{CDF}). The volume accumulates principally in BTC/USD and BTC/USDT, and then on ETH/USDT, BTC/ETH, and XRP/ETH.

\subsection{Fluctuation distributions}
\label{CDF}

\begin{figure}[!ht]
\centering
\includegraphics[width=1\textwidth]{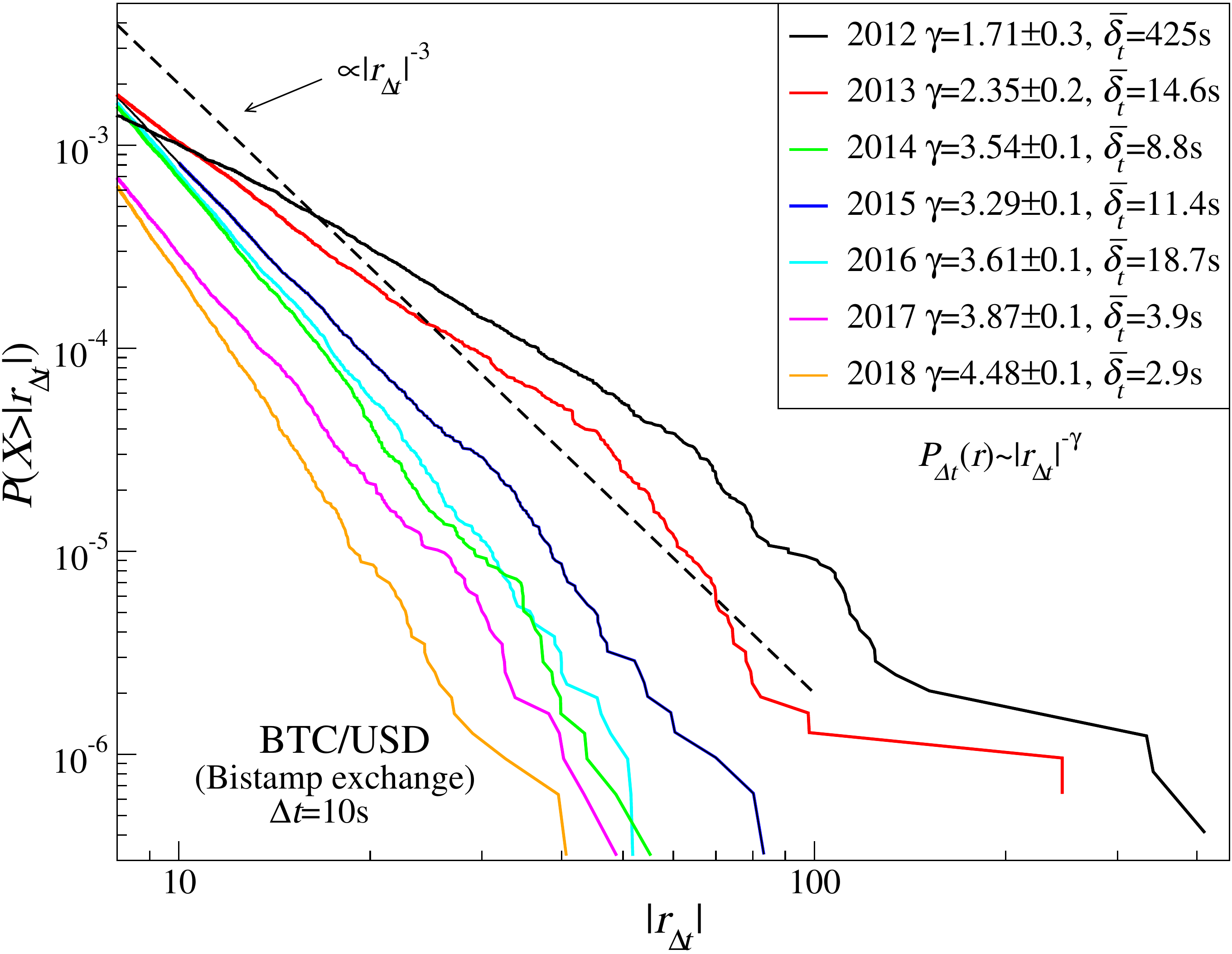}
\caption{Cumulative distribution of normalized absolute log-returns $r_{\Delta t=10\textrm{s}}$ for the bitcoin on the Bitstamp exchange. In the inset, the estimated $\gamma$ exponent is shown, with the average inter-transaction time over a given year.}
\label{fig:Bitstamp_rozklad}
\end{figure}

At the dawn of scientific interest in the financial markets, it was believed that a normal distribution~\cite{Bachelier} could approximate the PDF of the stock market returns. Later, it was discovered by Beno\^{i}t Mandelbrot that such distributions have significantly heavier tails and comply with the stable L\'{e}vy regime~\cite{Mandelbrot1963}. Consequently, the truncated L\'{e}vy distributions began to be used in this context~\cite{mantegna1994}. Currently, it is accepted that, in mature financial markets, the CDFs of the absolute normalized returns in the bulk part develop fat tails of the form
\begin{equation}
P(X>r_{\Delta t}) \sim |r_{\Delta t}|^{-\gamma},
\label{inversecubic}
\end{equation} 
where $r_{\Delta t}=(R_{\Delta t}-\mu)/ \sigma$ with $\mu$ and $\sigma$ denote, respectively, sample mean and standard deviation. This relationship with different values of $\gamma$ has been reported for many different asset types, including shares, currencies, commodities, and cryptocurrencies~\cite{Lux1996,Gopi1999,Plerou1999,
drozdzepps,kwapien2012,Begusic2018,DrozdzBTC2018,gebarowski2019,Watorek2019}. 

In particular, high-frequency stock-market data is frequently described by an inverse cubic dependence
~\cite{gopi1998,Gopi1999,gabaix2003}, i.e., the one with $\gamma \approx 3$. A well-known fact is that the tail thickness of the return PDFs/CDFs decreases as $\Delta t$ increases from second to minute, hour, and longer scales~\cite{Drozdz2007,Rak2013,futinternet2019}. A normal or exponential distribution can already approximate the return distributions for daily data. On the other hand, the return distributions for $\Delta t$ over the range of a few seconds are closely matched by the power-type functions~\cite{Drozdz2007}. This can easily be explained by the central limit theorem, provided the returns are independent: as $\Delta t$ increases, the distribution should converge to a normal distribution~\cite{Drozdz2003}.

In general, it often happens that the fluctuations of a power-law type are associated with some critical phenomenon that occurs in the dynamics of a given system. However, in the financial market case, it was pointed out~\cite{Cristelli2011} that such behaviour of the return distributions might be related solely to the finite-size effects in the number of active agents and, therefore, any particular form of a power-law PDF may be viewed as accidental and non-universal~\cite{alfi2009a,alfi2009b}, opposite to what was believed earlier~\cite{gabaix2003}.

\begin{figure}[!ht]
\centering
\includegraphics[width=1\textwidth]{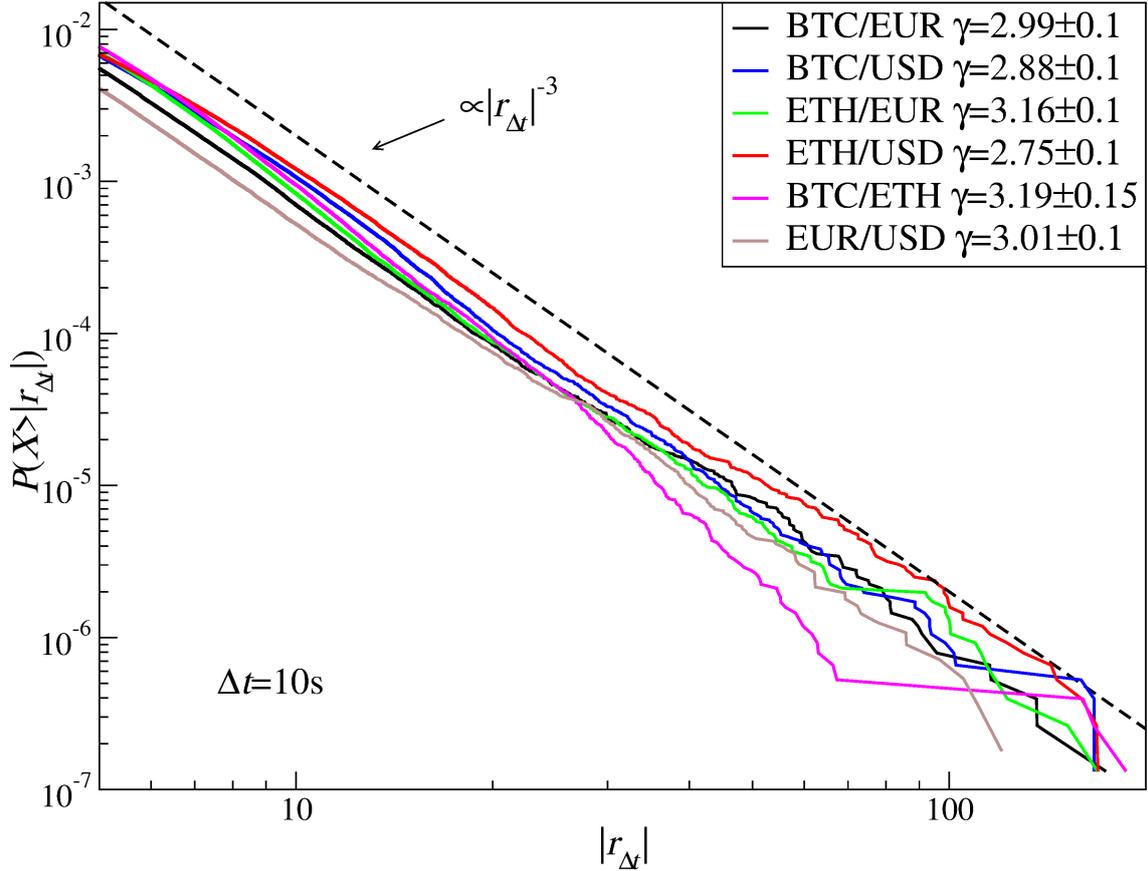}
\caption{Cumulative distribution of the normalized absolute log-returns $r_{\Delta t=10\textrm{s}}$ for the exchange rates BTC/EUR, BTC/USD, ETH/EUR, ETH/USD and BTC/ETH on Kraken, alongside EUR/USD, from July 2016 to December 2018. In the inset, the estimated exponent $\gamma$ is shown.}
\label{fig:Kraken10s_rozklad}
\end{figure}

In order to calculate the cumulative probability distribution function (CDF), the tick-by-tick datasets from Bitstamp and Kraken were transformed to time-series of normalized returns $r_{\Delta t}(t_i)$, $i=1,...,T$ at a sampling frequency $\Delta t=10$s. For the Bitstamp BTC/USD rate, it provides one with $T=21 \times 10^6$ returns. As the analyzed period of 2012-2018 features diversified dynamics, the statistical properties of the returns have to be considered separately for each year. Figure~\ref{fig:Bitstamp_rozklad} presents the CDFs of the absolute normalized returns. A power-type relation of the distribution tail (Eq.(\ref{inversecubic})) is observed for each year, at least over a part of the available return magnitudes, with the widest scaling range apparent in 2012. However, one has to notice that even in this latter case the power-law scale range does not exceed 1.5 decade, while its width gradually decreases each year and it falls well below one decade in 2018. This suggests that it is safer to consider these CDFs as only carrying some signatures of the power-law distributions and not being precisely the power-law ones.

There is evolution of the exponent $\gamma$ from lower ($\gamma_{\textrm{2012}}=1.7 \pm 0.3$) towards higher values ($\gamma_{\textrm{2018}}=4.5 \pm 0.1$), and the opposite trend for the estimation error. Furthermore, there is a relation between the shortening of the average inter-transaction time and the growing tail index $\gamma$, hallmarking a phenomenon observed in other financial markets as well. This relation can be interpreted as defining the internal time-flow of the market: the faster the trading and the shorter the inter-transaction intervals, the faster is the time-flow experienced by the assets and, consequently, the shorter are time-scales over which particular effects are observed~\cite{Drozdz2007}. The acceleration of market evolution is, therefore, a natural consequence of both the technological advances and the increasing market-participant influx that conjointly lead to higher transaction frequency. For example, for $\Delta t=10$s, the return CDF tails for the BTC/USD, which were initially in the L\'evy-stable regime with $\gamma < 2$, became gradually thinner with $\gamma\approx 3.5$ in 2014 and $\gamma\approx 4.5$ in 2018 (figure~\ref{fig:Bitstamp_rozklad}). At present, the inverse-cubic scaling has to be addressed for a sampling frequency of $\Delta t<10$s.

A comparison between the CDFs obtained for different exchange rates is presented in Figure~\ref{fig:Kraken10s_rozklad} based on five return time-series from Kraken, namely BTC/USD, BTC/EUR, ETH/USD, ETH/EUR, and BTC/ETH ($\Delta t=10$s, $T=7.6 \times 10^6$), alongside the EUR/USD return time-series from Dukascopy ($\Delta t=10$s, $T=5.6 \times 10^6$). Here, the data covers the entire available interval, from mid-2016 through end-2018. Except for BTC/ETH, the distributions show a near power-law decay over about one decade, with the scaling exponent dwelling around $\gamma \approx 3$. The values are slightly below this level for the BTC/USD and ETH/USD; these are the rates that, in a first part of the period under consideration, experienced the smallest trading frequency, with $\bar{\delta_t} > 100$s. The BTC/EUR and ETH/EUR returns are characterized by $\gamma$ that is the closest to 3, in parallel with the much more liquid Forex pair, EUR/USD, while the BTC/ETH crypto-crypto rate shows a slightly faster decay with $\gamma\approx 3.2$.

\begin{figure}[!ht]
\centering
\includegraphics[width=1\textwidth]{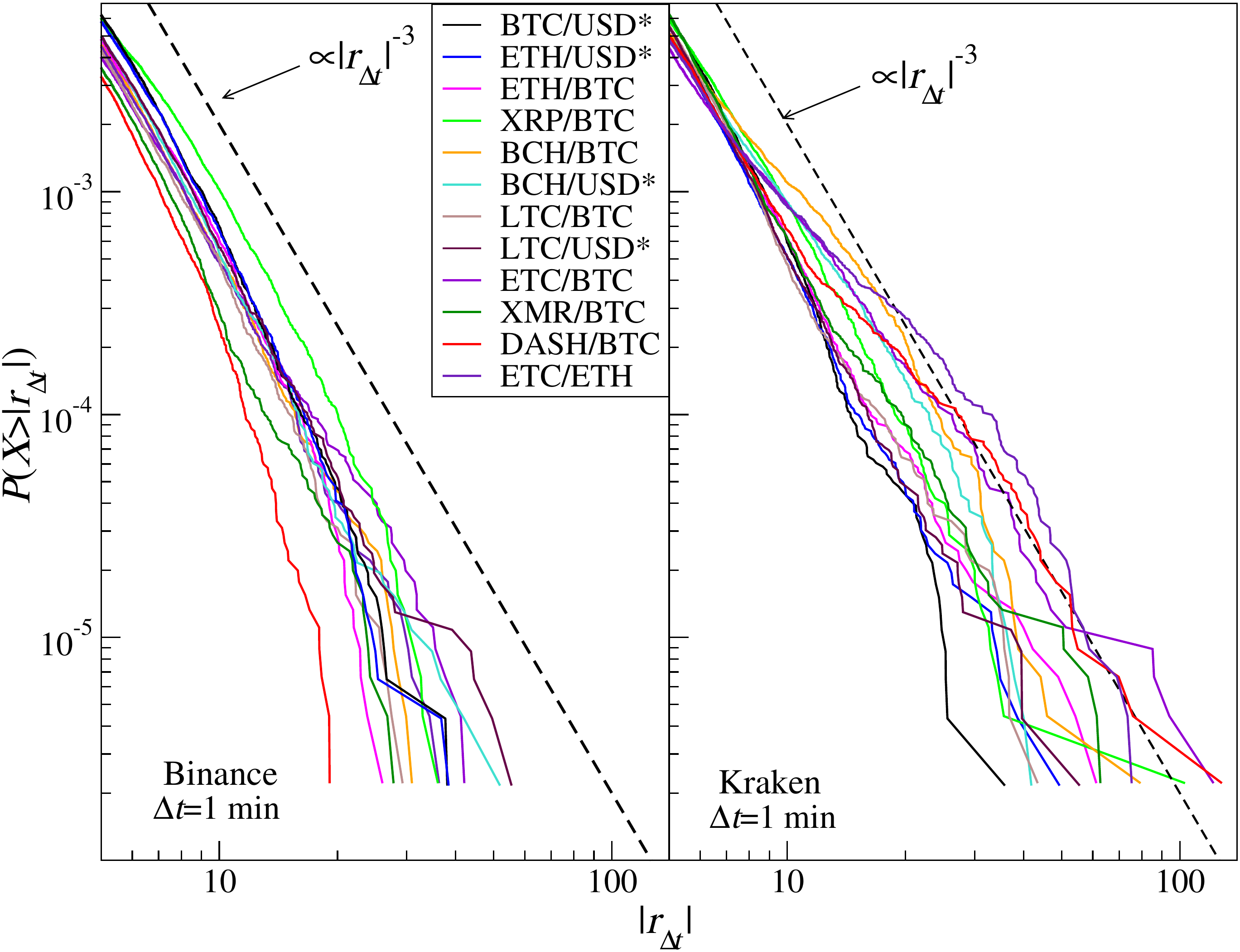}
\caption{Comparison of the cumulative distributions of normalized absolute log-returns $r_{\Delta t=1\textrm{min}}$ for the exchange rates listed simultaneously on Binance and Kraken (USD$^{*}$ means USD in the case of Kraken, and USDT in the case of Binance).}
\label{fig:Kraken_Binance_rozklad}
\end{figure}

The results for the two different platforms, Kraken and Binance, are collated in Figure~\ref{fig:Kraken_Binance_rozklad} and Table~\ref{tab:rozkladBiKR}. By decreasing the sampling frequency to $\Delta t=1$ min (Jan-Dec 2018, $T=463,000$), the fluctuations of BTC and ETH can be compared with those of a few among the least-liquid cryptocurrencies, namely, XRP, BCH, LTC, XMR, DASH, ETC (all cryptocurrencies alongside their tickers are listed in the Appendix~\ref{BiKrdodatek}). The exchange rates on Binance are characterized by a higher trading frequency (i.e., shorter average length $\langle t_0 \rangle$ and lower average number $N_0$ of non-trading periods, Table~\ref{tab:rozkladBiKR}) compared to their counterparts on Kraken (see also Fig.~\ref{fig:Kraken_Binance_WartV}). This affects the values of the $\gamma$ exponent: while on both platforms, the exchange rate returns form CDFs with $\gamma\approx 3$, the CDF tails for the exchange rate returns on Binance tend to decay faster than their counterparts on Kraken. This pattern is also discernible in Figure~\ref{fig:Kraken_Binance_rozklad} and is in line with the observation that the trading frequency is related to the inner time-pace of a market. The less liquid exchange rates from Kraken have a slower internal time and, consequently, their rates of return distributions are characterized by thicker tails and poorer scaling (i.e., larger estimation error). By contrast, the CDFs for the rates with a higher trading frequency and volume resemble those characterizing the mature financial markets.

\begin{table}[ht!]
\centering
\caption{Estimated $\gamma$ exponent, average length of non-trading periods $\langle t_0 \rangle$ (i.e., zero-returns sequences), number of non-trading periods $N_0$, average volume-per-minute in USD, for the cryptocurrency pairs listed simultaneously on Binance and Kraken in 2018 (USD$^{*}$ means USD in the case of Kraken, and USDT in the case of Binance).}
\small

\begin{tabular}{|c|c|c|c|c|c|c|c|c|}
\hline
        & \multicolumn{2}{c|}{\boldmath$\gamma$} & \multicolumn{2}{c|}{\boldmath$\langle t_0 \rangle$} & \multicolumn{2}{c|}{\boldmath$N_0$} & \multicolumn{2}{c|}{\boldmath $\langle W \rangle$ \textbf{[USD]}} \\ \hline
\textbf{Name}      & \textbf{Bi}         & \textbf{Kr}        & \textbf{Bi}           & \textbf{Kr} & \textbf{Bi}         & \textbf{Kr}        & \textbf{Bi}         & \textbf{Kr}        \\ \hline
BTC/USD$^{*}$  & 3.45$\pm$0.1   & 3.63$\pm$0.15    & 1.07        & 1.61        & 8559       & 55360      & 196331     & 33998     \\ \hline
ETH/USD$^{*}$  & 3.39$\pm$0.1 & 3.4$\pm$0.1     & 1.10        & 1.78        & 16184      & 67807      & 61430
      & 19711
     \\ \hline
BTC/ETH  & 3.30$\pm$0.1   & 3.18$\pm$0.2     & 1.09        & 2.62        & 12134      & 82074      & 60824
      & 5250
      \\ \hline
XRP/BTC  & 2.91$\pm$0.1   & 2.99$\pm$0.15    & 1.20        & 3.50        & 41043      & 72556      & 31510
      & 2763
      \\ \hline
BCH/BTC  & 3.36$\pm$0.1   & 2.35$\pm$0.15   & 1.25        & 4.39        & 29973      & 68059      & 16299
     & 1251
      \\ \hline
BCH/USD$^{*}$  & 3.36$\pm$0.1   & 2.61$\pm$0.15     & 1.40        & 3.83        & 40922      & 75427      & 15893
     & 2151
      \\ \hline
LTC/BTC  & 3.42$\pm$0.1 & 3.34$\pm$0.15     & 1.17        & 5.09        & 34360      & 62675      & 13110
     & 940
       \\ \hline
LTC/USD$^{*}$  & 3.23$\pm$0.15    & 3.41$\pm$0.1     & 1.29        & 3.48        & 45208      & 79983      & 11793
      & 1725
     \\ \hline
ETC/BTC  & 3.02$\pm$0.1 & 2.31$\pm$0.2     & 1.56        & 6.34        & 76691      & 55647      & 10164
      & 709
       \\ \hline
XMR/BTC  & 3.68$\pm$0.15  & 3.21$\pm$0.15    & 1.37        & 6.58        & 47852      & 54832      & 4176
       & 614
       \\ \hline
DASH/BTC & 4.04$\pm$0.15   & 2.52$\pm$0.2    & 1.46        & 7.58        & 55684      & 49134      & 2777
      & 416
       \\ \hline
ETC/ETH  & 3.34$\pm$0.1  & 2.18$\pm$0.2     & 2.30        & 9.15        & 82458      & 41861      & 1333
      & 266
       \\ \hline
Average  & 3.38$\pm$0.11  & 2.93$\pm$0.16   & 1.35        & 4.66        & 40922      & 63785      & 35470
      & 5816
     \\ \hline
\end{tabular}
\label{tab:rozkladBiKR}
\end{table}

Summarizing, it can be argued that, for a given exchange rate, in order to meet the characteristics typical of mature financial markets like the inverse-cubic dependence, a sufficiently high trading frequency is required. From the vantage point of the $\Delta t=10$s, for the most liquid BTC/USD exchange rate on the Bitstamp platform, this was achieved in 2014; by contrast, on the Kraken the exchange the rates involving some of the least-liquid cryptocurrencies appear to be still in a ``pre-cubic'' immature phase.

\subsection{Temporal autocorrelations}
\label{StatAutoCorr}

Another standard characteristic used to describe the financial returns is the autocorrelation function
\begin{equation}
C(\tau) = \langle r_{\Delta t}(t) r_{\Delta t}(t-\tau) \rangle _t,
\label{ACFwzor}
\end{equation}
where $\langle \cdot \rangle_t$ denotes the time-average. A typical behaviour of $C(\tau)$ for the financial returns is its immediate decay either to zero or to below zero~\cite{Fama,Gopi1999,Palagyi1999,Xu2003,drozdzepps}. However, the autocorrelation function of volatility (i.e., absolute value of the returns) decays with an increasing $\tau$ according to a power law~\cite{DING1993,Gopi1999,ekonofizyka,Cont2001,kutner2004,drozdzepps}. The decay range corresponds to the average width of the volatility clusters~\cite{drozdz2010}. This phenomenon, which can be observed across all markets, is a manifestation of long-term memory~\cite{kwapien2012}: large fluctuations are statistically followed by large ones and small fluctuations by small ones~\cite{Mandelbrot1963}. The long-range autocorrelation, which is a linear effect in volatility, becomes a nonlinear correlation from the perspective of returns. Importantly, this implies that the returns can reveal multiscaling (see Section\ref{nonlinear}).\par

\begin{figure}[!ht]
\centering
\includegraphics[width=1\textwidth]{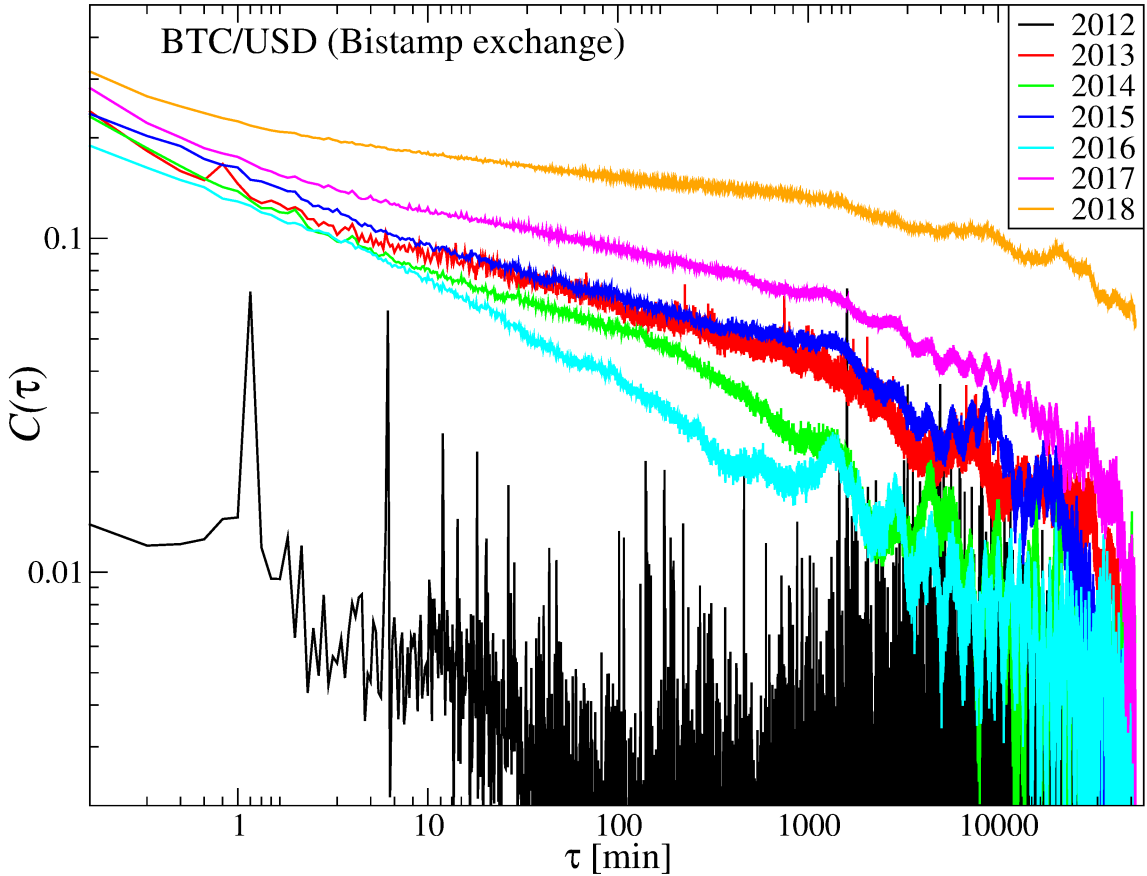}
\caption{Autocorrelation function of the normalized absolute log-returns $r_{\Delta t=10\textrm{s}}$ for bitcoin on the Bitstamp exchange over subsequent years.}
\label{fig:Bitstamp_ACF}
\end{figure}

\begin{figure}[!ht]
\centering
\includegraphics[width=1\textwidth]{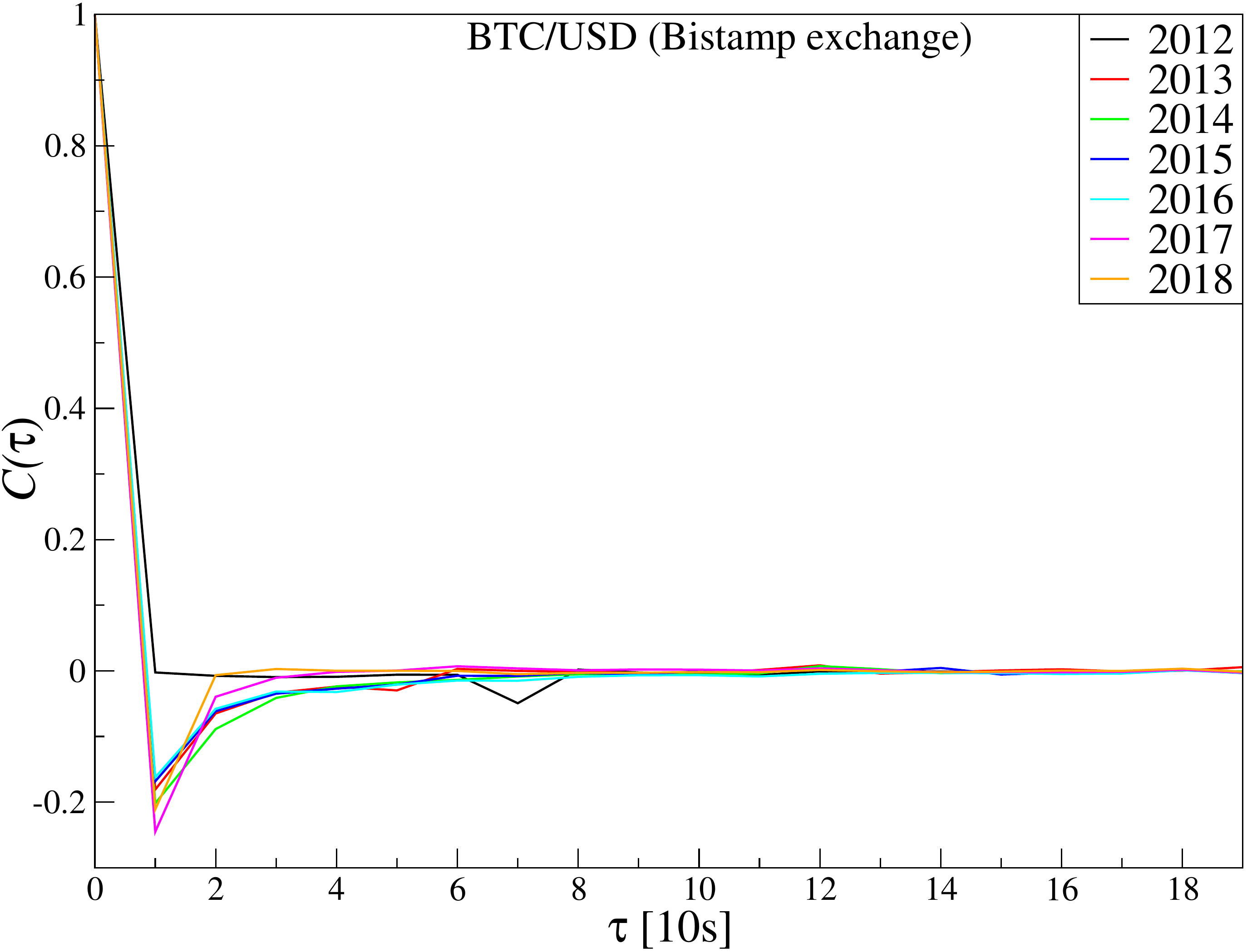}
\caption{Autocorrelation function of the normalized log-returns $r_{\Delta t=10\textrm{s}}$ for bitcoin on the Bitstamp exchange over subsequent years.}
\label{fig:Bitstamp_ACFr}
\end{figure}

It should be borne in mind that the autocorrelation function is a well-defined quantity only insofar as the data are drawn from a stationary process. Thus, if there is a trend in the data, it has to be removed beforehand. This also includes trivial periodic patterns. In the high-frequency financial data, one observes, for example, significant seasonal trends in volatility, including daily trends, weekly trends, and some longer-period trends. It is well-established that a satisfactory method of trend removal entails dividing each return corresponding to a time-point $t_j$ by the standard deviation of the returns calculated for the same moment over a sufficient number of trading days, i.e., $r_{\Delta t}(t_j) = R_{\Delta t}(t_j) / \sigma_R(t_j)$, where $j$ indexes the intervals $\Delta t$ starting from the market opening.\par

Figure~\ref{fig:Bitstamp_ACF} shows the autocorrelation function calculated for the BTC/USD volatility time-series split into annual parts. The quotes from the Bitstamp platform were first converted to absolute logarithmic returns, then detrended to remove seasonality. Similar to the CDF tails, a significant shift in the autocorrelation is observed between 2012 and the subsequent years. In the former, $C(\tau)$ is only short-range correlated, with memory disappearing after a few minutes. However, long-range memory has a power-law form: accordingly, $C(\tau)\sim \tau^{-\beta}$ started to emerge already in 2013 and was systematically built up to reach substantial strength even for intervals as long as $\tau>10^4$ min by 2018. The effect of increasing the volatility autocorrelation time can be explained in terms of the increasing number of transactions carried out in a time unit.\par

The long-range, linear autocorrelation in volatility corresponds to non-linear correlations on the level of signed returns. In turn, non-linear correlations are responsible for the multifractal effects (see Sect.~\ref{nonlinear}).\par

On the other hand, the autocorrelation function calculated for the original, signed returns does not indicate any linear memory effects for $\Delta t=10$s (Figure~\ref{fig:Bitstamp_ACFr}). A negative value of $C(1)$ is a statistical artifact related to isolated price jumps followed by price regression to a former level; these events are frequent on sampling frequencies comparable to the average inter-transaction interval. The observed autocorrelation function behavior for both returns and volatility resembles their counterparts on the other mature markets.\par

\begin{figure}[!ht]
\centering
\includegraphics[width=1\textwidth]{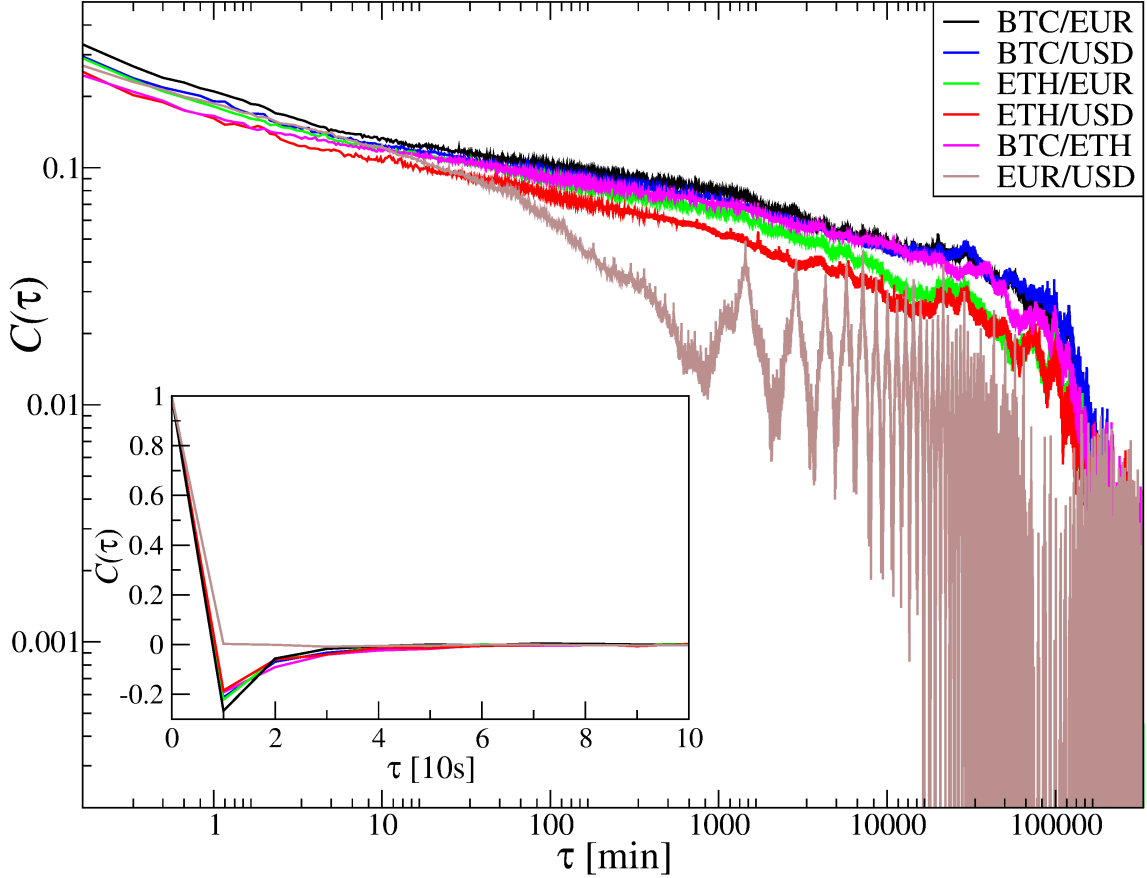}
\caption{Autocorrelation function of the normalized absolute log-returns $r_{\Delta t=10\textrm{s}}$ and normalized log-returns $r_{\Delta t=10\textrm{s}}$ (inset) for exchange rates BTC/EUR, BTC/USD, ETH/EUR, ETH/USD, BTC/ETH on Kraken exchange, and for EUR/USD from July 2016 to December 2018.}
\label{fig:Kraken10_ACF}
\end{figure}

\begin{figure}[!ht]
\centering
\includegraphics[width=1\textwidth]{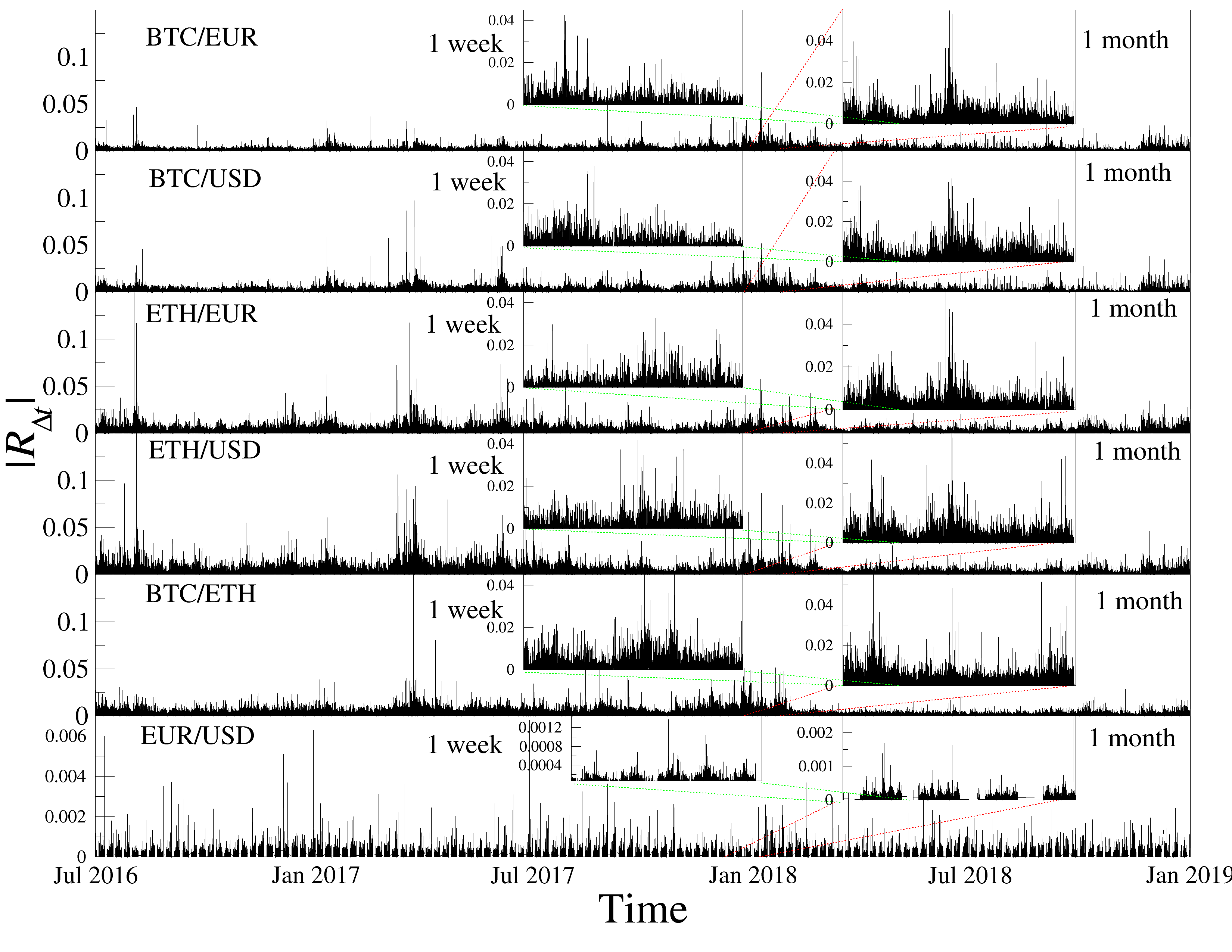}
\caption{Absolute values of log-returns $R_{\Delta t=10\textrm{s}}$ for the exchange rates BTC/EUR, BTC/USD, ETH/EUR, ETH/USD and EUR/USD from July 2016 to December 2018. The insets show magnified selected periods.}
\label{fig:Kraken10_volat}
\end{figure}

\begin{figure}[!ht]
\centering
\includegraphics[width=1\textwidth]{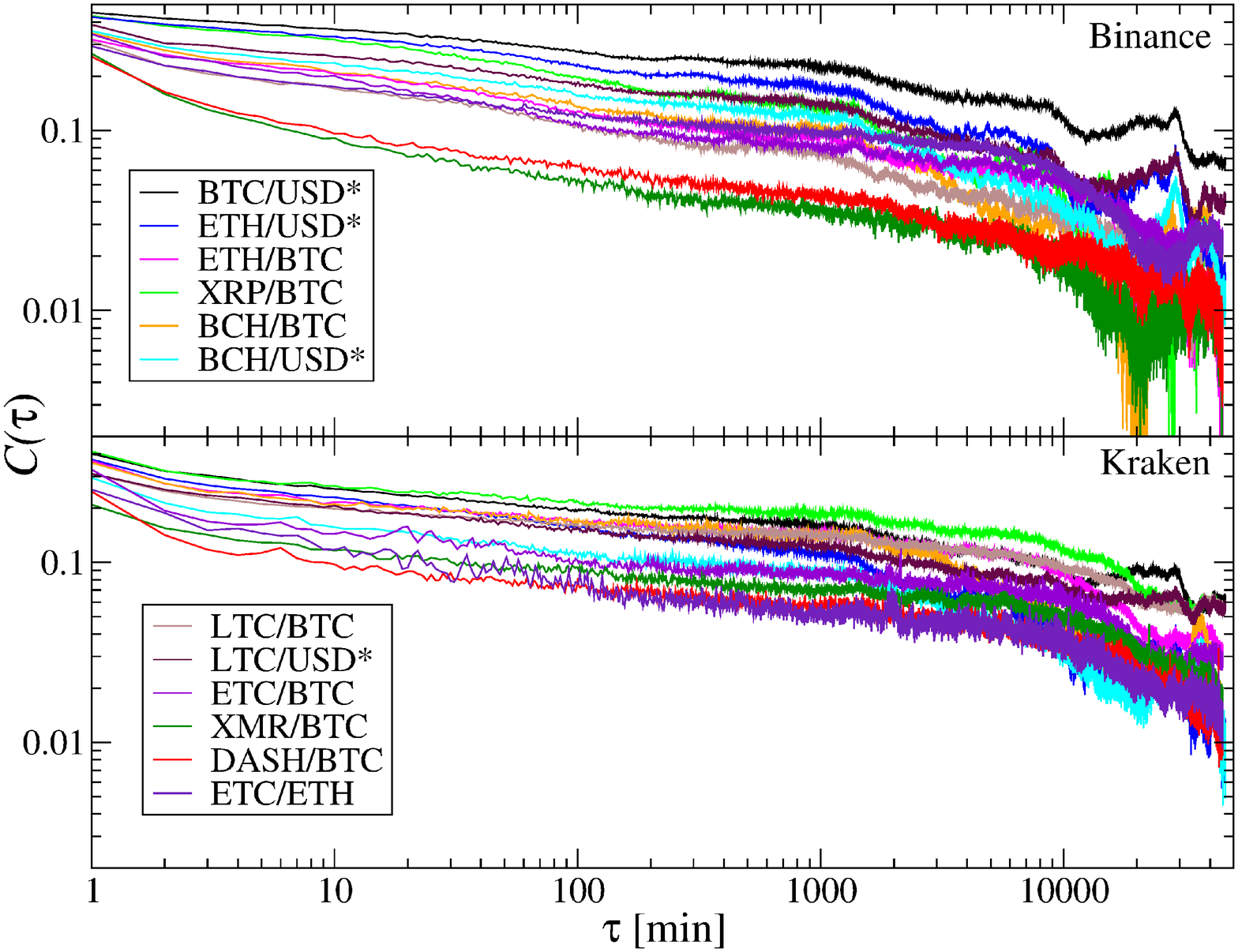}
\caption{Comparison of the autocorrelation functions of normalized absolute log-returns $r_{\Delta t=1\textrm{min}}$ for exchange rates listed simultaneously on Binance and Kraken (USD$^{*}$ means USD in the case of Kraken, and USDT in the case of Binance).}
\label{fig:Kraken_Binance_ACF}
\end{figure}

The volatility autocorrelations function for the BTC and ETH expressed in USD and EUR, as well as for BTC/ETH (Kraken), are displayed in Figure~\ref{fig:Kraken10_ACF}. For all cryptocurrency exchange rates under consideration, it decays with $\tau$ approximately according to a power-law up to about a week. However, in the case of EUR/USD (Dukascopy) the power-law regime breaks after already about one day, which can be accounted for by a shorter volatility-cluster length. This effect can be seen in Figure~\ref{fig:Kraken10_volat}, where $|R_{\Delta t}(t_i)|$ with $\Delta t=10$s are shown for six exchange rates over the analyzed period. For the rates involving cryptocurrencies, the cluster length is an order of magnitude larger than for EUR/USD. Five cycles of high/low volatility during a week can be distinguished for the EUR/USD, while only three cycles per month are visible for the cryptocurrencies (insets in Fig.~\ref{fig:Kraken10_volat}). Again, this can be explained by a higher trading frequency on the Forex than on Kraken. The average inter-transaction time is a few seconds on Kraken, whereas Forex transactions take place at least an order of magnitude faster.\par

Similarly to Bitstamp BTC/USD data, and indeed to all mature financial markets, the autocorrelation function for the returns (inset in Fig.~\ref{fig:Kraken10_ACF}) drops immediately to zero for the EUR/USD and even below zero for the cryptocurrency rates, indicating that no linear correlation exist for the signed returns.\par

The difference in average trading frequency between Binance and Kraken affects not only the return CDFs but also the volatility autocorrelation function. In both cases, one observes its power-law decay (Figure~\ref{fig:Kraken_Binance_ACF}), but on Binance, it is faster: it breaks down already after about 1,000 minutes as opposed to about 10,000 on Kraken. In the case of signed returns, the autocorrelation function exhibits a typical behavior: it immediately drops to negative values and then oscillates around zero (not shown).

\section{Multiscale correlations}
\label{nonlinear}

The commonly used correlation measures in time-series analysis, such as the autocorrelation function, the spectral density~\cite{spectral2005} and the Pearson correlation coefficient~\cite{Pearson1895}, only provide information about linear interdependencies in the data. However, a typical feature of financial markets is the occurrence of nonlinear correlations~\cite{kwapien2012,multirev}, which are the primary source of multifractality in their temporal dynamics (an additional factor reinforcing multifractality is related to the ``fat tails'' of the probability density function)~\cite{Kwapien2005,Zhou2009,Barunik2012,Grech2013,Rak2018,multirev}. Multifractality is an attribute of many complex systems wherein a sufficiently large number of elements interact according to nonlinear laws~\cite{kwapien2015}. This kind of system organization is a well-evident characteristic of the financial markets, which in this study are represented by the cryptocurrency market.

In Section~\ref{StatAutoCorr}, it was demonstrated that the power-law decay of the autocorrelation function estimated for absolute values of the log-returns reveals nonlinear correlations. Therefore, signatures of multiscaling can also be anticipated in the cryptocurrency market. In the following section, advanced methods for the quantitative description of nonlinear correlations based on the multifractal methodology will be presented: namely, the multifractal detrended fluctuation analysis (MFDFA), its generalization for the cross-correlation case, known as multifractal cross-correlation analysis (MFCCA), and the $q$-dependent detrended cross-correlation coefficient $\rho(q,s)$. The analyses of the time-series data from the cryptocurrency market were carried out through these methods.

\subsection{Detrending based multifractal methodology}
\label{detrendned}

Historically, the first method for analyzing fractal time-series is the rescaled range analysis (R/S), proposed by the British hydrologist Harold Edwin Hurst~\cite{Hurst}. In his original work, Hurst introduced and applied the R/S algorithm to study the recurrent flooding on the river Nile. In the case of nonstationary data, to correctly quantify correlations in a signal, it is necessary to identify and remove trends, which is impossible to attain using the R/S algorithm. To address this fundamental shortcoming, in 1995, Peng et al. proposed the detrended fluctuation analysis (DFA), which can be applied to nonstationary data~\cite{Peng94,kantelhardt01}. Within the DFA method, the trend represented by the $n$-th degree polynomial is subtracted. This procedure leads to the removal of nonstationarity from the data. However, the choice of the polynomial degree is crucial to obtain reliable results. If the polynomial degree is too high, the procedure could destroy the fluctuation structure. On the other hand, with too low a degree, the polynomial may not remove nonstationarity effectively~\cite{kantelhardt01,oswiecimka2013}. In most cases, a second-order polynomial is used for the analysis of financial data. Alternatives have been discussed in~Ref. \cite{oswiecimka2013}.

Multifractal detrended fluctuation analysis (MFDFA) is a more recent development of the DFA algorithm, covering the case of nonlinear correlations case~\cite{kantelhardt02}. Using the MFDFA, the nonlinear organization in a time-series is quantitatively described within a fractal formalism.  In recent years, cross-correlation versions of the DFA and MFDFA methods have also been proposed. The generalization of the DFA to the cross-correlation case is known as the detrended cross-correlation analysis (DCCA) method~\cite{podobnik2008}, whereas multifractal detrended cross-correlation analysis (MF-DXA) is an extension of the MFDFA~\cite{Zhou2009}.

In the latter case, since the cross-correlation fluctuation function may assume positive and negative values, proposed formulas, related to calculation of the moments, lead to erroneous conclusions. To avoid this effect, many authors resort to taking the absolute values of the cross-correlation fluctuations functions, which, in turn, risks leading to the artifactual detection of multifractal cross-correlations even for uncorrelated signals~\cite{oswiecimka2014}. A solution to this problem is offered by the multifractal cross-correlation analysis (MFCCA) algorithm, which has been proposed in~Ref. \cite{oswiecimka2014}.

An alternative to the DFA-based methods of quantifying multiscale correlations is provided by the algorithms using wavelet transforms like, e.g., the wavelet transform modulus maxima (WTMM)~\cite{Muzy91,Muzy94,Arneodo95} and the multifractal cross-wavelet transform (MFXWT)~\cite{Jiang2017} for the auto- and cross-correlation cases, respectively. However, this is MFCCA that is currently deemed to be the most reliable method for studying the multiscale correlations in the time series~\cite{oswiecimka2006,drozdz2010}.

The robustness of the time-series analysis carried out by means of the multifractal methodology has fuelled considerable popularity of the corresponding methods, and their application throughout highly diversified scientific fields~\cite{kwapien2015} such as biology~\cite{Eke2012,Maiorino2015},
chemistry~\cite{Grout1998,Udovichenko2002}, physics~\cite{Lafouti2014,Mali2015}, geophysics~\cite{Telesca2005,Witt2013}, hydrology~\cite{KOSCIELNYBUNDE2006,Zhang2009}, quantitative linguistics~\cite{Ausloos2012,Drozdz2016}, medicine~\cite{Echeverria2003,Chen2018}, meteorology~\cite{Rak2016,Adarsh2020}, music~\cite{Su2006,Roeske2018}, psychology~\cite{Garas2012,Ihlen2013} and especially finance~\cite{Ausloos2002,Calvet2002,Turiel2005,Dutta2016,Grech2016,Zhao2017,Klamut2018,multirev}. 

Notwithstanding the broad application of the multifractal methodology, it must be noted that their proper application and drawing conclusions require certain experience, otherwise some serious misinterpretation of results might occur. One needs to realize that not all indications of multiscaling found in data may be considered an actual manifestation of multifractality in the strict mathematical sense. It was shown in literature that multiscaling can also be a finite-sample effect especially if the analyzed data has a fat-tail PDF~\cite{drozdz2010}. In many instances of empirical data, including the financial data, multiscaling is observed over rather a moderate range of available scales like, e.g., one decade, while at the remaining scales other effects can be seen, like an exponential cut-off of the PDF tails or the outlier events, which do not fit into any fractal organization of the data. The multifractal methodology allows one for a reliable and convenient quantitative description of the temporal organization of nonstationary signals, including the heterogeneity of their singularities, while only sometimes it also allows for undoubted identification of a genuine multifractal structure.

\subsubsection{MFCCA as a generalization of MFDFA and DCCA}
\label{MFCCA}

The multifractal cross-correlation analysis (MFCCA)~\cite{oswiecimka2014} allows for quantitative description of both the scaling properties of data and the degree of multiscale detrended cross-correlation between two time series. Given two time-series $x_i$, $y_i$, where $i=1,2...T$, corresponding profiles are calculated according to \begin{equation}
X(j) =\sum_{i=1}^j[x_{i}-\langle x\rangle] ,\quad
Y(j) =\sum_{i=1}^j[y_{i}-\langle y\rangle],
\end{equation}
where $\langle x\rangle$ and $\langle y\rangle$ denote the means of the $x_{i}$ and $y_{i}$ time-series, respectively.

Then, the profiles $X(j)$ and $Y(j)$ are divided into $2M_s$ separate segments $\nu$ with length $s$, starting at the beginning and at the end of the time-series, where $M_s=\floor{T/s}$. In each segment $\nu$, a trend is identified by fitting a polynomial of degree $m$, separately for the series $X$ -- $P^{(m)}_{X,\nu}$ i $Y$ -- $P^{(m)}_{Y,\nu}$. As said, for financial time-series a typical choice is $m=2$~\cite{oswiecimka2006}. After removing the trend, cross-covariance is calculated
\begin{multline}
F_{xy}^{2}(\nu,s)=\frac{1}{s}\sum_{k=1}^{s}\lbrace
(X((\nu-1)s+k)-P^{(m)}_{X,\nu}(k)) \\ \times
(Y((\nu-1)s+k)-P^{(m)}_{Y,\nu}(k))\rbrace,
\label{Fxy2}
\end{multline}
for segment $\nu=1,...,M_s$ and   
\begin{multline}
F_{xy}^{2}(\nu,s)=\frac{1}{s}\sum_{k=1}^{s}\lbrace
(X(T-(\nu-M_s)s+k)-P^{(m)}_{X,\nu}(k)) \\ \times
(Y(T-(\nu-M_s)s+k)-P^{(m)}_{Y,\nu}(k))\rbrace,
\label{Fxy2T}
\end{multline}
for $\nu=M_s+1,...,2M_s$, and subsequently it is used to calculate the $q$-th order covariance function~\cite{oswiecimka2014}
\begin{equation}
F_{xy}^{q}(s)=\frac{1}{2M_s}\sum_{\nu=1}^{2M_s} {\rm
sign}(F_{xy}^{2}(\nu,s))|F_{xy}^{2}(\nu,s)|^{q/2},
\label{Fq}
\end{equation}
where ${\rm sign}(F_{xy}^{2}(\nu,s))$ denotes the sign of $F_{xy}^{2}(\nu,s)$. Fractal detrended cross-correlation between two time-series $x_i$ i $y_i$ is manifested in the scaling relationship
\begin{equation}
F_{xy}^{q}(s)^{1/q}=F_{xy}(q,s) \sim s^{\lambda(q)}, 
\label{Fxy}
\end{equation}
where $q\neq0$ and $\lambda(q)$ are the scaling exponents. The parameter $q$ acts as a filter which strengthens or suppresses the covariance calculated on segments having length $s$. For positive $q$, segments containing large fluctuations have a predominant impact on the sum in Eq.(\ref{Fq}), whereas for negative $q$, segments with small fluctuations dominate in $F_{xy}^{q}(s)$. In the presence of multifractal cross-correlation, a dependence of $\lambda(q)$ on $q$ is observed, whereas for the monofractal case, $\lambda(q)$ is independent of $q$.

Multifractal analysis of a single time-series by means of MFDFA~\cite{kantelhardt02} corresponds to a special case of the MFCCA procedure, wherein $x_i=y_i$. Then Eq.(\ref{Fq}) is reduced to
\begin{equation}
F(q,s)=\Big[\frac{1}{2M_s}\sum^{2M_s}_{\nu=1}{[F^2(\nu,s)]^{\frac{q}{2}}}\Big]^{\frac{1}{q}}.
\label{F}
\end{equation}
As in Eq.(\ref{Fxy}), fractality manifests itself in a power-law relation
\begin{equation}
F(q,s) \sim s^{h(q)},
\label{Hq}
\end{equation}
where $h(q)$ is the generalized Hurst exponent, which for $q=2$ corresponds to standard Hurst exponent~\cite{Halsey87}. In the case of a monofractal time series, $h(q)$ is constant, whereas for a multifractal one, $h(q)$ depends on $q$. In the latter case, the time series reveals a sort of hierarchical organization of the returns~\cite{drozdz2015}.

The singularity spectrum $f(\alpha)$ can be calculated by a Legendre transform:
\begin{equation}
\alpha=h(q)+qh'(q), \quad f(\alpha)=q[\alpha-h(q)]+1,
\label{spektrum}
\end{equation}
where $\alpha$ is the H\"older exponent, and $f(\alpha)$ is fractal dimension of the time series points with singularities $\alpha$. In the case of multifractals, the shape of the singularity spectrum characteristically resembles an inverted parabola.

Width of the singularity spectrum
\begin{equation}
\Delta \alpha = \alpha _\textrm{max}  - \alpha _\textrm{min},
\label{Dm}
\end{equation}
where $\alpha _\textrm{min}$ and $\alpha _\textrm{max}$ are the minimum and maximum values of $\alpha$, is a quantity intended to measure the diversity of singularities present in time series. Even if the scaling range in $s$ of the fluctuation functions $F(q,s)$ is limited to a decade or so and, thus, no actual multifractal structure exists in correlation of the time series, $f(\alpha)$ remains a convenient quantity to assess the detrended correlation heterogeneity.

Another important feature of the $f(\alpha)$ spectrum is its asymmetry (skewness), which can be measured by the asymmetry coefficient~\cite{drozdz2015}
\begin{equation} 
A_{\alpha}=\frac{\Delta \alpha _L - \Delta \alpha _R}{\Delta \alpha _ L + \Delta \alpha _R},
\label{Ass}
\end{equation}
where $\Delta \alpha _L = \alpha _0 - \alpha _\textrm{min}$, $\Delta \alpha _R = \alpha _\textrm{max} - \alpha _0$, and $\alpha _0$ corresponds to the maximum value $f(\alpha)$ observed for $q=0$. A positive coefficient $A _\alpha$ denotes a left-sided asymmetry of the spectrum. This corresponds to a more developed multiscaling (stronger correlations) at the level of large fluctuations. In turn, a negative value of $A _\alpha$ denotes a right-sided asymmetry, implicating small fluctuations as the dominant source of multiscaling.

To assess the strength of multifractal cross-correlation, the scaling exponent $\lambda(q)$ and the average of generalized Hurst exponents calculated for each time series separately
\begin{equation} 
h_{xy}(q) = (h_x(q)+h_y(q))/2
\label{hqxy}
\end{equation} 
have to be compared. Thus, the difference between $\lambda(q)$ and $h_{xy}(q)$ is calculated
\begin{equation}
d_{xy}(q) = \lambda(q) - h_{xy}(q).
\label{dqxy}
\end{equation}
For perfect multiscale cross-correlation, $d_{xy}(q)=0$, whereas for desynchronization between the time-series, $d_{xy}(q)\neq 0$~\cite{futinternet2019,Watorek2019}. Additional notions about $d_{xy}(q)$ are given in Subsect.~\ref{efektW}.

\subsubsection{Detrended cross-correlation coefficient}
\label{rho}

The fluctuation functions as defined by Eq.(\ref{Fxy}) can also be used as a basis for determining the $q$-dependent detrended correlation coefficient ($q$DCCA)~\cite{kwapien2015}
\begin{equation}
\rho(q,s) = {F_{xy}^q(s) \over \sqrt{ F_{xx}^q(s) F_{yy}^q(s) }},
\label{rhoq}
\end{equation}
where $F_{xx}$ and $F_{yy}$ are calculated from Eq.(\ref{F}). This measure allows one for describing quantitatively the cross-correlations between two time-series $x_i$ and $y_i$ after removal of a polynomial trend at different time scales $s$. Besides, the parameter $q$ allows the detrended fluctuation magnitude range for which the signals $x_i$ and $y_i$ are the most strongly correlated. Such two-dimensional decomposition of the overall correlations into the components corresponding to particular magnitudes and scales provides a significant advantage over standard correlation measures since correlations between empirical time-series are usually not uniformly distributed over fluctuations of different size and different time scales~\cite{kwapien2017}. $\rho(q,s)$ can be used for investigating non-stationary time series that are not necessarily fractal~\cite{kwapien2015}. For $q>0$ the coefficient $\rho(q,s)$ assumes values from $-1$ to $1$~\cite{kwapien2015} and their interpretation is the same as the Pearson correlation coefficient~\cite{Pearson1895}. For $q<0$ small fluctuations have the dominant contribution to $\rho(q,s)$ and vice-versa for $q>0$. In the case of $q=2$, the coefficient $\rho(q,s)$ is reduced to $\rho_{\textrm{DCCA}}$~\cite{zebende2011}.

\subsubsection{Relationship between $\lambda(q)$ and $\rho(q,s)$}
\label{efektW}

In the previous section, it was shown that the coefficient $\rho(q,s)$ quantifies the correlations with respect to a time scale $s$. Throughout the literature on financial cross-correlations, the Epps effect~\cite{Epps,kwapien2004,Toth2009,drozdzepps}, which involves an increasing level of correlation as the time-interval $\Delta t$ between the returns $R_{\Delta t}$ increases (a higher correlation level for hourly returns compared to the minute returns), is well-evident. Similarly, for the coefficient $\rho_{\textrm{DCCA}}$ and its more general version $\rho(q,s)$, different correlation levels appear across scales~\cite{Ma2013,Ma2014,Pal2014,Reboredo2014,kwapien2015,Barunik2016,Li2016,Yang2016,Hussain2017,Zhao2017,Ferreira2019}.

Since $\lambda(q)$ and $\rho(q,s)$ are derived from the same fluctuation function given by Eq.(\ref{Fxy}), they are linked by a meaningful relationship. If the condition of the scaling of the fluctuation functions for both cross-correlation $F_{xy} (q,s)=a_{xy}(q)s^{\lambda(q)}$ and single time-series $F_{xx}(q,s)=a_x(q)s^{h_x(q)}$ and $F_{yy}(q,s)=a_y(q)s^{h_y(q)}$ is fulfilled, proportionality coefficients $a_{xy}(q), a_x(q), a_y(q)$ can be determined. Based on the relation (\ref{rhoq}) and the properties of $\rho(q,s)$ (for $q\ge 0$, $-1\le \rho(q,s)\le 1$), the following relation can be written~\cite{kwapien2015}:
\begin{equation}
 F_{xy}^q(s) \le \sqrt{ F_{xx}^q(s) F_{yy}^q(s) },
\label{eq::cauchy-schwarz-like}
\end{equation}
and after substitution:
\begin{equation}
(a_{xy}(q))^qs^{q\lambda(q)}\leq (a_x(q)a_y(q))^{q/2}s^{q(h_x(q)+h_y(q))/2}.
\label{eq::ca}
\end{equation}
For $q>0$ this leads to:
\begin{equation}
\lambda(q) \leq \log_s(\frac{\sqrt{a_x(q)a_y(q)}}{a_{xy}(q)}) + \frac{h_x(q)+h_y(q)}{2}.
\label{eq:ca1}
\end{equation}
To satisfy Eqs.(\ref{eq::cauchy-schwarz-like}) and (\ref{eq::ca}), for $q>0$ the proportionality coefficients must fulfill the relation $a_{xy}(q) \leq \sqrt{a_x(q)a_y(q)}$ \cite{oswiecimka2014}. Hence, $\log_s(\frac{\sqrt{a_x(q)a_y(q)}}{a_{xy}(q)})$ is positive whereas $d_{xy}(q)$, which is the difference between $\lambda(q)$ and $h_{xy}(q)$, can be either positive or negative~\cite{oswiecimka2014}. For positive $d_{xy}(q)$, i.e., $\lambda(q) > (h_x(q)+h_y(q))/2$, $F_{xy} (q,s)$ grows faster with increasing $s$ than $\sqrt{F_{xx}^q(s) F_{yy}^q(s)}$. Thus, $\rho(q,s)$ also increases with increasing $s$. The reverse situation occurs when $d_{xy}(q)$ is negative~\cite{Watorek2019}. In that case, $\rho(q,s)$ decreases with increasing $s$. For $d_{xy}(q)=0$, $\rho(q,s)$ does not change with increasing $s$.

The above result holds true under a condition that both time series have finite length. For $s \to \infty$ only $\lambda(q) \leq  h_{xy}(q)$) is allowed~\cite{He2011,Kristo2015}. These relationships will be studied for empirical data in the following sections.

\subsection{Analysis of the Hurst exponent in the cryptocurrency market}
\label{Hurst}

In section~\ref{StatAutoCorr}, it was shown that the autocorrelation function for absolute returns (volatility) decays according to a power-law relation (``long memory'')~\cite{lilo2004}. A well-known measure that describes long memory quantitatively is the Hurst exponent~\cite{Hurst}. It quantifies the kind of self-similarity in the time-series alongside the associated degree of persistence. $H=0.5$ denotes the absence of autocorrelation, that is, the successive fluctuations are uncorrelated. $0.5 \le H \le 1$ imply a persistent time-series with positive autocorrelation: a price change in the same direction as the previous one is more probable than in the opposite direction. On the other hand, $H < 0.5$ denotes an antipersistent time series with negative autocorrelation: a price change in the opposite direction is more likely than in the same direction as the previous one.

Mature financial markets typically feature Hurst exponents dwelling around $0.5$~\cite{Ausloos2000,Matteo2005}, which effectively implies the impossibility of predicting the direction of the next price change at any time point. Contrariwise, significant deviations from this level are often fount in emerging markets~\cite{Matteo2005,Cajueiro2006}, hallmarking market inefficiency~\cite{Tabak2006}. $H \to 0.5$ can, therefore, be taken as an indication of market maturation~\cite{Matteo2003,DrozdzBTC2018}. Furthermore, a drop below $0.5$ over a short time-interval may be a predictor of an upcoming trend change~\cite{Grech2004,Stawiarski2016}. Throughout this study, the Hurst exponent is calculated using the MFDFA method for $q=2$ (Eq.(\ref{Hq})).

\begin{figure}[!ht]
\centering
\includegraphics[width=0.9\textwidth]{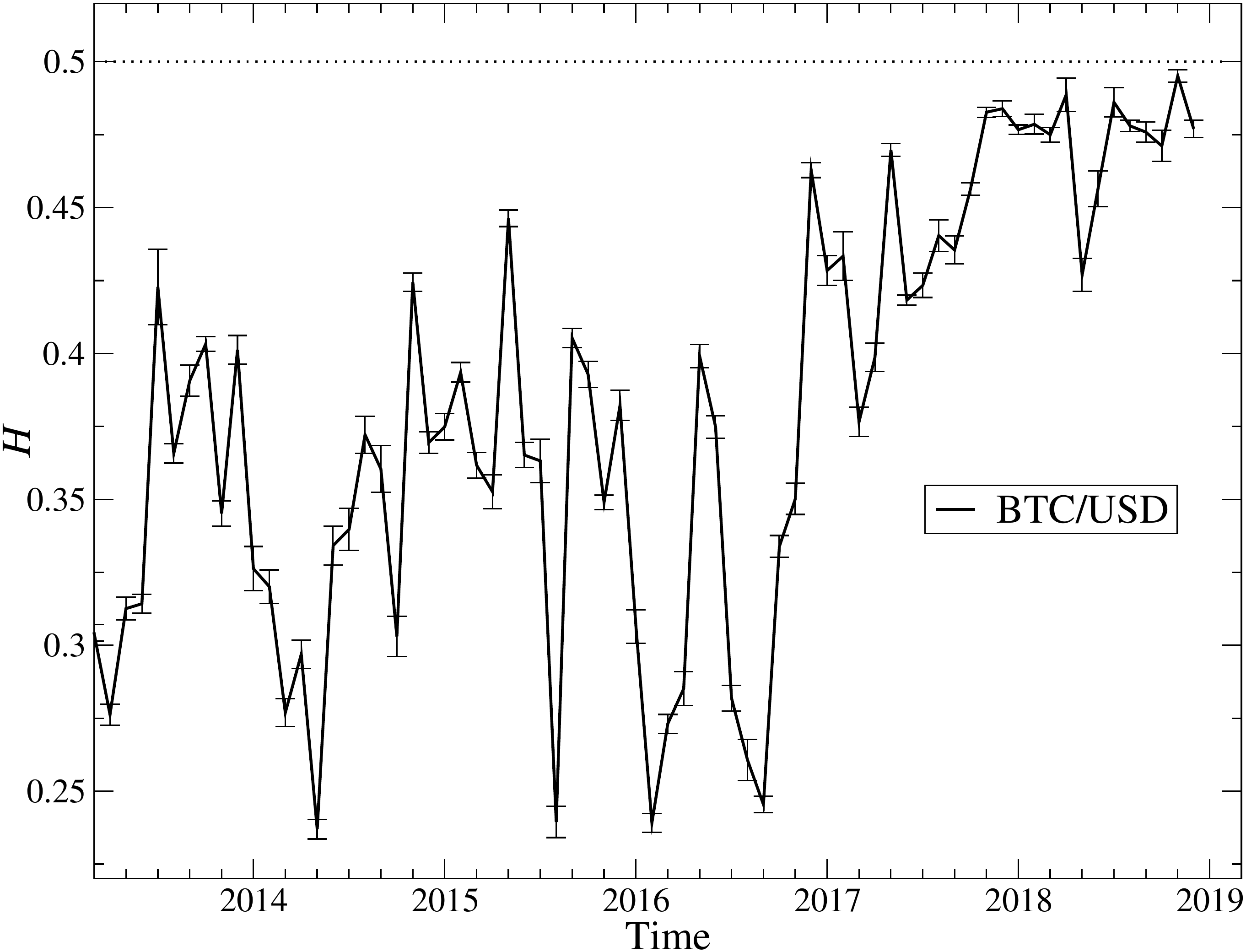}
\includegraphics[width=0.9\textwidth]{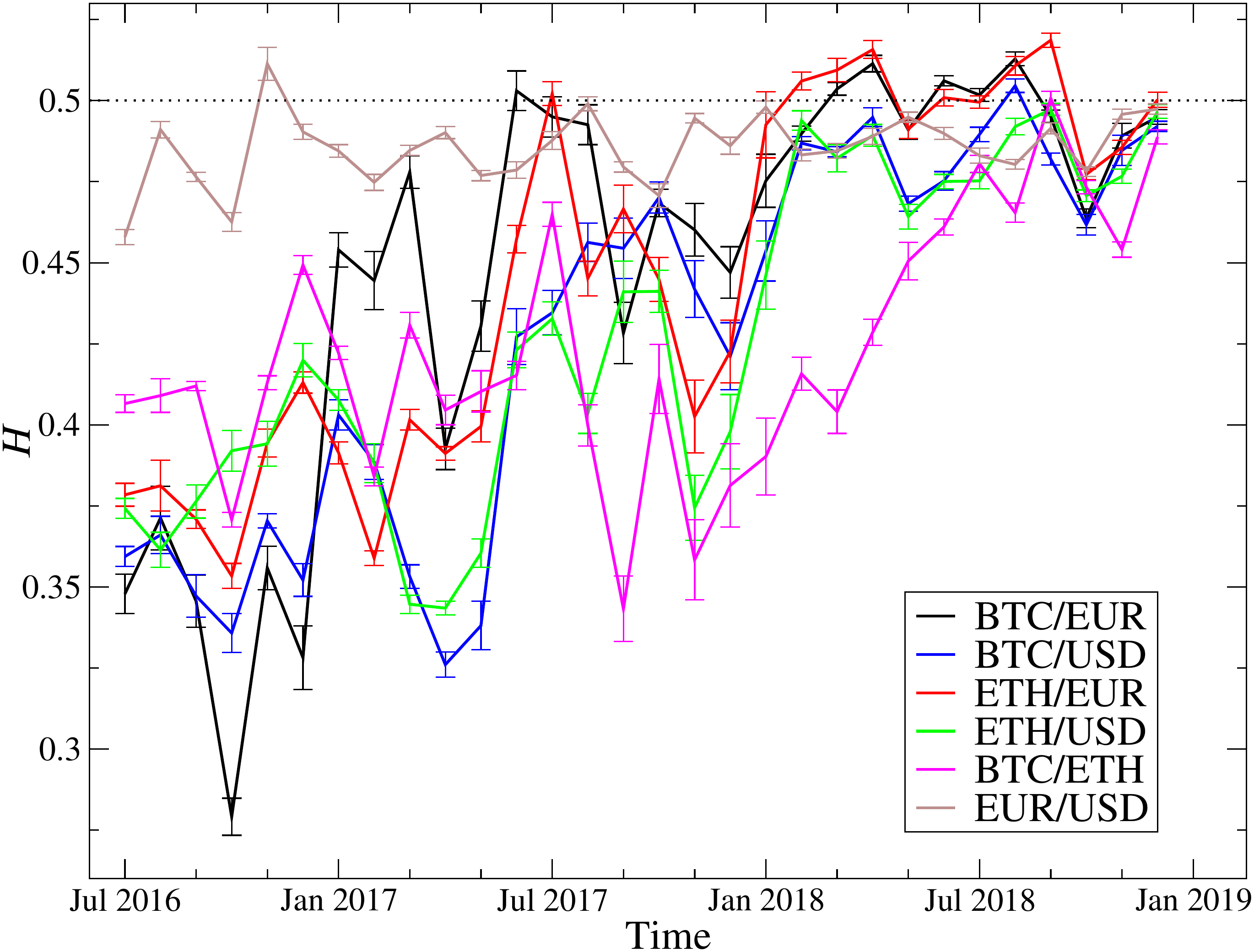}
\caption{Hurst exponent for the $\Delta t=10$s returns calculated over monthly windows for different exchange rates. (Top) The BTC/USD exchange rate from Bitstamp (March 2013--December 2018). (Bottom) BTC/EUR, BTC/USD, ETH/EUR, ETH/USD, and BTC/ETH from Kraken (July 2016--December 2018), together with EUR/USD, for comparison. In both panels error bars indicate the standard error of linear regression.}
\label{fig:Bitstamp_Hurst}
\end{figure}

The temporal evolution of the Hurst exponents over the 10-seconds log-returns of BTC/USD exchange rate ($R_{\Delta t=10\textrm{s}}$) from the Bitstamp exchange, from March 2013 to December 2018, is shown in Fig.~\ref{fig:Bitstamp_Hurst} (top panel). Earlier periods (see Fig.~\ref{fig:BTCbitstamp}) were characterized by considerably longer inter-transaction times and, hence, a larger number of zero returns. The Hurst exponent was calculated for separate monthly windows, each containing about 260,000 observations, and the results are shown in Fig.~\ref{fig:Bitstamp_Hurst} (top panel). By the end of 2016, the exponent had attained values indicating antipersistence ($H<0.5$), which is typical for risky, growing markets. Alongside increased trade frequency (i.e., shortening of the inter-transaction times), $H$ approached 0.5 in 2018. This is in line with the results obtained for the return PDFs and autocorrelation functions in Sect.~\ref{Dane stat}, wherein 2018 was also indicated as the year during which the BTC/USD exchange rate attained characteristics closest to those typically observed for Forex.

The Hurst exponent evolution for more exchange rates: BTC/EUR, BTC/USD, ETH/EUR, ETH/USD, and BTC/ETH can be shown based on data from the Kraken platform. This is done, however, in expense of a shorter interval available: July 2016--December 2018. The returns were constructed with $\Delta t=10$s and $H$ was calculated over monthly windows -- Fig.~\ref{fig:Bitstamp_Hurst} (bottom panel). Like in the case of BTC/USD from Bitstamp, the exchange rates from Kraken were associated with $H<0.5$ until mid-2017. Then from 2018 the Hurst exponents increased to a 0.5 level for all the exchange rates involving USD and EUR, while BTC/ETH remained in the antipersistent regime until mid-2018. This shift of $H$ is parallel to the shortening of the inter-transaction times on Kraken that started in April 2017. For comparison, $H$ for the EUR/USD exchange rate is also shown -- Fig.~\ref{fig:Bitstamp_Hurst} (bottom panel). This demonstrates clearly the cryptocurrency market development~\cite{futinternet2019}.

\begin{table}[ht!]
\centering
\caption{The Hurst exponents derived for the exchange rates listed on Binance and Kraken in 2018. The estimated errors are of the order of the third decimal place, therefore, they are not reported. (USD$^{*}$ means US dollar in the case of Kraken and USDT in the case of Binance).}
\begin{tabular}{|c|c|c|}
\hline
        & \multicolumn{2}{c|}{\boldmath$H$} \\ \hline
\textbf{Name}   & \textbf{Binance}    & \textbf{Kraken}    \\ \hline
BTC/USD$^{*}$  & 0.47       & 0.48      \\ \hline
ETH/USD$^{*}$  & 0.48       & 0.49      \\ \hline
ETH/BTC  & 0.51       & 0.48      \\ \hline
XRP/BTC  & 0.46       & 0.45      \\ \hline
BCH/BTC  & 0.48       & 0.43      \\ \hline
BCH/USD$^{*}$  & 0.48       & 0.48      \\ \hline
LTC/BTC  & 0.47       & 0.45      \\ \hline
LTC/USD$^{*}$  & 0.47       & 0.47      \\ \hline
ETC/BTC  & 0.47       & 0.45      \\ \hline
XMR/BTC  & 0.46       & 0.43      \\ \hline
DASH/BTC & 0.46       & 0.44      \\ \hline
ETC/ETH  & 0.42       & 0.41      \\ \hline
Average & 0.47       & 0.45      \\ \hline
\end{tabular}
\label{tab:BiKrHurst}
\end{table}

Table~\ref{tab:BiKrHurst} shows values of the Hurst exponent calculated for the returns ($\Delta t=1$ min) of a few equivalent exchange rates that were obtained independently from two platforms: Binance and Kraken. The time series cover the whole year 2018. On average, larger exponents can be observed for the cryptocurrency pairs listed on Binance, where $H$ is closer to $0.5$. This goes in parallel with a higher trading frequency on this platform than on Kraken (see Sect.~\ref{Dane stat}). The smallest Hurst exponent was obtained on both platforms for ETC/ETH, which is characterized by the lowest trading frequency correlating also with relatively small Hurst exponents for Kraken's BCH/BTC, XMR/BTC, and DASH/BTC (in each case $H<0.5$).

\subsection{Multiscaling of the exchange rates}
\label{MFDFA}

The Hurst exponent discussed in the previous subsection can account for linear correlations only. A more subtle characteristics of the time series complexity is the occurrence of multiscale autocorrelations. They can be described by the generalized Hurst exponent family $h(q)$ with $q$ that belongs to some predefined range related to the properties of the return PDF tails. $h(q)$ can be calculated with help of MFDFA according to Eq.(\ref{Hq}). A complementary information can be found in the singularity spectrum $f(\alpha)$, whose width given by Eq.(\ref{Dm}) can be considered as a measure of the time series complexity or diversity of its internal structure~\cite{drozdz2010,drozdz2018,multirev}.

\subsubsection{Temporal evolution of the multiscale autocorrelations}
\label{BTCMFDFA}

\begin{figure}
\centering
\includegraphics[width=1\textwidth]{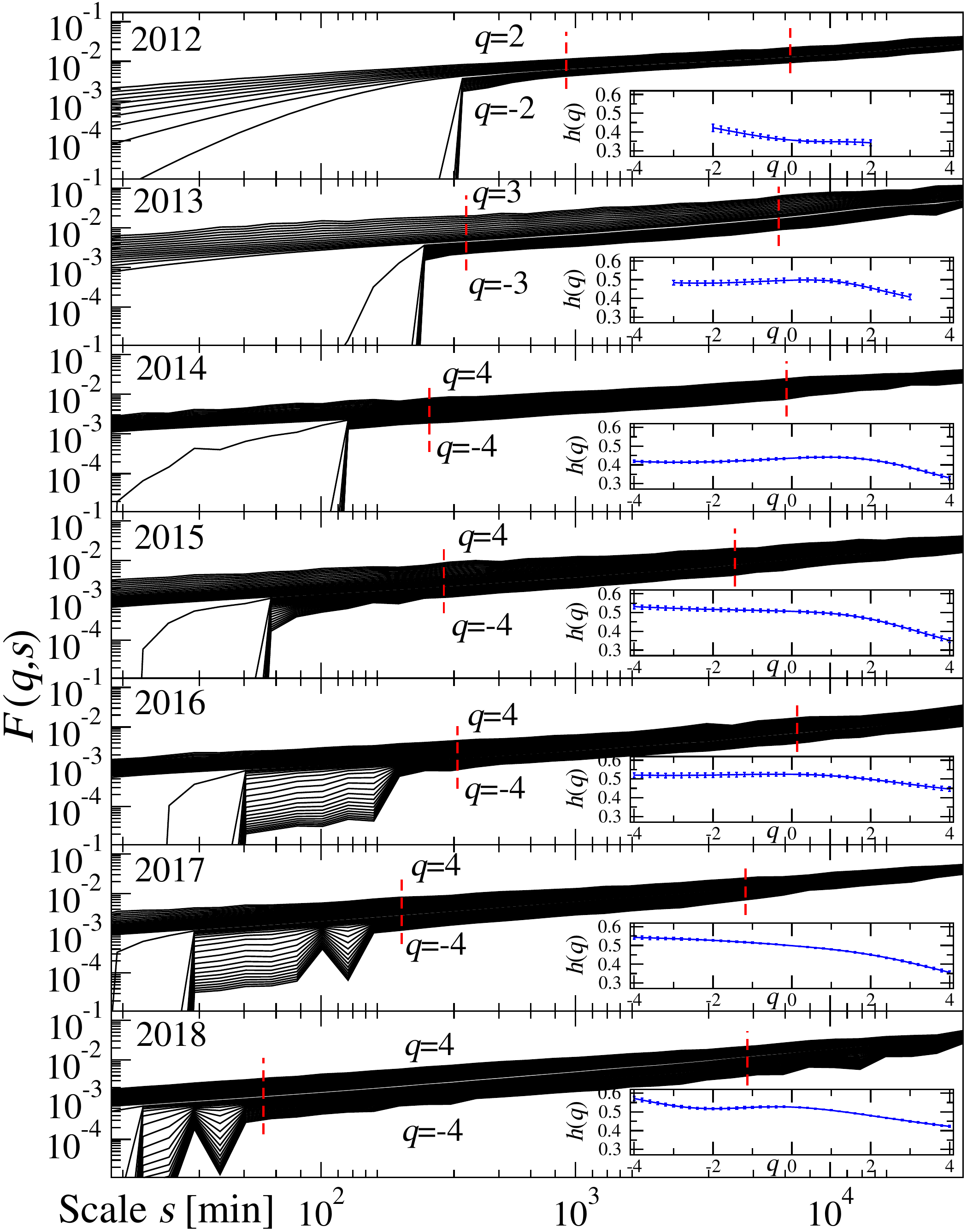}
\caption{Fluctuation functions $F(q,s)$ ($\Delta q=0.2$) calculated for the BTC/USD exchange rate on Bitstamp over subsequent annual intervals from 2012 to 2018. The range of scales selected for the calculation of $h(q)$ is indicated by the dashed lines. Inset presents the generalized Hurst exponent $h(q)$ calculated from the scale range marked with dashed lines. Error bars represent the standard error of linear regression.}
\label{fig:Bitstamp_Fq}
\end{figure}

Figure~\ref{fig:Bitstamp_Fq} presents the fluctuation functions $F(q,s)$ defined by Eq.~\ref{F} and the generalized Hurst exponents $h(q)$ calculated for the $\Delta t=10$s return time series of the BTC/USD exchange rate divided into annual intervals 2012-2018 (the Bitstamp data). A systematic improvement of the $F(q,s)$ scaling quality can be observed each subsequent year, especially for $q<0$ (small returns). This improvement is closely related to the gradual shortening of the inter-transaction times on this exchange rate, thus shortening also the zero-return sequences that influence the algorithm output. In 2012 such intervals without trading reached 600 minutes, thus the range of satisfactory scaling $F(q,s)$ is very short in this case. There was no dependence of $h(q)$ on negative $q$s until 2017 (see insets in Fig.~\ref{fig:Bitstamp_Fq}). Thus, a power-law dependence of $F(q,s)$ on $s$ during earlier years reflects solely the organization of the larger fluctuations. It can also be recalled that the CDF tail for BTC/USD increased from $\gamma\approx 2$ in 2012 to over 3 from 2014 on (Sect.~\ref{Dane stat}). This result led to limiting the $q$ range in $h(q)$ from [-2,2] for 2012, through [-3,3] for 2013, to [-4,4] for 2014 and the subsequent years (because of the non-existent higher-order moments of the corresponding PDFs).

\begin{figure}[!ht]
\centering
\includegraphics[width=1\textwidth]{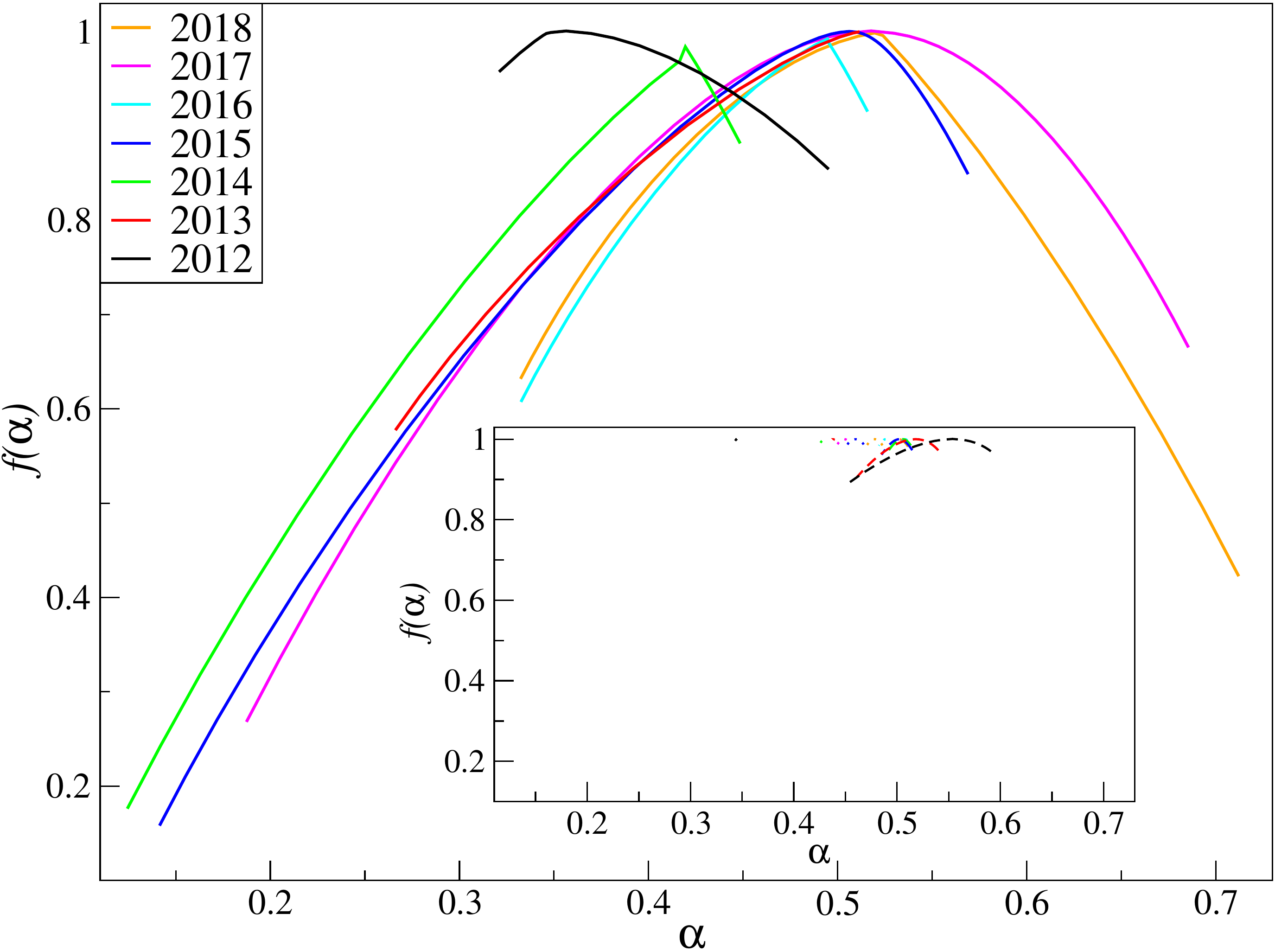}
\caption{Singularity spectra $f(\alpha)$ calculated for the BTC/USD exchange rate returns over subsequent annual periods with $q\in [-4,4]$, except for 2012 ($q\in[-2,2]$) and 2013 ($q\in[-3,3]$). Inset shows $f(\alpha)$ for the Fourier surrogates (dotted line) and the shuffled surrogates (dashed line).}
\label{fig:Bitstamp_spektra}
\end{figure}

Figure~\ref{fig:Bitstamp_spektra} presents the singularity spectra $f(\alpha)$ for the same individual years as Fig.~\ref{fig:Bitstamp_Fq}. From 2013 until 2017 the spectra are asymmetric with strong dominance of the left wing corresponding to large returns and only in 2018 the right wing corresponding to small returns becomes significant. In the years 2013-2016 the asymmetry coefficient $A_{\alpha}$ assumed values between 0.7 and 0.8. The spectrum was widest in 2017 attaining width $\Delta \alpha\approx 0.5$ with $A_{\alpha}\approx 0.34$, while in 2018 it became symmetric with $A_{\alpha}\approx -0.03$ and $\Delta \alpha\approx 0.41$. Based on these values, $f(\alpha)$ in 2018 can be categorized as multifractal with well-developed hierarchical organization for both the small and large returns. This means that, along with increasing transaction frequency on BTC/USD, the small returns ceased to behave like noise and the nonlinear correlations eventually built up also in this case. Such a symmetry between the fluctuations of different magnitude is a well-grounded observation on Forex~\cite{drozdzepps,Rak2015} and other mature markets~\cite{multirev,Watorek2019}. The results for 2012 and 2013 are characterized by the largest error, which is related to a short range of scaling in $s$ that may be used to assess the scaling exponents $h(q)$ and $f(\alpha)$. Moreover, taking into account the lack of long-range nonlinear correlations in 2012 (see Sect.~\ref{StatAutoCorr}), the time-series for this period can be considered as monofractal (if fractal at all).

The observed multiscale characteristics disappear when considering the Fourier surrogates of the data~\cite{schreiber2000} or the shuffled time series. In the former case, nonlinear correlations are destroyed through the Fourier-phase randomization of the original data and only linear correlations survive. Their singularity spectra are narrow and located at the values of $\alpha$ that match the Hurst exponent $H$ of the original time series (see Fig.~\ref{fig:Bitstamp_spektra}). In the latter case, shuffling destroys all correlations (i.e., linear and nonlinear) but preserves the return distributions. The spectra $f(\alpha)$ are also narrow and located roughly at $\alpha=0.5$. It should be pointed out that the slightly wider spectra featured for 2012 and 2013 are a numerical artifact caused by the heavy tails of the corresponding PDFs (see Refs.~\cite{drozdz2010,Zhou2012,Grech2013,Rak2018} for more details).
\begin{figure}[!ht]
\centering
\includegraphics[width=1\textwidth]{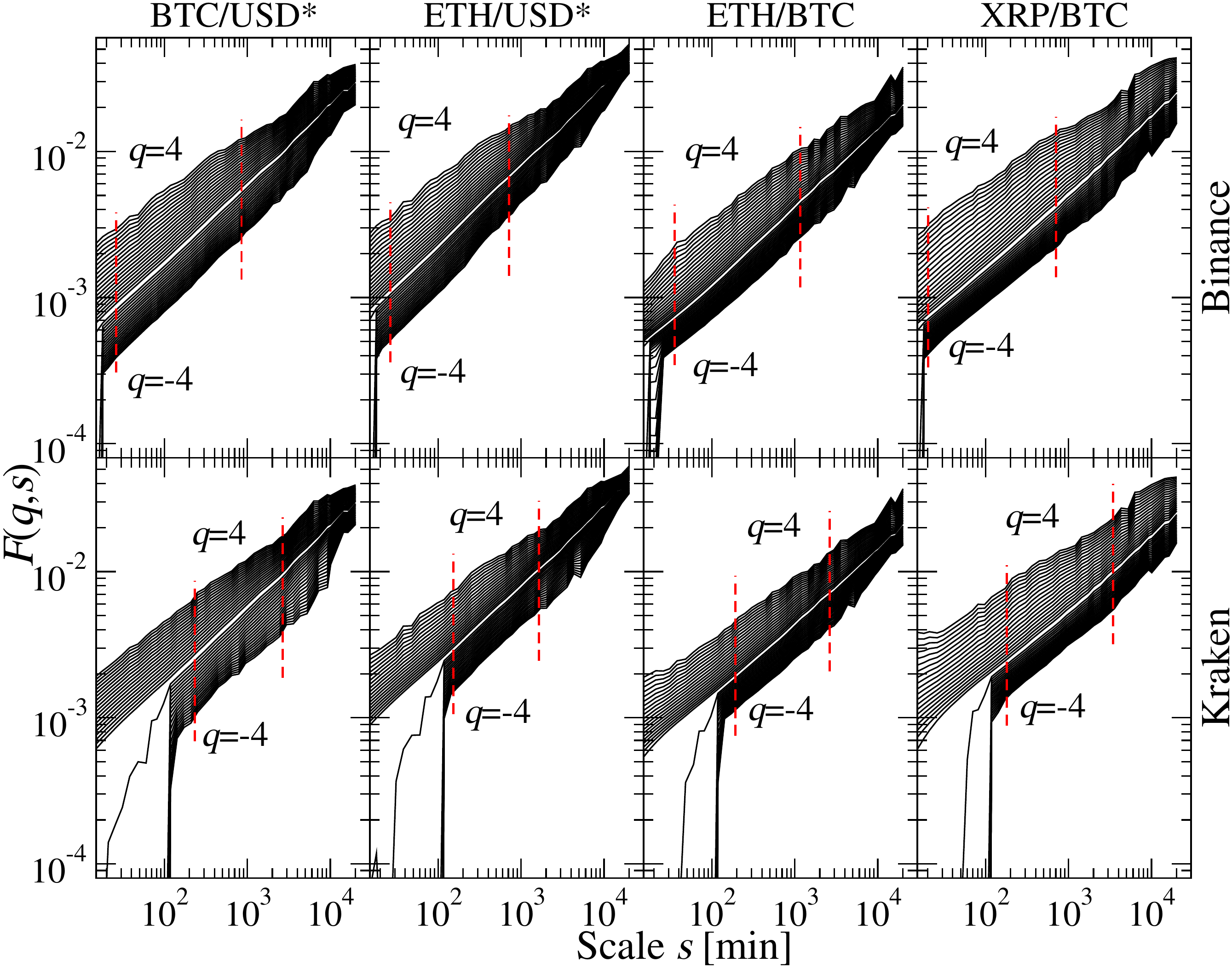}
\caption{Fluctuation functions $F(q,s)$ ($\Delta q=0.2$) calculated for the exchange rates BTC/USDT, ETH/USDT, BTC/ETH and XRP/BTC on Binance (upper panel) and Kraken (lower panel) in 2018 (USD$^{*}$ means USD in the case of Kraken, and USDT in the case of Binance).}
\label{fig:KrBi_Fq}
\end{figure}

\begin{figure}[!ht]
\centering
\includegraphics[width=1\textwidth]{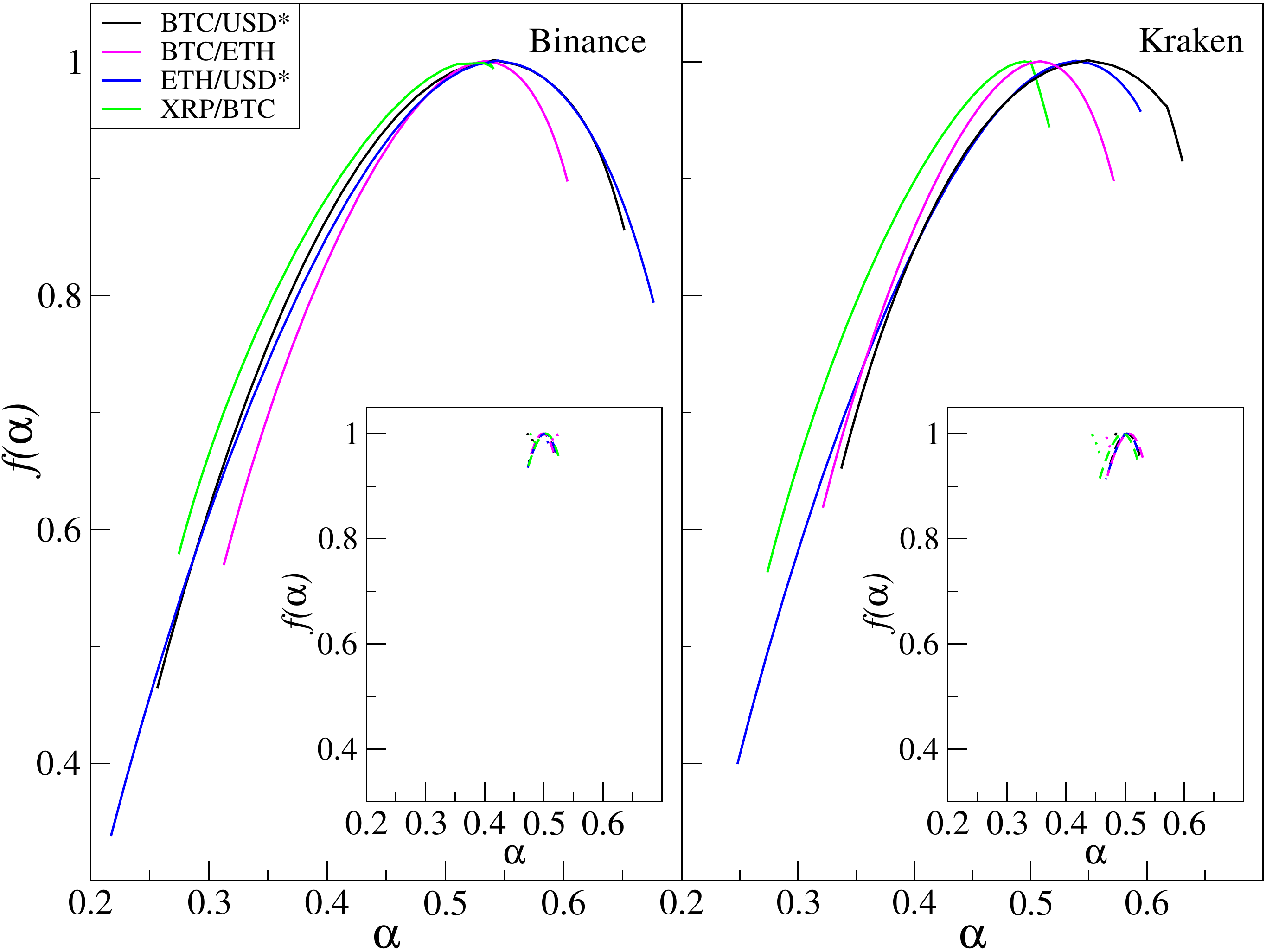}
\caption{Singularity spectra calculated for the exchange rates BTC/USD$^{*}$, ETH/USD$^{*}$, BTC/ETH and XRP/BTC on Binance (left panel) and Kraken (right panel) in 2018 for $q\in[-4,4]$(USD$^{*}$ means USD in the case of Kraken, and USDT in the case of Binance). In the inset, returns spectra for the Fourier surrogates (dotted line) and the random shuffling surrogates (dashed line) are shown.}
\label{fig:KrBi_spektra}
\end{figure}

In order to look at a broader picture, one may also compare the multiscale properties of BTC/USD with other exchange rates or even the same exchange rates from different trading platforms. Fig.~\ref{fig:KrBi_Fq} exhibits the fluctuation functions $F(q,s)$ calculated from the $\Delta t=1$ min returns of the most liquid exchange rates listed on Binance and Kraken:  BTC/USD$^{*}$, ETH/USD$^{*}$, BTC/ETH, and XRP/BTC, where USD$^{*}$ means either US dollar (Kraken) or USDT (Binance). The data covers 2018. Since the related PDFs are heavy-tailed with $\gamma\approx 3$, a range of $q$ was restricted to [-4,4]~\cite{oswiecimka2005}. The functions $F(q,s)$ are indefinite for small $s$ and negative $q$ because the considered exchange rates were traded with a moderate frequency, especially on Kraken, where non-trading intervals could be as long as 100 min. Fortunately, the scaling quality for $s>100$ min is comparable between both exchanges. The dotted lines in Figure~\ref{fig:KrBi_Fq} mark the range of scales $s$, for which the singularity spectra presented in Fig.~\ref{fig:KrBi_spektra} were estimated. These spectra are characterized by a left-sided asymmetry on both platforms. For Binance they are slightly wider and more symmetric than for Kraken, which suggests that the nonlinear organization of the small fluctuations is better developed on the former platform, owing to a higher trading frequency there. Like in the case of BTC/USD from Bitstamp, the observed multiscale effects disappear here when the Fourier and the shuffled surrogates are considered. Other exchange rates listed in Tab.~\ref{tab:rozkladBiKR} (Sect.~\ref{StatAutoCorr}), which are quoted simultaneously on Binance and Kraken, cannot be compared between the two because of a low trading frequency on Kraken and the PDF heavy tails of the corresponding time series. However, in general the returns for these exchange rates are monofractal and they have already been characterized by the Hurst exponent (Tab.~\ref{tab:BiKrHurst}).

\subsubsection{Multiscale characteristics for currencies and cryptocurrencies}
\label{MFDFAKraken}

\begin{figure}[!ht]
\centering
\includegraphics[width=0.8\textwidth]{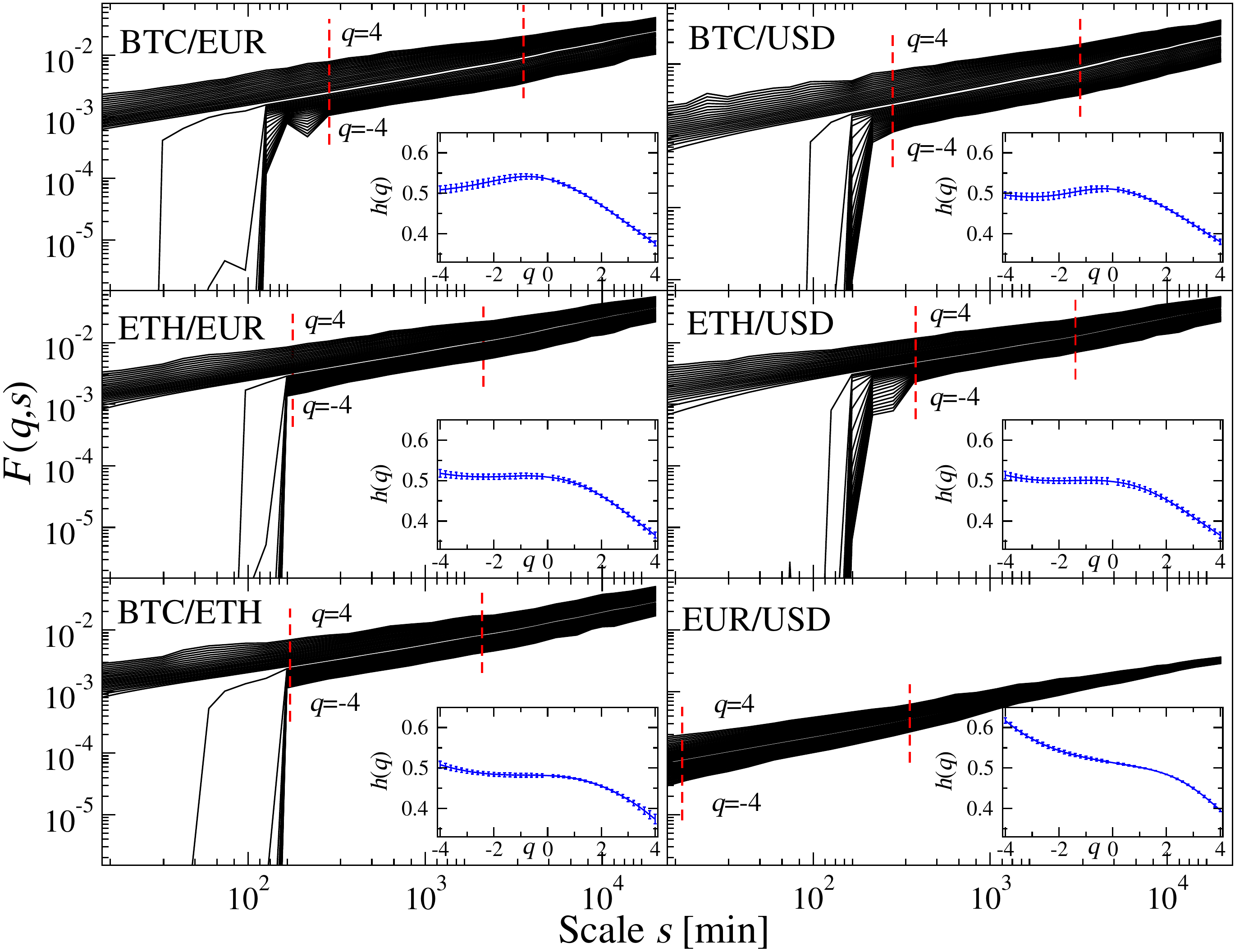}
\caption{Fluctuation functions $F(q,s)$ ($\Delta q=0.2$) calculated for returns of the exchange rates BTC/EUR, BTC/USD, ETH/EUR, ETH/USD and BTC/ETH from the Kraken exchange, and EUR/USD from the Forex market. The calculations include returns recorded from July 2016 to December 2018. Insets present the generalized Hurst exponent $h(q)$ estimated from the range of scales indicated by the dashed lines. Error bars represent the standard error for linear regression.}
\label{fig:Kraken10s_Fq}
\end{figure}

Figure~\ref{fig:Kraken10s_Fq} shows $F(q,s)$ and $h(q)$ calculated or five the most liquid cryptocurrency pairs: BTC/EUR, BTC/USD, ETH/EUR, ETH/USD, and BTC/ETH from Kraken and, for a benchmark, the fiat currency pair EUR/USD from Forex. The returns with $\Delta t=10$s cover an interval from July 2016 to December 2018. The scale range used to calculate $h(q)$ is denoted by vertical dashed lines in each case. A maximum $s$ used to estimate $h(q)$ should not be longer than the length of the average cluster of volatility assessed with the help of autocorrelation function (Sect.~\ref{StatAutoCorr}), otherwise the estimated $h(q)$ is not credible~\cite{drozdz2010}. Similar to BTC/US from Bitstamp discussed in Sect.~\ref{BTCMFDFA}, on Kraken the long periods without transactions lead to $F(q,s)$ with undefined values for the negative $q$s and for the small scales $s$. This effect does not occur in the case of EUR/USD, because Forex is a much more liquid market.

A lack of dependence of $h(q)$ on $q$ for the negative $q$s and a clear dependence for $q>0$ (insets of Fig.~\ref{fig:Kraken10s_Fq}) indicates that a heterogeneous hierarchical organization can only be seen for the medium and large returns, while the small returns are rather homogeneous. In contrast, $h(q)$ for EUR/USD is symmetric, proving that the fluctuations of all magnitudes are responsible for the multiscale organization of the data. It appears noteworthy that a symmetric $f(\alpha)$ characterized also BTC/USD in 2018.

\begin{figure}[!ht]
\centering
\includegraphics[width=0.8\textwidth]{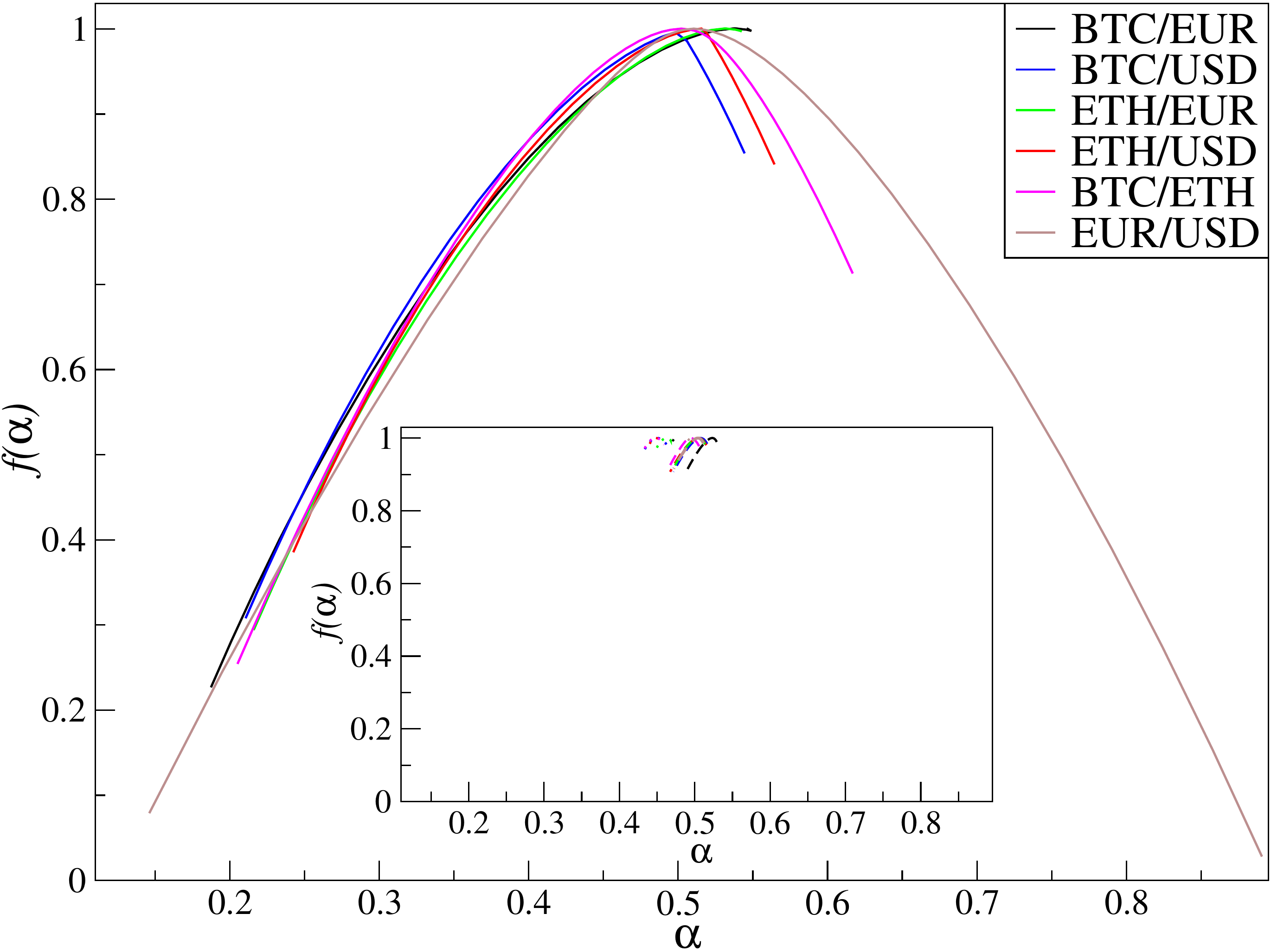}
\caption{Singularity spectrum calculated for the exchange rates BTC/EUR, BTC/USD, ETH/EUR, ETH/USD and BTC/ETH on the Kraken exchange and EUR/USD, from July 2016 to December 2018, for $q\in [-4,4]$. In the inset, spectra for Fourier surrogates (dotted line) and random shuffling surrogates (dashed line) are given for the returns.}
\label{fig:Kraken10s_spektra}
\end{figure}

The multiscale properties characterizing the cryptocurrency exchange rates from Kraken and EUR/USD from Forex are illustrated in Figure~\ref{fig:Kraken10s_spektra}, where their singularity spectra are compared. The spectrum for EUR/USD is the widest with $\Delta \alpha\approx 0.75$ and the most symmetric with $A_{\alpha}\approx -0.02$. The widest spectrum for Kraken, representing the crypto-crypto pair BTC/ETH, is a result of the longest right-side wing of $f(\alpha)$ with $\Delta \alpha\approx 0.42$. This spectrum is still considerably narrower if compared to that for EUR/USD, however. The other cryptocurrency/fiat exchange rates are also characterized by the left-sided asymmetry: $A_{\alpha}\in [0.7,0.9]$ and $\Delta \alpha\approx 0.35$ in each case. As a consequence, the small returns quantified by the right wing of $f(\alpha)$ behave as if they were monofractal noise. This results confirms that the Forex exchange rates show more complex dynamics than the cryptocurrency exchange rates.

The observed multiscale effects disappear entirely if the non-linear correlations are selectively destroyed via the Fourier surrogates or if the returns are shuffled. In the latter case, the impact of the heavy tails on the singularity spectra is evident (inset of Fig.~\ref{fig:Kraken10s_spektra}). The spectra for the shuffled data are a little broader if compared to the case of the Fourier surrogates. However, they are narrower than their counterparts derived for BTC/USD from Bitstamp in 2012 and 2013, which stems from the thinner return PDF tails on Kraken.

Summing up, complexity of the BTC/USD exchange rate returns has evolved in time and, by 2017, it approached the complexity observed typically on Forex and even attained that level in 2018. Similar conclusion could be drawn for the exchange rates listed on Kraken, where ETH/USD shows the multiscale properties similar to BTC/USD. An exchange rate without any fiat currency involved -- BTC/ETH -- also reveals complexity typical for Forex. The exception that allows for distinguishing between the cryptocurrency exchange rates and the Forex ones is a less evident hierarchical organization of the small fluctuations in the former case. This is because a high trading frequency is needed to develop multiscaling at the level of small fluctuations. The impact of trading frequency is also seen when comparing the same exchange rates listed on different trading platforms. On Binance, where the frequency is larger, the return time-series show the multiscale properties that are closer to those typically observed on Forex than their counterparts listed on Kraken.

\section{Multiscale cross-correlations}
\label{sect::NonlinearCrossCorrel}

Throughout the two preceding sections, it was shown that the cryptocurrency exchange rate fluctuations are characterized by heavy tails of the return probability distributions and by volatility clustering related to nonlinear temporal correlations in the returns and responsible for multifractality. Additional properties are revealed when one considers multivariate data consisting of two or more exchange rates. In Sec.~\ref{sect::MultiscaleCrossBTCETH}, cross-correlations among the cryptocurrencies and fiat currencies will be analyzed using the previously-presented methods based on detrending, namely the MFCCA (Sect.~\ref{MFCCA}) and $\rho(q,s)$~(\ref{rho}). These allow for a quantitative description of nonlinear cross-dependencies over different time-scales and fluctuation magnitudes, conferring a significant advantage over classical linear correlation~\cite{Pearson1895}. As for linear cross-correlations, the multifractal characteristics of individual time-series does not imply that such the same are also multifractally cross-correlated. Multifractal cross-correlations of the triangular arbitrage opportunity data will be discussed in Sect.~\ref{sect::TriangularArbitrage}. The results reported in this section are based on three data sets:
\begin{itemize}
\item \textit{Kraken-2}: BTC and ETH quotations expressed in USD and EUR taken from Kraken and covering from Jul. 2016 to Dec. 2018, transformed to return time-series with $\Delta t=10$s,
\item \textit{Kraken-11}: Quotations of BCH, BTC, DASH, ETC, ETH, LTC, REP, USDT, XMR, XRP, ZEC expressed in USD and EUR, taken from Kraken and covering the interval Jan. 2018 to Dec. 2018, transformed to return time-series with $\Delta t=1$ min (here, only the most liquid pairs were considered), and
\item \textit{Binance-12}: Crypto-crypto exchange rates obtained from BAT, BCH, BNB, BTC, ETH, ICX, IOTA, LTC, LSK, NEO, USDT, XLM taken from Binance and covering the same interval, also transformed to return time-series with $\Delta t=1$ min.
\end{itemize}

See Appendix~\ref{BiKrdodatek} for the full names and statistical properties of all cryptocurrencies considered in this section.\par

\subsection{Multiscale cross-correlations for the BTC and ETH}
\label{sect::MultiscaleCrossBTCETH}

\begin{figure}[!ht]
\centering
\includegraphics[width=1\textwidth]{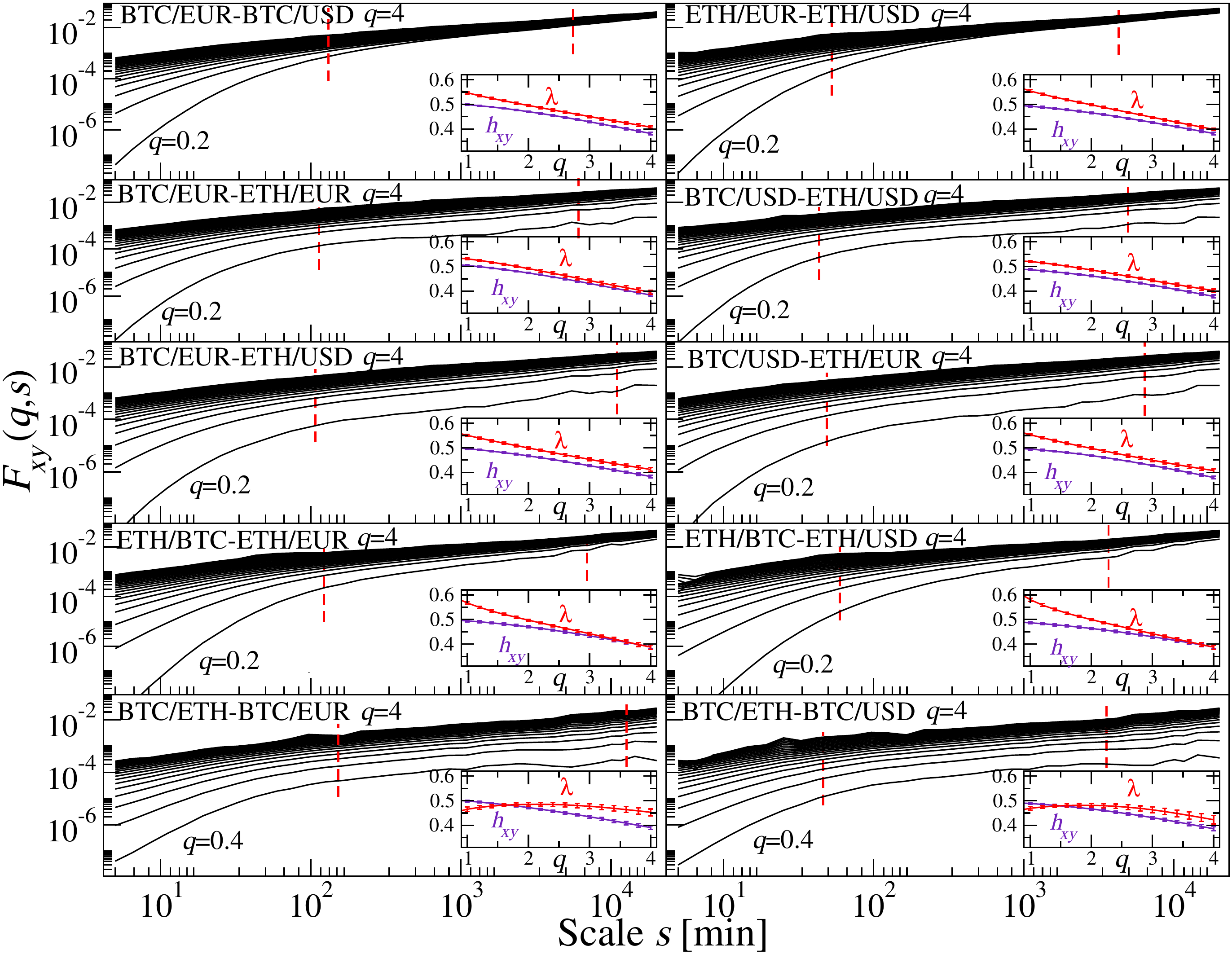}
\caption{(Main) Bivariate fluctuation function $F_\textrm{xy}(q,s)$ for $q>0$ ($\Delta q=0.2$) calculated for the exchange rates BTC/EUR, BTC/USD, ETH/EUR, ETH/USD, and BTC/ETH covering the interval from Jul. 2016 to Dec. 2018 on Kraken. Inset shows the scaling exponent $\lambda(q)$ and the average generalized Hurst exponent $h_\textrm{xy}(q)$ calculated over the scale range marked with the dashed lines. Error bars represent the standard error of linear regression.}
\label{fig:Kraken10s_Fqxy}
\end{figure}

The bivariate fluctuation functions $F_\textrm{xy}(q,s)$ calculated~(Eq.(\ref{Fq})) for the exchange rates involving cryptocurrencies show scaling behaviour over a range of scales. Figure~\ref{fig:Kraken10s_Fqxy} (main panels) illustrates this observation by showing $F_\textrm{xy}(q,s)$ for ten pairs of the exchange rate return time-series, representing the most liquid cryptocurrencies BTC and ETH (the \textit{Kraken-2} dataset), ordered from top to bottom with decreasing scaling quality. 
In this case, scaling signifies that the considered exchange rates are cross-correlated at various levels of their hierarchical organization. The most convincing scaling is seen for the bitcoin expressed in EUR and USD, as well as for the ethereum in the same fiat currencies (top panels). Contrariwise, the weakest scaling is associated with correlations for the exchange rates involving BTC and a fiat currency, or only cryptocurrencies, namely BTC/ETH (two bottom panels).\par

However, the scaling of $F_\textrm{xy}(q,s)$ can only be observed for medium and large returns ($q>0$) whereas, for small returns ($q<0$), the function becomes negative and there is no multiscale cross-correlation. This finding agrees with the multifractal properties of the individual exchange rates reported in Sec.\ref{nonlinear}, for which similar properties were also observed for medium and large returns, with their spectra accordingly featuring a strong left-side asymmetry. Indeed, this effect is typical for mature financial markets~\cite{Rak2015,Zhao2017,gebarowski2019,Watorek2019}.\par

\begin{figure}[!ht]
\centering
\includegraphics[width=1\textwidth]{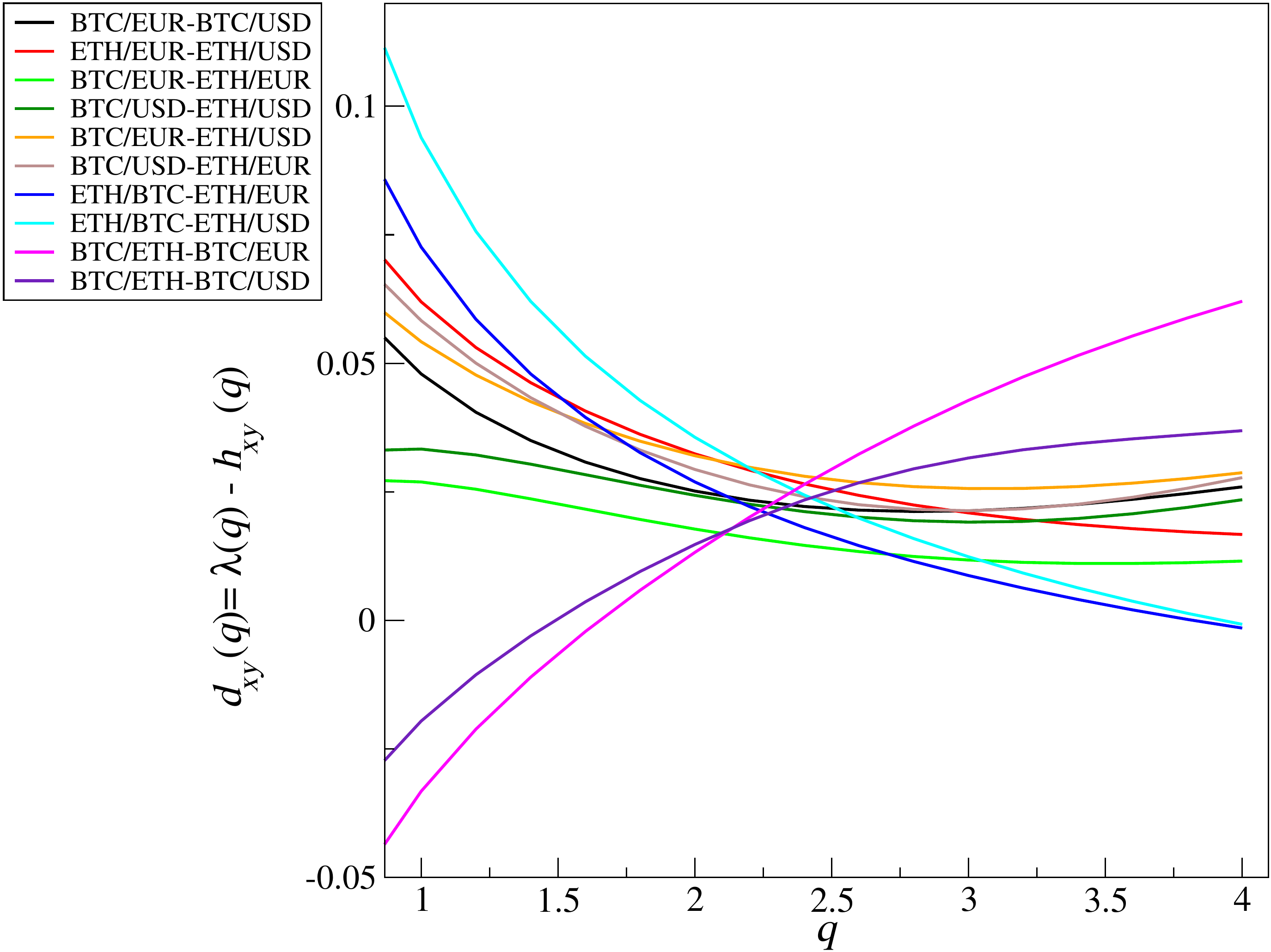}
\caption{Difference between the scaling exponent $\lambda(q)$ and the average generalized Hurst exponent $h_\textrm{xy}(q)$ for the exchange rates shown  in Fig.~\ref{fig:Kraken10s_Fqxy}.}
\label{fig:Kraken10s_lambda_hsrednie}
\end{figure}

\begin{figure}[!ht]
\centering
\includegraphics[width=1\textwidth]{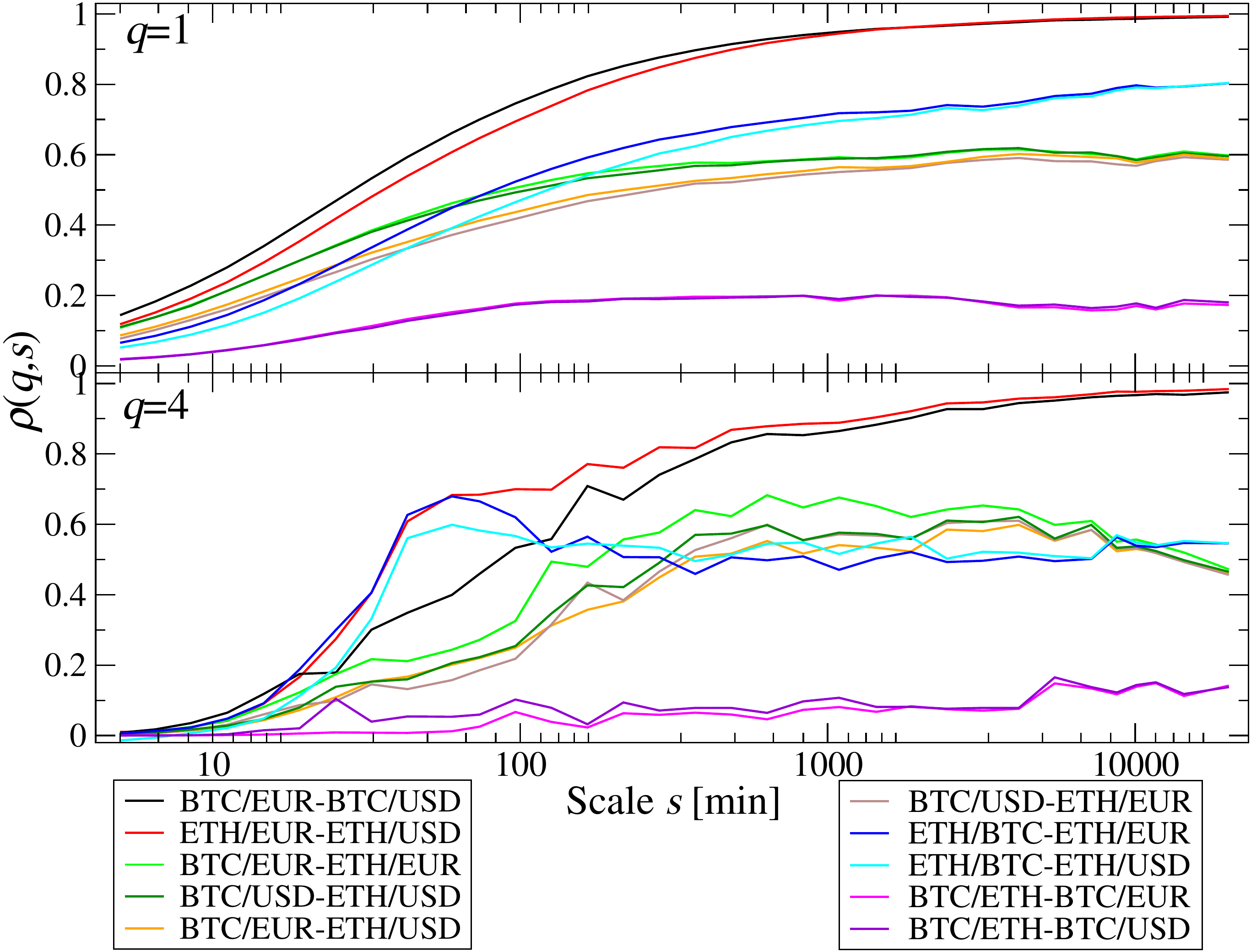}
\caption{Detrended cross-correlation coefficient $\rho(q,s)$ calculated for the exchange rate pairs from Fig.~\ref{fig:Kraken10s_Fqxy}, for $q=1$ (top) and $q=4$ (bottom).}
\label{fig:Kraken10s_pq1all}
\end{figure}

The dependencies of the scaling exponent $\lambda(q)$ and average generalized Hurst exponent $h_\textrm{xy}(q)$ defined, respectively, by Eq.(\ref{Fxy}) and Eq.(\ref{hqxy}) in Sect.\ref{detrendned}, are shown in Fig.~\ref{fig:Kraken10s_Fqxy}. Through varying $q$, $\lambda(q)$ reflects the multifractal character of cross-correlations between all 10 exchange rates, at least for $q \geqslant 1$. The weakest dependence on $q$ and, thus, the least-developed multifractality could be seen for BTC/ETH--BTC/EUR and BTC/ETH--BTC/USD, which are also characterized by the weakest scaling. The cross-correlations between these exchange rate pairs are also characterized by a negative difference $d_\textrm{xy}(q) = \lambda(q) - h_\textrm{xy}(q)$;  this is in contrast with the other pairs, for which this is positive and depends appreciably on $q$ (Figure~\ref{fig:Kraken10s_lambda_hsrednie}). The significant magnitude of $d_\textrm{xy}(q)$ observed for some exchange rate pairs implies a strong dependence of the cross-correlation on the scale $s$ (see Sect.~\ref{efektW}). Indeed, since $d_\textrm{xy}(q)> 0$ for the majority of pairs, the detrended cross-correlation coefficient $\rho(q,s)$ increases substantially with $s$ for both medium and large returns, with this effect being stronger for $q=1$ than for $q=4$ (Figure~\ref{fig:Kraken10s_pq1all}). The same can be explained by the occurrence of large returns in the individual exchange rates, significantly contributing variance on small scales and counteracting the build-up of cross-correlations (i.e., on longer scales these are averaged out). Since for $q=4$ such returns are amplified more intensely than for $q=1$, it becomes obvious that $\rho(q,s)$ for small $s$ is closer to zero in the former case. However, this $q$-dependence is less evident here than on mature financial markets~\cite{kwapien2015,Zhao2017,Watorek2019}, in particular the Forex, for which $\rho(q=1,s)/\rho(q=4,s) \approx 2$~\cite{gebarowski2019} (see Sect.~\ref{sect::Forex} for further details).\par

The maximum level of $\rho(q,s)$ depends on the exchange rate pair; for the dataset considered here, high values are attained by the pairs such that the same cryptocurrency is expressed in terms of the fiat currencies, USD and EUR, for which $\rho(q,s) \to 1$. This situation can be explained by the considerably lower volatility of EUR/USD as compared to the cryptocurrencies (Fig.~\ref{fig:Kraken_serie}). By contrast, the lowest level of $\rho(q,s)$ is found for the pairs involving the same cryptocurrency expressed in a fiat currency and another cryptocurrency, namely BTC/ETH--BTC/EUR and BTC/ETH--BTC/USD. The remaining pairs are associated with intermediate levels of $\rho(q,s)$. Moreover, BTC/ETH--BTC/EUR and BTC/ETH--BTC/USD together with ETH/BTC--ETH/EUR and ETH/BTC--ETH/USD show a sign change of $d_\textrm{xy}(q)$ (Fig.~\ref{fig:Kraken10s_lambda_hsrednie}) when transitioning from $q=1$ to $q=4$, which hallmarks a change in the monotonic behaviour of $\rho(q,s)$ for $s \geqslant 300$.\par 

\begin{figure}[!ht]
\centering
\includegraphics[width=1\textwidth]{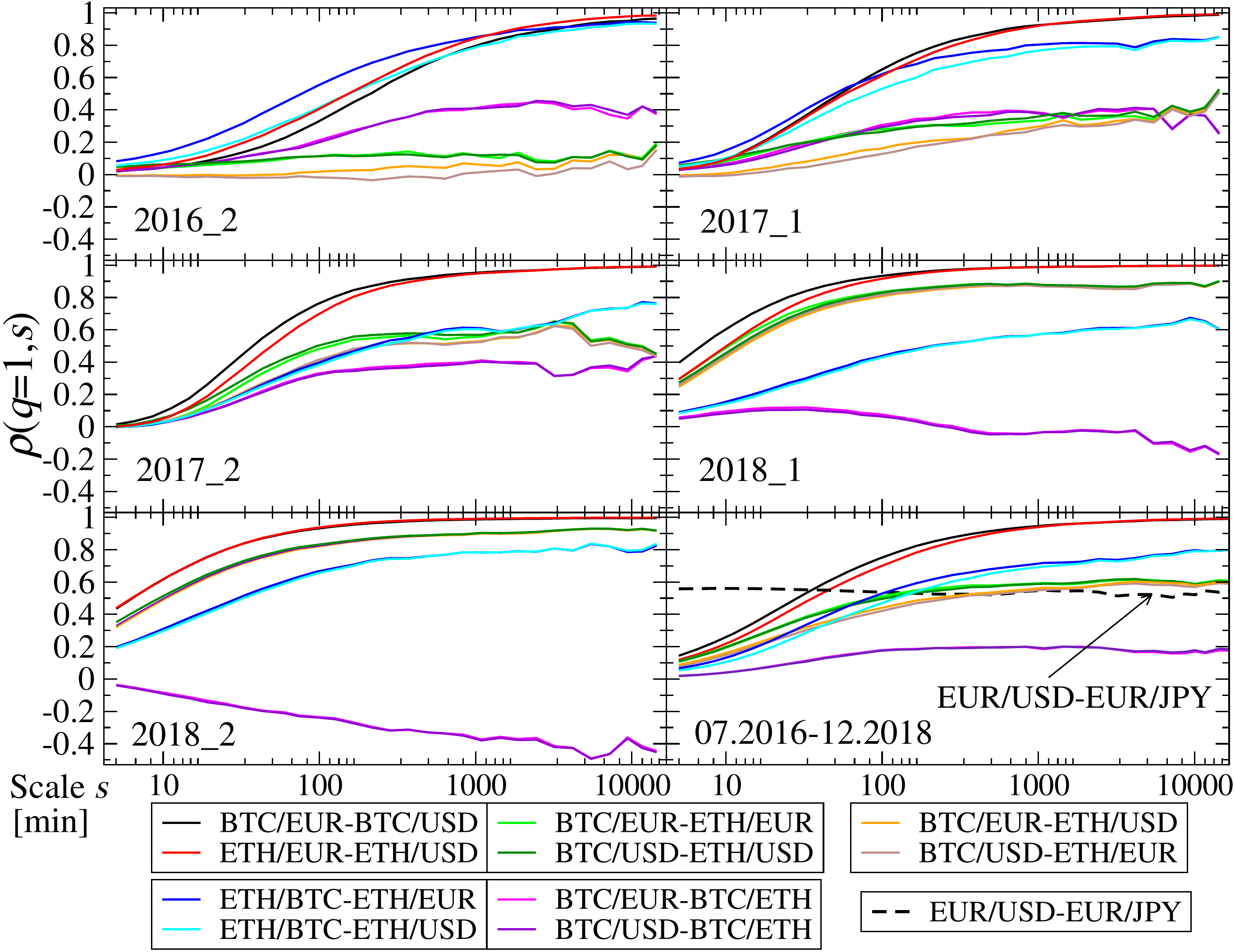}
\caption{Detrended cross-correlation coefficient $\rho(q=1,s)$ calculated for the exchange rate pairs shown in Fig.~\ref{fig:Kraken10s_Fqxy} in semi-annual windows from Jul. 2016 to Dec. 2018, shown alongside together the pair EUR/USD--EUR/JPY from the Forex.}
\label{fig:Kraken10s_pq1sub}
\end{figure}

The \textit{Kraken-2} dataset covers the period Jul. 2016 through Dec. 2018, over which the cryptocurrency trading steadily accelerated, leading to a consistent reduction in the inter-transaction times. The coefficients $\rho(q=1,s)$ calculated over a half-year periods are presented in Figure~\ref{fig:Kraken10s_pq1sub}. During the interval from Jul. 2016 to Dec. 2017, the correlations on the smallest time-scales remained close to zero, and $\rho(q=1,s)$ increased markedly with $s$. This markedly changed during 2018, when $\rho(q,s)$ became higher than before for small $s$, and its maximum value was attained faster. The weakest correlations between BTC/ETH--BTC/EUR and BTC/ETH--BTC/USD observed in Fig.~\ref{fig:Kraken10s_pq1all} can be accounted for the fact that, following the crash of the cryptocurrency market in Jan 2018, $\rho(q=1,s)$ flipped sign because both the BTC and ETH decreased with respect to the EUR and USD (Fig.~\ref{fig:Kraken_Binance_WartV}). However, the decrease observed for the ETH is larger and faster than for the BTC, so BTC/ETH rate can be seen to rise, weakening the global (i.e., whole-period) correlations between BTC/ETH and BTC/EUR, and between BTC/ETH and BTC/USD. For comparison, it should be noted that $\rho(q=1,s)$ does not depend on $s$ in the case of a Forex exchange rate cross-correlation, such as EUR/USD--EUR/JPY (bottom right panel of Fig.~\ref{fig:Kraken10s_pq1sub}), which is typical in low-volatility periods wherein no extraordinary event affects the Forex systematically~\cite{gebarowski2019}.\par

\begin{figure}[ht!]
\includegraphics[width=1\textwidth]{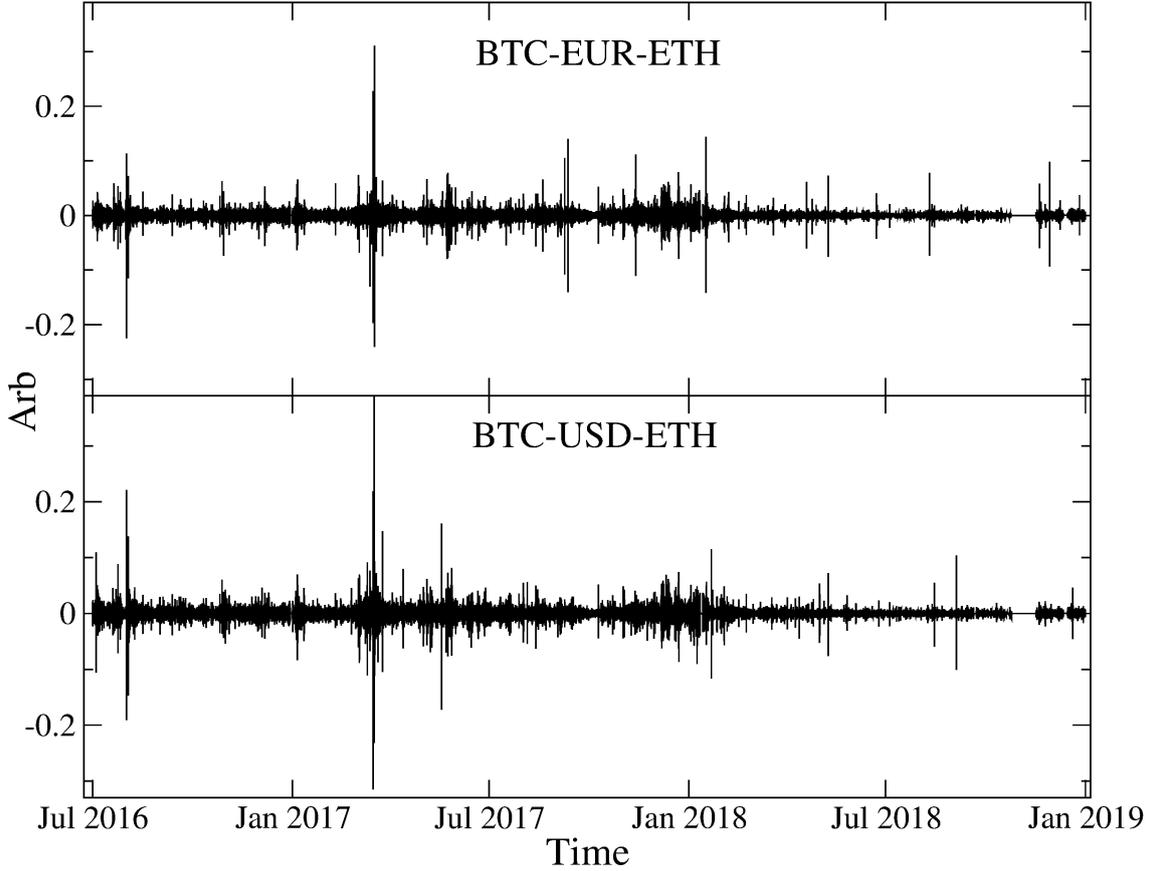}
\caption{Arbitrage opportunity time-series for the exchange rates that form a triangle relation, including the BTC/EUR, BTC/ETH, ETH/EUR and BTC/USD, BTC/ETH, ETH/USD.}
\label{fig:KrArbit}
\end{figure}

Until the end of 2017, insofar as two exchange rates hinged around a common base currency (e.g., BTC/EUR--ETH/EUR and BTC/USD--ETH/USD), one would observe a stronger cross-correlation compared to the situation of different base currencies (e.g., BTC/EUR--ETH/USD and BTC/USD--ETH/EUR); an example is visible in Fig.~\ref{fig:Kraken10s_pq1sub}. In 2018, however, similar cross-correlation levels could already be seen between the two cases. This observation can be explained in terms of the market synchronization that takes place over longer time-scales; it is also related to a significant slowing in the frequency and size of the triangular arbitrage opportunities over the period from Mar. 2018 to Dec. 2018, as visible in Figure~\ref{fig:KrArbit}.\par

Three (crypto)currencies are deemed to be in a triangle relationship if it is possible to make a transaction cycle through three exchange rates and return to the initial (crypto)currency~\cite{Fenn2009,gebarowski2019}. Given the five exchange rates from \textit{Kraken-2}, two triangles are available: EUR $\rightleftarrows$ BTC $\rightleftarrows$ ETH $\rightleftarrows$ EUR and USD $\rightleftarrows$ BTC $\rightleftarrows$ ETH $\rightleftarrows$ USD, where the double arrows indicate the two possible exchange directions~\cite{Fenn2009}. The arbitrage opportunity occurs at a time $t$ if
\begin{equation}
\frac{1}{\textrm{BTC/EUR}_\textrm{{ask}}(t)}\cdot\textrm{BTC/ETH}_\textrm{{bid}}(t)\cdot\textrm{ETH/EUR}_\textrm{{bid}}(t) > 1
\label{arbitEUR1} 
\end{equation}
or
 \begin{equation}
\frac{1}{\textrm{ETH/EUR}_\textrm{{ask}}(t)}\cdot\frac{1}{\textrm{BTC/ETH}_\textrm{{ask}}(t)}\cdot\textrm{BTC/EUR}_\textrm{bid}(t) > 1.
\label{arbitEUR2} 
\end{equation}
The formulas are, evidently, unchanged when considering the USD in place of the EUR. Because the bid price for the same exchange rate needs to be lower than the ask price, such arbitrage opportunities are rather rare. For example, on Forex they occur only during events that elevate the volatility~\cite{gebarowski2019}.\par

In the case of a dataset without the bid and ask prices, it is impossible to calculate the exact arbitrage opportunity as one has to replace the exchange rates in Eq.(\ref{arbitEUR1}) and Eq.(\ref{arbitEUR2}) with logarithmic prices at the end of each interval of length $\Delta t$:
 \begin{equation}
R_{\Delta t}\textrm{(A/B},t)=\log(\textrm{A}(t+\Delta t)/\textrm{B}(t+\Delta t))-\log(\textrm{A}(t)/\textrm{B}(t)),
\label{arbit_zwroty} 
\end{equation}
where A/B is the exchange rate between the (crypto)currencies A and B, and $\Delta t=10\textrm{s}$. For the above-mentioned triangles, the following arbitrage cases are possible:
\begin{equation}
\begin{split}
\textrm{Arb}_{\Delta t}(\textrm{EUR-BTC-ETH},t) = & -R_{\Delta t}(\textrm{BTC/EUR},t) + R_{\Delta t}(\textrm{BTC/ETH},t)\\
& + R_{\Delta t}(\textrm{ETH/EUR},t),
\end{split}
\label{arbit_zwrotyEUR} 
\end{equation}
\begin{equation}
\begin{split}
\textrm{Arb}_{\Delta t}(\textrm{USD-BTC-ETH},t) = & -R_{\Delta t}(\textrm{BTC/USD},t) + R_{\Delta t}(\textrm{BTC/ETH},t)\\
& + R_{\Delta t}(\textrm{ETH/USD},t).
\end{split}
\label{arbit_zwrotyUSD} 
\end{equation}
If Arb(EUR-BTC-ETH) $\neq 0$ or Arb(USD-BTC-ETH) $\neq 0$, there exists a possibility of carrying out the arbitrage. Time-series of the arbitrage opportunity are shown in Fig.~\ref{fig:KrArbit}. Towards the end of 2018, their frequency and size decreased, which could be associated with a more intense synchronization of the exchange rates engendered by the inter-transaction time shortening on Kraken. This situation also translates into an increased cross-correlation level over the shortest time-scales during the same period.\par

\subsection{Multiscale cross-correlations on the different trading platforms}
\label{sect::CrossCorrelations}

This section reports the primary results concerning the exchange-rate cross-correlations of liquid cryptocurrencies on the two trading platforms, Kraken (the \textit{Kraken-11} dataset) and Binance (the \textit{Binance-12} dataset). More specifically, both the cross-correlations within one platform and the inter-platform ones were considered. Only BTC and ETH (Kraken) and BTC, ETH, USDT, and BNB (Binance) were liquid enough to serve as bases. For each platform, a symmetric $q$-correlation matrix $C_{ij}^{(q,s)}$, where $1 \leqslant i,j \leqslant N$, can be created for each scale $s$, whose entries correspond to $\rho(q,s)$ calculated for two exchange rates, with $N=31$.\par

\subsubsection{Intra-platform multiscale cross-correlations}
\label{sect::IntraPlatform}

\begin{figure}[ht!]
\centering
\includegraphics[width=1\textwidth]{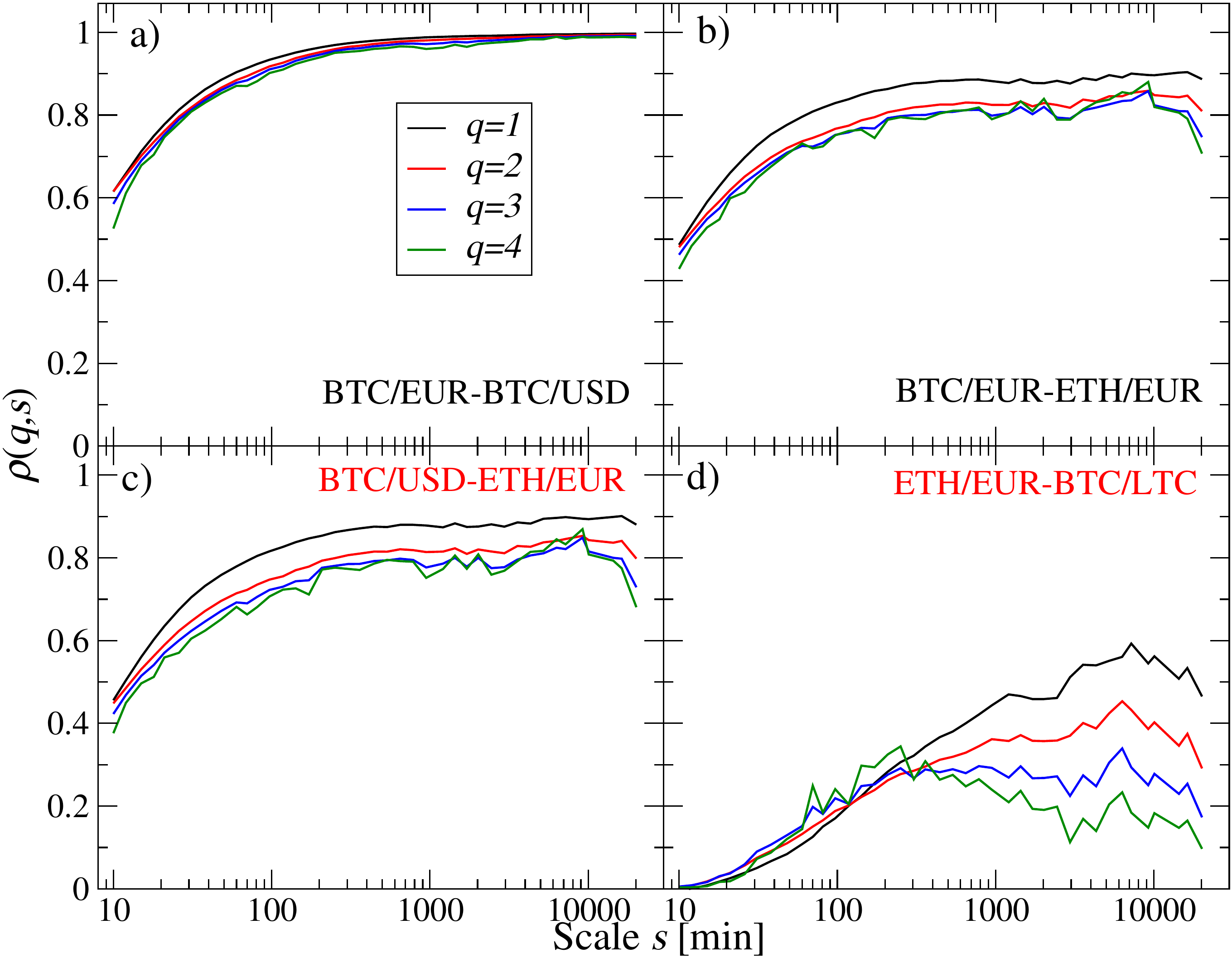}
\caption{Detrended cross-correlation coefficient $\rho(q,s)$ for $q>0$ calculated for the exchange rate pairs listed on Binance. Situations when the exchange rates do not have a common base currency are marked in red. (a) The same cryptocurrency in EUR and USD, (b) two cryptocurrencies expressed in the same base one, i.e., EUR or USD, (c) two cryptocurrencies expressed in EUR and USD, (d) no common base currency. Panels (a)-(d) are ordered according to decreasing correlation strength.}
\label{fig:Kraken_pq_ex}
\end{figure}

\begin{figure}[ht!]
\centering
\includegraphics[width=1\textwidth]{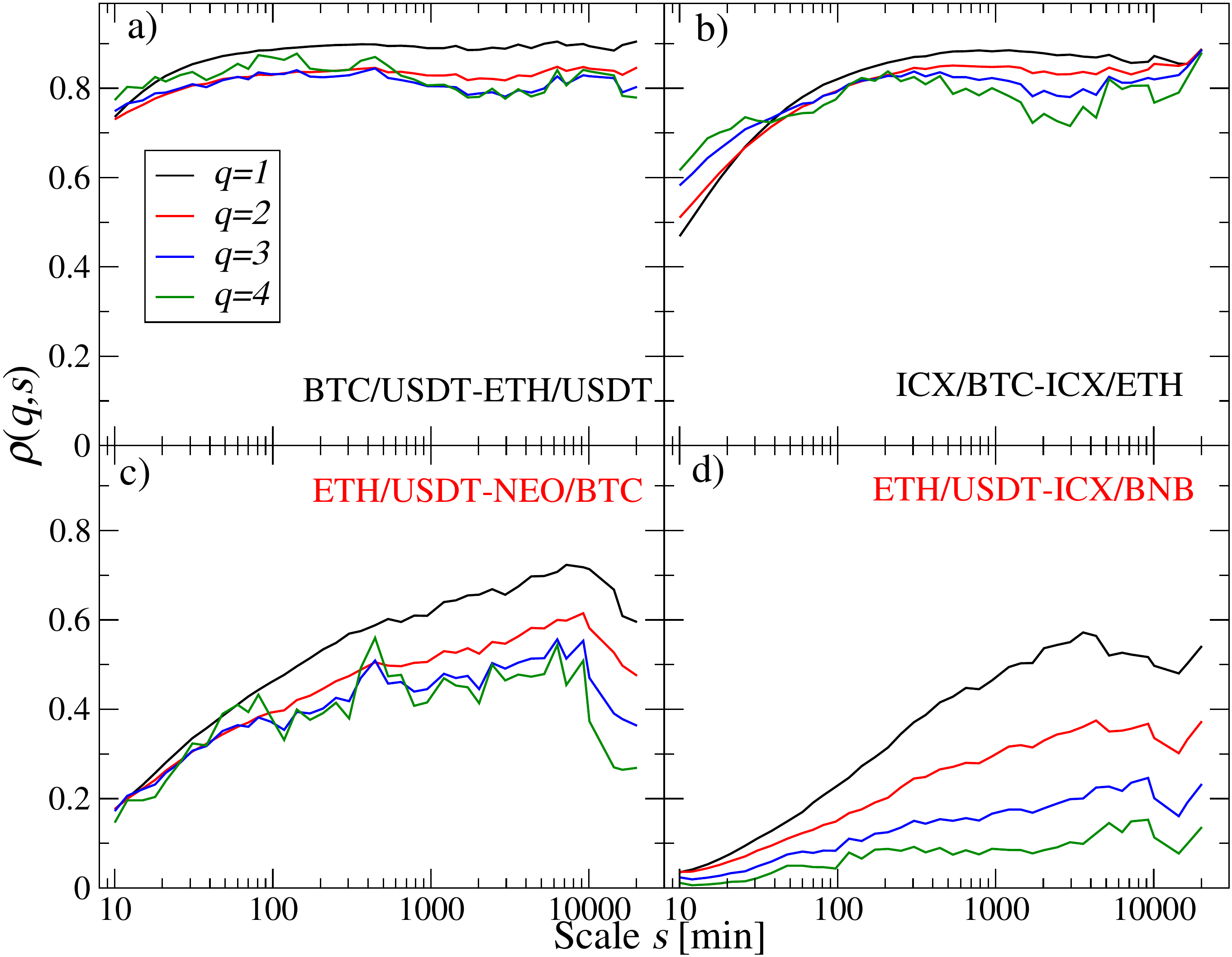}
\caption{Detrended cross-correlation coefficient $\rho(q,s)$ for $q>0$ on Binance calculated for exchange rate pairs delineating a triangle relation (top panels) and not being in such a relation (bottom panels).}
\label{fig:Binance_pq_ex}
\end{figure}

The representative behavior of $\rho(q,s)$ in four typical cases for each platform is shown in Fig.~\ref{fig:Kraken_pq_ex} (Kraken) and Fig.~\ref{fig:Binance_pq_ex} (Binance).

Among the exchange rates considered for Kraken, the strongest correlation, approaching unity for large $s$, is seen for BTC/EUR--BTC/USD (Fig.~\ref{fig:Kraken_pq_ex}(a)); this indicates that, on such scales, it is largely irrelevant whether the BTC is expressed in EUR or USD, since the traditional currencies are markedly less volatile than the cryptocurrencies. Over shorter scales, however, the cross-correlations are weaker for any setting of $q$, because the market is incompletely synchronized and there occur arbitrage opportunities. Meanwhile, for a given $s$, $\rho(q,s)$ is larger for shorter average inter-transaction intervals. The cross-correlation strength for the cryptocurrencies expressed in the same fiat currency, EUR or USD, ranks second, with the largest value of $\rho(q,s) \approx 0.8$ ($s>200$ min) observed for the most liquid pair, namely BTC and ETH (Fig.~\ref{fig:Kraken_pq_ex}(b)).\par

The third case is a situation that characterizes Kraken in particular: a pair of exchange rates does not form a triangle relation (i.e., four (crypto)currencies are involved), but one cryptocurrency is expressed in USD whereas the other in EUR. If the cryptocurrencies are BTC and ETH (Fig.~\ref{fig:Kraken_pq_ex}(c)), the cross-correlation strength is only slightly lower than that in Fig.~~\ref{fig:Kraken_pq_ex}(b), and it is only for the smallest $s$. Again, this is connected to the small volatility of the fiat currencies. Finally, there is the case of exchange rates that also do not form a triangle, but involve only one fiat currency. Here, a maximum $\rho(q,s)$ is considerably smaller than in the previous cases; the most strongly correlated pair is ETH/EUR--BTC/LTC, with $\rho(q=1,s)<0.6$ and $\rho(q=4,s)<0.4$ even for the longest scales (Fig.~\ref{fig:Kraken_pq_ex}(d)). For the shortest scales considered ($s \approx 10$ min), such exchange rates are largely independent of each other. A notable feature is the significant dependence of the cross-correlation strength on $q$.\par

Since on Binance cryptocurrencies are only traded among themselves, one cannot calculate $\rho(q,s)$ for BTC/X--BTC/Y, where X, Y are fiat currencies, and one cannot create a plot equivalent to Fig.~\ref{fig:Kraken_pq_ex}(a). Nevertheless, it is possible to use Tether (USDT) as a proxy for the US dollar, and thus create a plot equivalent to Fig.~\ref{fig:Kraken_pq_ex}(b). Indeed, Fig.~\ref{fig:Binance_pq_ex}(a) shows that the cross-correlation between the exchange rates that involve the same ``fiat'' base, i.e., BTC/USDT--ETH/USDT, is the strongest among all the rates considered on Binance. By replacing the USDT with ICX, a less liquid cryptocurrency, the cross-correlations decrease for small $s$, while remaining on a level similar to that for USDT for $s>200$ min (Fig.~\ref{fig:Binance_pq_ex}(b)).\par

If the exchange rates do not share the same base, they do not form a triangle, and $\rho(q,s)$ is substantially smaller than in panels (a) and (b) over all $q$ and all scales (Fig.~\ref{fig:Binance_pq_ex}(c)-(d)). Since, over small $s$, trading on different exchange rates does not happen to be synchronized, especially for the exchange rates that do not form any triangle (no arbitrage possibility), the cross-correlations are much weaker than over large $s$, from the perspective of which trading is more synchronous. The liquidity of a cryptocurrency liquidity plays a key role here: for the more liquid exchange rates, e.g., BTC/NEO in Fig.~\ref{fig:Binance_pq_ex}(c), the maximum level of $\rho(q,s)$ is higher compared to the less liquid ones, e.g., ICX/BNB in Fig.~\ref{fig:Binance_pq_ex}(d).\par

\begin{figure}[ht!]
\centering
\includegraphics[width=1\textwidth]{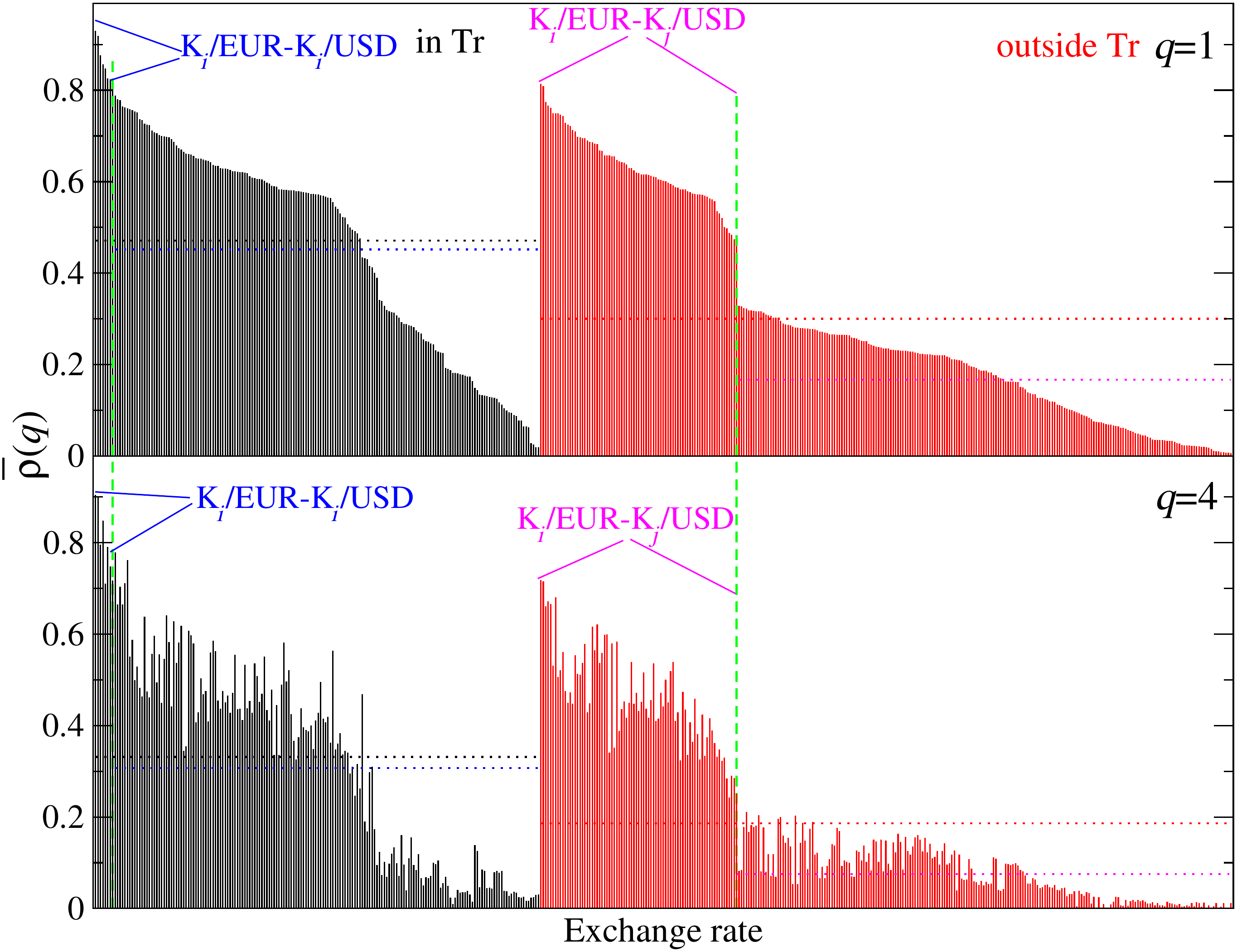} 
\caption{Absolute detrended cross-correlation coefficient $\overline{\rho}(q)$ averaged over all the scales $s$, calculated for the exchange rate pairs from Kraken. The exchange rates related by a triangle relationship (in Tr) are shown on the left (black), while the exchange rate pairs that do not form any triangle (outside Tr) are shown on the right (red). The coefficients $\overline{\rho}(q)$ are sorted in descending order for the top panel, and this order is preserved for $q=4$ in the bottom panel. Horizontal lines indicate the average value of $\overline{\rho}(q)$ for a group of exchange rates in a triangle relationship, i.e., K$_i$/USD--K$_i$/EUR, and without such a relation, i.e., K$_i$/EUR--K$_j$/USD, where K$_i$, K$_j=$ BCH, BTC, DASH, ETC, ETH, LTC, XMR, XRP, and ZEC, K$_i \ne$ K$_j$).}
\label{fig:KrakenPqpoq}
\end{figure}

Cross-correlations between the exchange rates of the form K$_i$/EUR--K$_j$/USD, where K$_i$, K$_j$ ($1 \leqslant i,j \leqslant N$) denote cryptocurrencies, are on average stronger than those involving only a single fiat currency X, i.e.,  K$_i$/X--K$_j$/K$_l$, or no fiat currency at all, i.e., K$_i$/K$_l$--K$_j$/K$_m$, where $1 \leqslant l,m \leqslant N, i \neq l, j \neq m$. This is shown in Fig.~\ref{fig:KrakenPqpoq} for all possible exchange rate pairs from \textit{Kraken-11}, split into the triangle-related (black) and triangle-unrelated (red) ones. Each vertical bar represents the absolute detrended cross-correlation coefficient averaged over all scales: $\overline{\rho}(q)={1 \over s} |\sum_{s=s_\textrm{min}}^{s_\textrm{max}} \rho(q,s)|$, and in descending order. By comparing the triangle cross-correlations with the non-triangle ones, it becomes apparent that a primary factor responsible for the magnitude of $\overline{\rho}(q)$ is the number of exchange rates that revolve around a fiat currency: the exchange rates of two cryptocurrencies tend to be less correlated with the other exchange rates than those involving the USD or EUR. Notably, this property is largely preserved if a different value of $q$ is selected (e.g., compare the top and bottom panels of Fig.~\ref{fig:KrakenPqpoq}).\par

\begin{figure}
\centering
\includegraphics[scale=0.498]{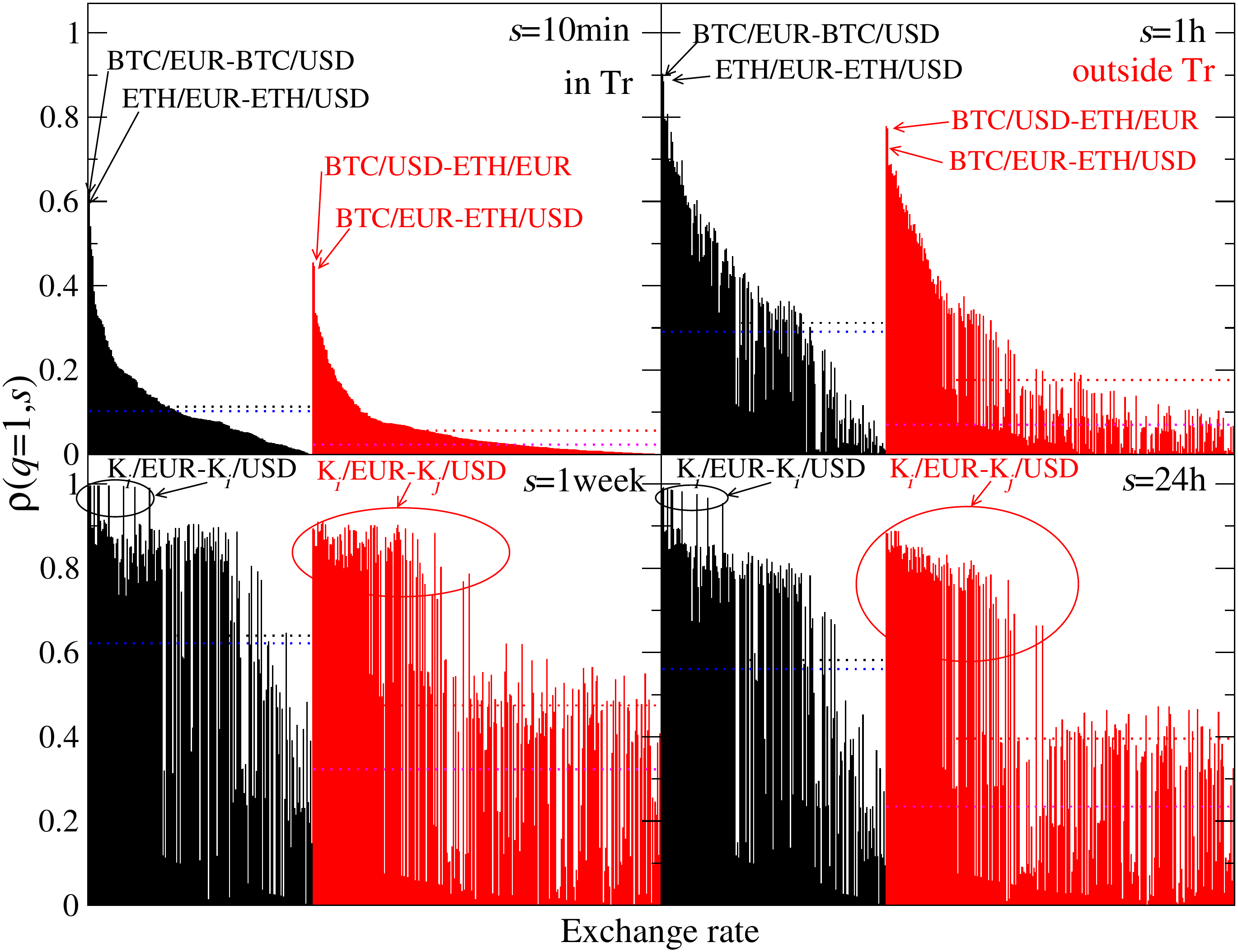}\\ 
\includegraphics[scale=0.498]{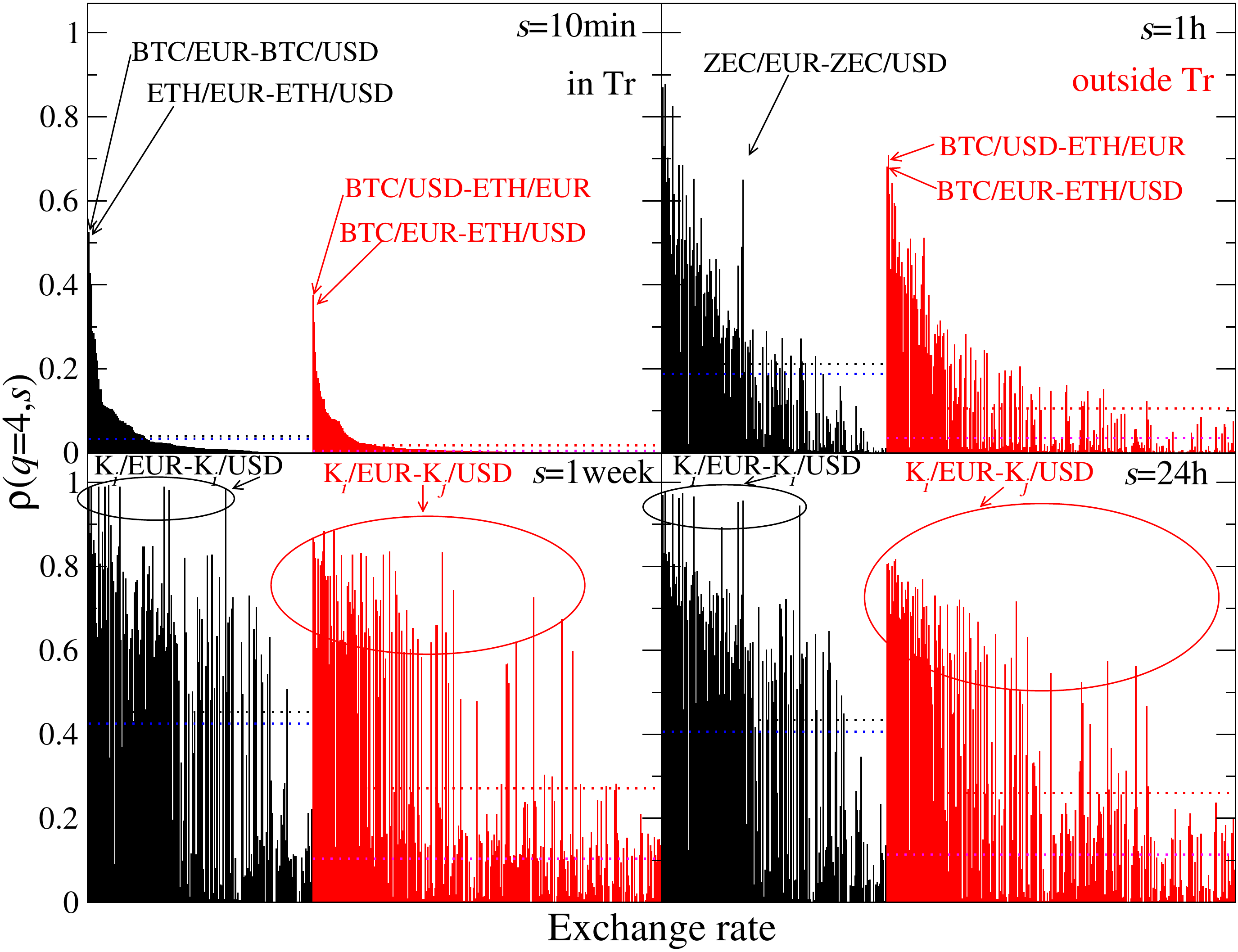} 
\caption{Absolute detrended cross-correlation coefficient $\rho(q,s)$ for sample time-scales $s$, with $q=1$ (medium returns) and $q=4$ (large returns) calculated for the exchange rate pairs from Kraken connected by a triangle relationship (in Tr) and without any triangle relation (outside Tr). The coefficients $\rho(q,s)$ are sorted in descending order for the $s=$10 min, and this order is preserved for the remaining scales. Horizontal lines indicate the average value of $\rho(q,s)$ for a group of exchange rates forming a triangle, i.e., K$_i$/EUR--K$_i$/USD, and without such a relation, i.e., K$_i$/EUR--K$_j$/USD (K$_i$,  K$_j$ like in Fig.~\ref{fig:KrakenPqpoq}, K$_i \ne$ K$_j$).}
\label{fig:KrakenPqpos}
\end{figure}

Fig.~\ref{fig:KrakenPqpos} shows $|\rho(q,s)|$ calculated for representative individual scales (i.e., $s=10$ min, $s=1$~h, $s=1$ day, and $s=7$ days) instead of $\overline{\rho}(q)$, and arranged in the same manner as Fig.~\ref{fig:KrakenPqpoq}, for $q=1$ (top panel) and $q=4$ (bottom panel). Typically, the magnitude of $\rho(q,s)$ increases with raising $s$ for a given exchange rate pair with the largest cross-correlation, close to unity, being reached for K$_i$/EUR--K$_i$/USD in accordance with Fig.~\ref{fig:Kraken_pq_ex}(a). Thus, on long temporal scales, it becomes inconsequential whether a cryptocurrency is expressed in EUR or USD. This effect is visible not only for the most liquid cryptocurrencies, e.g., the BTC and ETH, but also for the less liquid ones. However, for $s=10$ min, only the most liquid exchange rates feature $\rho(q,s) \gg 0$. By increasing $q$ and focusing on the larger returns, one observes that only the exchange rates having the form K$_i$/EUR--K$_j$/USD differ from 0 significantly on the shortest scale $s=10$ min; these exchange rates also increase most markedly in $s$.\par

\begin{figure}[ht!]
\centering
\includegraphics[scale=0.52]{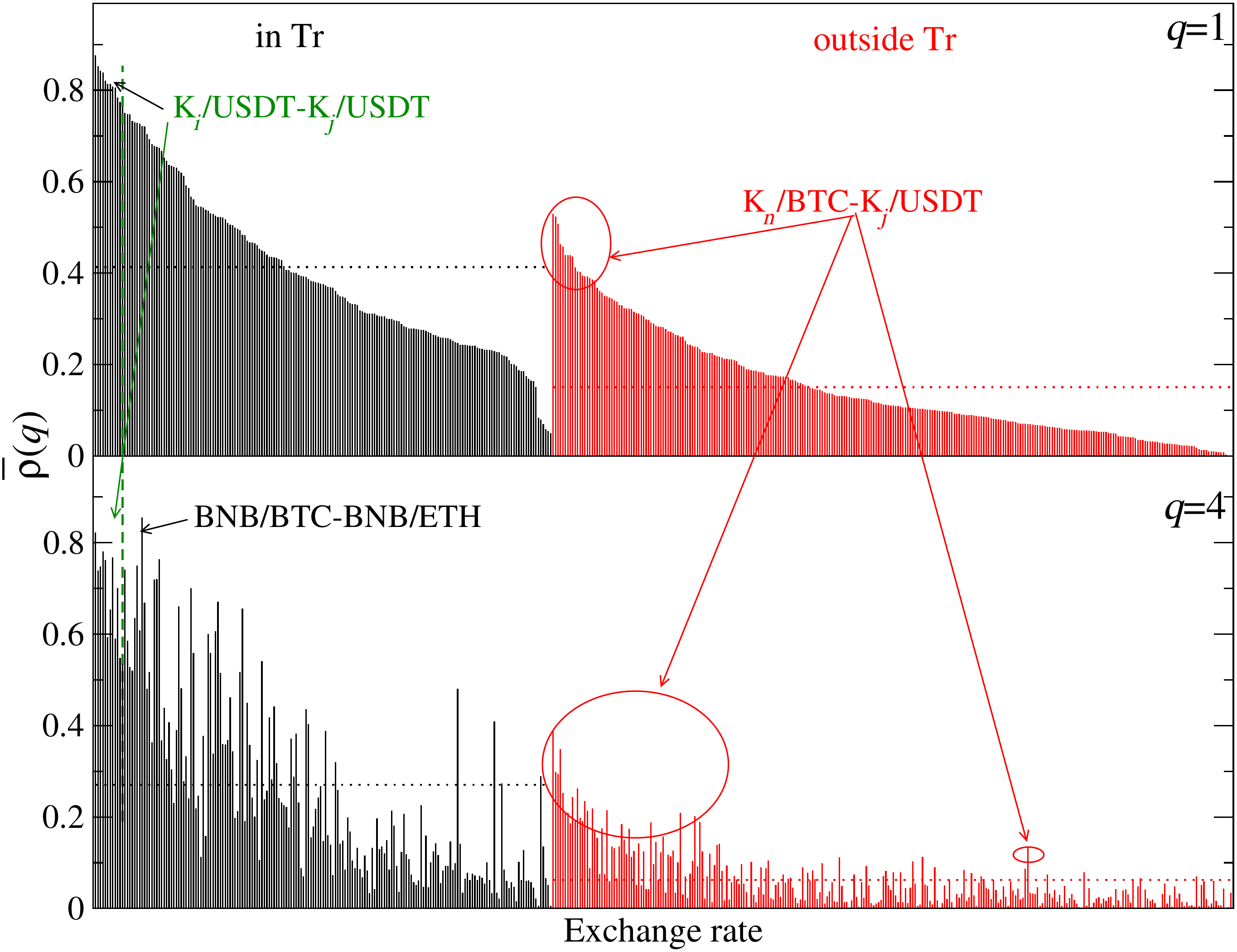}
\caption{Absolute detrended cross-correlation coefficient $\overline{\rho}(q)$ averaged over all scales $s$, calculated for the exchange rate pairs from Binance. The exchange rates related by a triangle relationship (in Tr) are shown on the left (black), while the exchange rate pairs that do not form any triangle (outside Tr) are shown on the right (red). The coefficients $\overline{\rho}(q)$ are sorted in descending order in the top panel, and this order is preserved for $q=4$ in bottom panel. Horizontal lines indicate the average value of $\overline{\rho}(q)$ for a group of exchange rates with a common base currency (i.e., in a triangle relationship): K$_i$/USDT--K$_j$/USDT (where K$_i$, K$_j=$ BCH, BNB, BTC, ETH, LTC, and NEO, K$_i$ $\ne$ K$_j$) and without such a relation: K$_l$/BTC--K$_j$/USDT (where K$_l=$ BAT, BCH, BNB, ETH, ICX, IOTA, LTC, LSK, NEO, and XLM, K$_l$ $\ne$ K$_j$).}
\label{fig:BinancePqpoq}
\end{figure}

The same key factor, namely the average number of transactions in a time interval, determines the cross-correlation strength on the Binance platform (the \textit{Binance-12} dataset). Given the absence of any exchange rate involving fiat currency, for $q=1$ the strongest cross-correlations are those observed for the exchange rate pairs involving two different cryptocurrencies alongside a common base, USDT: K$_i$/USDT--K$_j$/USDT (the triangle-relation case with K$_i$, K$_j$=BCH, BNB, BTC, ETH, LTC, NEO, K$_i \neq$ K$_j$), and those involving two cryptocurrencies, USDT and the most liquid BTC: K$_l$/BTC--K$_j$/USDT (the non-triangle-relation case with K$_l$=BAT, BCH, BNB, ETH, ICX, IOTA, LTC, LSK, NEO, XLM, K$_l \neq$ K$_j$); see Fig.~\ref{fig:BinancePqpoq}. However, the agreement between the magnitude of $\overline{\rho}(q)$ for $q=1$ and $q=4$ is less clear on Binance than on Kraken, because of higher volatility of the cryptocurrencies as compared to USD and EUR.\par

\begin{figure}
\centering
\includegraphics[scale=0.46]{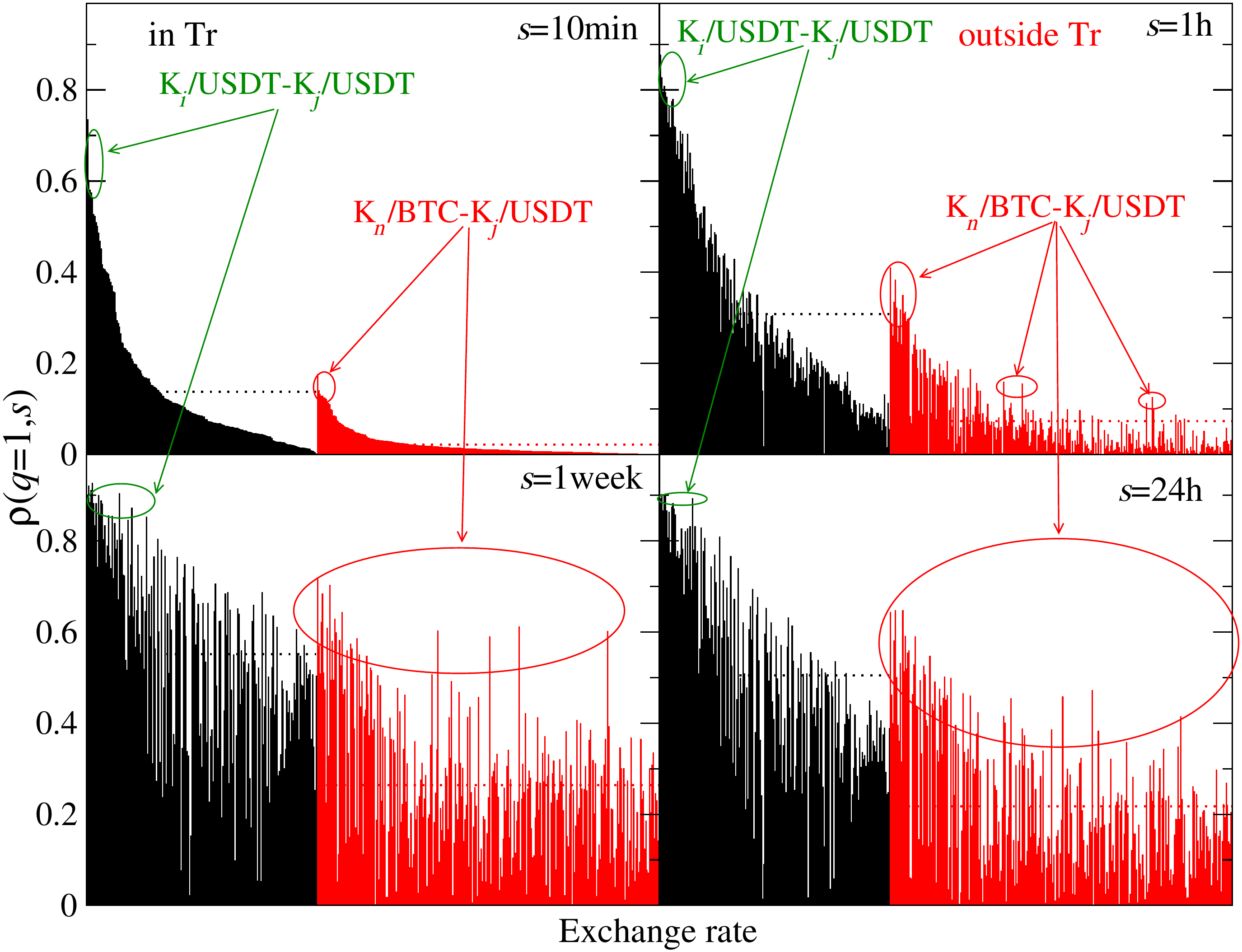}\\ 
\includegraphics[scale=0.46]{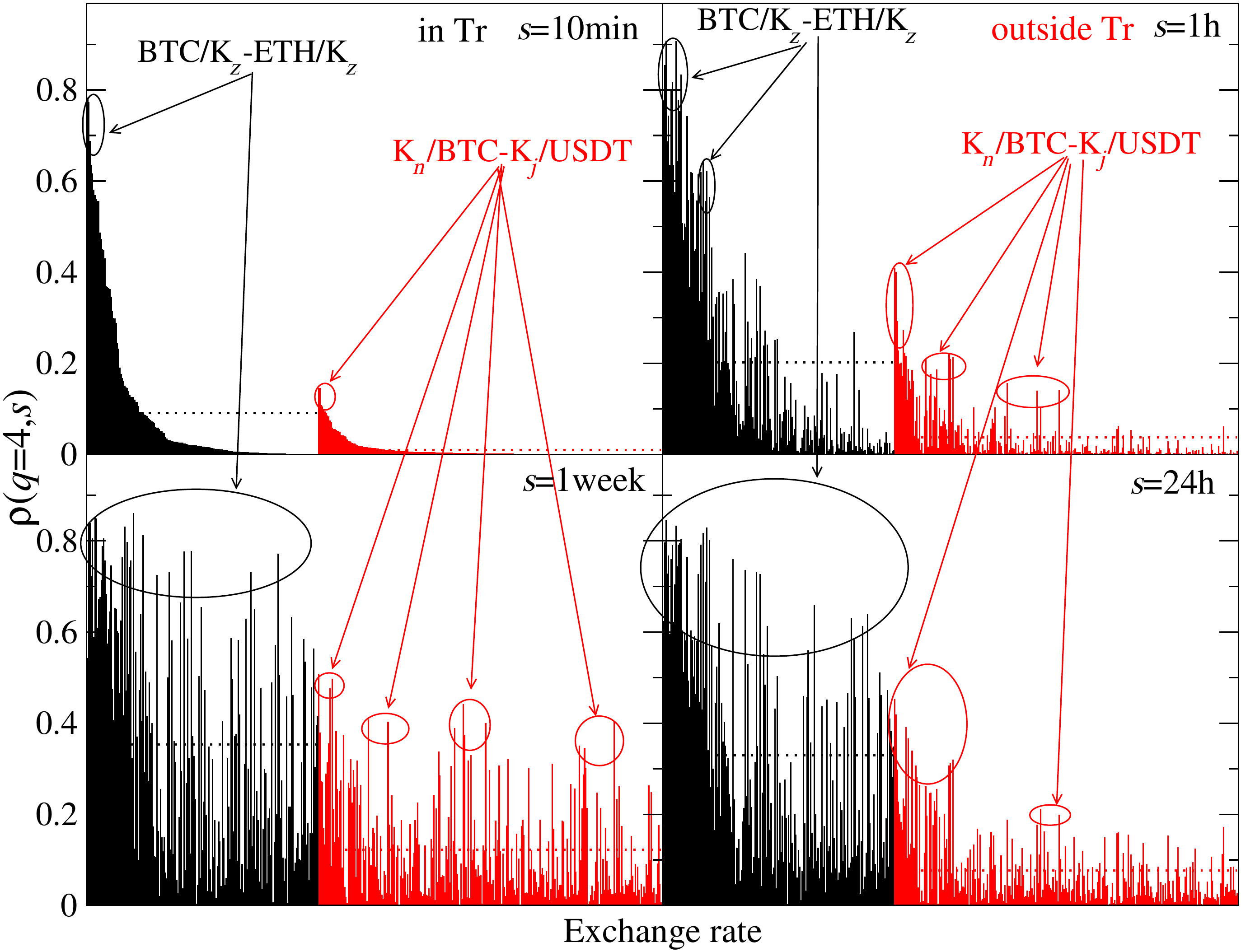} 
\caption{Absolute detrended cross-correlation coefficient $\rho(q,s)$ for representative time-scales $s$ and for $q=1$ (medium returns) and $q=4$ (large returns) calculated for the exchange rate pairs from Kraken connected by a triangle relationship (in Tr), and without any triangle relation (outside Tr). The coefficients $\rho(q,s)$ are sorted in descending order for the $s=$10 min, and this order is preserved for the remaining scales. Horizontal lines indicate the average value of $\rho(q,s)$ for a group of exchange rates forming a triangle: K$_i$/USDT--K$_j$/USDT for $q=1$ (where K$_i$, K$_j=$ BCH, BNB, BTC, ETH, LTC,  and NEO, K$_i \ne$ K$_j$) and BTC/K$_m$--ETH/K$_m$ for $q=4$ (where K$_m=$ BAT, BCH, BNB, ICX, IOTA, LTC, LSK, NEO, USDT, and XLM) and without such a relationship:  K$_l$/BTC--K$_j$/USDT for $q=1$ and $q=4$ (where K$_l=$ BAT, BCH, BNB, ETH, ICX, IOTA, LTC, LSK, NEO, and XLM, K$_l \ne$ K$_j$).}
\label{fig:BinancePqpos}
\end{figure}

On the level of individual scales, the average cross-correlations among the Binance exchange rates increase with $s$ (see Fig.~\ref{fig:BinancePqpos}), which is in agreement with Fig.~\ref{fig:Binance_pq_ex} and points to the analogous result obtained for the Kraken data (Fig.~\ref{fig:KrakenPqpoq}). In contrast with Kraken, however, the largest values of $\rho(q,s)$ do not change with $s$ in the case of the triangle-forming exchange rates K$_i$/USDT--K$_j$/USDT (for $q=1$) and BTC/K$_m$--ETH/K$_m$, where K$_i$ and K$_j$ denote the same cryptocurrencies as in the preceding paragraph, and K$_m$ denotes any cryptocurrency denoted by K$_l$ and additionally USDT (for $q=4$). On the other hand, the pairs K$_l$/BTC--K$_j$/USDT are the strongest cross-correlated ones among the triangle-unrelated exchange rates. On Binance, there is a considerably larger difference in $\rho(q,s)$ between the triangle-related and the triangle-unrelated rates on the short scale of $s=10$ min. In the former case (i.e., with a common base cryptocurrency), market synchronization occurs on shorter scales because of the triangular arbitrage possibility. In contrast, in the latter case (i.e., lack of a common base), there is no direct arbitrage possible, and a longer time is needed to synchronize the exchange rates.\par

Remarkably, since the cross-correlations among the exchange rates differ in magnitude, a cross-correlation hierarchy is formed, which can be expressed as a dendrogram by applying a hierarchical clustering algorithm~\cite{Mantegna1999,Drozdz2007curr,Gorski2008}. As this is defined over a metric space, each average $q$-coefficient $\overline{\rho_{ij}}(q)$ for the exchange rates $i$ and $j$ must be transformed to a respective distance $\delta_{ij}(q)$, defined as follows
\begin{equation}
\delta_{ij}(q) = \sqrt{2 \big( 1 - |\overline{\rho_{ij}}(q)| \big)}, \qquad q > 0,
\label{dij} 
\end{equation}
where the absolute value is taken to eliminate the impact of exchange direction. This definition of $\delta_{ij}$ implies that $0 \leqslant \delta_{ij} \leqslant \sqrt{2}$ with $\delta_{ij}=0$ for perfect cross-correlation between the exchange rates $i$ and $j$, and $\delta_{ij}=\sqrt{2}$ for statistically independent pairs.\par

\begin{figure}[ht!]
\includegraphics[width=1\textwidth]{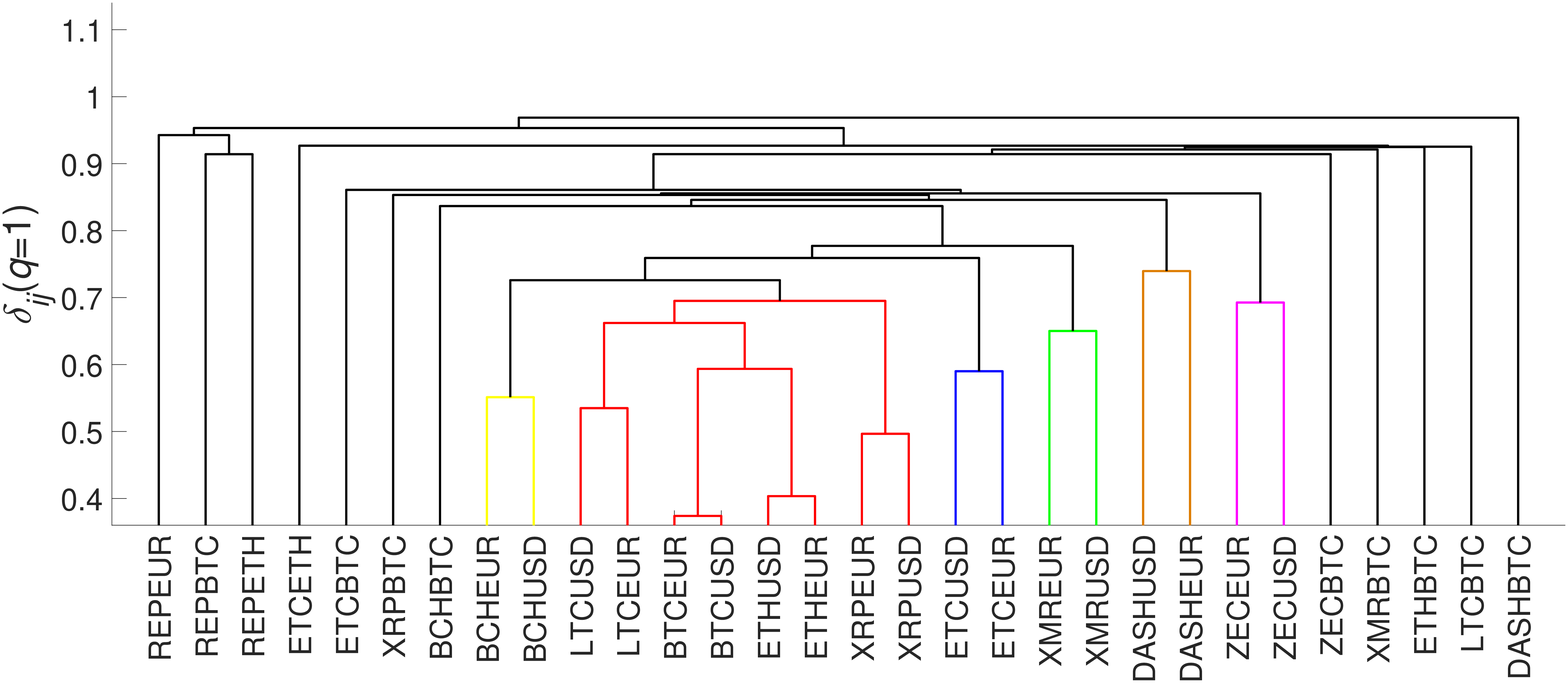}\\ 
\includegraphics[width=1\textwidth]{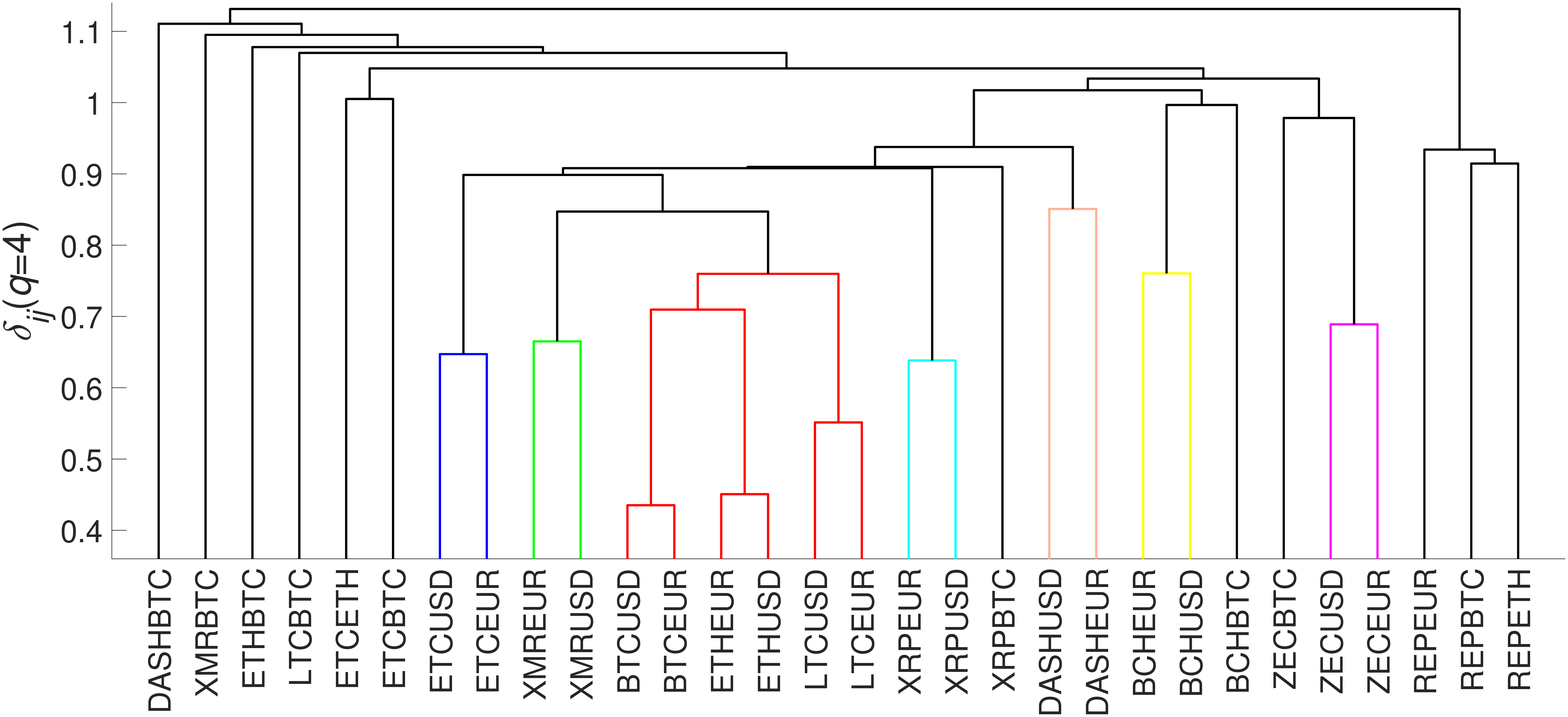} 
\caption{Dendrograms for Kraken, presenting a hierarchy of cross-correlated exchange rates obtained by agglomerative clustering based on metric distances calculated from the average cross-correlation coefficients $|\overline{\rho}(q)|$ for $q=1$ (top) and $q=4$ (bottom).}
\label{fig:KrakenPqhtree}
\end{figure}

\begin{figure}[ht!]
\includegraphics[width=1\textwidth]{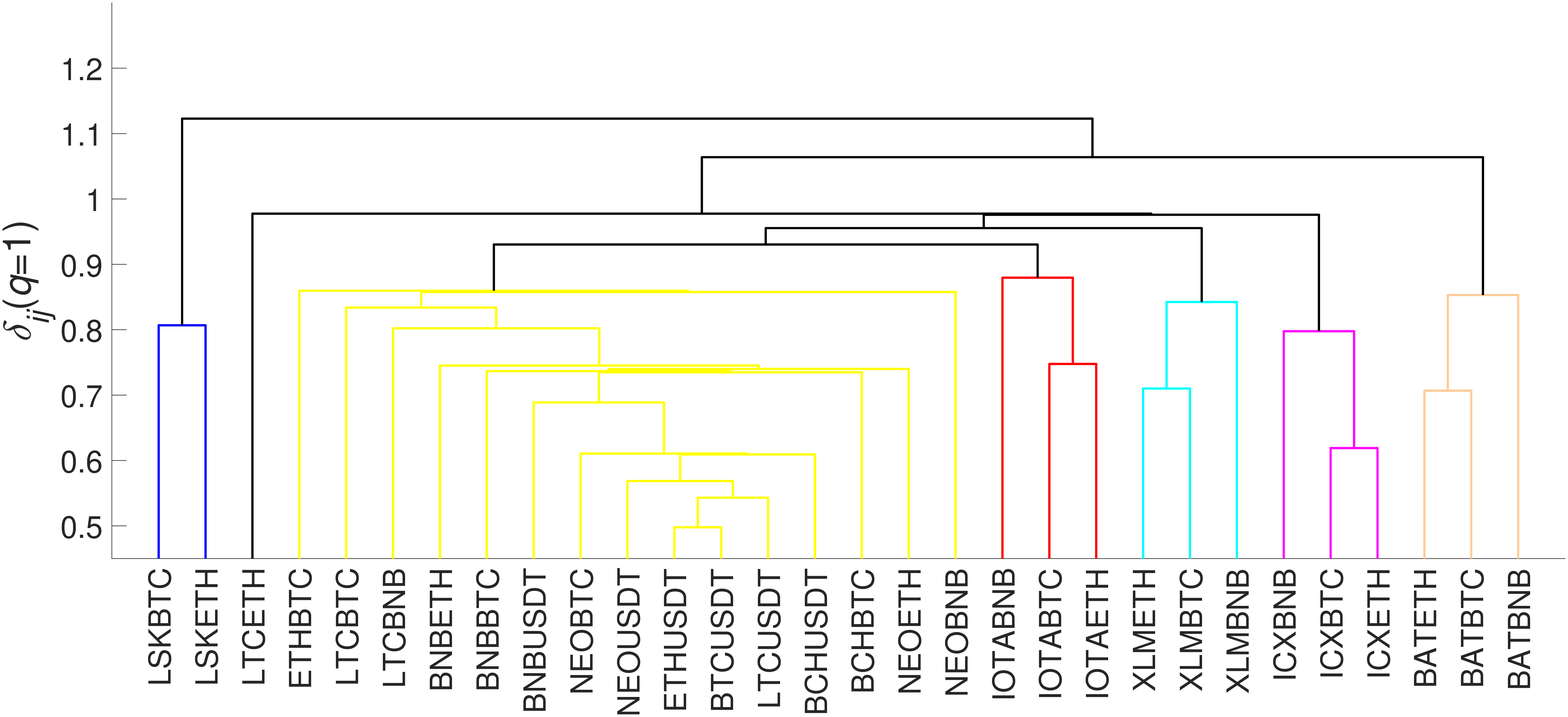}\\ 
\includegraphics[width=1\textwidth]{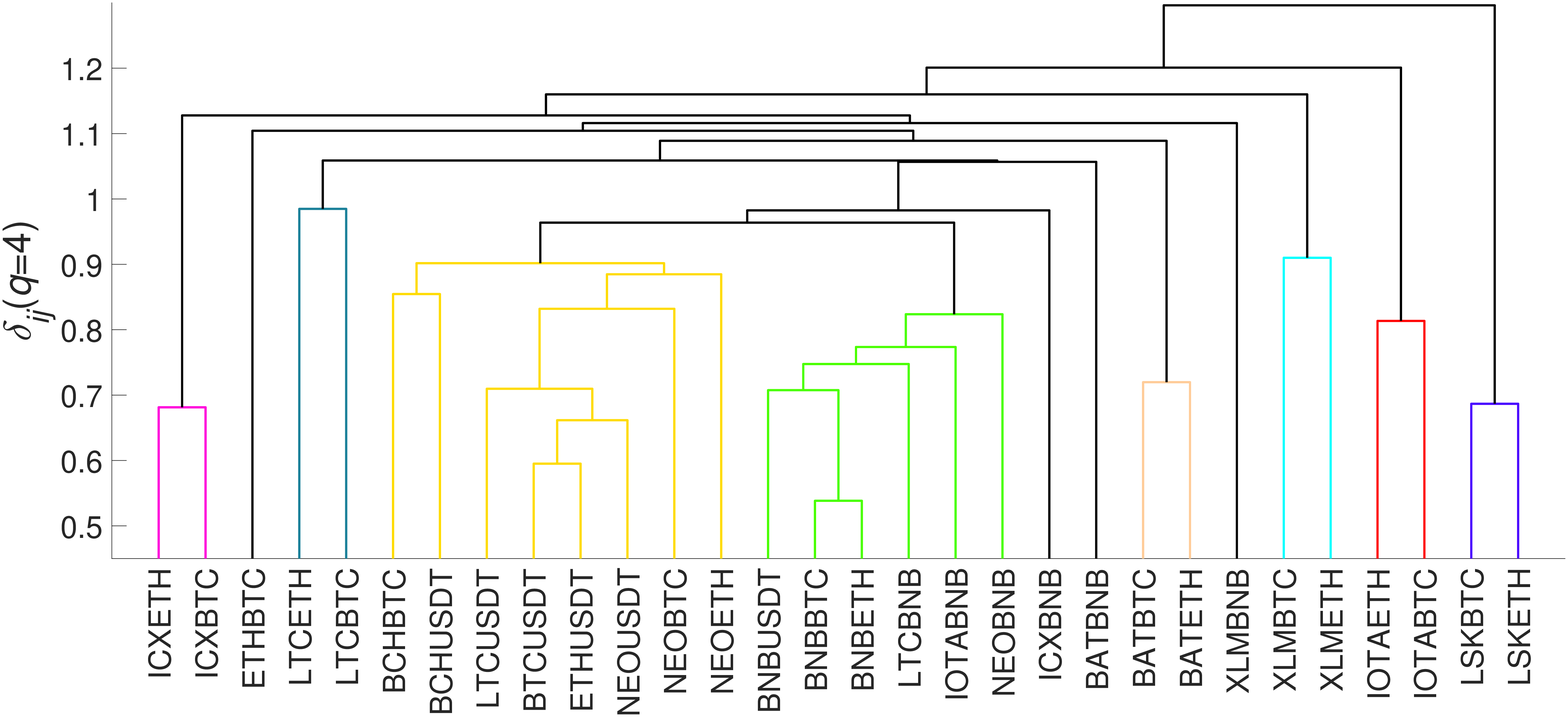} 
\caption{Dendrograms for Binance, mirroring Fig.~\ref{fig:KrakenPqhtree}.}
\label{fig:BinancePqhtree}
\end{figure}

The so-constructed dendrograms are presented in Fig.~\ref{fig:KrakenPqhtree} for the exchange rates from Kraken (except for USDT/USD, which is independent of the rest of the market), and in Fig.~\ref{fig:BinancePqhtree} for the exchange rates from Binance (except for the least correlated one, LSK/BNB). On both platforms, the distances increase with $q$ (i.e., substantial returns are less cross-correlated than typical ones) but, for a given $q$, they are typically larger on Binance; actually, this could already be observed in Fig.~\ref{fig:KrakenPqpoq} and Fig.~\ref{fig:BinancePqpoq} based on $\overline{\rho}(q)$. On Kraken, clusters for $q=1$ and $q=4$ appear similar to each other, with the most liquid cryptocurrencies (i.e., BTC, ETH, LTC, and XRP) being separated the least if expressed in fiat currencies (the red cluster). The other exchange rates form clusters insofar as the same cryptocurrency is expressed in EUR and USD; hence, they essentially elucidate a correlated movement of USD and EUR viewed from a perspective of less liquid cryptocurrencies.\par

On Binance, two large clusters, based on USDT and BNB, can be distinguished for $q=4$. On the other hand, a cluster of cryptocurrencies expressed in BNB does not exist for $q=1$, and the exchange rates involving USDT merge with those involving the BTC, BNB, and ETH, as well as a few other rates, including BNB/USDT and NEO/BTC. Less liquid cryptocurrencies, such as LSK, ICX, BAT, XLM, and IOTA expressed in BTC, BNB, and ETH, form their own clusters. This arrangement signifies that, while the most liquid cryptocurrencies are strongly correlated within their group, the less liquid ones are more independent. For $q=4$, the main clustering factor is a common base. A significant fluctuation of the base cryptocurrency price implies that all other assets become correlated if they are expressed in this particular cryptocurrency. This effect is also observed on the Forex when some currency experiences a high volatility period (see Ref.~\cite{gebarowski2019} for examples about the Swiss franc during the Swiss National Bank interventions in 2011 and 2015 British pound after the Brexit referendum result was announced in 2016).\par

\subsubsection{Inter-platform multiscale cross-correlations}
\label{sect::InterPlatform}

In this subsection, cross-correlations for the selected exchange rate pairs listed simultaneously on both Kraken and Binance are directly compared. In this context, the USD and USDT are treated as the same currency and denoted by USD$^{*}$.\par

\begin{figure}
\centering
\includegraphics[scale=0.52]{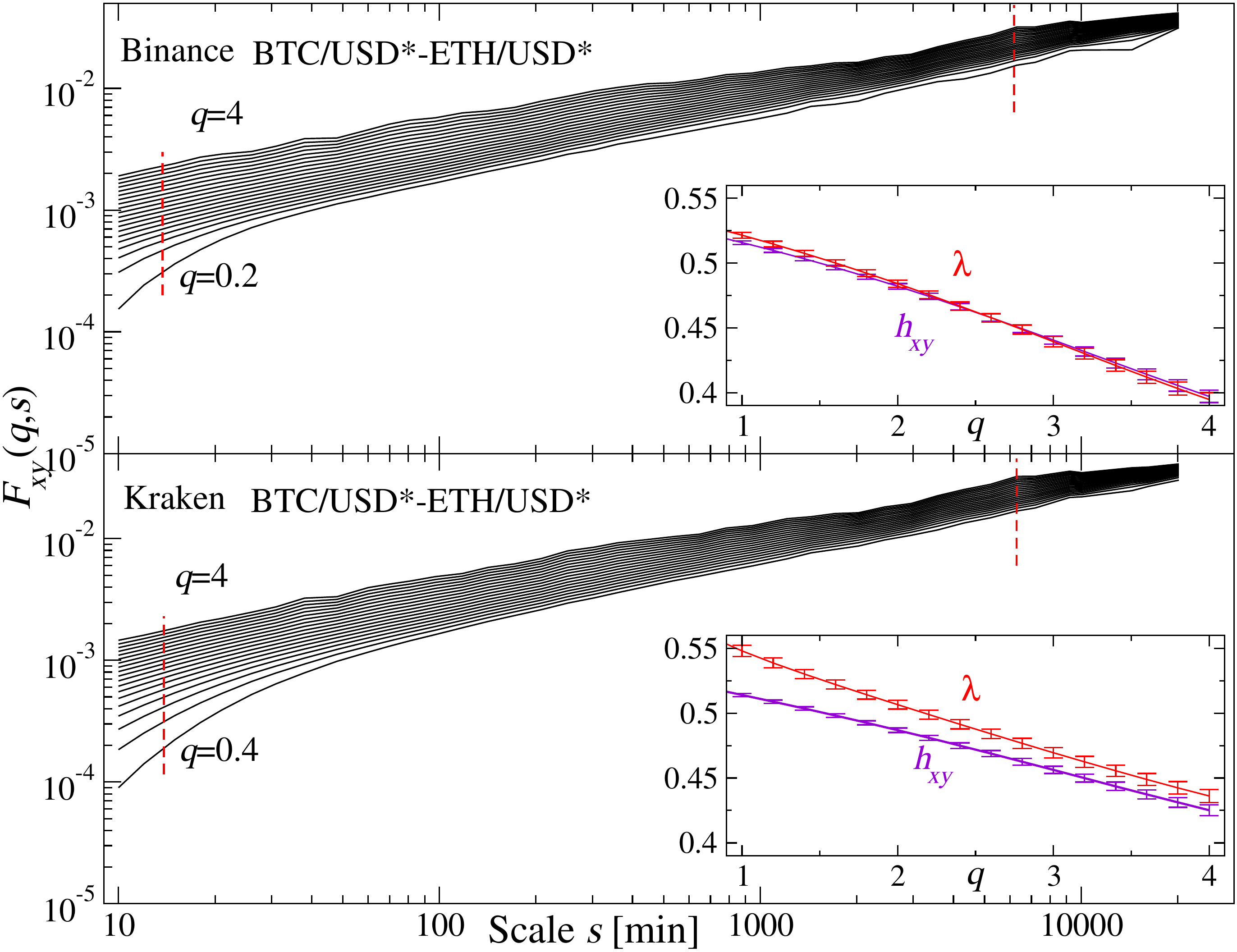}\\
\includegraphics[scale=0.52]{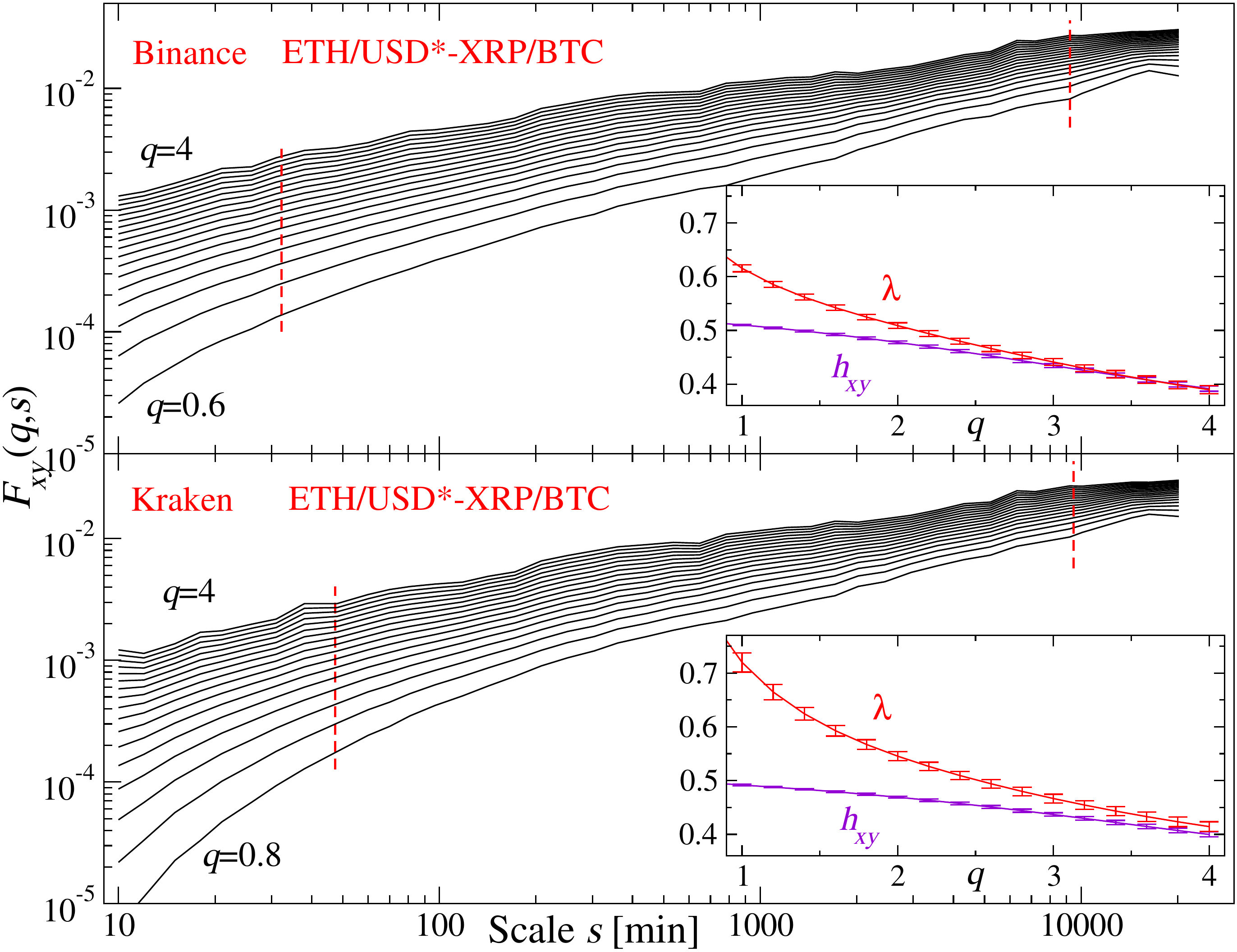}
\caption{Bivariate fluctuation function $F_\textrm{xy}(q,s)$ for $q>0$ ($\Delta q=0.2$), calculated for the exchange rate pairs BTC/USD$^*$--ETH/USD$^*$ (a triangle relationship, black) and ETH/USD$^*$--XRP/BTC (without a triangle relationship, red) listed simultaneously on Binance and Kraken. Insets present scaling exponent $\lambda(q)$ and the average generalized Hurst exponents $h_\textrm{xy}(q) = (h_x(q)+h_y(q))/2$ estimated from the scale range marked with dashed lines in main panels.}
\label{fig:KrBi_Fqxy}
\end{figure}

Figure~\ref{fig:KrBi_Fqxy} shows the bivariate fluctuation function $F_\textrm{xy}(q,s)$ for the most correlated exchange rates in a triangle relationship, namely the BTC/USD$^*$--ETH/USD$^*$ (top panel), and without a triangle relationship, namely the ETH/USD$^*$--XRP/BTC (bottom panel). As in Sect.~\ref{sect::MultiscaleCrossBTCETH}, $F_\textrm{xy}(q,s)$ is also presented here only for $q>0$, and shows a consistent power-law dependence on scale. By contrast, for negative $q$ corresponding to small returns, $F_\textrm{xy}(q,s)$ fluctuates around zero, denoting the absence of cross-correlation among such returns (not shown because of the log-log plot). A difference between the platforms can be appreciated in insets of Figure~\ref{fig:KrBi_Fqxy}, which present $h_\textrm{xy}(q)$ and $\lambda(q)$ as defined by Eq.(\ref{Fxy}) and Eq.(\ref{hqxy}) in Sect.~\ref{detrendned}. The difference $d_\textrm{xy}(q)$ between these quantities is considerably larger on Kraken than on Binance, which suggests that, on the latter, the exchange rates synchronize over shorter time-scales. Nevertheless, on both platforms the exchange rates without a triangle relation show a later synchronization onset (larger $d_\textrm{xy}(q)$) compared to the exchange rates in a triangle relationship, especially for $q \in [1,3]$.\par

\begin{figure}[ht!]
\includegraphics[width=1\textwidth]{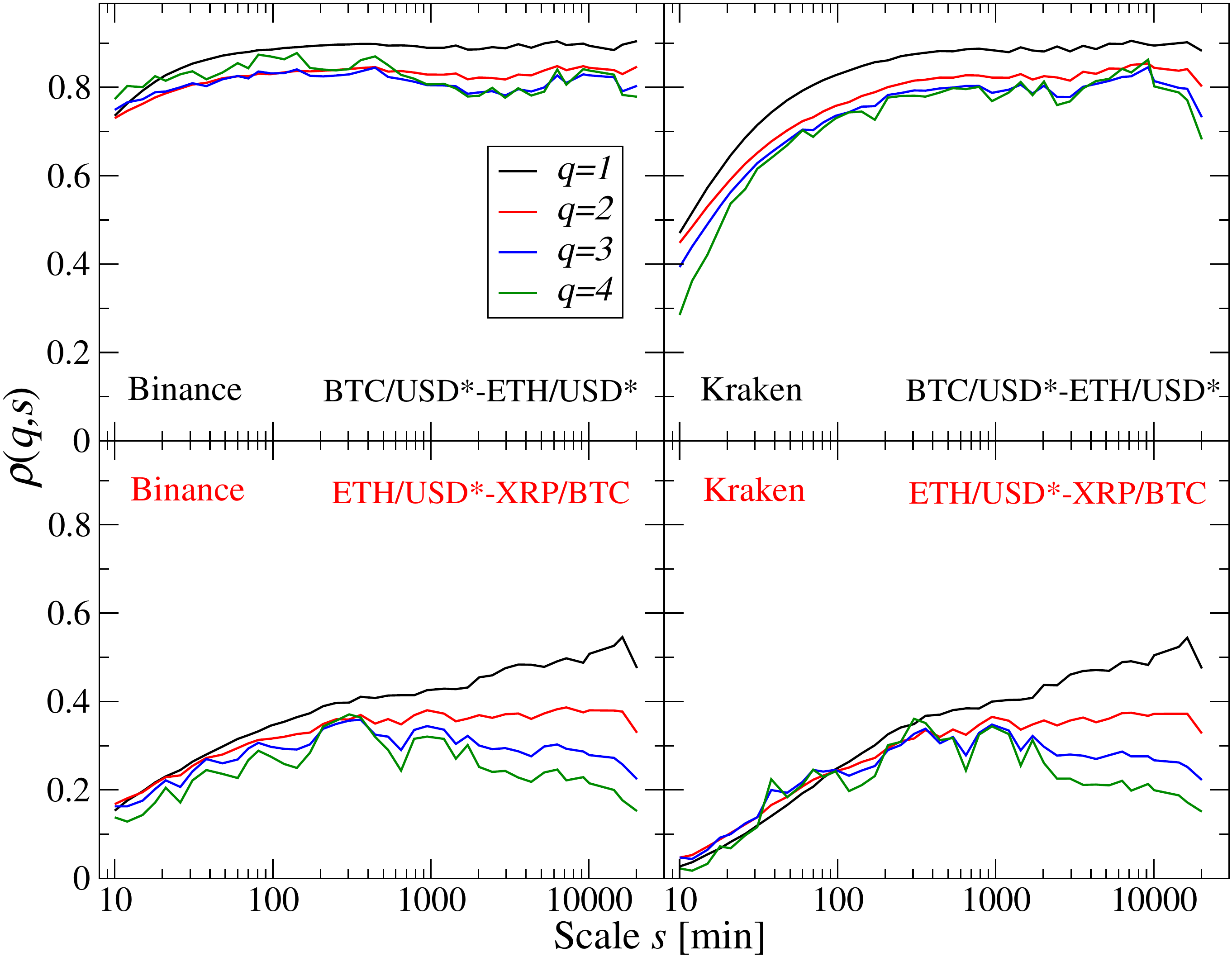} 
\caption{Comparison of the detrended correlation coefficient $\rho(q,s)$ calculated for selected exchange rate pairs that listed simultaneously on Binance and Kraken. Top panels: representative exchange rates forming a triangle relationship; bottom panels: representative exchange rates that do not fulfill this condition.}
\label{fig:KBiKrcomp}
\end{figure}

Given that $d_\textrm{xy}(q)$ corresponds directly to the functional behavior of $\rho(q,s)$, this quantity is shown in Fig.~\ref{fig:KBiKrcomp} for the same exchange rate pairs as in Fig.~\ref{fig:KrBi_Fqxy}. Over the shortest scales, on the order of $s=10$ min, and for both BTC/USD$^*$--ETH/USD$^*$ and ETH/USD$^*$--XRP/BTC, $\rho(q,s)$ is smaller in the case of Kraken: for ETH/USD$^*$--XRP/BTC, the cross-correlation even approaches zero. Moreover, $\rho(q,s)$ substantially increases with $s$ (however, only in terms of its overall behavior, as $d_\textrm{xy}(q)$ is not sensitive to scale) on Kraken, which corresponds to positive $d_\textrm{xy}(q)$ (insets in Fig.~\ref{fig:KrBi_Fqxy}). Differently, $\rho(q,s)$ practically does not depend on $s$ for BTC/USD$^*$--ETH/USD$^*$ on Binance, which goes in parallel with $d_\textrm{xy}(q) \approx 0$ in this case. On the other hand, in the case of the exchange rates that do not form a triangle (ETH/USDT--XRP/BTC) on Binance, $d_\textrm{xy}(q) \approx 0$ for $q>3$, which agrees with a weak average dependence of $\rho(q,s)$ on $s$ in Fig.~\ref{fig:KBiKrcomp}. The correlation strength on both platforms converges to a common level at the scale $s \approx 300$ min, which defines the time needed to attain full synchronization of different exchange rates.\par

\begin{table}[ht!]
\centering
\caption{Detrended cross-correlation coefficient $\langle \rho(q,s) \rangle$ averaged over the 66 exchange rate pairs listed simultaneously on Binance and Kraken for representative values of $q$ and $s$. Both the rates forming a triangle (Tr, white rows) and those not related in this manner (P, red rows) are included.} 
\begin{tabular}{|c|c|c|c|c|c|}
\hline
                                 & \multicolumn{2}{c|}{\boldmath$\langle \rho(q=1,s) \rangle$}                                     &                       & \multicolumn{2}{c|}{\boldmath$\langle \rho(q=4,s)\rangle$}                                     \\ \hline
\textbf{Time scale} \boldmath$s$             & \textbf{Binance}            & \textbf{Kraken}                           &                         & \textbf{Binance}  & \textbf{Kraken}   
 \\ \hline
Tr $s$=10min                       & 0.163                        & 0.092                        &                         & 0.117                        & 0.029                        \\ \hline
\rowcolor[HTML]{FD6864} 
{\color[HTML]{000000} P $s$=10min} & {\color[HTML]{000000} 0.045} & {\color[HTML]{000000} 0.029} & {\color[HTML]{000000} } & {\color[HTML]{000000} 0.021} & {\color[HTML]{000000} 0.005} \\ \hline
Tr $s$=1h                          & 0.309                        & 0.229                        &                         & 0.208                        & 0.162                        \\ \hline
\rowcolor[HTML]{FD6864} 
P $s$=1h                           & 0.120                        & 0.086                        &                         & 0.061                        & 0.070                        \\ \hline
Tr $s$=24h                         & 0.490                        & 0.479                        &                         & 0.262                        & 0.259                        \\ \hline
\rowcolor[HTML]{FD6864} 
P $s$=24h                          & 0.299                        & 0.287                        &                         & 0.120                        & 0.123                        \\ \hline
Tr $s$=1 week                     & 0.573                        & 0.570                        &                         & 0.276                        & 0.276                        \\ \hline
\rowcolor[HTML]{FD6864} 
P $s$=1 week                      & 0.381                        & 0.380                        &                         & 0.139                        & 0.137                        \\ \hline
\end{tabular}
\label{tab:sredniecorrKrBi}
\end{table}

Table~\ref{tab:sredniecorrKrBi} presents $\langle\rho(q,s)\rangle$ averaged over all possible exchange rate pairs listed on both platforms for sample values of $q$ and $s$. In almost every case, the triangle-related pairs produce larger $\langle\rho(q,s)\rangle$ than the unrelated ones. While for shorter time-scales (i.e., $s=10$ min and $s=1$ hour) the cross-correlations are stronger on Binance than Kraken, for longer time-scales (i.e., $s=1$ day and $s=1$ week) the correlation levels converge, as was already been pointed out above (Fig.~\ref{fig:KBiKrcomp}).\par

\begin{figure}[ht!]
\includegraphics[width=1\textwidth]{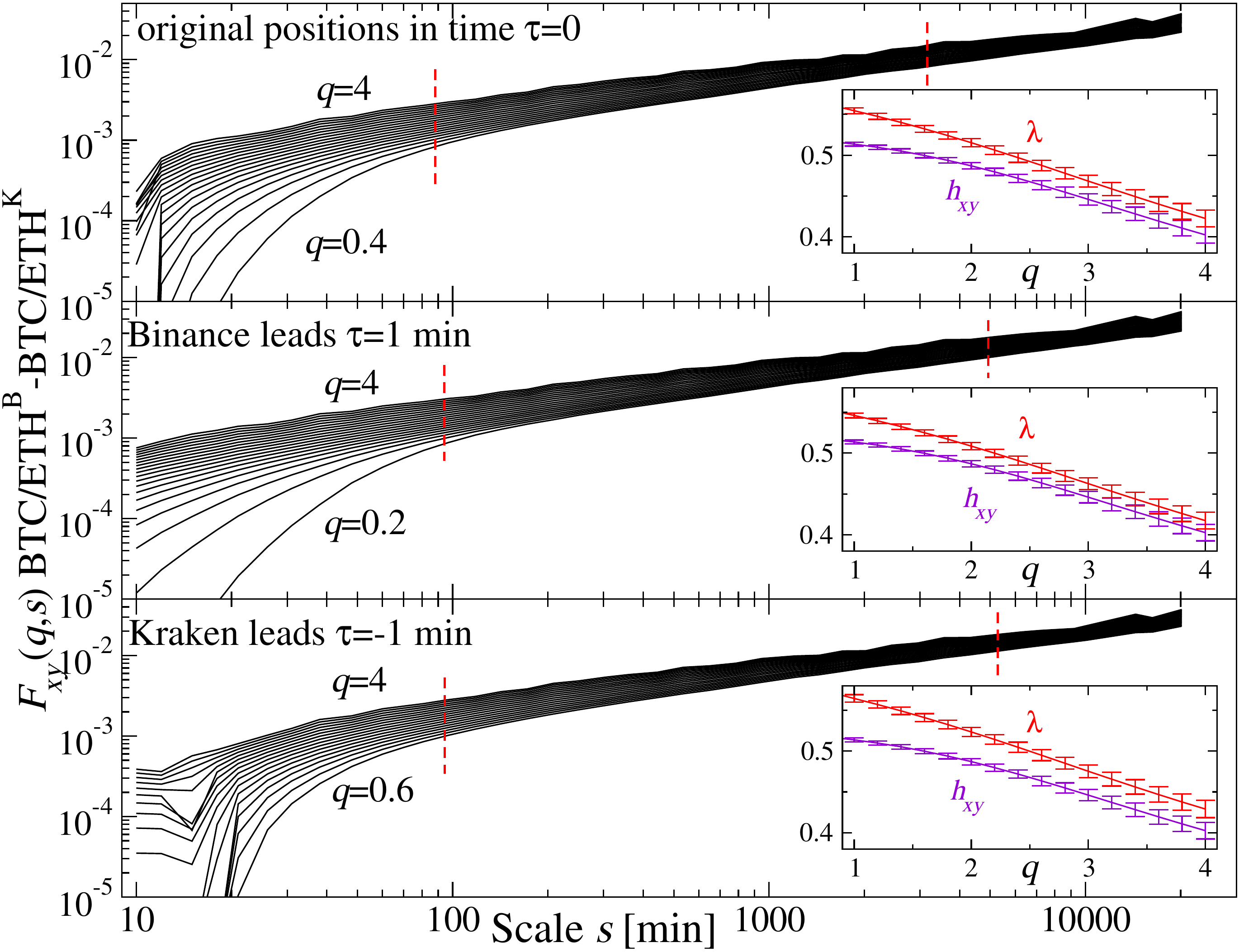} 
\caption{Bivariate fluctuation function $F_\textrm{xy}(q,s)$ for $q>0$ ($\Delta q=0.2$) calculated for BTC/ETH$^{(\textrm{B})}$--BTC/ETH$^{(\textrm{K})}(\tau)$, where $\tau$ denotes time lag, B labels the exchange rate listed on Binance and K labels the exchange rate listed on Kraken. Three sample situations are shown: synchronous time-series from both platforms ($\tau=0$, top), Binance ahead of Kraken by $\tau=1$ min (middle), and Kraken ahead of Binance by $\tau=1$ min (bottom). Insets present the scaling exponent $\lambda(q)$ and the average generalized Hurst exponent $h_\textrm{xy}(q)$ calculated over the scale range marked with dashed lines; error bars represent the standard error of linear regression.}
\label{fig:BTCETH_Fq}
\end{figure}

Since there are exchange rates listed on both platforms, pairs may be formed not only within the same platform but also across them, for example, drawing the first from Binance and the second from Kraken. As events on Binance may occur ahead of Kraken or \textit{vice versa}, an additional parameter has to be introduced, namely a delay $\tau$, whose sign denotes which exchange rate is delayed: here, $\tau>0$ signifies that an exchange rate representing Binance leads, whereas $\tau<0$ means this exchange rate is lagged with respect to Kraken. Figure~\ref{fig:BTCETH_Fq} displays the fluctuation function $F_\textrm{xy}(q,s,\tau)$ for a sample exchange rate involving the most liquid cryptocurrencies BTC/ETH. Different temporal relations between the time-series are considered: $\tau=0$ (top), $\tau=1$ min (middle), and $\tau=-1$ min (bottom). The scaling quality of $F_\textrm{xy}(q,s,\tau)$ is maximized and the difference $d_\textrm{xy}(q)$ is minimized if Binance leads by a minute (middle panel). The weakest scaling and the largest $d_\textrm{xy}(q)$ can be seen in the opposite case when Kraken leads by a minute (bottom panel). The equal-time case ($\tau=0$) is located in between these extremes.\par

\begin{figure}[ht!]
\includegraphics[width=1\textwidth]{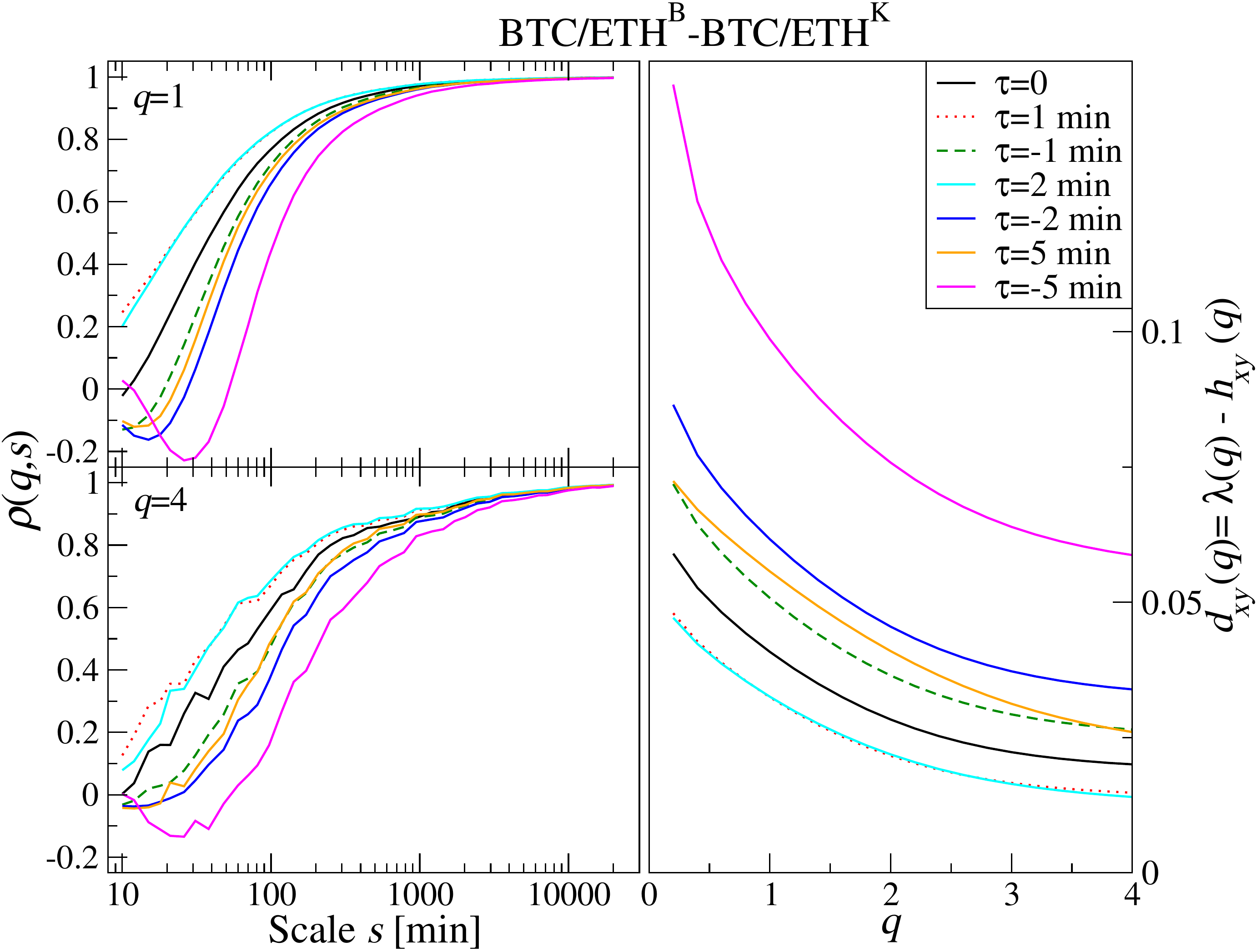}
\caption{Three synchronization situations: (1) original positions in time, (2) Binance ahead of Kraken by $\tau$, (3) Kraken ahead of Binance by $\tau$ Left: correlations $\rho(q,s)$ between BTC/ETH exchange rate on the Binance and Kraken exchanges. Left: corresponding differences $d_\textrm{xy}(q)$.}
\label{fig:BTCETH_rhoq}
\end{figure}

The cross-correlation strength for the pair BTC/ETH$^{(\textrm{B})}$--BTC/ETH$^{(\textrm{K})}$, expressed in terms of $\rho(q,s,\tau)$ for $\tau=\pm 1$ min, $\tau=\pm 2$ min, and $\tau= \pm 5$ min, is shown in Fig.~\ref{fig:BTCETH_rhoq} (left). While, for all considered lags $\tau$, the function $\rho(q,s,\tau)$ converges to unity over sufficiently long scale $s$, which is a desired effect in the case of the equivalent exchange rates, the strongest cross-correlation at the shortest scales of $s\approx 10$ min is observed for $\tau=1$ min and $\tau=2$ min. This finding implies that the BTC/ETH movements on Binance are, on average, advanced by 1-2 min with respect to Kraken. For $\tau=0$, the cross-correlation is weaker as well as for $\tau<0$ and $\tau=5$ min. This picture is stable across different positive values of $q$ (top left and bottom left), suggesting that the effect is present on both typical and large returns. A leading position of Binance is by no means surprising, as the BTC/ETH volume traded and associated trading frequency are higher than on Kraken. These results, based on $\rho(q,s,\tau)$, are supported by the characteristics of $d_\textrm{xy}(q)$, which is the smallest for $\tau=1$ min and $\tau=2$ min, as visible in Fig.~\ref{fig:BTCETH_rhoq} (right).\par

\begin{figure}
\centering
\includegraphics[width=1\textwidth]{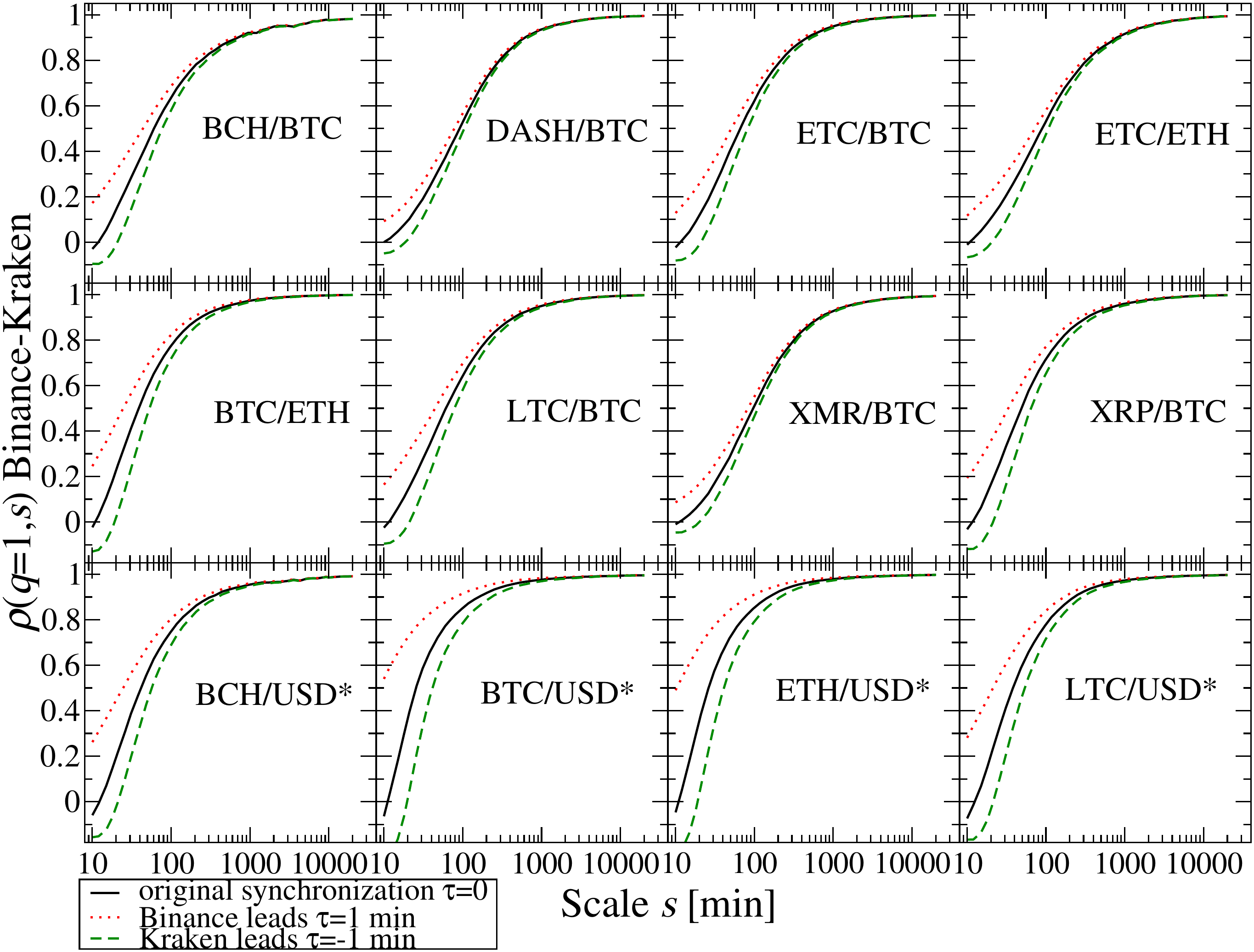}\\
\includegraphics[width=1\textwidth]{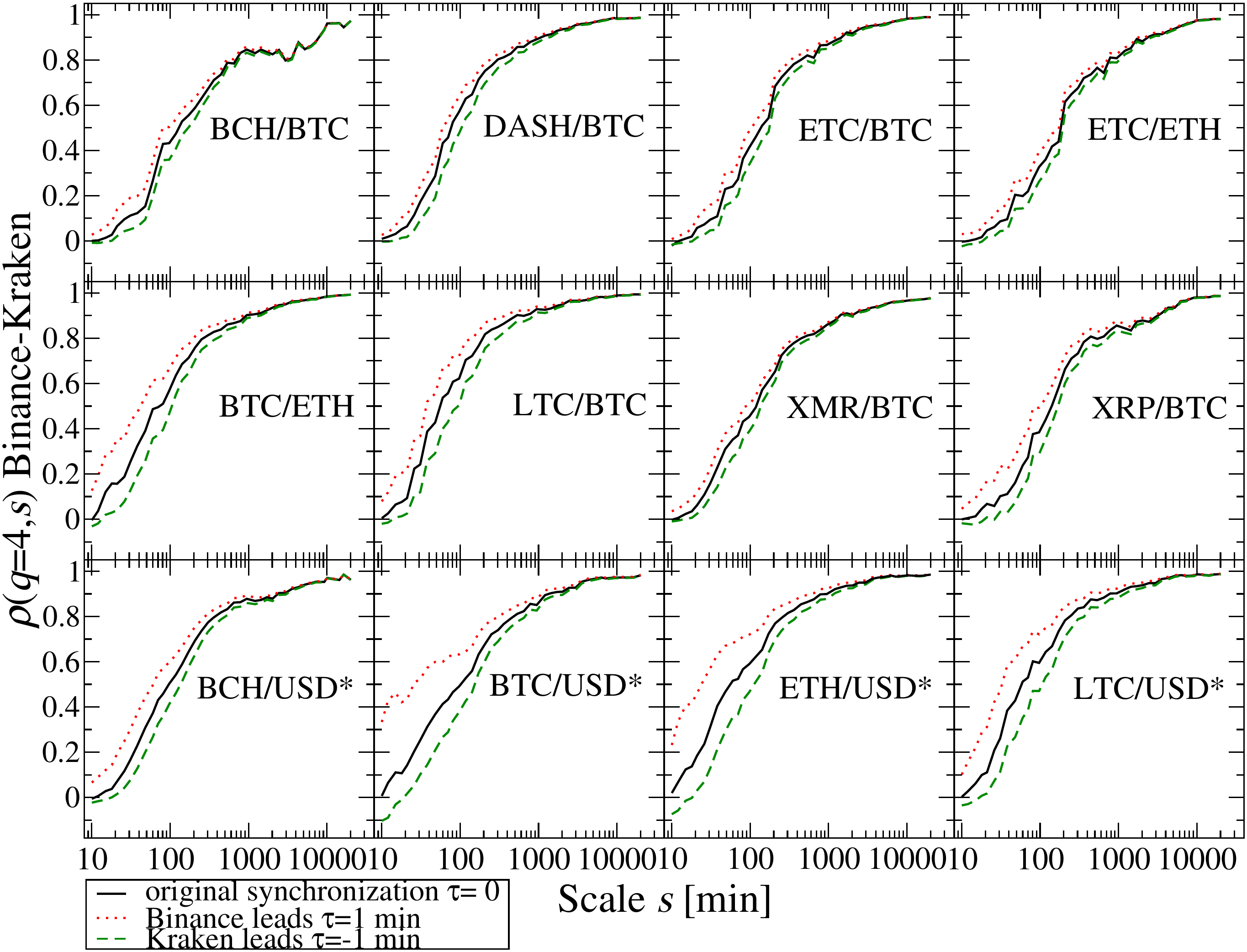}
\caption{Correlations $\rho(q,s)$ between the exchange rates listed simultaneously on Binance and Kraken in three synchronization situations: (1) original positions in time, (2) Binance ahead of Kraken, $\tau=1$ min, (3) Kraken ahead of Binance, $\tau=-1$ min; USD$^*$ stands for USDT on Binance and USD on Kraken.}
\label{fig:KrBipq1_4}
\end{figure}
The leading role of Binance is also visible for the remaining 11 cross-platform exchange rate pairs by means of $\rho(q,s,\tau)$ calculated for $\tau=-1$ min, $\tau=0$, and $\tau=1$ min (Fig.~\ref{fig:KrBipq1_4}). While, in each case with $q=1$, Binance is advanced by 1 min (the largest value of $\rho(q,s,\tau)$), the exact cross-correlation strength depends on the exchange rate liquidity, with the most liquid rates being BTC/USD$^*$ and ETH/USD$^*$. Like for BTC/ETH, the cross-correlation strength increases with $s$ and converges to unity, more slowly for $q=4$ than for $q=1$. Furthermore, for $q=4$ (large returns), not all exchange rates synchronize more markedly when $\tau=1$ min is used; in this regard, it is possible that, for the less liquid exchange rates, the market needs a longer lag to sense synchronization of large fluctuations. This effect could be accounted for by the higher trading frequency on Binance. In other words, when a transaction is completed on it, and there is no parallel trade on Kraken, the next transaction on Kraken will take into account what already has happened on Binance.\par

\begin{figure}[ht!]
\includegraphics[width=1\textwidth]{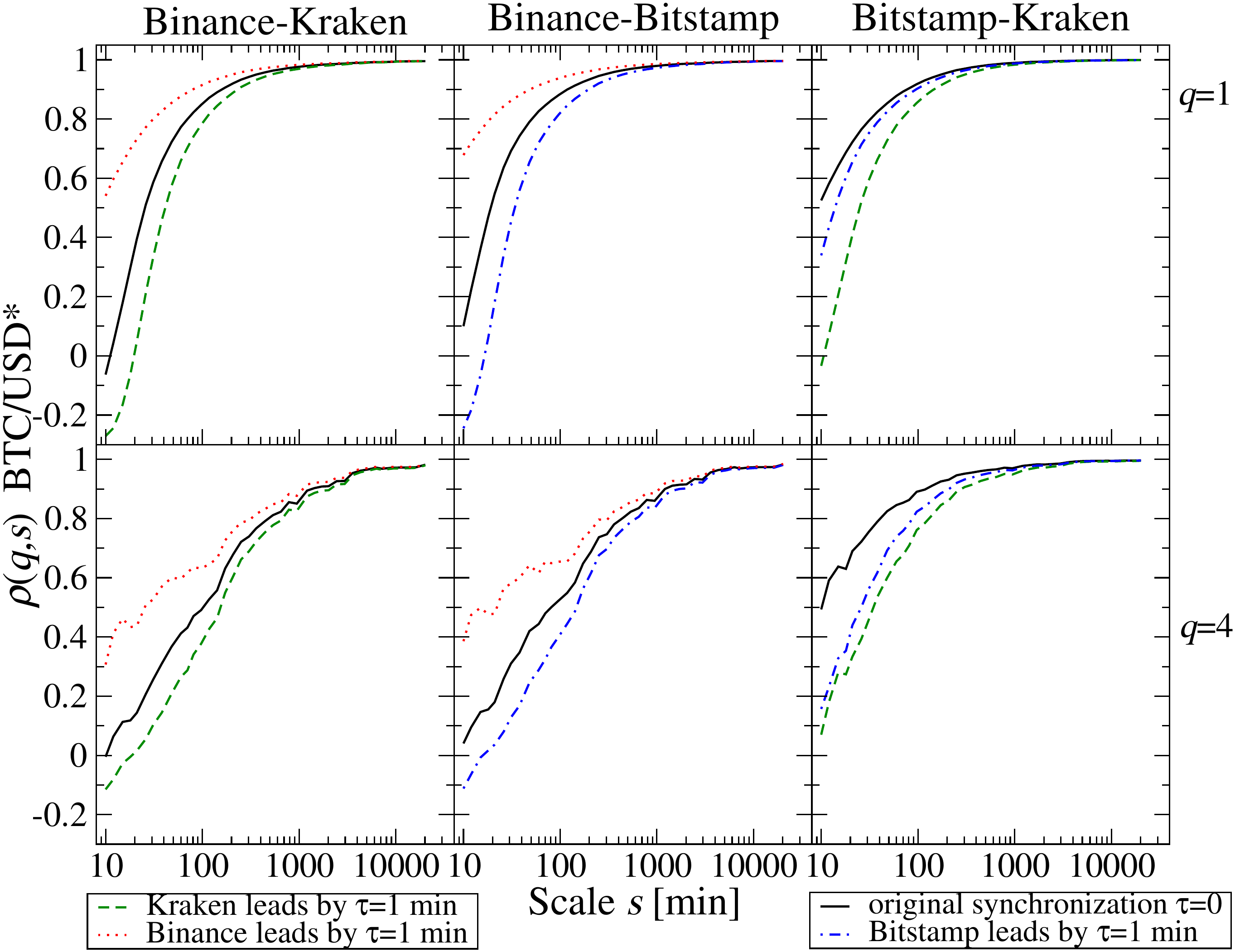}
\caption{Detrended cross-correlation coefficient $\rho(q,s,\tau)$ for the BTC/USD exchange rate listed on Binance, Kraken, and Bitstamp. All exchange rate pairs are shown, namely Binance vs. Kraken (left), Binance vs. Bitstamp (middle), Bitstamp vs. Kraken (right) for $\tau=-1$ min (the second platform leads), $\tau=0$ (both platforms are synchronous), and $\tau=1$ min (the first platform leads).}
\label{fig:BTCpqBiKrBit}
\end{figure}

Another platform specialized in cryptocurrency trading is Bitstamp. The detrended cross-correlation coefficient $\rho(q,s,\tau)$ calculated from the BTC/USD$^*$ returns (USD$^*$ stands for USD on Bitstamp) with $\Delta t=1$ min traded simultaneously on two out of three platforms: Binance, Kraken, and Bitstamp, is plotted in Fig.~\ref{fig:BTCpqBiKrBit} for data spanning the entire year 2018. For Bitstamp vs. Kraken data, the effect of the asymmetric information spreading is absent, and for any $q>0$, the cross-correlations are strongest for $\tau=0$. On the other hand, comparing Binance and Bitstamp data, the situation resembles that of Binance vs. Kraken, i.e., Binance advances by 1 min. This can be explained in terms of the trading frequency, which is comparable between Bitstamp and Kraken, but lower than on Binance.\par

\subsubsection{Arbitrage on the cryptocurrency market}
\label{sect::TriangularArbitrage}

\begin{figure}[ht!]
\includegraphics[width=1\textwidth]{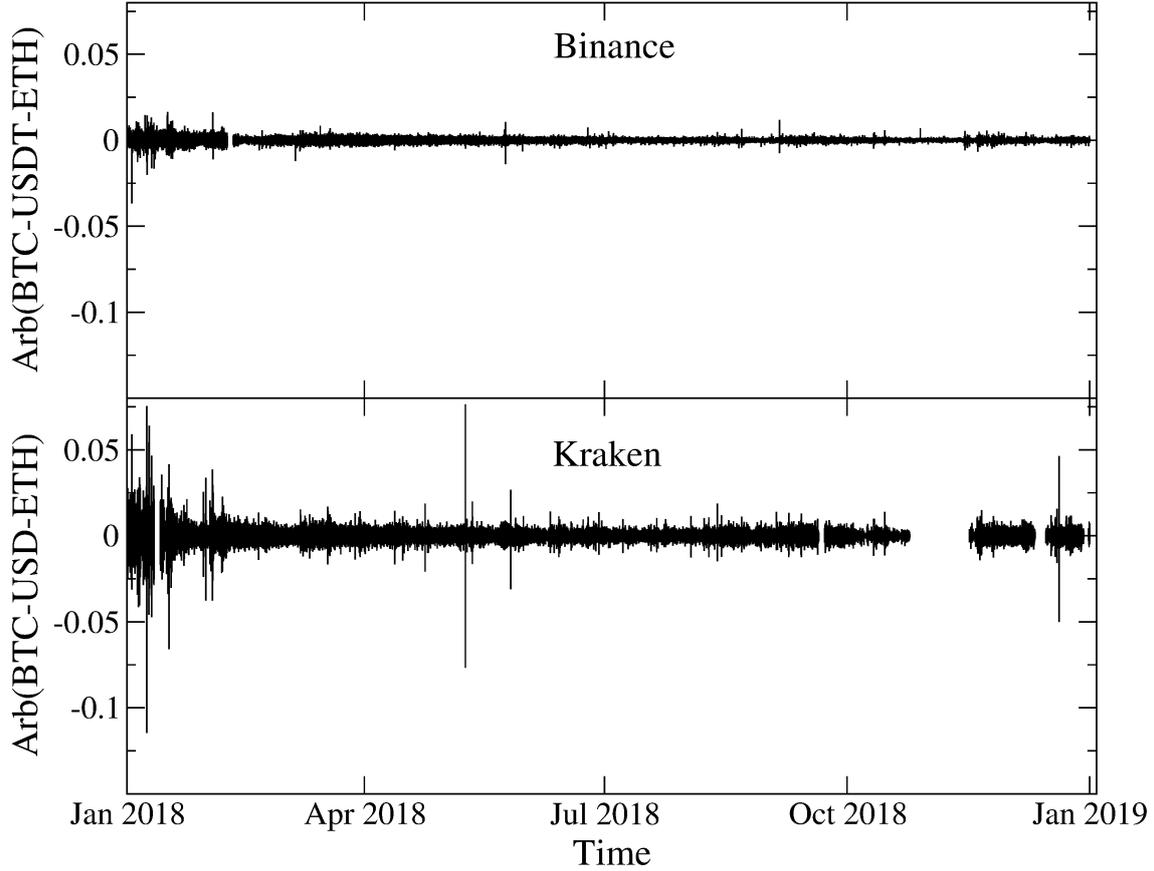} 
\caption{Arbitrage opportunity time-series for sample exchange rates forming a triangle relationship: BTC/USD$^*$, BTC/ETH, ETH/USD$^*$ from Binance (top) and Kraken (bottom). USD$^*$ stands for USDT on Binance and USD on Kraken.}
\label{fig:BTCETHUSDTarbitraz}
\end{figure}

Exchange rates on the Binance platform are characterized by higher volume and trading frequency compared to their counterparts on Kraken (see Tab.~\ref{tab:rozkladBiKR}) in Sect.~\ref{CDF}. Due to the less frequent trade, information needs more time to diffuse, directly translating into triangular arbitrage opportunities. In an analogy with Eq.(\ref{arbit_zwrotyUSD}), these are defined as an exchange rate return sum
\begin{equation}
\textrm{Arb}_{\Delta t}(\textrm{A-B-C},t)=R_{\Delta t}(\textrm{A/B},t)+R_{\Delta t}(\textrm{B/C},t)+R_{\Delta t}(\textrm{C/A},t),
\label{arbit_zwrotyAB} 
\end{equation}
where $R(\textrm{A/B},t)$ is calculated according to Eq.(\ref{arbit_zwroty}) and A, B, C are cryptocurrencies (in the case of Kraken, additionally including the USD). Time-series of the arbitrage opportunities at the return time sampling frequency $\Delta t=1$ min on Binance and Kraken (independently) for BTC/USD$^*$, BTC/ETH, and ETH/USD$^*$ are presented in Fig.~\ref{fig:BTCETHUSDTarbitraz}. The largest arbitrage opportunities occurred in parallel on both exchanges at the beginning of 2018, i.e., during a period of highest volatility, which accompanied the cryptocurrency market crash. Their magnitudes, however, were systematically larger on Kraken.\par

Tab.~\ref{tab:okazjearbitrazoweKrBi} lists the average $\langle \textrm{Arb}_{\Delta t}\rangle$ and the maximum Arb$_{\Delta t}^\textrm{max}$ values of the arbitrage opportunity for all triangles that can be created simultaneously on both platforms based on the available data (the \textit{Kraken-11} and \textit{Binance-12} datasets). Both $\langle \textrm{Arb}_{\Delta t}\rangle$ and Arb$_{\Delta t}^\textrm{max}$ are markedly lower on Binance. The smallest values of these quantities occur for a triangle including the most-traded exchange rates, namely, the BTC/USD$^*$, BTC/ETH, and ETH/USD$^*$. Triangular arbitrage opportunities for the exchange rates listed separately on Binance and Kraken are provided in the Appendix~\ref{BiKrdodatek} in Tables~\ref{tab:BiArbitall} and~\ref{tab:KrArbitall}.\par

\begin{table}[ht!]
\centering
\caption{Average and maximum triangular arbitrage opportunity on Binance and Kraken in 2018; USD$^*$ stands for USDT on Binance and USD on Kraken.}
\begin{tabular}{|c|c|c|c|c|}
\hline
     & \multicolumn{2}{c|}{\textbf{Kraken}} & \multicolumn{2}{c|}{\textbf{Binance}} \\ \hline
 \textbf{Triangle}        & \textbf{Average}         & \textbf{Max}        & \textbf{Average}          & \textbf{Max}         \\ \hline
BTC-USD$^*$-ETH & 0.0017  & 0.1147  & 0.0006   & 0.0367  \\ \hline
BTC-USD$^*$-BCH & 0.0028  & 0.4701  & 0.003   & 0.0793  \\ \hline
BTC-USD$^*$-LTC & 0.0019 & 0.2016  & 0.0009   & 0.093  \\ \hline
ETC-ETH-BTC & 0.0019  & 0.246  & 0.0014  & 0.063 \\ \hline
\end{tabular}
\label{tab:okazjearbitrazoweKrBi}
\end{table}
\begin{figure}[ht!]
\includegraphics[width=1\textwidth]{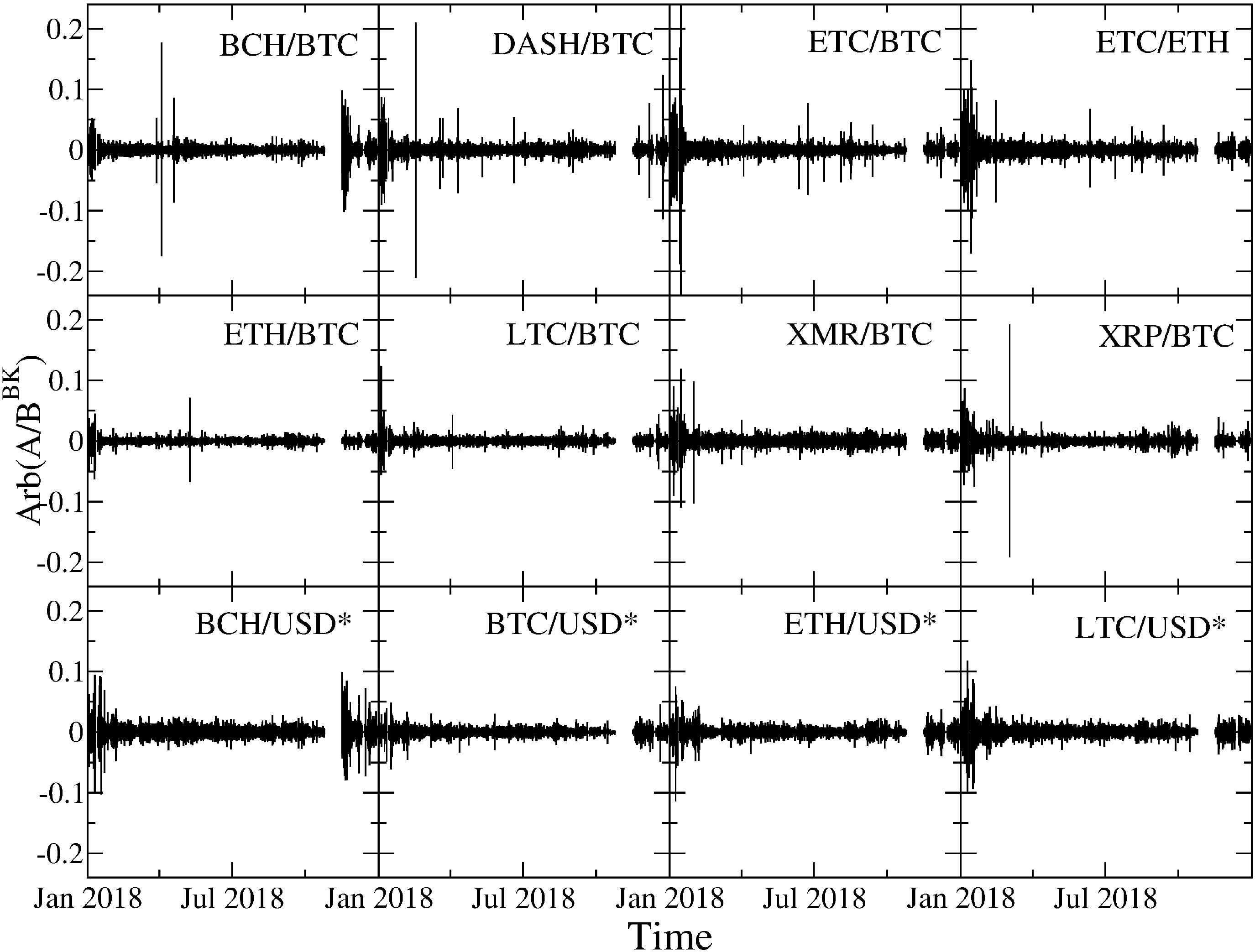}
\caption{Arbitrage opportunity time-series for the same exchange rates listed independently on Binance (B) and Kraken (K). USD$^*$ stands for USDT on Binance and USD on Kraken.}
\label{fig:BiKr_Arbit}
\end{figure}
The leading role of Binance results in the cross-platform correlations that was discussed in Sect.~\ref{sect::InterPlatform}, and is also identifiable in the arbitrage opportunity data comprising 12 exchange rates listed simultaneously on it and Kraken. In this context, time-series are defined by
\begin{equation}
\textrm{Arb}_{\Delta t}^\textrm{(BK)}(\textrm{A/B},t)=R_{\Delta t}\textrm{(A/B}^\textrm{(B)},t)-R_{\Delta t}\textrm{(A/B}^\textrm{(K)},t),
\label{arbit_zwrotyBiKr} 
\end{equation}
where $R_{\Delta t}(\textrm{A/B},t)$ is given by Eq.(\ref{arbit_zwroty}), B stands for Binance, and K for Kraken. As expected, the largest opportunities occur on the least liquid rates, namely, DASH/BTC, ETC/BTC, and ETC/ETH (Fig.~\ref{fig:BiKr_Arbit}). Conversely, on the most liquid rates such as BTC/USD$^*$, ETH/USD$^*$, and BTC/ETH, arbitrage is a rare event at the sampling frequency of $\Delta t=1$ min, and its magnitude is considerably lower.\par

It appears as though the effect of the synchronization delays experienced by the different trading platforms occurs over time-scales that are unique for the cryptocurrency market and vastly longer than the other markets, like the Forex or the stock market. This is due to the decentralization, slower information spreading, and lower liquidity of the cryptocurrency market. It translates into numerous arbitrage opportunities, both for exchange rate pairs and triangles. Also, there is no reference price of a cryptocurrency, unlike in the Forex where Reuters offers the reference rates. Similar effects were analyzed by applying the same methodology to the stock markets~\cite{oswiecimka2014} and oil markets~\cite{Watorek2019}, but the results were not as conclusive as for the cryptocurrency market.\par

To summarize this part, it can be stated that (1) the more frequently traded cryptocurrencies are, the faster is the synchronization of their fluctuations. This leads to high cross-correlation values over short time-scales and their weak dependence on $s$, which translates into a small difference in $d_\textrm{xy}(q)$. (2) The cross-correlations between the exchange rates forming a triangle are more robust, especially over time-scales on the order of a minute, compared to those not related in this way. (3) Large fluctuations are more weakly cross-correlated than medium ones. However, this difference is smaller than in the Forex, which will be addressed in Sect.~\ref{sect::Forex}. (4) The arbitrage opportunities in a triangle relationship are sparser on the frequently-traded exchange rates, and their magnitude is smaller.\par

\subsection{Multiscale cross-correlations and triangular arbitrage on Forex}
\label{sect::Forex}

\begin{figure}[!ht]
\centering
\includegraphics[width=1\textwidth]{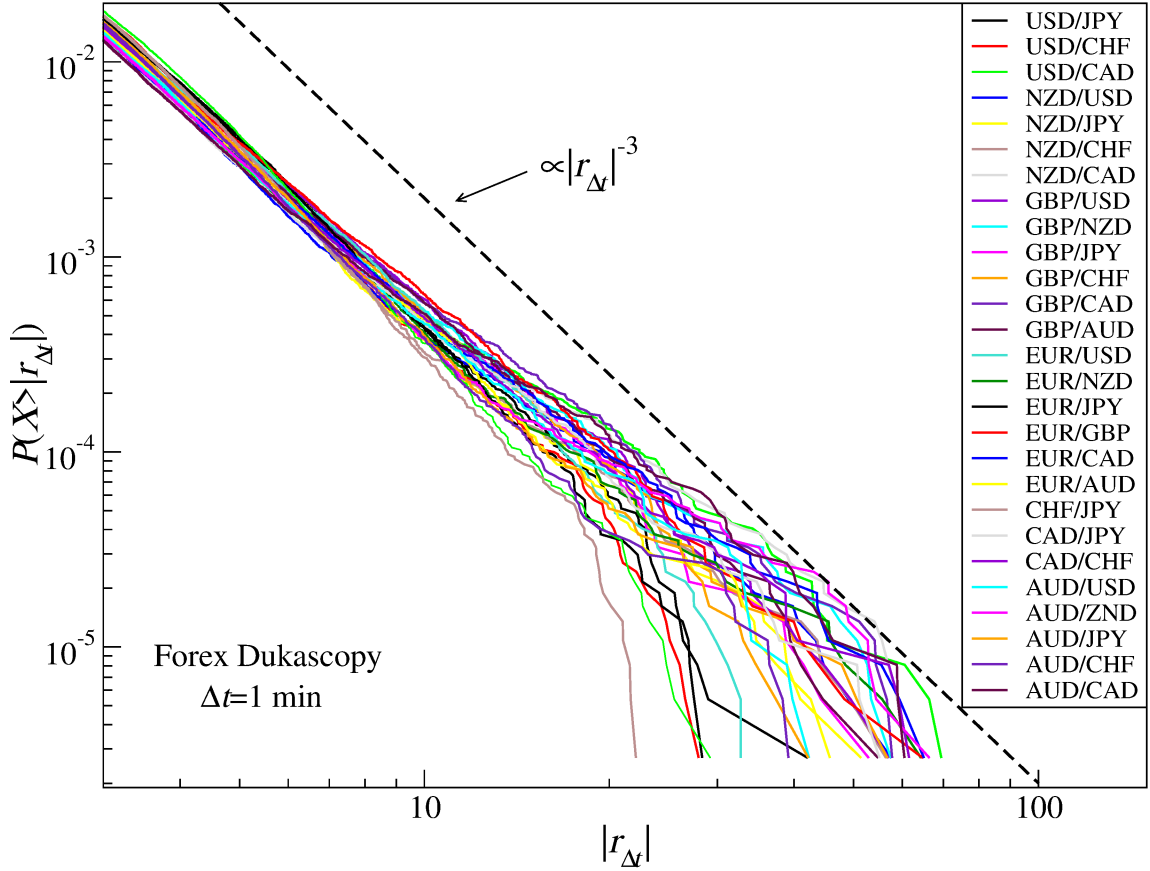}
\caption{Cumulative distributions of the absolute normalized returns $r_{\Delta t}$ with $\Delta t=1$ min representing the main currency exchange rates on the Forex in 2018. The inverse cubic law is shown as a slanted line.}
\label{fig:Forex_rozklady}
\end{figure}

The Forex differs from the cryptocurrency markets in the volume traded, transaction frequency, liquidity, and factors related to its external connections with economy and politics. Predictably, its multiscale properties are therefore different. To illustrate this, a selection of the eight most liquid major currencies are considered in this section, namely the Australian dollar (AUD), Canadian dollar (CAD), Swiss franc (CHF), Euro (EUR), British pound (GBP), Japanese yen (JPY), New Zealand dollar (NZD), and US dollar (USD). Each currency can be expressed in the remaining ones, giving a total of 28 exchange rates (all of them are explicitly listed in Figure~\ref{fig:Forex_rozklady}). The representative data were drawn from the Forex broker Dukascopy~\cite{Dukascopy}, and cover the year 2018.
As for the cryptocurrencies, the returns $R_{\Delta t}(A/B,t)$ are calculated with $\Delta t=1$ min. For most of the exchange rates, the inverse cubic law is met by the return CDFs when $\gamma>2.5$ (Fig.~\ref{fig:Forex_rozklady}). Therefore, the bivariate fluctuation function $F_\textrm{xy}(q,s)$ may be calculated for $q\in [-4,4]$ without concerns about moment divergence.\par

\begin{figure}[!ht]
\centering
\includegraphics[width=1\textwidth]{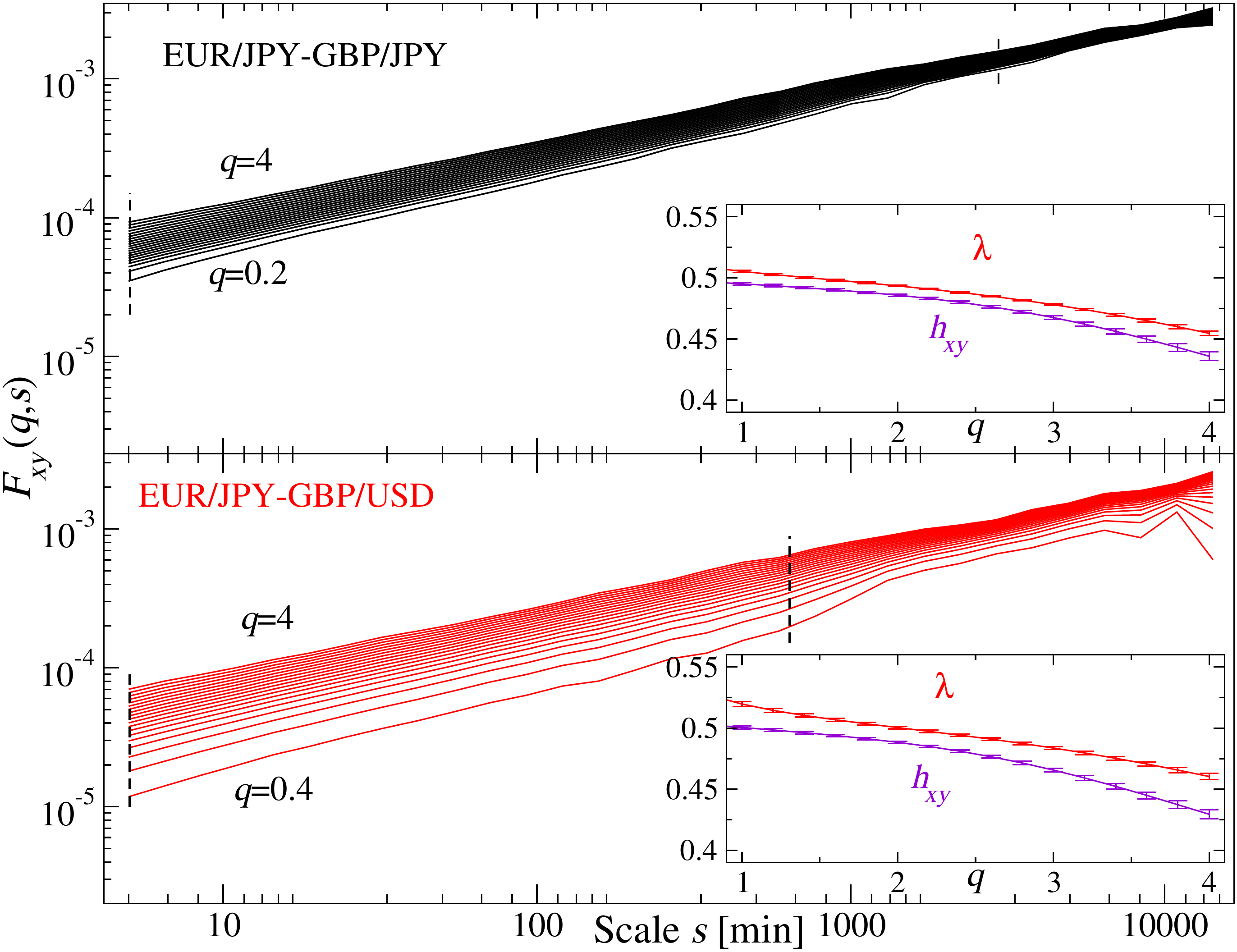}
\caption{(Main) Bivariate fluctuation function $F_\textrm{xy}(q,s)$ for $q>0$ ($\Delta q=0.2$) calculated for the exchange rate pairs EUR/JPY--GBP/JPY (a triangle relation, top) and EUR/JPY--GBP/USD (no triangle, bottom). The inset shows the scaling exponent $\lambda(q)$ and average generalized Hurst exponent $h_\textrm{xy}(q)$ calculated over the scale range marked with dashed lines.}
\label{fig:Fx_przyklad}
\end{figure}
\begin{figure}[ht!]
\includegraphics[width=1\textwidth]{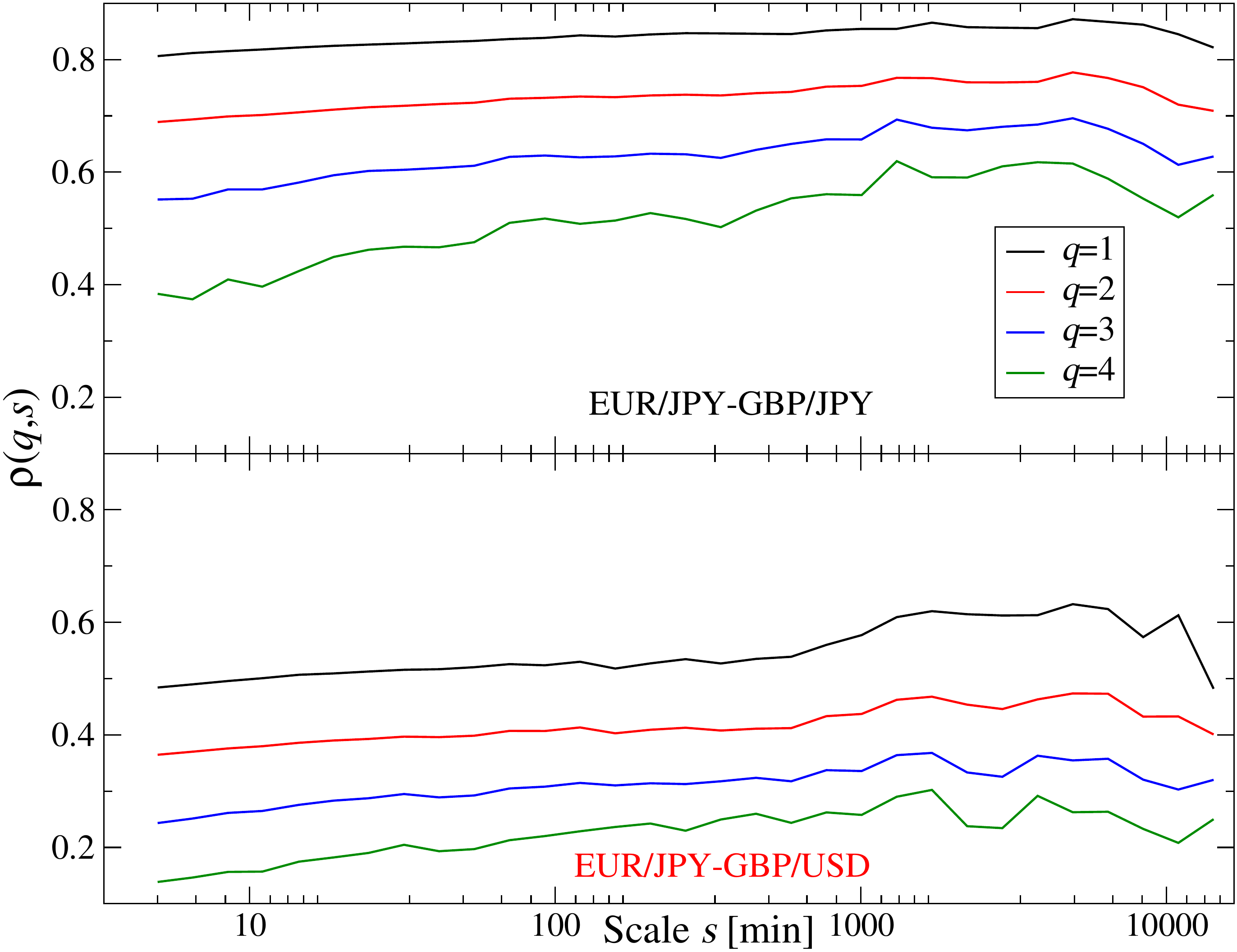} 
\caption{Detrended cross-correlation coefficient $\rho(q,s)$ for the exchange rate pairs connected by a triangle relation (EUR/JPY--GBP/JPY, top) and those not forming a triangle (EUR/JPY--GBP/USD, bottom). Compare $\rho(q,s)$ in the cryptocurrency market in Fig.~\ref{fig:Kraken_pq_ex} and Fig.~\ref{fig:Binance_pq_ex}.}
\label{fig:FxPq}
\end{figure}

The bivariate fluctuation function $F_\textrm{xy}(q,s)$ reveals better scaling for the triangle-related exchange rates (i.e., a common base currency, top panel), which translates into a smaller distance between $\lambda(q)$ and $h_{xy}(q)$ compared to the exchange rates without a triangle relation; Fig.~\ref{fig:Fx_przyklad} illustrates this based on two sample cases (i.e., a triangle relation - top, without triangle - bottom). In both cases, the difference $d_\textrm{xy}(q)$ is maximal for $q=4$, unlike the cryptocurrencies, for which the largest $d_\textrm{xy}(q)$ arises for $q=1$ (Fig.~\ref{fig:Kraken10s_lambda_hsrednie}). In general, this difference is smaller on the Forex than the cryptocurrency markets, since the cross-correlation strength $\rho(q,s)$ depends less markedly on $s$ (Fig.~\ref{fig:FxPq}). This effect can be attributed to the larger liquidity of the Forex exchange rates because, on the cryptocurrency platforms, $\rho(q,s)$ does not depend on the scale for the most liquid pair, BTC and ETH, expressed in a common base currency. The most pronounced difference between cryptocurrencies and fiat currencies is the much higher values of $\rho(q,s)$ for the Forex exchange rates without a triangle relationship. In the case of cryptocurrencies, $\rho(q,s)$ approaches zero on the shortest scales for the out-of-a-triangle rates. On the Forex, $\rho(q,s)$ is also lower in this case than for triangle-related exchange rates (Fig~\ref{fig:FxPq}), but not as substantially as in the cryptocurrency market. Large returns are, in general, considerably less cross-correlated on the Forex than on the cryptocurrency market: namely, there is a substantial dependence of $\rho(q,s)$ on $q$ for both the in-triangle and out-of-triangle exchange rates (Fig.~\ref{fig:FxPq}).\par

\begin{figure}
\centering
\includegraphics[scale=0.499]{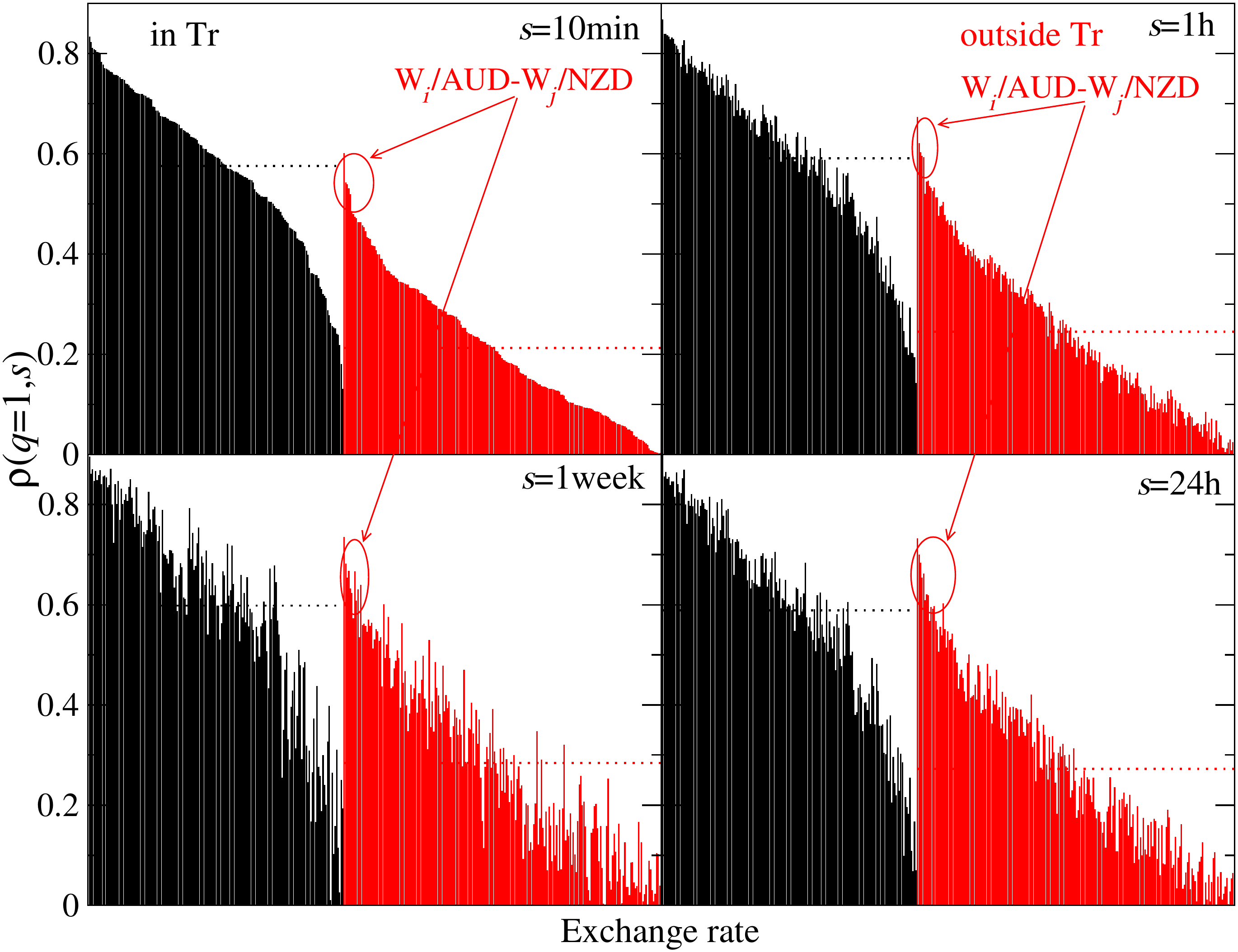}\\
\includegraphics[scale=0.499]{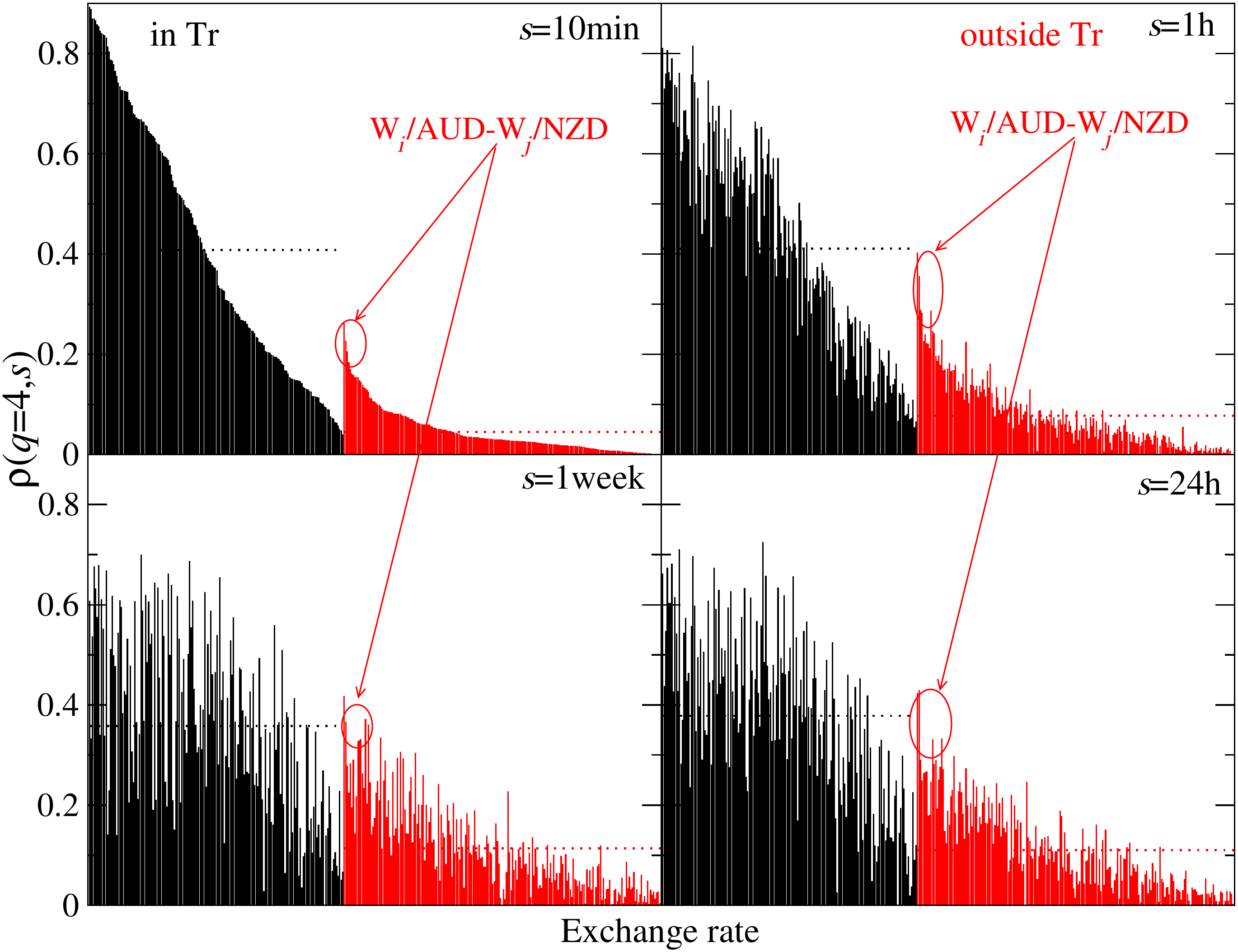}
\caption{Absolute detrended cross-correlation coefficient $|\rho(q,s)|$ calculated for over exchange rate pairs for $q=1$ (top 4 panels) and $q=4$ (bottom 4 panels). Sample time-scales are shown: $s=10$ min, $s=1$ h, $s=1$ day, and $s=1$ week. Exchange rate pairs connected by a triangle relationship (in Tr, black) and without such a relation (outside Tr, red) are distinguished. Horizontal line in each panel indicates the average value of  $|\rho(q,s)|$ calculated for each group separately. Values of $|\rho(q,s)|$ are sorted descending for $s=10$ min and these locations are preserved in the remaining 3 panels. The following pairs are distinguished: W$_i$/AUD--W$_j$/NZD, where W$_i$, W$_j=$ CAD, CHF, EUR, GBP, JPY, and USD, W$_i \ne$ W$_j$.}
\label{fig:Fx_Pq_pos}
\end{figure}

Fig.~\ref{fig:Fx_Pq_pos} shows $|\rho(q,s)|$ for all exchange rates under consideration over four scales, namely $s=10$ min, $s=1$ h, $s=1$ day, and $s=1$ week. For $q=1$, the coefficient $|\rho(q,s)|$ only weakly depends on scale, while such a dependence is more pronounced for $q=4$. The exchange rates in a triangle relation are considerably more strongly cross-correlated than those not related in this way (black vs. red plots). Since the AUD and NZD represent countries whose economies are greatly coupled, the exchange rate pairs of the form W$_i$/AUD--W$_j$/NZD, where W$_i$, W$_j$ denote all the other currencies, behave effectively as if W$_i$ and W$_j$ were expressed in a common base currency, which significantly amplifies the corresponding $|\rho(q,s)|$.\par

For $q=4$, there is a large difference in the cross-correlation strength between the in-triangle and the outside-triangle exchange rate pairs, at both the individual and average levels; $|\rho(q=4,s)|$ is also smaller than for $q=1$. Furthermore, a moderate scale dependence can be appreciated for the outside-of-triangle exchange rates. The shifts in the ranked positions of $|\rho(q,s)|$ with increasing $s$ are substantial, but this does not lead to an increase in the average cross-correlation levels (horizontal lines in Fig.~\ref{fig:Fx_Pq_pos}), unlike the cryptocurrencies showing $\rho(q,s) \approx 0$ for $s=10$ min (see Fig.~\ref{fig:KrakenPqpos} and Fig.~\ref{fig:BinancePqpos} in Sect.~\ref{sect::IntraPlatform}). A difference in trading frequency between the cryptocurrencies and the fiat currencies can plausibly account for this observation.\par

\begin{figure}[!ht]
\centering
\includegraphics[width=1\textwidth]{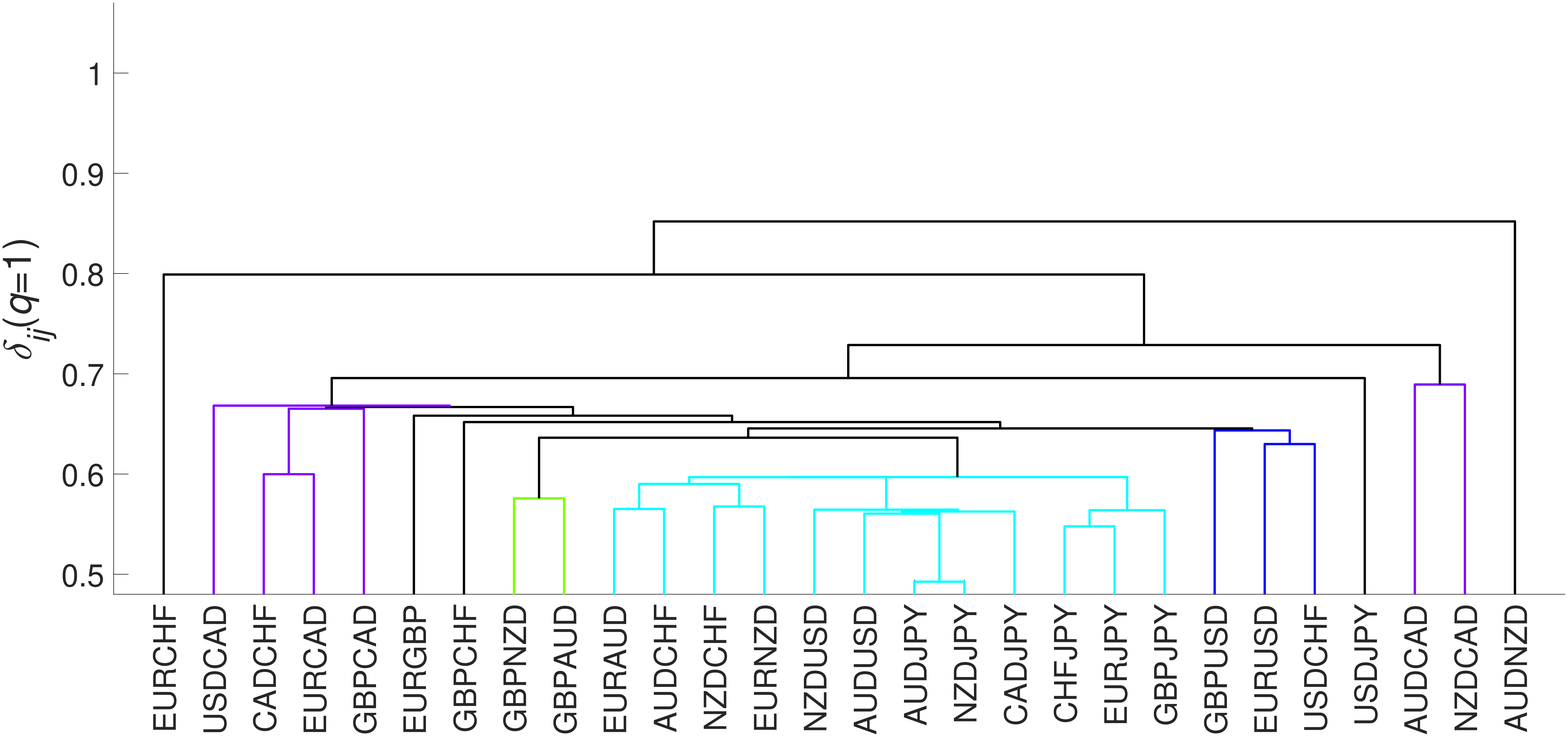}\\
\includegraphics[width=1\textwidth]{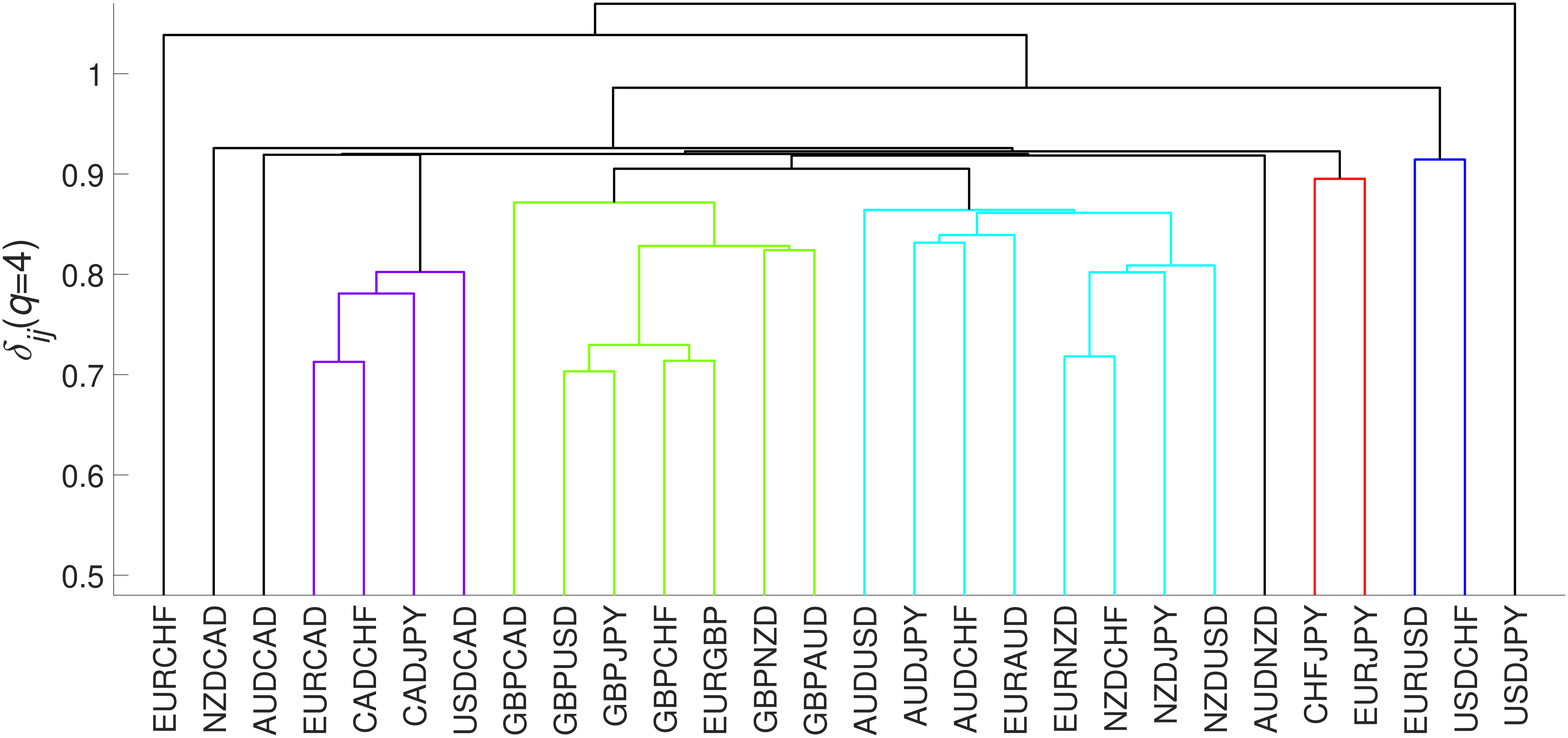}
\caption{Dendrograms for 28 exchange rates created with the hierarchical clustering algorithm from the average detrended cross-correlation coefficient $|\overline{\rho}(q)|$ for $q=1$ (top) and $q=4$ (bottom).}
\label{fig:Fx_Htreeq}
\end{figure}

A cross-correlation hierarchy of the 28 exchange rates calculated with the hierarchical clustering algorithm from $\overline{\rho}(q) = {1 \over s} |\sum_{s=s_\textrm{min}}^{s_\textrm{max}} \rho(q,s)|$ shows that the large fluctuation cross-correlations ($q=4$) form more pronounced clusters consisting of the exchange rates involving a common base currency, namely CAD, GBP, AUD, and NZD (Fig.~\ref{fig:Fx_Htreeq}). This arises because each independent movement of a base currency with respect to the other currencies contributes to their evolution, which can be viewed as their cross-correlation insofar as they are expressed in this particular base currency. The effect is most influential during the extreme volatility periods that typically accompany important economic and political events~\cite{gebarowski2019}. For typical fluctuations ($q=1$), the subdivision into clusters is weaker. The most numerous cluster encompasses exchange rates not related by a triangle relation, namely, EUR/AUD--NZD/CHF and AUD/JPY--NZD/USD. This indicates that, for a sampling interval of $\Delta t=1$ min, the Forex is largely synchronized, and information is transmitted quickly, even outside triangle relations.\par

\begin{figure}[!ht]
\centering
\includegraphics[width=1\textwidth]{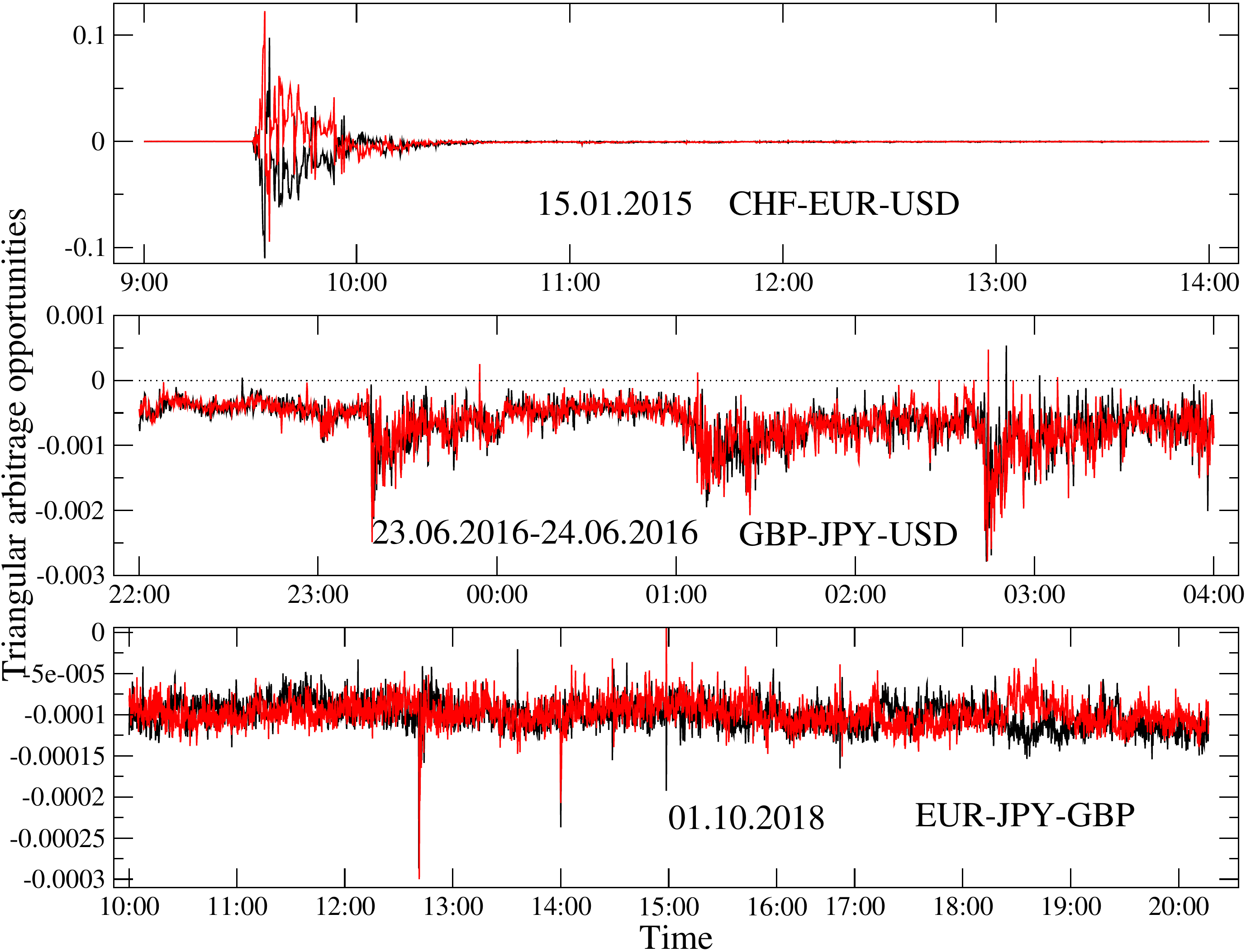}
\caption{Triangular arbitrage opportunity time-series for CHF-EUR-USD on Jan 15, 2015 (the CHF/EUR fluctuation limit lifted, top), GBP-JPY-USD on Jun 23-24, 2016 (the Brexit referendum, middle), and EUR-JPY-GBP on Oct 1, 2018 (a sample day without turmoil, bottom). Unity was subtracted from each signal. Because of the bid and ask price spread, an arbitrage opportunity arises when one of two relations (black or red) is above zero.}
\label{fig:Fx_arbitrage}
\end{figure}
\begin{figure}[!ht]
\centering
\includegraphics[width=1\textwidth]{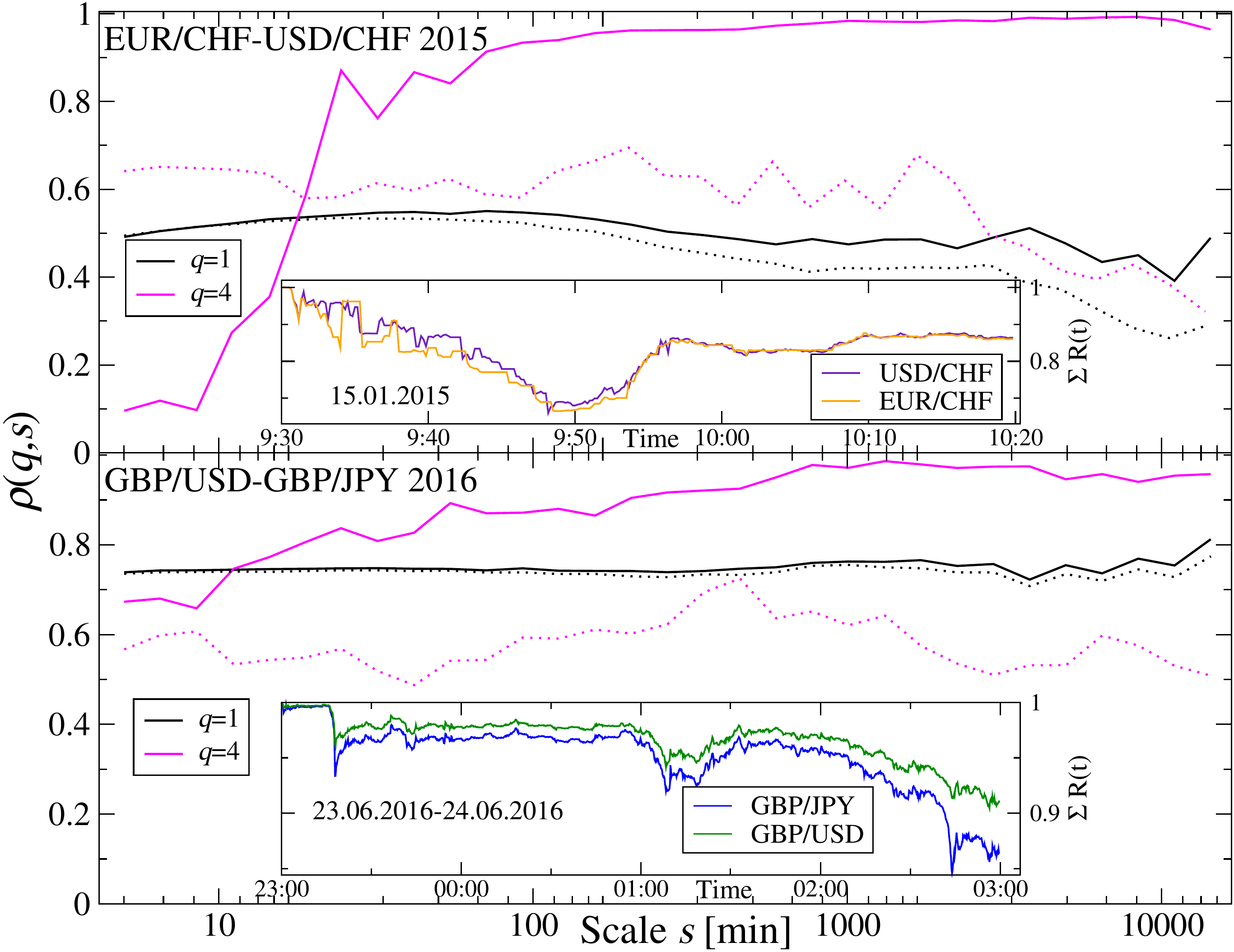}
\caption{(Main) Detrended cross-correlation coefficient $\rho(q,s)$ (solid line) calculated for EUR/CHF--USD/CHF in 2015 (top) and GBP/USD--GBP/JPY in 2016 (bottom). Dashed lines mark $\rho(q,s)$ calculated after removing the volatile periods shown in insets. (Inset) Triangle-related exchange rates expressed as cumulative returns $R_{\Delta t}$ for $\Delta t=1$ min (starting from unity) on days of extreme volatility: Jan 15, 2015 (the SNB intervention) and Jun 23-24, 2016 (the Brexit referendum).}
\label{fig:Fx_extreme_events}
\end{figure}
Together with the higher cross-correlation levels over the shortest scales $s$, this is associated with a higher liquidity and trading frequency on the Forex. An associated effect is a much rarer occurrence of the triangular arbitrage opportunities. In Fig.~\ref{fig:Fx_arbitrage}, deviations from a triangle relation calculated for the bid and ask prices in analogy to Eq.(\ref{arbitEUR1}) and Eq.(\ref{arbitEUR2}) are shown for three triangles: CHF-EUR-USD, GBP-JPY-USD, and EUR-JPY-GBP during time-intervals of particular market instability, namely, stemming from the Swiss National Bank's policy change on EUR/CHF, and the Brexit referendum (on the Forex, an arbitrage opportunity can occur only during elevated volatility periods~\cite{gebarowski2019}). Periods of maximum volatility are presented in detail in Fig.~\ref{fig:Fx_extreme_events} (insets). During them, the individual exchange rates involving the CHF and GBP changed so rapidly that the others could not keep up, allowing the formation of arbitrage possibilities on the triangle-related exchange rates. For comparison, a day without any volatility-amplifying event from 2018 is also shown in Fig.~\ref{fig:Fx_arbitrage}. The main plots in Fig.~\ref{fig:Fx_extreme_events} display $\rho(q,s)$ calculated for the above-mentioned exchange rate pairs with $q=1$ and $q=4$. The latter case shows that large returns during the turmoil periods are responsible for significant decorrelation of the exchange rates, whereas the average returns remain cross-correlated as customary. If these periods are removed, the arbitrage opportunities disappear, and the cross-correlation strength increases.\par

These properties of the Forex, including the paucity of arbitrage opportunities, more frequent trading, and shorter time-scales at which market synchronization occurs, motivate the view that the cryptocurrency market remains a markedly less mature market by comparison. Among the cryptocurrencies, only the BTC and ETH, when expressed in USD (or its equivalent USDT) on the most significant trading platforms, reveal signatures similar to those typical of exchange rates on the Forex.

\subsection{COVID-19 impact on multiscale cross-correlations between the cryptocurrency market and traditional markets}
\label{duka_krypto}

\begin{figure}[ht!]
\centering
\includegraphics[width=1\textwidth]{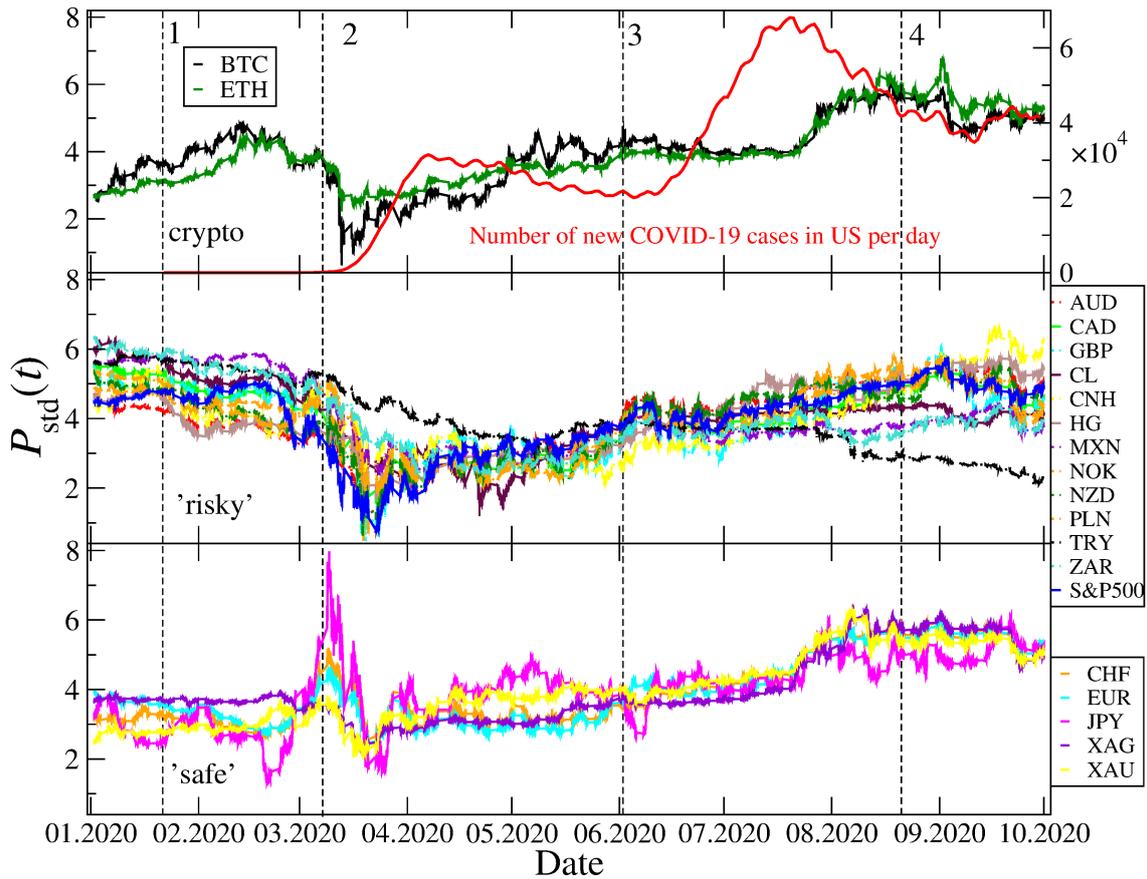}
\caption{Standardized price/value of BTC and ETH (top), AUD, CAD, GBP, CL, CNH, HG, MXN, NOK, NZD, PLN, TRY, ZAR, and S\&P500 (``risky'' assets, middle), CHF, EUR, JPY, XAG, and XAU (``safe'' assets, bottom) since Jan 1, 2020. For comparison, the daily new cases of Covid-19 in US are also shown (red line, top). Vertical dashed lines mark the events related to Covid-19 that caused significant detrended cross-correlations between the cryptocurrencies and other financial instruments: (1) the first identified case of Covid-19 in the United States (Jan 21, 2020), (2) global market panic (March 2020), (3) the 2nd wave of the US pandemic (June 2020), (4) the pandemic slowdown in the United States, all time highs on S\&P500 and Nasdaq (Aug 2020).}
\label{fig::normprice}
\end{figure}

Till beginning of 2020, it was generally believed that, viewed by many as a ''safe haven'', the cryptocurrency market was rather detached from the other financial markets and exhibited its individual dynamics~\cite{ji2018,urquhart2019,shahzad2019,wang2019,Conlon2020,shahzad2019a}. Whether this is still a valid hypothesis can be verified by calculating the generalized detrended cross-correlation coefficient $\rho(q,s)$ given by Eq.(\ref{rhoq}) for time series of 1-min log-returns of BTC/USD and ETH/USD and 20 conventional assets, among which are 13 traditional currencies: Australian dollar (AUD), euro (EUR), British pound (GBP), New Zealand dollar (NZD), Canadian dollar (CAD), Swiss franc (CHF), Chinese offshore yuan (CNH), Japanese yen (JPY), Mexican peso (MXN), Norwegian krone (NOK), Polish zloty (PLN), Turkish lira (TRY), and South African rand (ZAR), 3 stock market indices: Dow Jones Industrial Average (DJIA), NASDAQ100, S\&P500, and 4 commodities: gold (XAU), crude oil (CL), silver (XAG), and copper (HG), all expressed in US dollar. Fig.~\ref{fig::normprice} shows temporal evolution of their standardized prices or values
\begin{equation}
P_{\rm std}(t)={P(t)-\langle P(t) \rangle \over \sigma_{P}}+4
\end{equation}
between Jan 2018 and Sep 2020 (Monday-Friday). This interval includes a market panic related to the Covid-19 pandemic outburst~\cite{estrada2020}. The data comes from Dukascopy~\cite{Dukascopy}.

The analysis was carried out for several different time scales $s$ but, for the sake of clarity, only two scales are considered here: the shortest available scale $s=10$ min and the scale corresponding roughly to a trading day in the NYSE stock market $s=360$ min (an actual trading day lasts for 390 min). For the return-magnitude filtering parameter $q$ also two values are discussed here: $q=1$, which does not favor any specific magnitude range, and $q=4$, which amplifies the large return contribution.

\begin{figure}[ht!]
\centering
\includegraphics[width=1\textwidth]{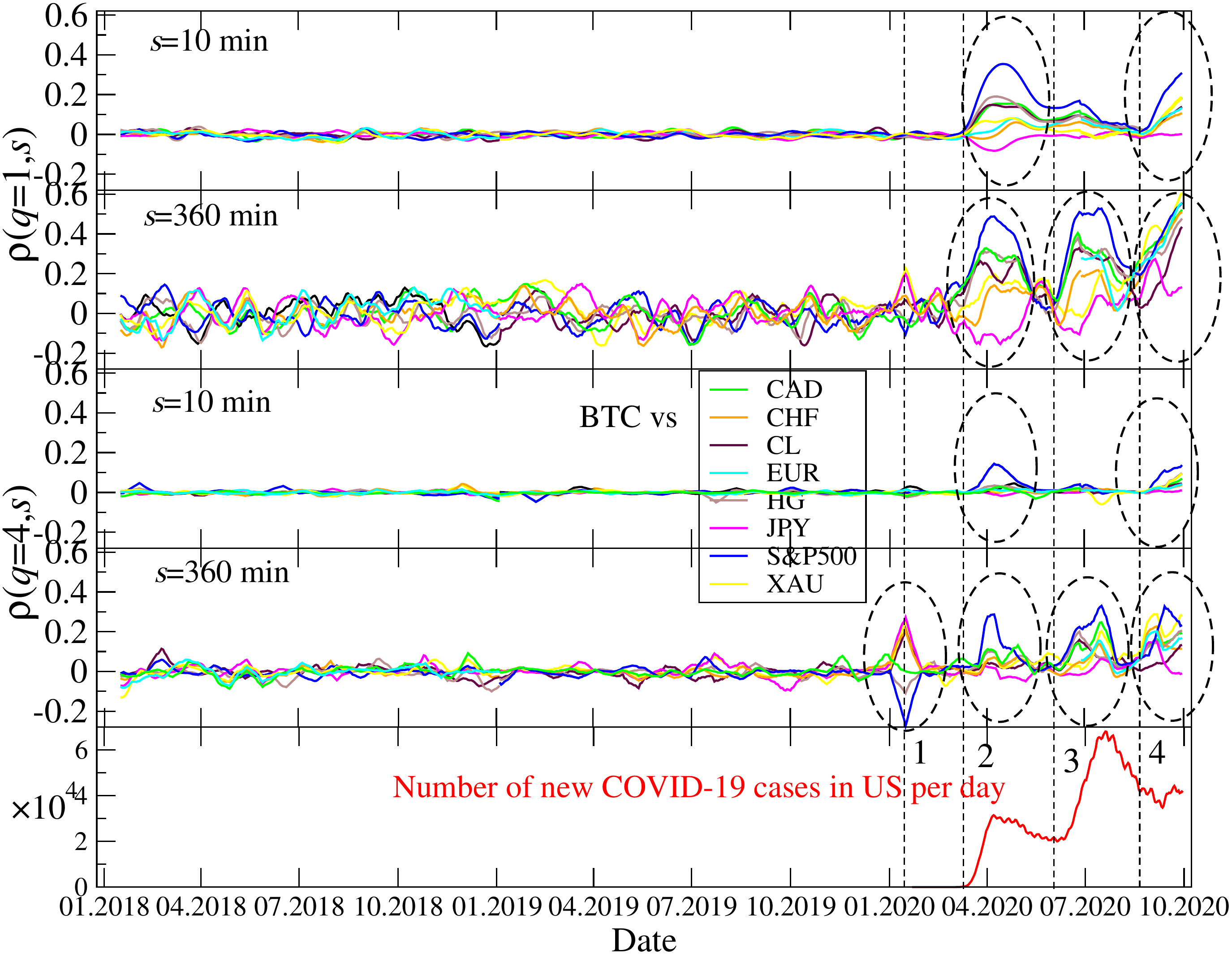}
\caption{Evolution of the generalized detrended cross-correlation coefficient $\rho(q,s)$ calculated in 10-day-wide moving windows with a 1-day step for the returns of the BTC/USD exchange rate and the selected conventional assets expressed in US dollar: CAD, CHF, CL, EUR, HG, JPY, S\&P500, and XAU. $q=1$ does not favour any specific magnitude range (top and upper middle), while $q=4$ amplifies the large return contribution (middle-middle and lower middle). $s=10$ min is the shortest available time scale in this analysis (top and middle-middle), while $s=360$ min is approx. a trading day in the NYSE stock market (upper middle and lower middle). Periods with the statistically significant values of $\rho(q,s)$ are marked with dashed ellipses (see Fig.~\ref{fig::normprice} for their description.}
\label{fig::rhoq.BTC}
\end{figure}

\begin{figure}[ht!]
\centering
\includegraphics[width=1\textwidth]{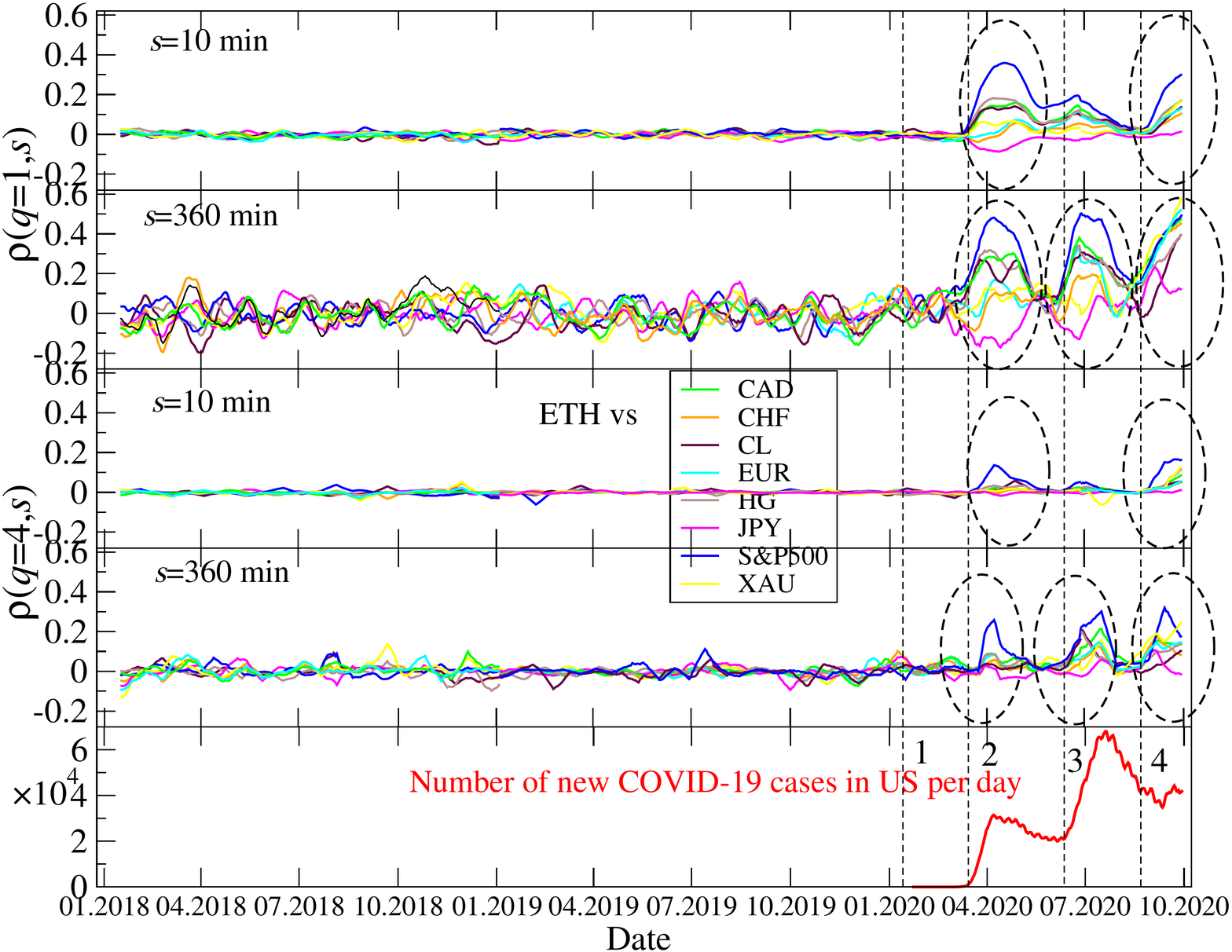}
\caption{Evolution of the generalized detrended cross-correlation coefficient $\rho(q,s)$ calculated in 10-day-wide moving windows with a 1-day step for the returns of the ETH/USD exchange rate and the selected conventional assets expressed in US dollar: CAD, CHF, CL, EUR, HG, JPY, S\&P500, and XAU. See caption to Fig.~\ref{fig::rhoq.BTC} for more details.}
\label{fig::rhoq.ETH}
\end{figure}

Figs.~\ref{fig::rhoq.BTC} and~\ref{fig::rhoq.ETH} display $\rho(q,s)$ for a combination of the $s$ and $q$ values described above. In each panel $\rho(q,s)$ for BTC/USD (Fig.~\ref{fig::rhoq.BTC}) or ETH/USD (Fig.~\ref{fig::rhoq.ETH}) and each of 8 other assets: CAD, CHF, CL, EUR, HG, JPY, S\&P500, and XAU are shown. Some of the assets listed in the beginning of this section have been omitted since their cross-correlation with other assets is similar to one of the assets that are shown: NASDAQ100 and DJIA are similar to S\&P500, XAG is similar to XAU, the currencies: AUD, NZD, ZAR, CHN, MXN, EUR, GBP, NOK, and PLN are similar to CAD. On the other hand, TRY has been omitted because for almost all the time it was uncorrelated with BTC and ETH.

In accordance with the initial hypothesis, the cryptocurrency market and the traditional assets used to reveal no significant cross-correlation in 2018 and 2019. However, this situation altered in 2020 with 4 specific periods of the statistically significant cross-correlations that can be distinguished. All these periods are related to the Covid-19 pandemic. The first one (see the corresponding number in in Figs.~\ref{fig::normprice}--\ref{fig::rhoq.ETH}) was associated with a drop of S\&P500 and the other US stock market indices triggered by the first identified local case of Covid-19 on Jan 21, 2020. During this time BTC continued its upward trend, which led to negative $\rho(q,s)$ for BTC and S\&P500 for $\rho(q=4,s=360{\rm min})$. At that time, BTC was also positively correlated with some other financial instruments considered safe: JPY, CHF, and gold. Interestingly, the cross-correlation between ETH and the other assets were still insignificant. One could argue that, at that initial Covid-19 phase, BTC fulfilled its planned role as a hedge for risky assets.

This statement is no longer true since the Covid-19-related market panic of March 2020, when almost all financial assets expressed in USD, including the cryptocurrencies, sharply lost their value. The positive cross-correlation between BTC and the traditional assets were the most evident during that time. $\rho(q,s)$ was larger for $s=360$ min than for $s=10$ min, but nevertheless both were significant. Typical for the cryptocurrency market as it has been reported in Sect.~\ref{sect::CrossCorrelations}, such behaviour can be explained by the fact that the cross-correlations need time to build up. In the case of large returns ($q=4$), BTC was positively cross-correlated with S\&P500 mainly, while it was more independent in respect to the other assets. However, generally ($q=1$) BTC is positively cross-correlated also with copper, crude oil, gold, and CAD. There was no substantial cross-correlation between BTC and CHF or EUR, and $\rho(q,s)$ was negative for BTC vs. JPY. From this point of view, BTC became a risky asset. In March 2020 one could observe similar behaviour of $\rho(q,s)$ also in the case of ETH, because the global financial market panic did not spare any cryptocurrency.

After transient restoration of the financial markets in April and May, the fear returned in June 2020 with the beginning of the 2nd Covid-19 wave (period 3 in Figs.~\ref{fig::normprice}--\ref{fig::rhoq.ETH}). It was accompanied by strong positive cross-correlation of BTC and ETH with the traditional assets except JPY that was observed mostly for $s=360$ min. $\rho(q,s)$ was smaller for $s=10$ min, but still positive and significant for $q=1$. What is even more interesting is the behaviour of the detrended cross-correlations in the 4th considered period that started in late Aug 2020 after Covid-19 slowing down in the United States. Since that point BTC and ETH have been positively correlated with all the considered traditional assets, even with JPY, on all considered time scales. BTC, ETH, and the other assets expressed in USD gained value; S\&P500 and NASDAQ100 even reached their all-time highs. This surprisingly collective behaviour of the considered financial instruments may be connected with weakness of USD, in which they are expressed. This is visible especially for EUR, which is now one of the most strongly correlated assets with BTC and ETH for $s=360$ min, while it was rather weakly correlated before.

It seems that the events connected with Covid-19 triggered the emergence of cross-correlations between the major cryptocurrencies and the traditional markets. These cross-correlations occurred not only during the sharp market falls earlier in 2020, but also during the still lasting recovery phase. Therefore it may be concluded that the cryptocurrencies have ceased to be an island detached from the traditional markets and become a connected part of the world's financial markets. This can be a sign of positive verification of the cryptocurrencies during one of the most turbulent periods in the recent years. However, this observation is still fresh, therefore a question of whether it is transient or stable remains open. Especially in the context of a highly inflationary policy adopted by the Federal Reserve and other central banks, which is about substantial increasing of the monetary base (Fig.~\ref{fig:monetarybase}).

\section{Correlation matrix and network analysis of cross-correlations}
\label{Matrixcorr}

\begin{figure}[!ht]
\centering
\includegraphics[width=1\textwidth]{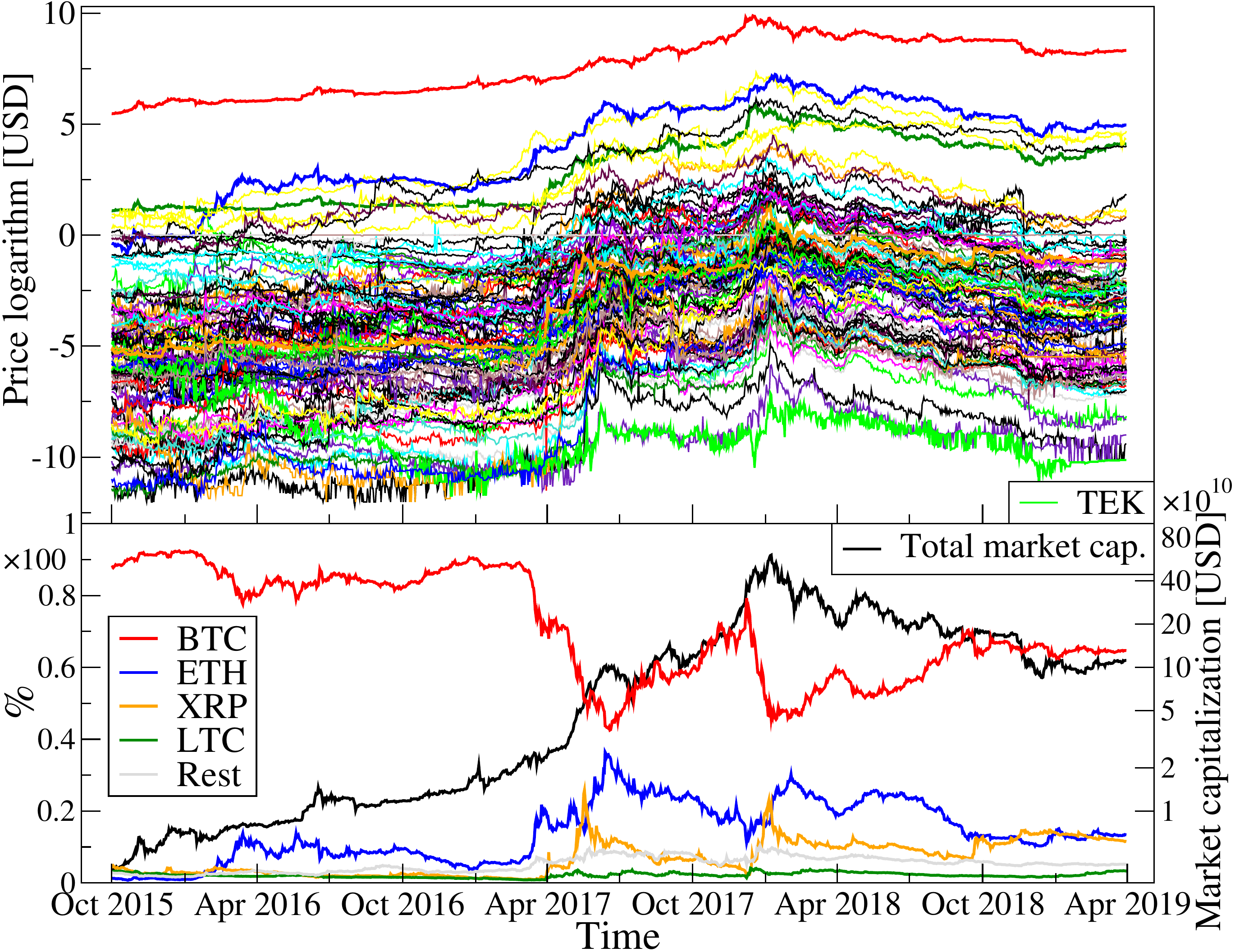}
\caption{(Top) Cryptocurrency price evolution over the period 10/2015--03/2019; 100 cryptocurrencies with the largest market capitalization are shown. (Bottom) Total market capitalization share of BTC, ETH, XRP, LTC, and the remaining cryptocurrencies.}
\label{fig:dailycena}
\end{figure}

This section discuss a multivariate approach to cross-correlations among the cryptocurrency prices. A data set consists of $N=100$ cryptocurrency daily quotes expressed in USD covering the period 10/2015-03/2019. They form a set of time series of length $T=1278$. A complete list of cryptocurrency tickers and names can be found in the Appendix~\ref{listamac}. The data source is CoinMarketCap website~\cite{coinmarket} that collects quotations from all active cryptocurrency trading platforms (273 in March 2019) and evaluates the cryptocurrency prices on an ongoing basis. Prices of the same cryptocurrency from different trading platforms are weighted by their transaction volume with a larger volume corresponding to a larger weight. Fig.~\ref{fig:dailycena} shows the individual cryptocurrency price logarithm time series (top panel) and cryptocurrency capitalization (bottom panel), defined as a product of price and the number of issued cryptocurrency units.

By the turn of 2017 and 2018, a systematic increase in the entire market capitalization can be observed. It was followed by a bear market continuing throughout 2018. There were also changes in the individual cryptocurrency market share: by March 2017, BTC capitalization accounted for over $80\%$ of the total cryptocurrency market value, then it fell to around $40\%$ in May 2017, only to rise again to $80\%$ in December 2017. The next two cryptocurrencies in terms of their capitalization, ETH and XRP, moved in the opposite direction. The market share of ETH increased in the first half of 2017, which was related to the ICO-mania discussed earlier. The ICO offers at that time were released on the Ethereum platform mainly and that raised the ETH valuation. In May 2017, the shares of BTC and ETH in the entire market capitalization differed by just a few per cent. The second period of growing ETH importance fell on tip of the ICO-bull market: ETH peaked in January 2018, almost a month after the BTC peak. Fig.~\ref{fig:dailycena} presents also a market share of the 4th largest cryptocurrency: LTC. The remaining cryptocurrencies did not account for more than $10\%$ of the entire market capitalization. All the cryptocurrencies considered here account for $95\%$ of the market capitalization.

\subsection{Correlation matrix}
\label{corrM}

The analysis starts from a set of time series of logarithmic returns $R_{\Delta t}(\textrm{K}_j/\textrm{X},t_l) = \log P (\textrm{K}_j/\textrm{X},t_l + \Delta t) - \log P(\textrm{K}_j/\textrm{X},t_l)$, with K$_j$ denoting cryptocurrencies ($j=1,...,N$), X denoting either a fiat currency (USD) or a cryptocurrency, $\Delta t$ being sampling interval, and $l=1,...,T$ indexing consecutive sampling moments. The log-returns can be normalized
\begin{equation}
r_{\Delta t} (\textrm{K}_j/\textrm{X},t_l) = {R_{\Delta t} (\textrm{K}_j/\textrm{X},t_l)-\langle R_{\Delta t}(\textrm{K}_j/\textrm{X},t_l) \rangle_l \over \sigma(R_{\Delta t}(\textrm{K}_j/\textrm{X}))},
\label{g}
\end{equation}
where $\sigma(\cdot)$ is standard deviation of $R_{\Delta t}(\textrm{K}_j/\textrm{X},t_l)$. Correlation matrix for a base (crypto)currency X is given by
\begin{equation}
{\bold C}^\textrm{(X)} = {1 \over T} {\bold M}^{(\textrm{X})} {\tilde {\bold M}}^{(\textrm{X})},
\label{C}
\end{equation}
where ${\bold M}^{(\textrm{X})}$ is a data matrix of size $(N-1) \times T$ and $\sim$ denotes matrix transpose. A separate matrix can be constructed for each of $N$ available base (crypto)currencies X. Each element of $\textbf{C}^{(\textrm{X})}$ is a Pearson correlation coefficient for a pair of exchange rates K$_{j_1}$/X--K$_{j_2}$/X with $j_1 \neq j_2$. As a real symmetric matrix, $\textbf{C}^{(\textrm{X})}$ has $N-1$ real eigenvalues $\lambda^{(X)}_i$ and the same number of the associated eigenvectors ${\bf v}^{(X)}_i$:
\begin{equation}
{\bf C}^{(\textrm{X})} {\bf v}_i^{(\textrm{X})} = \lambda^{(\textrm{X})}_i {\bf v}_i^{(\textrm{X})},
\label{eigenequation}
\end{equation}
where the eigenvalues are numbered in the ascending order: $0 \le \lambda^{(\textrm{X})}_{1} \le \lambda^{(\textrm{X})}_{2} \le$...$\le \lambda^{(\textrm{X})}_{(N-1)}$ and $\textrm{Tr} {\bf C}^{(\textrm{X})} = N-1$. Through their components $v^{(\textrm{X})}_{ij}$, the eigenvectors ${\bf v}^{(\textrm{X})}_i$ can be associated with eigensignals $z^{(\textrm{X})}_i(t)$ that are linear combinations of the original return time series:
\begin{equation}
z^{(\textrm{X})}_i(t)= \sum_{j=1}^{N-1} v^{(\textrm{X})}_{ij} r_{\Delta t}(\textrm{K}_j/\textrm{X},t).
\label{eigensignal}
\end{equation}
These eigensignals may be associated with respective orthogonal portfolios~\cite{markowitz}.

A convenient reference for the empirical correlation matrices ${\bf C}^{(\textrm{X})}$ is the Wishart ensemble of random matrices~\cite{randommatrix}, whose universal properties describe uncorrelated data. Matrices $\bf W$ of the Wishart ensemble are constructed as the correlation matrices from time series of \textit{i.i.d.} random variables with the Gaussian distribution $N(0,\sigma)$~\cite{wishart1928}. Analytical form of p.d.f. describing an eigenvalue distribution of $\bf W$ is known:
\begin{equation}
\phi_\textrm{W}(\lambda)={1 \over N} \sum_{i=1}^N \delta(\lambda - \lambda_i) = {Q \over 2 \pi \sigma_{\textrm{W}}^2} {\sqrt{(\lambda_{+}-\lambda)(\lambda-\lambda_{-})} \over \lambda},
\label{rhoW}
\end{equation}
\begin{equation}
\lambda_{\pm} = \sigma_{\textrm{W}}^2 (1 + 1/Q \pm 2 \sqrt{1 \over Q}),
\label{lambdaW}
\end{equation}
where $\lambda\in[\lambda_-,\lambda_+]$ and $Q=T/N$, $T$ id the length of the time series and $N$ is the number of time series. The above relationship is called the Marchenko-Pastur distribution that is strictly valid in the infinite limit $T, N \to \infty$~\cite{Marchenko1967}. A comparison between the Marchenko-Pastur distribution and the empirical matrix eigenvalue distribution can show how noise and collectivity impacts the cryptocurrency market.

A typical feature of mature financial markets is a dominant share of noise effects in the cross-correlations among financial instruments. In this case an eigenvalue distribution of $\textbf{C}^{(\textrm{X})}$ can be approximated by the Marchenko-Pastur distribution for a majority of the eigenvalues~\cite{Marchenko1967}. At the same time, there is usually at least one eigenvalue outside the Marchenko-Pastur distribution~\cite{Laloux1999,Plerou1999a,Drozdz2001,Drozdz2002a,Utsugi2004,Drozdz2002,Plerou2002,Kwapien2006}. This eigenvalue or these eigenvalues, together with their associated eigenvectors, reflect how collective are the price movements on a given market~\cite{Drozdz2000,kwapien2012}.

\subsubsection{Correlation matrix elements}
\label{matrixelements}

\begin{figure}
\centering
\includegraphics[width=1\textwidth]{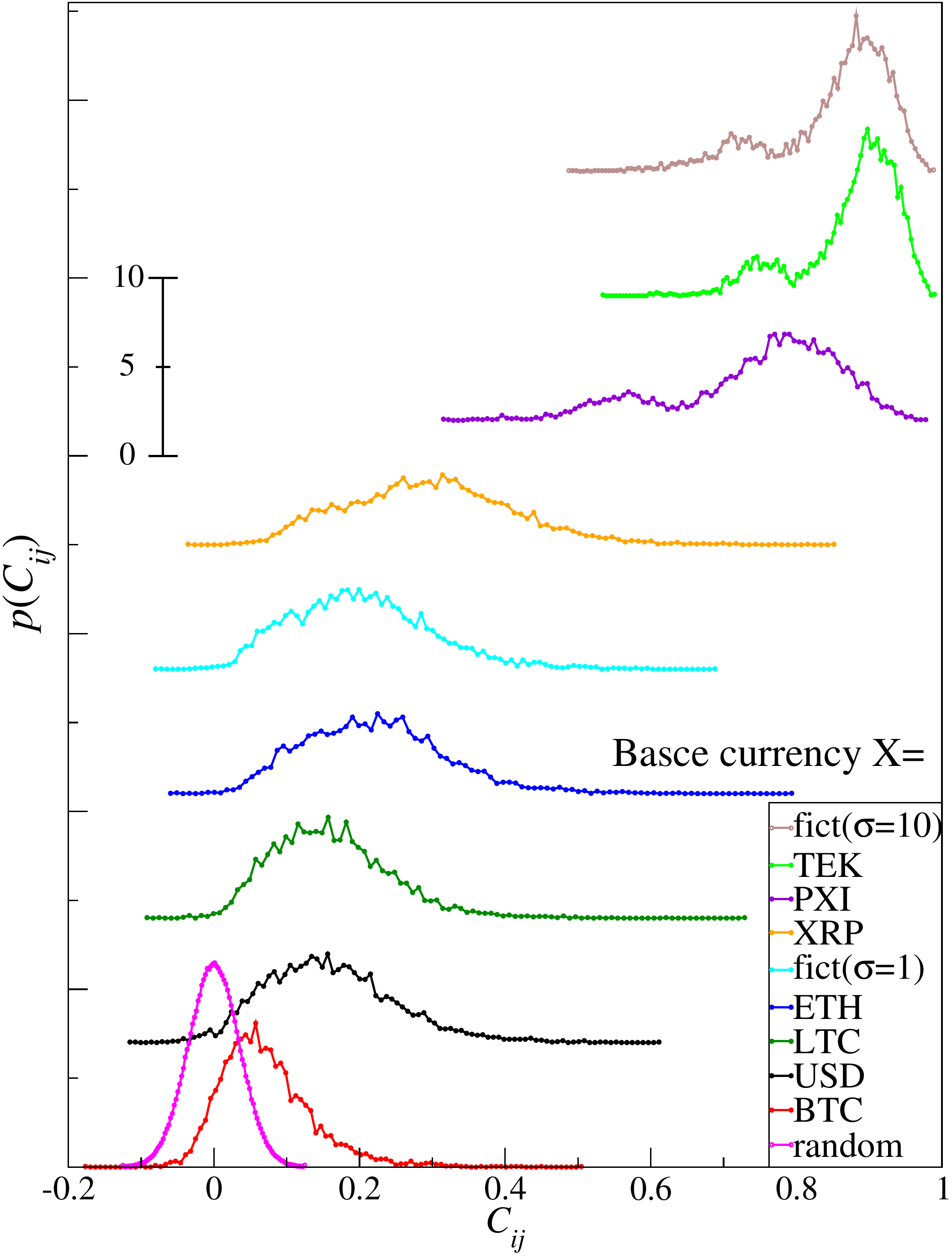}
\caption{PDFs calculated from the off-diagonal elements $C^{(\textrm{X})}_{ij}$ for six selected cryptocurrencies and USD as base currencies. Additionally, the random matrix case and the fictitious base currency are shown (for $\sigma=1$ and $\sigma=10$). For better visualization, the consecutive cases are shifted upwards.}
\label{fig:elements}
\end{figure}

In a generic case of uncorrelated signals, the off-diagonal entries of ${\bf C}^{(\textrm{X})}$ are Gaussian-distributed with zero mean~\cite{randommatrix}. However, in the case of empirical correlation matrices, there may be a shift in this distribution (positive or negative mean) and presence of heavy tails~\cite{Drozdz2000,Drozdz2001}. On the cryptocurrency market, the off-diagonal matrix entry distribution largely depends on a base currency as Fig.~\ref{fig:elements} documents for several representative cases. For all base currencies, this distribution differs from a random case. The smallest deviation is noticeable for X=BTC, which means that the market observed from a BTC perspective is the least correlated. The more peripheral (lower capitalization) a base cryptocurrency is chosen, the more the ${\bf C}^{\textrm{(X})}$ distribution is shifted to the right with an extreme case of TEK. In this case, all the exchange rate pairs involving TEK as the base give $C_{ij}^{\textrm{TEK}} > 0.5$ and the distribution mean is located close to 0.9. For such a peripheral cryptocurrency, the market is the most collective. This is even more evident for an artificial cryptocurrency (fict in Fig.~\ref{fig:elements}), whose each exchange rate $\textrm{K}_j/\textrm{fict} = \textrm{K}_j/\textrm{USD} \cdot \textrm{USD}/\textrm{fict}$ is driven by a geometric Brownian motion ($\mu=0, \sigma=1$) of $\textrm{USD}/\textrm{fict}$. In this situation the distribution is closer to the random matrix case than any other one. The fictitious cryptocurrency volatility $\sigma$ is responsible for this behavior: when it increases, the off-diagonal elements are shifted to the right. For $\sigma=10$ the distribution for X=fict looks similar to that obtained for X=TEK.

\subsubsection{Correlation matrix eigenvalues}

\begin{figure}
\centering
\includegraphics[scale=0.36]{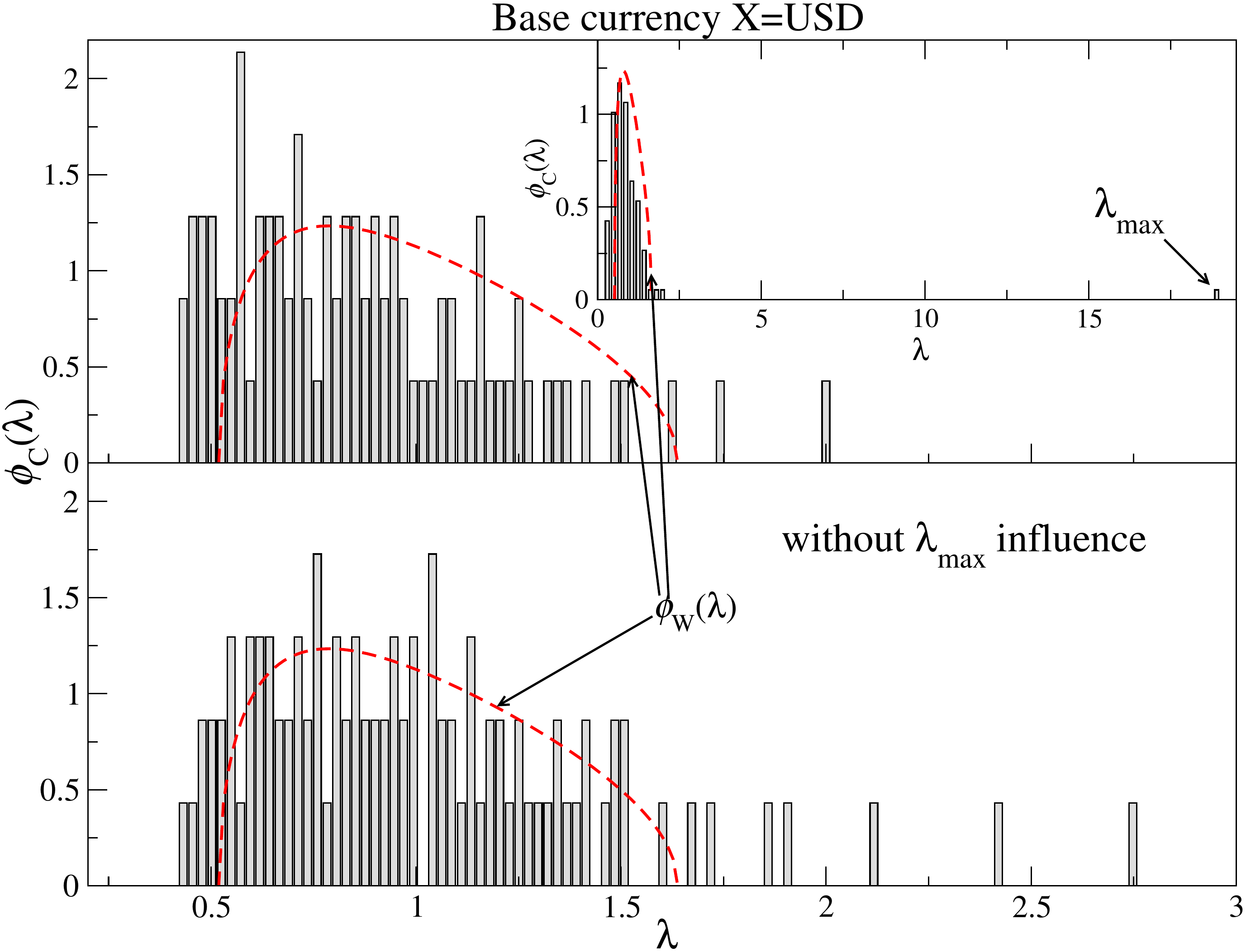}\\
\includegraphics[scale=0.36]{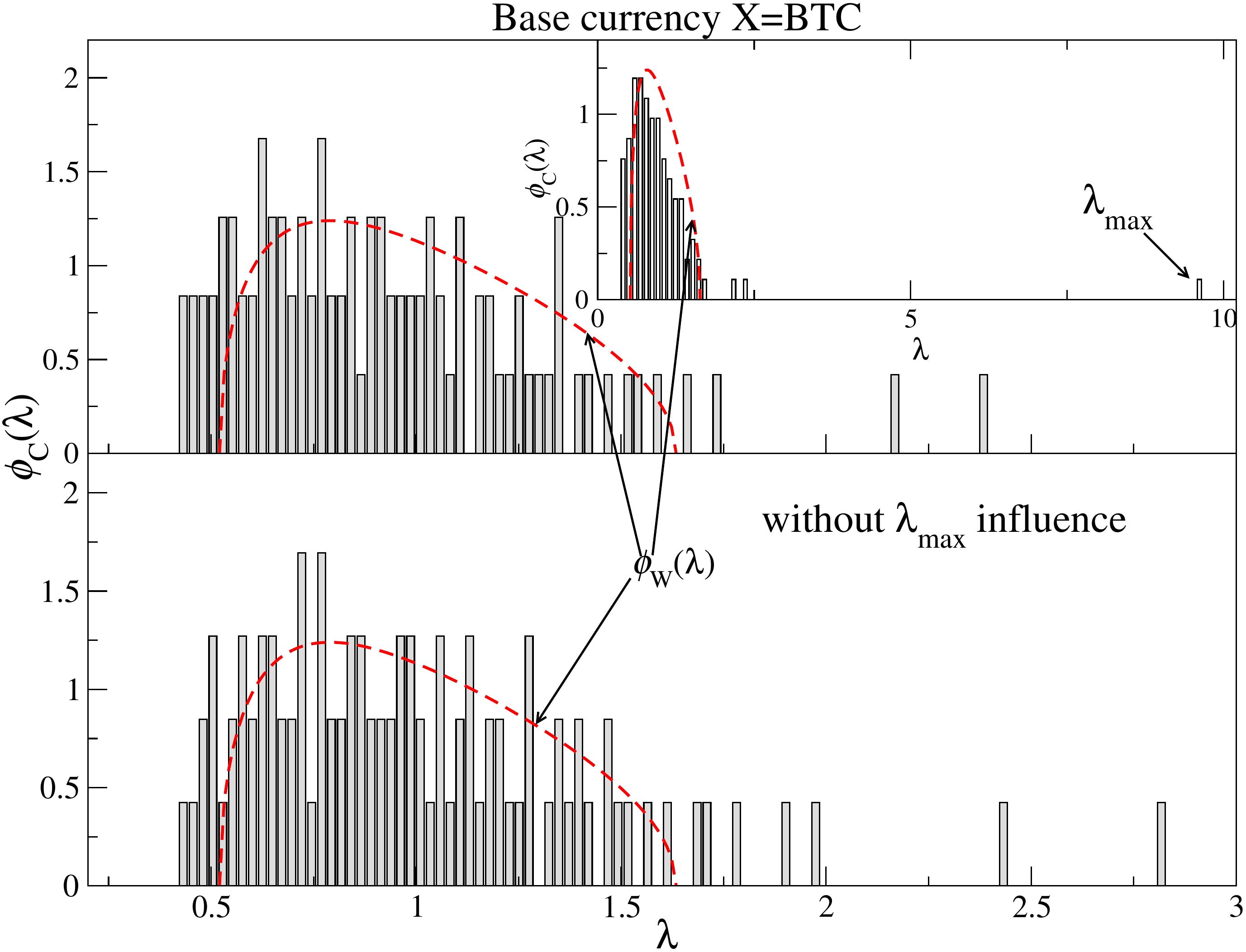}\\
\includegraphics[scale=0.36]{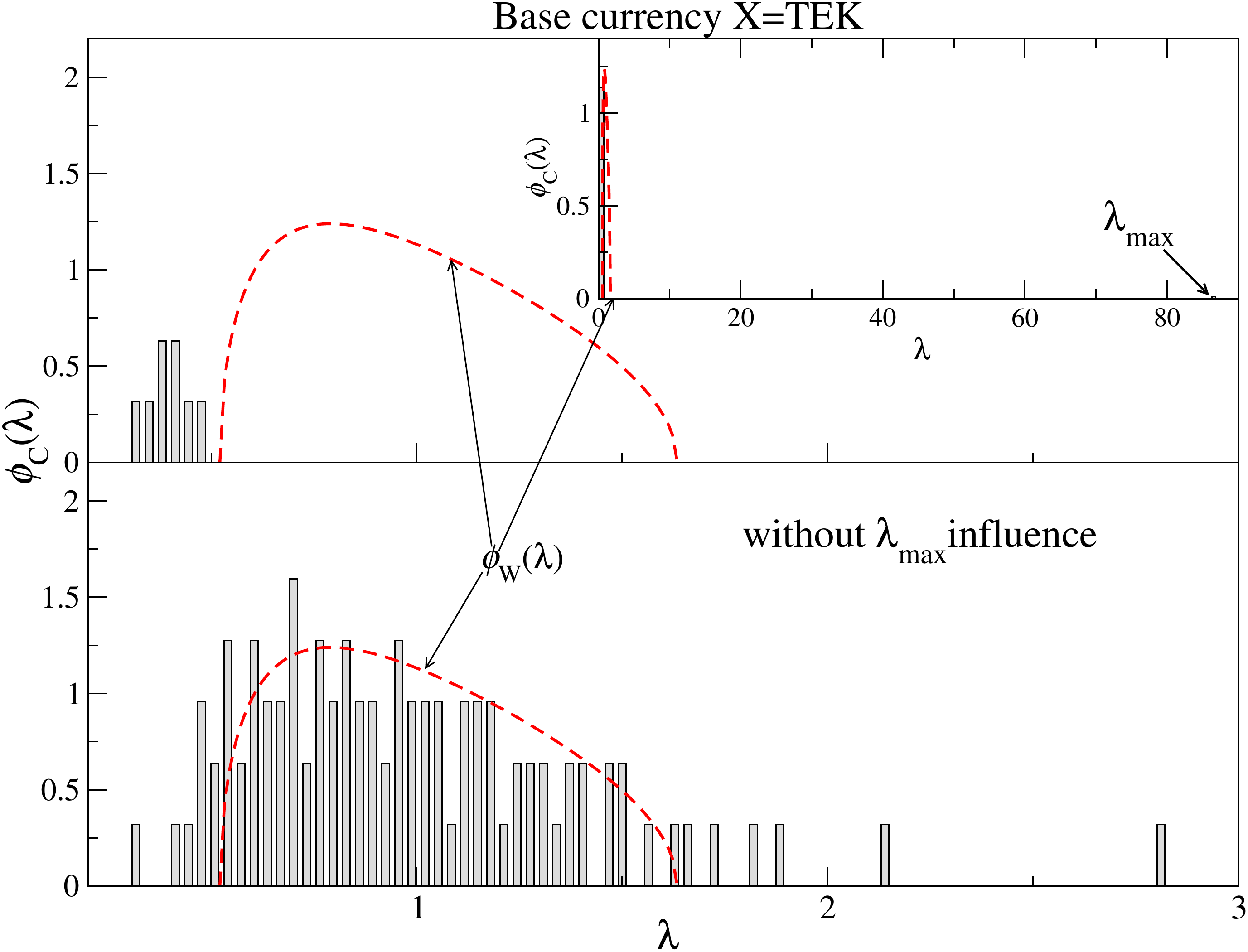}
\caption{PDFs $\phi_\textrm{C}(\lambda)$ for X=USD (top), X=BTC (middle), and X=TEK (bottom). In each case, the bulk of eigenvalues is shown in the upper graph, the entire eigenvalue distribution is shown in inset, and the residual eigenvalue distribution after removing $\lambda_{\textrm{max}}$ is shown in the lower graph. Dashed line denotes a fitted Marchenko-Pastur distribution $\phi_\textrm{W}(\lambda)$.}
\label{fig:spectra}
\end{figure}

PDFs describing the eigenvalue distribution $\phi_C(\lambda)$ for three base currencies: USD and the two extreme case cryptocurrencies: BTC and TEK are shown in Fig.~\ref{fig:spectra}. The following characteristics describing the market cross-correlation structure can be observed: (1) a market factor $\lambda_{\textrm{max}}^{(\textrm{X})}$~\cite{Plerou2002,Kwapien2006}, which shows the average level of cross-correlations (2) a distance between $\lambda_{\textrm{max}}$ and the rest of eigenvalues, which determines a share of the genuine cross-correlations, and (3) the eigenvalue bulk location in the Wishart region $\phi_\textrm{W}(\lambda)$~\cite{chaos2020}. Value of $\lambda_\textrm{max}^{(\textrm{X})}$ corresponds to a right-side shift of the average off-diagonal element of $\textbf{C}^{(\textrm{X})}$ in Fig.~\ref{fig:elements}. The smallest shift of $\lambda_{\textrm{max}}^{(\textrm{X})} \approx 9.7$ occurs for X=BTC. Most of the remaining eigenvalues in this case are located within the limits $[\lambda_{-},\lambda_{+}]$ set by Eq.(\ref{lambdaW}). For X=USD a value of $\lambda_{\textrm{max}}^{\textrm{USD}} \approx 19$ is higher, which means that the cryptocurrency market is more collective if expressed in USD than in BTC. The largest value of $\lambda_{\textrm{max}}^{\textrm{TEK}}$ is obtained when the cryptocurrencies are expressed in TEK. $\lambda_{\textrm{max}}^{\textrm{TEK}}$ is responsible for more than $90\%$ of the matrix trace, which effectively ``squeezes'' the other eigenvalues to around zero. This leads them to ``push out'' the noise-related eigenvalues even though they still specify random cross-correlations~\cite{haken1987}.

\begin{figure}
\centering
\includegraphics[scale=0.37]{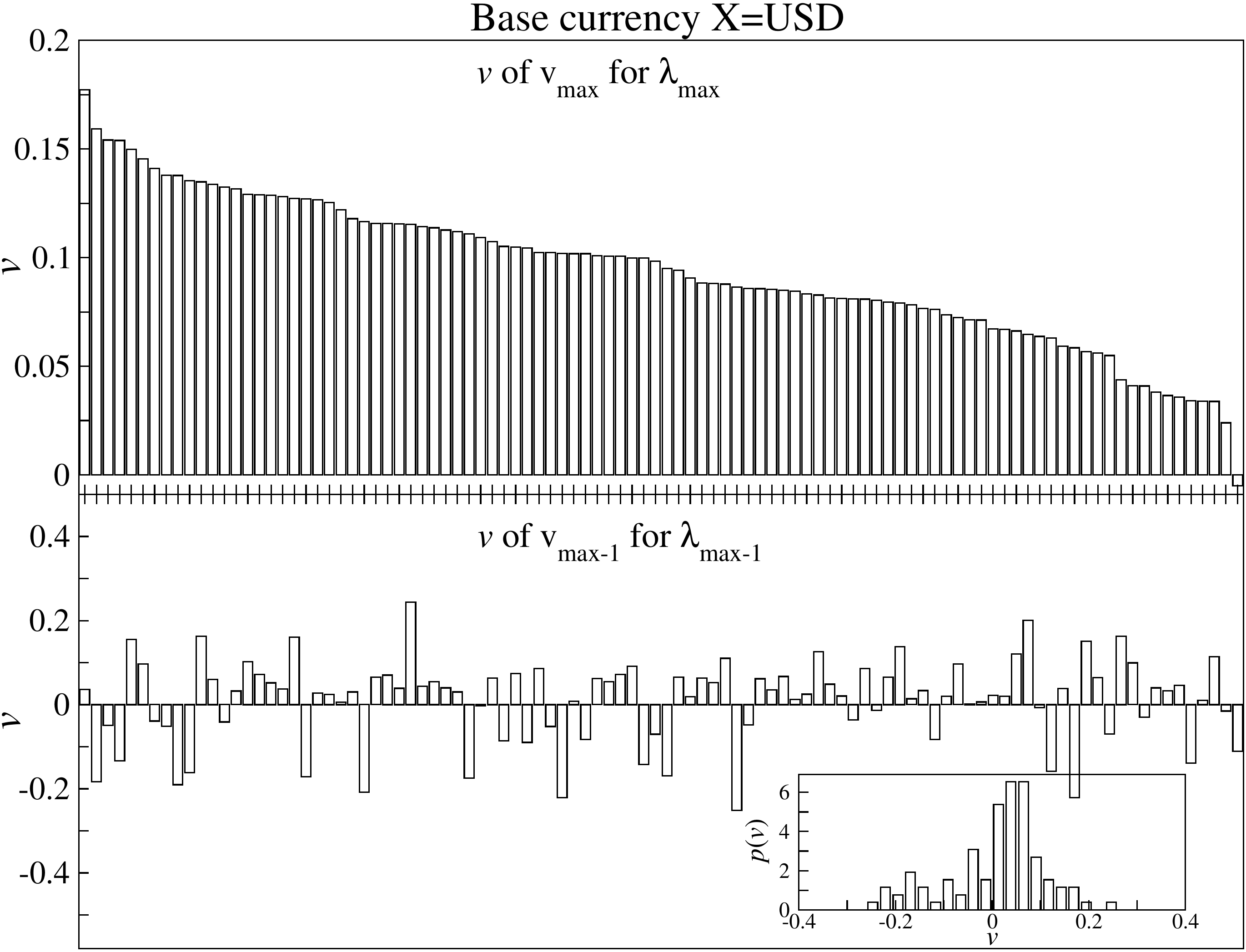}\\
\includegraphics[scale=0.37]{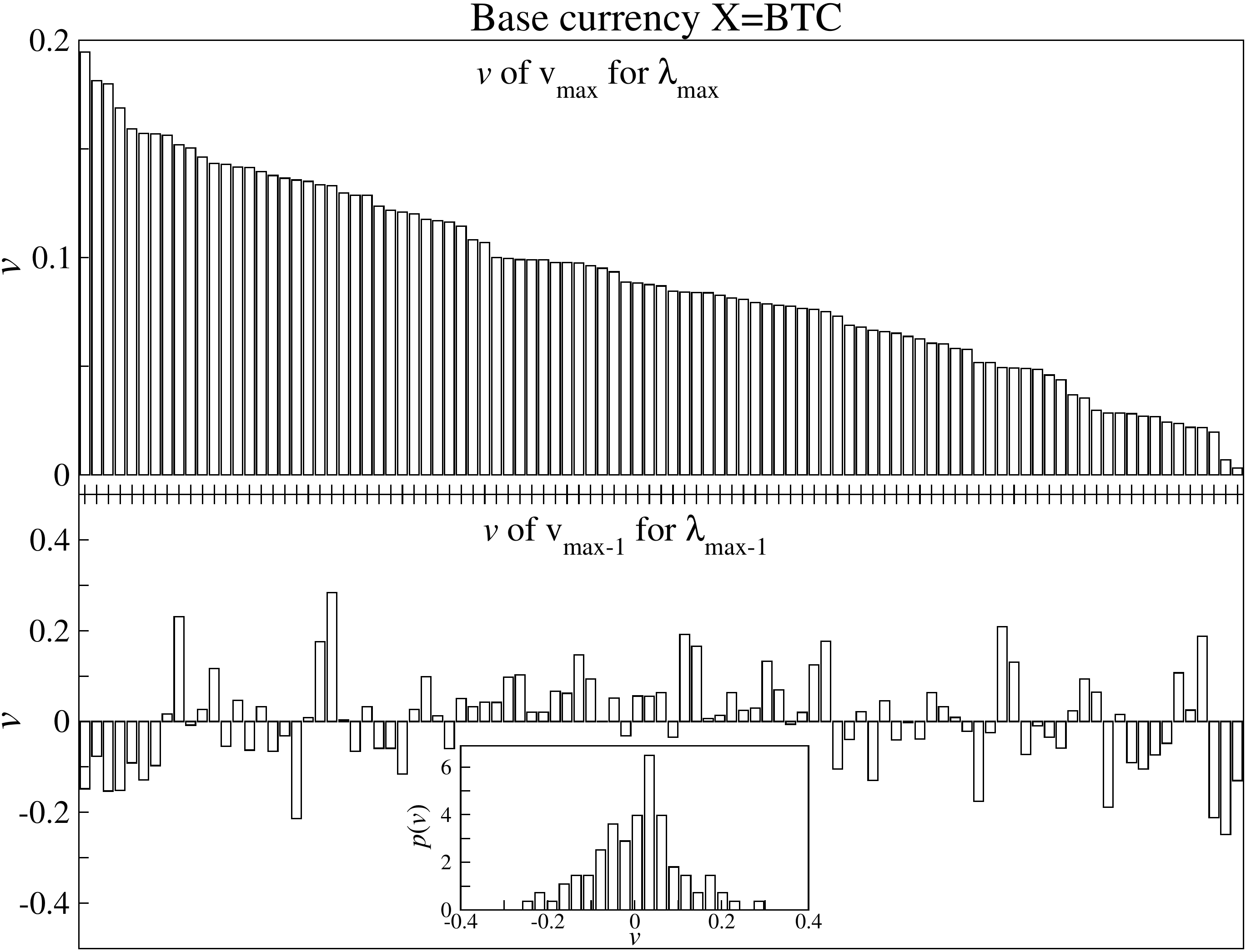}\\
\includegraphics[scale=0.37]{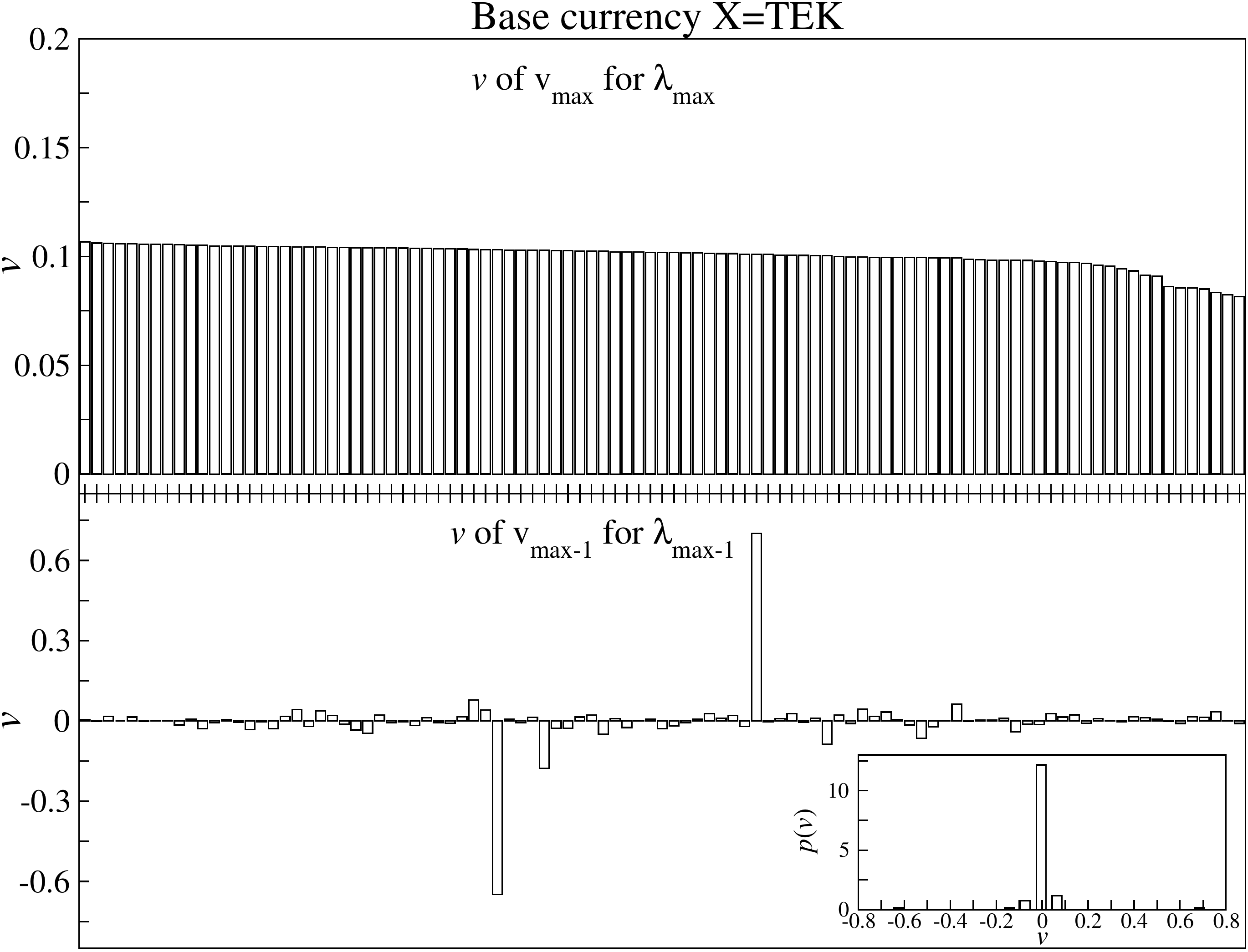}
\caption{Eigenvector components corresponding to $\lambda_{\textrm{max}}^{(\textrm{X})}$ (top) and $\lambda_{\textrm{max-1}}^{(\textrm{X})}$ (bottom panel) from the correlation matrix $C^{(\textrm{X})}_{ij}$ for X=USD, BTC and TEK. Insets presents the distribution of the eigenvector components.} 
\label{fig:expansion}
\end{figure}

A complementary indicator of collectivity is the eigenvector components ${\bf v}_{\textrm{max}}^{(\textrm{X})}$ corresponding to $\lambda_\textrm{max}^{(\textrm{X})}$ (Fig.~\ref{fig:expansion}). Through a significant contribution of the same sign from all time series (all eigenvector components), they reflect the collective nature of the respective eigenvector. Here, the ``squeezing'' effect is also visible while preserving the eigenvector ${\bf v}_{\textrm{max-1}}^{(\textrm{X})}$ corresponding to the second largest eigenvalue $\lambda_{\textrm{max-1}}^{(\textrm{X})}$. They are shown in Fig.~\ref{fig:expansion} (bottom panels). For X=BTC and X=USD, the eigenvector components have a distribution close to a normal one, while for X=TEK the distribution is dominated by a few components only.

\begin{figure}[ht!]
\centering
\includegraphics[width=1\textwidth]{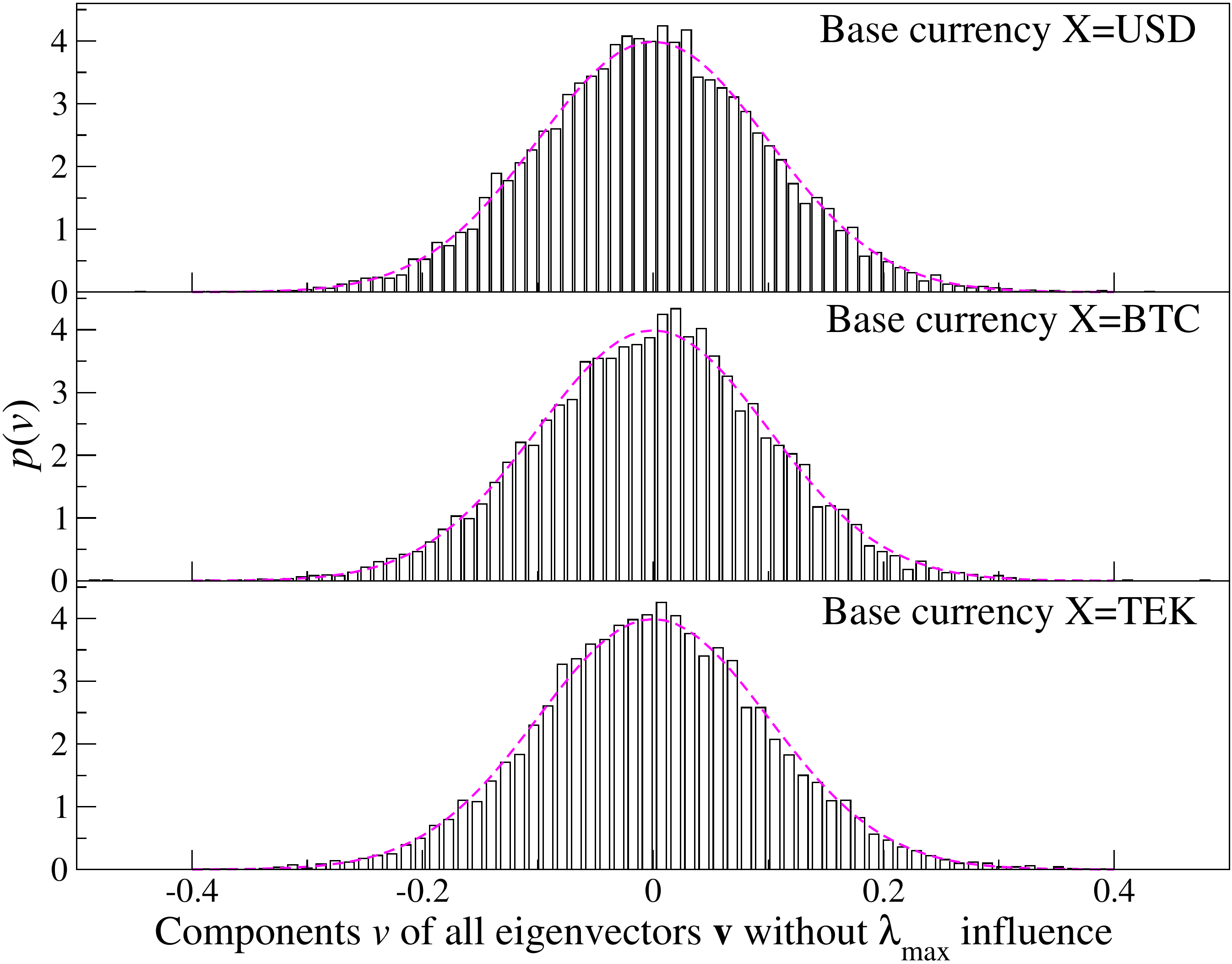} 
\caption{Eigenvector component PDFs for all eigenvectors $\textbf{v}_i^{(\textrm{X})}$ of the residual correlation matrix $\textbf{R}^{(\textrm{X})}$ for X=USD (top), X=BTC (middle), and X=TEK (bottom). In order to obtain $\textbf{R}^{(\textrm{X})}$, the market factor related to $\lambda_\textrm{max}^{(\textrm{X})}$ has been removed from the eigensignals of $\textbf{C}^{(\textrm{X})}$. Dotted line shows a fitted normal distribution.}
\label{fig:expansion-all}
\end{figure}

In order to eliminate the ``squeezing'' effect, the influence of the largest eigenvalue $\lambda_{\textrm{max}}^{(\textrm{X})}$ has to be subtracted. This can be accomplished by fitting to each time series $g^{(\textrm{X})}_i(t)$ the eigensignal $z_{\textrm{max}}^{(\textrm{X})}(t)$ with the least-square method:
\begin{equation}
g^{(\textrm{X})}_i = a_i + b_i z^{(\textrm{X})}_{\textrm{max}}(t) + \epsilon^{(\textrm{X})}_i(t), 
\label{residual}
\end{equation}
where $a_i$ and $b_i$ are parameters. A residual correlation matrix $\textbf{R}^{(\textrm{X})}$ that no longer contains a market factor $z_\textrm{max}^{(\textrm{X})}$ is constructed from the residual time series $\epsilon^{(\textrm{X})}_i(t)$~\cite{Kwapien2006}. The eigenvalue PDFs after eliminating $\lambda_\textrm{max}^{(\textrm{X})}$ are presented in the lower panels of Fig.~\ref{fig:spectra}. In all the cases, the outlying eigenvalue has disappeared and the vast majority of eigenvalues fall within $[\lambda_{-},\lambda_{+}]$. This observation also applies to the extreme case of X=TEK. The eigenvectors of ${\bf R}^{(X)}$ are different in general from the eigenvectors of $\textbf{C}^{(\textrm{X})}$ (see Fig.~\ref{fig:expansion-all}). For each X their component distribution is approximated well by the normal distribution, exactly as the eigenvector component distribution for the random matrices $\textbf{W}$.

\subsubsection{The largest eigenvalue for various base currencies}
\label{drabinka}

\begin{figure}
\centering\includegraphics[angle=90,origin=c,width=1\textwidth]{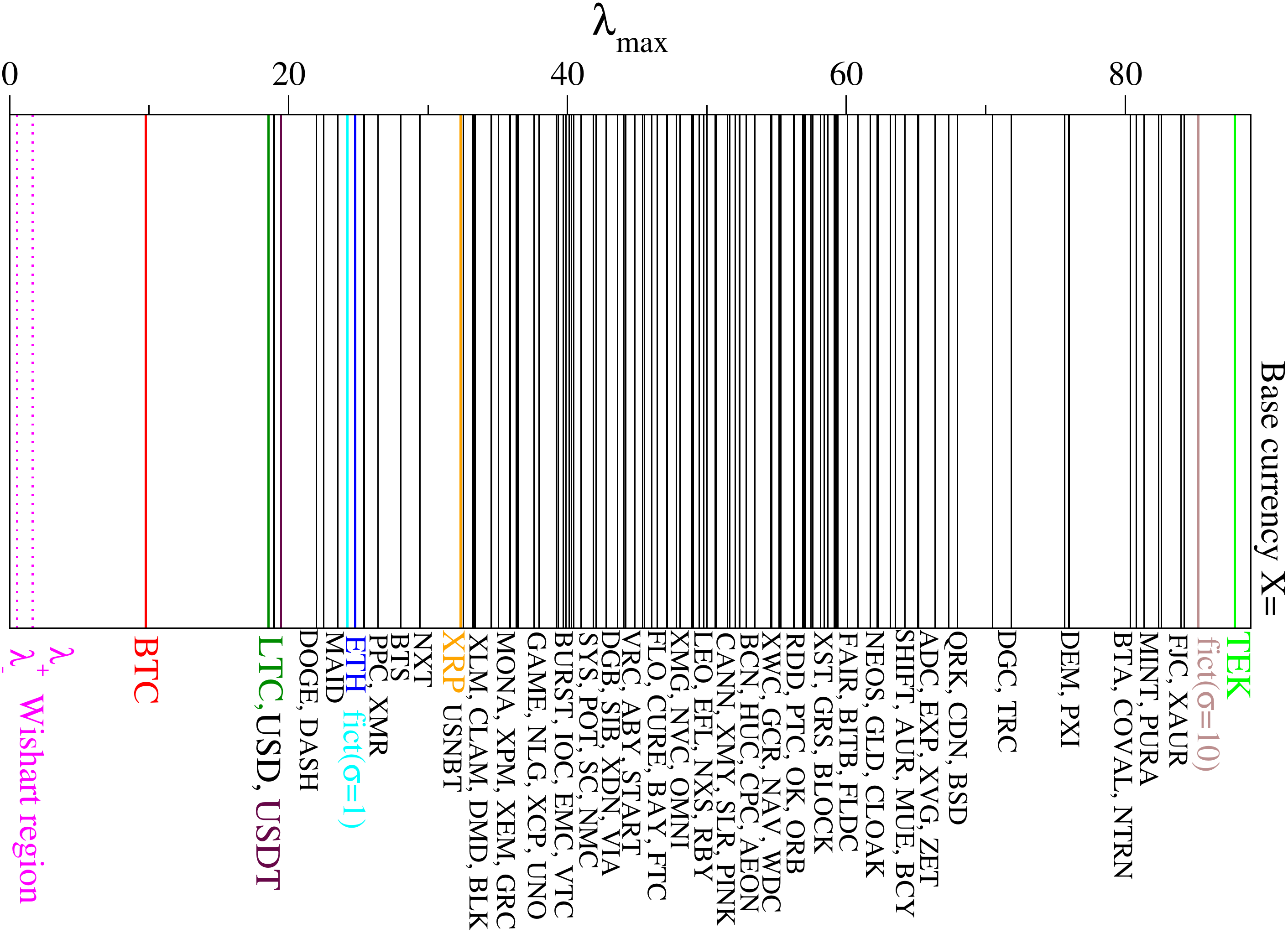} 
\caption{The largest eigenvalue $\lambda_\textrm{max}^{(\textrm{X})}$ of $\textbf{C}^{(\textrm{X})}$ for each of $N=100$ cryptocurrencies selected for the base (full names are listed in the Appendix~\ref{listamac}), as well as for USD and a fictitious currency (fict) with two different volatility levels $\sigma=1$ and $\sigma=10$. The limiting values $\lambda_{-}$ and $\lambda_{+}$ of the Wishart eigenvalue distribution $\phi_\textrm{W}(\lambda)$ given by Eq.~(\ref{lambdaW}) with $\sigma_\textrm{W}=1$ are denoted by dashed lines. $\lambda_{\textrm{max}}^{\textrm{USD}}$ and $\lambda_{\textrm{max}}^{\textrm{fict}}$ have been multiplied by 0.99 so that they can be comparable directly with the other eigenvalues.}
\label{fig:ladder}
\end{figure}

Fig.~\ref{fig:ladder} displays $\lambda_{\textrm{max}}^{(\textrm{X})}$ for all the considered cryptocurrencies treated as the base, together with a fictitious currency created like before (see~\ref{matrixelements}). This allows one to see what the entire market structure looks like from each cryptocurrency's perspective. The extremes X=BTC and X=TEK previously shown in detail are the lowermost and the uppermost ladder rungs here. A value of $\lambda_{\textrm{max}}^{(\textrm{X})}$ corresponding to a given base currency X should be interpreted as the relative importance of a given cryptocurrency.

A sharp change in the valuation of a peripheral cryptocurrency that is not correlated with the other cryptocurrencies and has its individual dynamics does not affect their behaviour, but nevertheless it does change values of all the cryptocurrencies expressed in it. As a result, the entire market's valuation shifts and the cryptocurrencies appear strongly cross-correlated and synchronous. In this case $\lambda_{\textrm{max}}^{(\textrm{X})}$ is large. On the other hand, a similar change in the valuation of an important cryptocurrency that have a number of other cryptocurrencies following it results in that the other cryptocurrencies can have their dynamics affected, too. Consequently, any valuation change when the other cryptocurrencies are expressed in this cryptocurrency turns out to be relatively small and the market appears less cross-correlated. 

For the fictitious currency (fict($\sigma=1$)) and USD, $\lambda_{\textrm{max}}^{(\textrm{X})}$ is relatively low, because despite having their dynamics independent from the cryptocurrency market, they are characterized by a low volatility level if compared to the cryptocurrencies. Therefore, their small, uncorrelated changes do not affect much the market valuation if they are this valuation base. A situation is completely different when volatility of the fictitious currency is raised to $\sigma=10$ (fict($\sigma=10$)). Then the largest eigenvalue reaches a high level above 80. A significant change of the fictitious currency value causes shifting of the cryptocurrency valuations expressed in it. Therefore, from an independent, highly volatile currency's perspective, the market seems substantially collective. The ``dragging'' effect of a rapid change in a base currency value is shown in section~\ref{sect::Forex} and in Ref.~\cite{gebarowski2019}. Large volatility of CHF and GBP caused on long time scales a cross-correlation increase among the exchange rates with one of these currencies being the base.

The above interpretation of $\lambda_{\textrm{max}}^{(\textrm{X})}$ -- the lower its value, the more important the base cryptocurrency X -- is also supported by capitalization volume. TEK, for which $\lambda_{\textrm{max}}^{\textrm{TEK}}$ is the largest, has one of the smallest capitalizations, while BTC has the largest one. In addition, TEK/USD shows a trend that differs from the other cryptocurrencies' trends
(Fig.~\ref{fig:dailycena}).

\subsubsection{Quasi-idempotence}

\begin{figure}[ht!]
\centering
\includegraphics[angle=90,origin=c,width=1\textwidth]{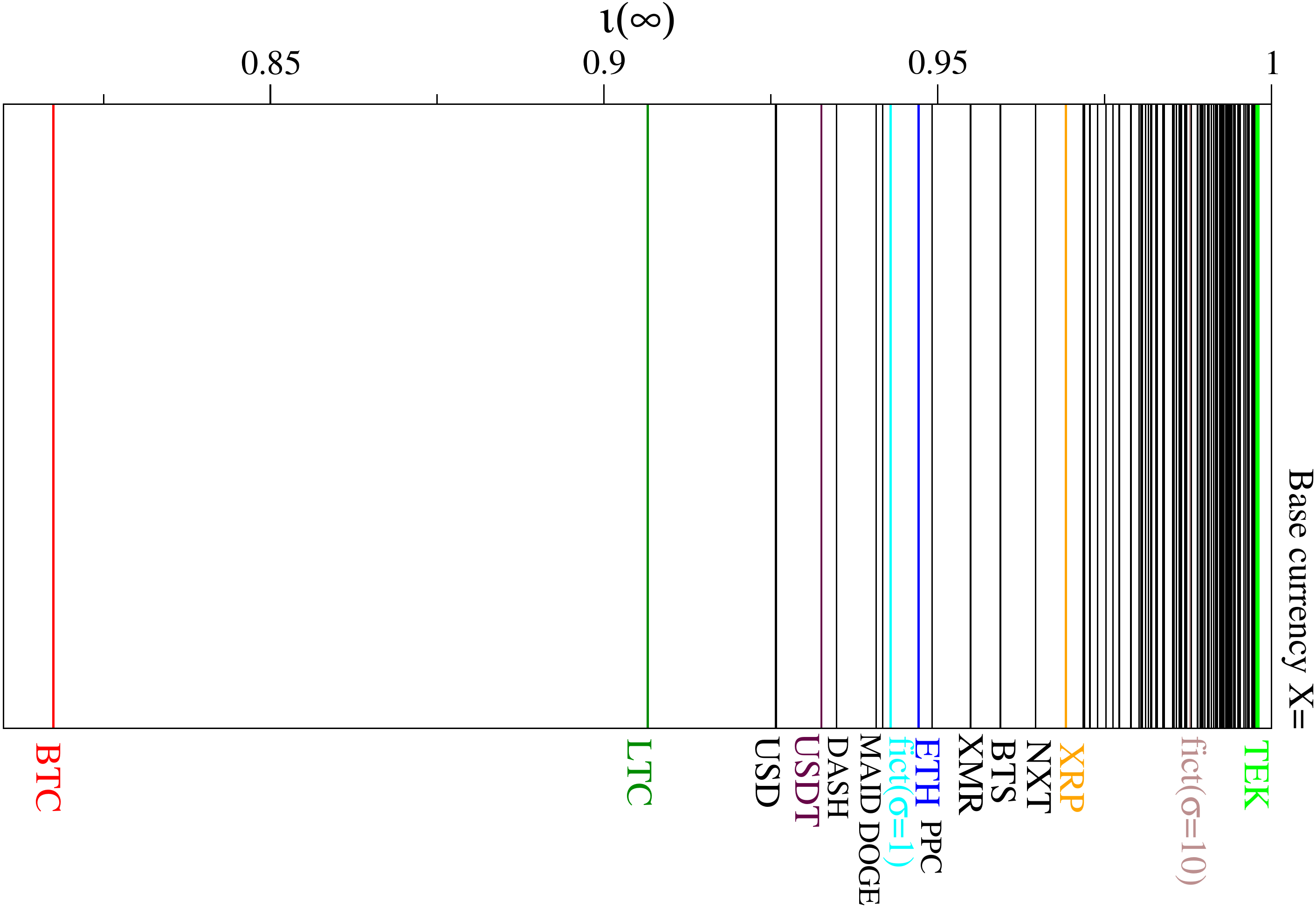} 
\caption{Quasi-idempotence $\iota^{\textrm{(X)}}(\infty)$ of $\textbf{C}^{(\textrm{X})}$ for each of $N=100$ cryptocurrencies, USD, and a fictitious currency (fict) selected to be the base -- X. For USD and fict, $\iota^{\textrm{(X)}}(\infty)$ has been multiplied by 0.99 so that it can be compared directly with $\iota^{\textrm{(X)}}(\infty)$ for the other base cryptocurrencies.}
\label{fig:IOTAladder}
\end{figure}

Another measure that can be applied to rank the importance of cryptocurrencies is quasi-idempotence~\cite{Minati2017}. The underlying notion is to assess how close an experimentally-obtained connectivity, synchronization, or correlation matrix is to the same squared infinitely many times. Considering a weighted network, the corresponding matrix is the least sensitive to squaring (closest to equilibrium) whenever the edge weights are simply reflections of the corresponding node strengths; in other words, when the connectivity between any two nodes is a reflection of their common participation in global connectivity. Quasi-idempotence, therefore, can be deemed an index at the same time of self-similarity and collectivity. Here, it was calculated according to
\begin{equation}
\iota^{\textrm{(X)}}(\infty)=r(\Xi(\mathbf{C}_{0}^{\textrm{(X)}}),\Xi(\mathbf{C}_{\infty}^{\textrm{(X)}}))
\end{equation}
where $\mathbf{C}_{0}^{\textrm{(X)}}$ and $\mathbf{C}_{\infty}^{\textrm{(X)}}$ denote the initial correlation matrix for a base (crypto)currency X and the same squared infinitely many times, practically obtained according to a truncation criterion while maintaining unit norm at each step to avoid divergence. $\Xi(\mathbf{C^{\textrm{(X)}}})$ yields a vector $\mathbf{c}$ containing the rearranged off-diagonal elements $C_{ij}^{\textrm{(X)}}=C_{ji}^{\textrm{(X)}}$ of $\mathbf{C}^{\textrm{(X)}}$ with $i>j$ and $r(\mathbf{x},\mathbf{y})$ denotes the element-wise Pearson correlation between the two vectors. As the measure is ill-defined in the presence of negative entries (due to possible non-convergence), anticorrelations were clipped to zero. While related to the largest eigenvalue analysis, this approach is mathematically distinct and entails a more complex dependency on the entire spectrum. Consequentially, as visible in Fig.~\ref{fig:IOTAladder}, it not only confirmed but demonstrated even more markedly the centrality of the Bitcoin, such as that the lowest values of $\iota^{\textrm{BTC}}(\infty)$ were consistently observed when this was taken as the base currency: such a finding underlines that the ``collective dynamics'' which endow the correlation matrix with a non-random structure largely stem from the influence of this currency. Similar dependencies can be detected in maturing culture neurons, physical networks of chaotic oscillators as well as simulations of stochastic Kuramoto networks~\cite{Minati2017}.

\subsubsection{Evolution of the market cross-correlation structure}
\label{EIGevolution}

\begin{figure}[ht!]
\includegraphics[width=1\textwidth]{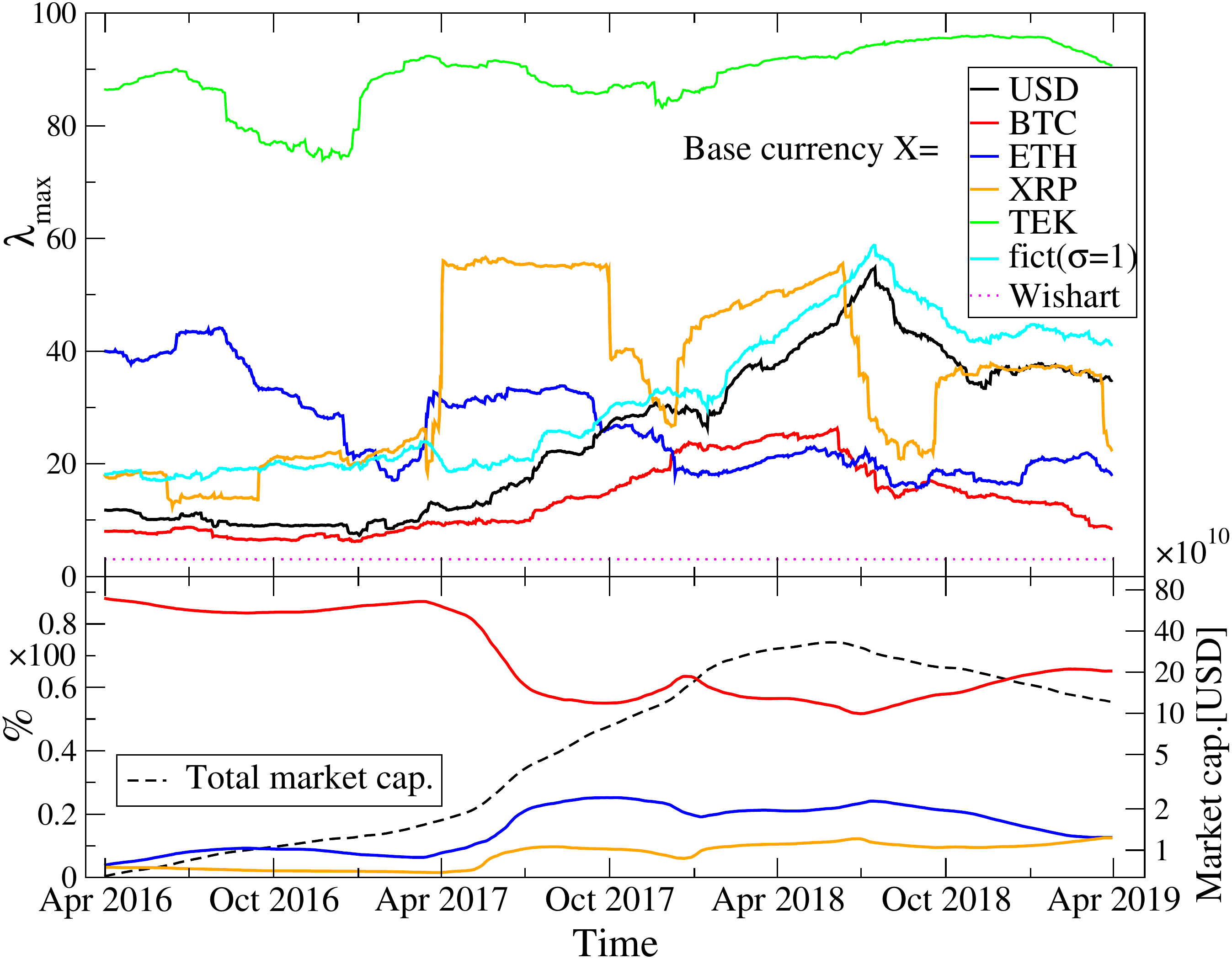} 
\caption{Top: $\lambda_{\textrm{max}}^{(\textrm{X})}(t)$ calculated from $\textbf{C}^{(\textrm{X})}(t)$ in a 182-day rolling window with a step of 1 day for X=BTC, X=ETH, X=XRP, X=TEK, and X=USD. A fictitious currency fict($\sigma=1$) is also shown. $\lambda_{\textrm{max}}^{\textrm{USD}}$ and $\lambda_{\textrm{max}}^{\textrm{fict}(\sigma=1)}$ have been multiplied by 0.99 so that they can be compared directly with the other eigenvalues. A dashed line marks $\lambda_{+}$ for the Wishart case. Bottom: market capitalization as a function of time (dashed line) and a market share of BTC, ETH, and XRP, calculated in a semi-annual rolling window.}
\label{fig:EIGtime-dependent}
\end{figure}

As characteristics of the individual cryptocurrency exchange rates evolve over time, a cross-correlation structure of the market can also change as the market develops. Top panel in Fig.~\ref{fig:EIGtime-dependent} presents $\lambda_\textrm{max}^{(\textrm{X})}(t)$ of $\textbf{C}^{(\textrm{X})}(t)$ for X=BTC, X=ETH, X=XRP, X=TEK, X=USD, and X=fict (constructed from a geometric Brownian motion with $\sigma=1$) calculated for various base currencies in a half-year (182 days) rolling window with a one-day step. Bottom panel of Fig.~\ref{fig:EIGtime-dependent} presents evolution of a market capitalization share of the most important cryptocurrencies calculated in the same way.

The extreme cases (X=BTC and X=TEK) do not change significantly over time. However, the increase in $\lambda_{\textrm{max}}^{(\textrm{X})}$ and, consequently, in the market cross-correlation strength is particularly strong for X=USD and X=fict($\sigma=1$), whose dynamics are independent of the cryptocurrency market. The largest growth can be seen in the semi-annual windows ending from January to August 2018 -- this interval includes a period of a strong bull market followed by a crash. This behaviour is consistent with an observation that stock markets are strongly collective during sharp declines~\cite{Drozdz2000,Preis2012,Sandoval2012}. After the cryptocurrency market calmed down in the second half of 2018, $\lambda_{\textrm{max}}^{(\textrm{X})}$ for USD and fict($\sigma=1$) decreased, but they still remain above the 2017 level.

If X=ETH one can see a systematic drop in $\lambda_{\textrm{max}}^{\textrm{ETH}}$ that is due to a capitalization increase of this cryptocurrency (Fig.~\ref{fig:EIGtime-dependent}). In the windows ending from December 2017 to June 2018 $\lambda_{\textrm{max}}^{\textrm{ETH}}$ was lower than $\lambda_{\textrm{max}}^{\textrm{BTC}}$, which was a temporary takeover of the dominant role on the cryptocurrency market by ETH. At that time ETH's share in the market capitalization increased while the BTC's share dropped, which was related to the ICO-mania mentioned earlier. The number of ICO offers was released mainly on the Ethereum platform which shot up a demand for ETH, rose its importance, and strengthened links with other cryptocurrencies. An additional factor that caused BTC to decline at the turn of 2017 and 2018 was that BTC experienced its valuation peak in December 2017, while the other major cryptocurrencies, including ETH, about a month later. From January 2018 a bear phase spread across the market, with BTC losing relatively less value, which translated into an increase in its market capitalization share. This was accompanied by a systematic decline in $\lambda_{\textrm{max}}^{\textrm{BTC}}$, which can be interpreted as a return of the bitcoin's dominant position.

If X=XRP one can see jumps in $\lambda_{\textrm{max}}^{\textrm{XRP}}$ (Fig.~\ref{fig:EIGtime-dependent}) that were associated with a 102\% increase in the XRP/USD rate valuation on 04/02/2017. Such a big change implied valuation change of all the cryptocurrencies expressed in XRP, leading to a collective behaviour of the entire market if observed from the XRP perspective. Inclusion of that day in a semi-annual rolling window translates directly into a sharp increase in $\lambda_{\textrm{max}}^{\textrm{XRP}}$. In contrast, a decline in $\lambda_{\textrm{max}}^{\textrm{XRP}}$ half a year later was caused by falling out of that day from the rolling window. This confirms the previously presented interpretation of the largest eigenvalue magnitude.

\subsection{Minimal spanning tree representation of the cryptocurrency market}

Based on the correlation matrix $\textbf{C}^{(\textrm{X})}$, one can also construct a network representation of the market. Nodes can represent the exchange rates and edges can be the cross-correlations between two exchange rates with weights equal to $C_{ij}^{(\textrm{X})}$. 

Minimal spanning tree (MST) is a connected subset of a weighted network that spans this network and minimizes a sum of the metric edge lengths. It was first applied to a cross-correlation analysis of a stock market in Ref.~\cite{Mantegna1999} and, since then, it has been successfully used to analyze cross-correlation topology of various markets such as: Forex~\cite{McDonald2005,Mizuno2006,Naylor2007,Gorski2008,Kwapien2009,Jang2011}, the stock market~\cite{Bonanno2000,MICCICHE2003,Onnela2003,Bonanno2004,EOM2009,tabak2010,Wilinski2013,Sensoy2014,kwapien2017}, the commodity market~\cite{Sieczka2009}, and the cryptocurrency market~\cite{stosic2018}. For the mature markets, a node degree PDF for MST can be approximated by a power-law tails~\cite{kwapien2012}. In the case of the stock and Forex markets, the tail scaling exponents assume values that are similar to the hierarchical models of complex networks~\cite{BOCCALETTI2006}.

Because the off-diagonal elements of $\textbf{C}^{(\textrm{X})}$ do not meet the metric conditions, it is necessary to transform edge weights into internode distances defined by
\begin{equation}
d^{(\textrm{X})}_{i,j} = \sqrt{2 \Bigl ( 1 - C^{(\textrm{X})}_{ij} \Bigr )}
\label{dij_MST} 
\end{equation}
that already fulfill these conditions~\cite{Mantegna1999}. Since the Pearson correlation coefficient takes values in $[- 1,1]$, the distance $d^{(\textrm{X})}_{i,j}$ is confined to $[0,2]$. With this measure, the MST graphs that represent market structure can be constructed by following the Prim's algorithm~\cite{Prim}.

\begin{figure}
\centering
\includegraphics[width=1\textwidth]{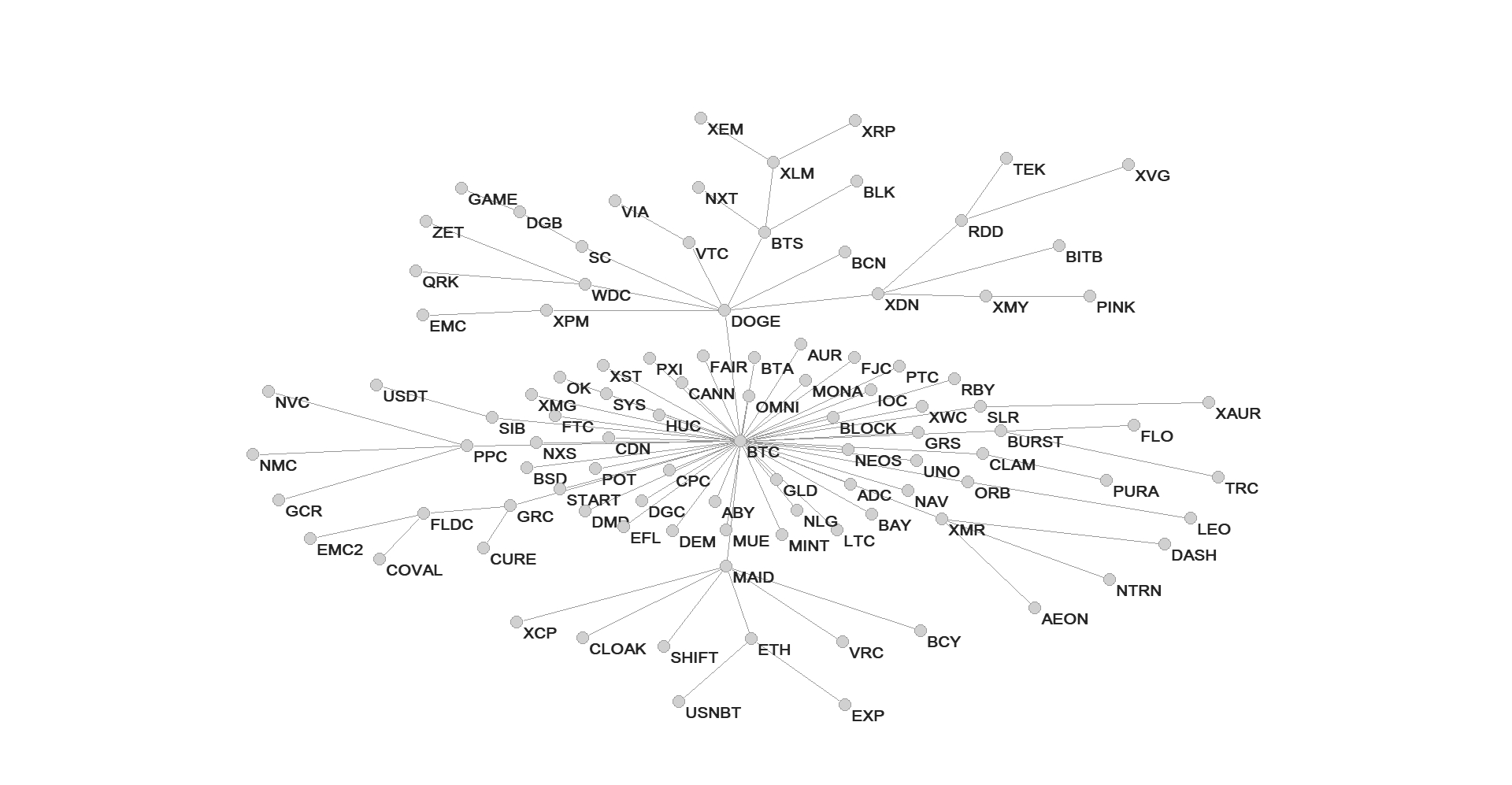} 
\caption{MST representing the cryptocurrency market constructed for USD playing the role of the base currency.}
\label{fig:MSTwUSD}
\end{figure}

Fig.~\ref{fig:MSTwUSD} presents a strongly centralized MST for X=USD, constructed from a set of $N=100$ X/USD exchange rates recorded daily over the period 10/2015--03/2019. $k_{\textrm{BTC}}^{\textrm{USD}}=51$ is multiplicity (degree) of a node corresponding to BTC/USD, while other nodes have multiplicity not exceeding 10. The node degree can be interpreted as a centrality measure expressing a relative importance of the exchange rate. The larger degree of a node corresponding to a given cryptocurrency, the more it affects the remaining cryptocurrencies~\cite{Kwapien2009}. Dominance of BTC in the structure of MST for USD as the base currency is consistent with the already-discussed results of the largest eigenvalue analysis (Fig.~\ref{drabinka}).

\begin{figure}
\centering
\includegraphics[width=1\textwidth]{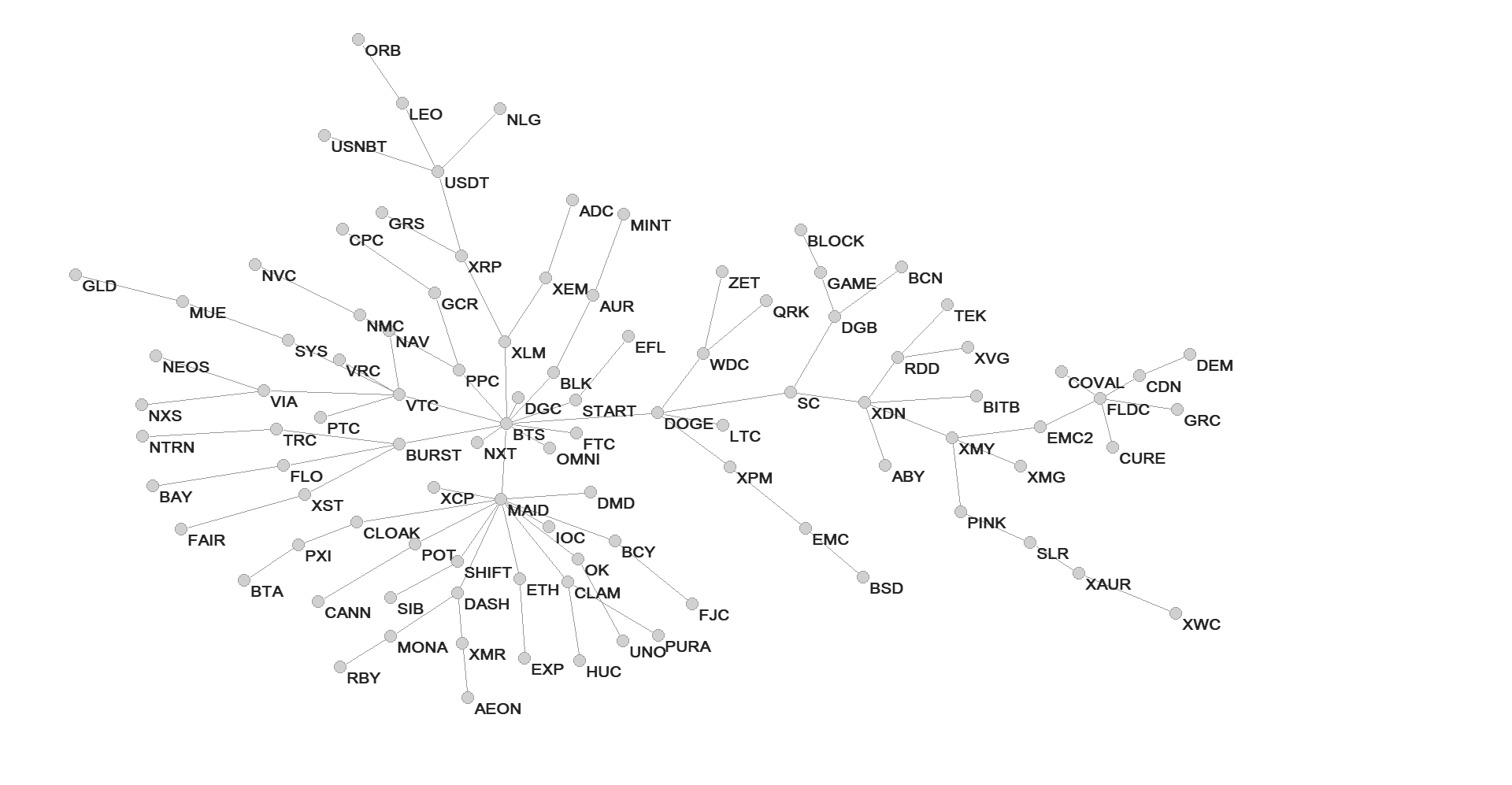} 
\caption{MST representing the cryptocurrency market constructed for BTC as the base currency.}
\label{fig:MSTwBTC}
\end{figure}

MST for X=USD, whose dynamics is not directly related to the cryptocurrency market, seems to best describe the market topology. However, as shown in Sect.~\ref{corrM}, the cross-correlation structure depends heavily on the base currency choice. Fig.~\ref{fig:MSTwBTC} shows MST for X=BTC, which looks profoundly different. It is much more decentralized and there is no dominant node. The largest degree  $k_{\textrm{max}}^{\textrm{BTC}}=12$ is observed for BTS/BTC and MAID/BTC. For the entire period under consideration, MSTs created for the base cryptocurrencies less important than BTC are even more centralized than that for X=USD: most nodes have a unit degree and they are connected to BTC. Results for all the cryptocurrencies used as the base ones are presented in Sect.~\ref{compEIGMST}.


\begin{figure}[ht!]
\centering
\includegraphics[width=1\textwidth]{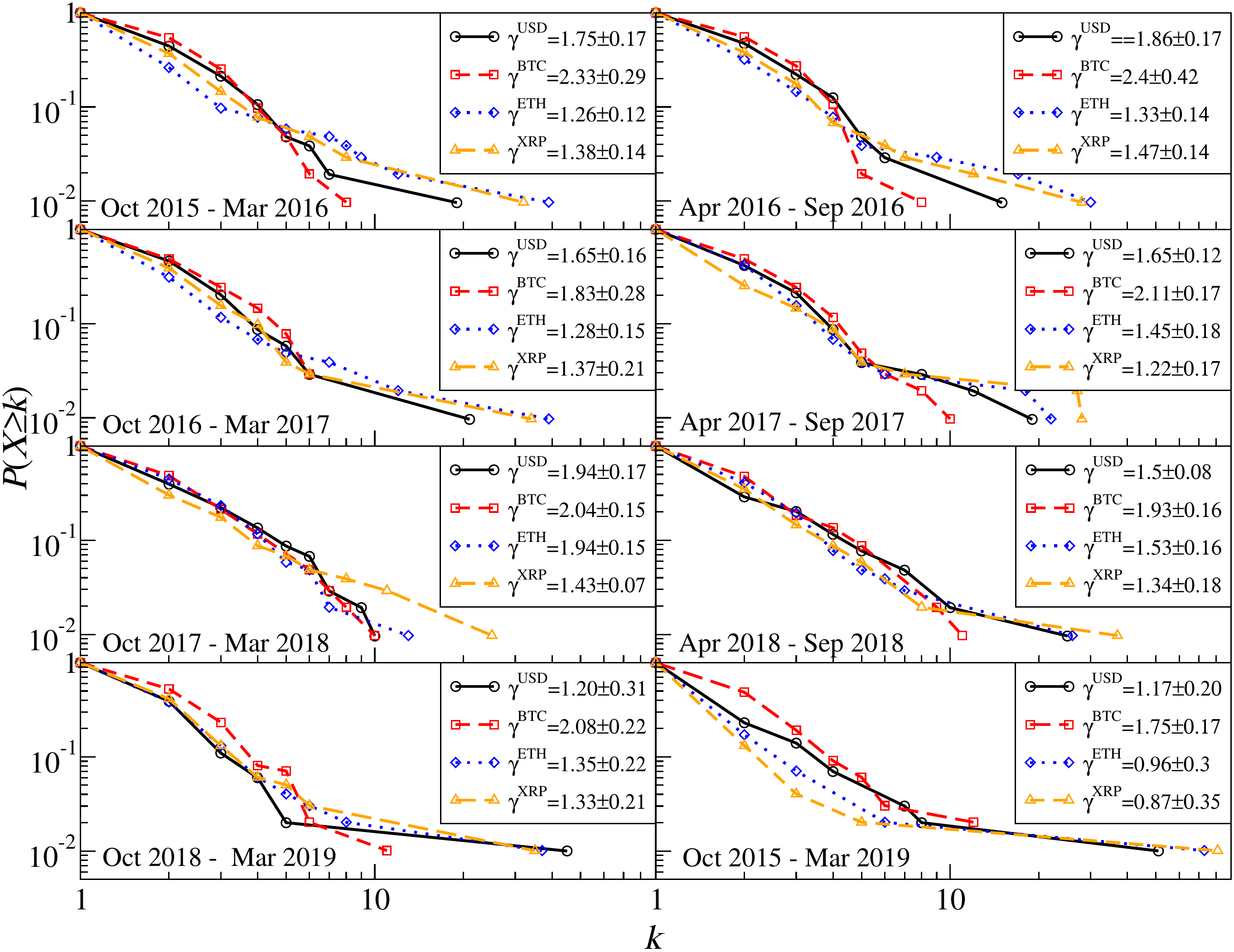} 
\caption{Node degree CDFs with the estimated scaling exponent $\gamma^{(\textrm{X})}$ for different choices of the base currency X=USD, X=BTC, X=ETH, and X=XRP in semi-annual windows and in the whole period 10/2015--03/2019 (bottom right).}
\label{fig:rozkladykrotnosci}
\end{figure}

On Forex, the node degree distributions are in agreement with the power-law model, which means that these trees are scale-free for some base currencies~\cite{Gorski2008,Kwapien2009}. Scaling of the node degree CDF can be parameterized by an exponent $\gamma^{(\textrm{X})}$:
\begin{equation}
P(X\geq k) \sim k^{-\gamma^{(\textrm{X})}},
\label{rozklkrot}
\end{equation} 
where $k$ is the MST node degree and X base cryptocurrency. Fig.~\ref{fig:rozkladykrotnosci} presents a node degree CDF $k^{(\textrm{X})}$ for three base cryptocurrencies with the largest capitalization: X=BTC, X=ETH, and X=XRP together with X=USD. Due to the changing market cross-correlation structure observed in Sect.~\ref{EIGevolution}, the results were also shown in semi-annual windows, apart from the entire-period case from 10/2015 to 03/2019.

For the whole considered period (bottom right panel in Fig.~\ref{fig:rozkladykrotnosci}) some trace of scaling can be identified only for X=BTC. In other cases a node with the largest degree, that is BTC, disrupts scaling, because the corresponding MSTs are highly centralized. In a majority of the semi-annual intervals, there is no satisfactory scaling quality, either. Only for X=BTC in some intervals, $\gamma^{\textrm{BTC}}$ oscillates around 2, which is a value associated with a decentralized network topology. A significant scaling quality improvement occurs in the interval 10/2017 to 3/2018. A related event is the BTC significance decline described in Sect.~\ref{EIGevolution} with the associated maximum node degree $k_{\textrm{max}}^{(\textrm{X})}$ decreasing for $X\neq$BTC. This made it possible to develop a hierarchical structure of the nodes with the scaling tails of the node degree CDFs. An example of such a situation is X=XRP with $\gamma^{\textrm{XRP}}=1.43\pm 0.07$ and a well-developed scale-free tail. Slightly worse scaling is seen for a subsequent window 04/2018--09/2018, where $\gamma^{\textrm{USD}}=1.5\pm 0.08$. In that interval a CDF scaling is also noticeable for X=USD. The last semi-annual window 10/2018--03/2019 displays a return of the BTC domination and, thus, a transition of the MST structure to a centralized form.
\begin{figure}[ht!]
\centering
\includegraphics[width=1\textwidth]{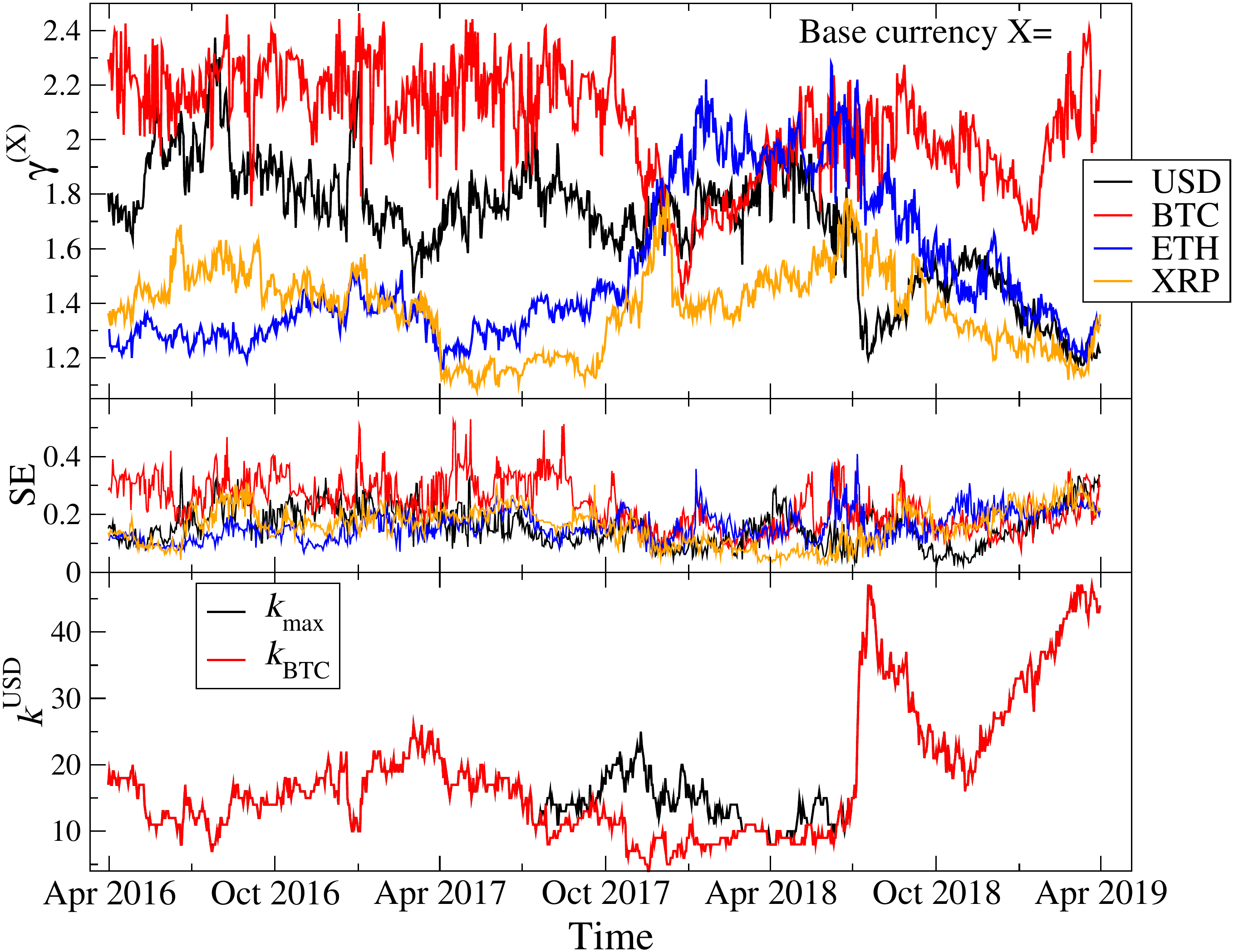} 
\caption{Top: scaling exponent of the node degree CDF $\gamma^{(\textrm{X})}(t)$ for X=USD, X=BTC, X=ETH, and X=XRP. Middle: standard error of a linear-regression-fitted $\gamma^{(\textrm{X})}$. Bottom: maximum node degree $k_{\textrm{max}}^{\textrm{USD}}$ for X=USD (black) and the BTC/USD node degree $k_{\textrm{BTC}}^{\textrm{USD}}$ (red).} 
\label{fig:rozklwierzch}
\end{figure}

Evolution of the scaling exponent $\gamma^{(\textrm{X})}(t)$ for the above-mentioned base currencies X is presented in Fig.~\ref{fig:rozklwierzch} (top panel). A semi-annual rolling window with a step of 1 day is applied. A scaling quality improvement measured by a power-law fit standard error can be observed in the rolling windows that end between 10/2017 and 5/2018 (middle panel) and it corresponds to the period of the interim BTC/USD market capitalization share and node degree decline (bottom panels of Fig.~\ref{fig:rozklwierzch} and Fig.~\ref{fig:rozklwierzch}). On contrary, in the windows that end between 8/2017 and 4/2018, there are nodes with a larger degree than BTC/USD. This is related to a strong bull market that started on all cryptocurrencies. During that time, the market was increasingly collective from the X=USD perspective as shown in Fig.~\ref{fig:EIGtime-dependent}. Between 11/2017 and 3/2018, $\gamma^{\textrm{ETH}}$ increased above 2, which corresponded to a dominant role takeover by ETH (see also Fig.~\ref{fig:EIGtime-dependent} for $\lambda_{\textrm{max}}^{(\textrm{X})}$ behaviour). This change was associated with the ICO-mania and the increasing capitalization and the ETH significance. Between 5/2018 and 7/2018, which was the final phase of the bull market, the market became decentralized and $\gamma^{\textrm{ETH}}$,  $\gamma^{\textrm{BTC}}$, and $\gamma^{\textrm{USD}}$ were above 1.8. However, $\gamma^{\textrm{XRP}}$ was equal to around 1.6 at that time, which is close to the theoretically derived value for a hierarchical network model: $\gamma^{\textrm{hier}}\approx 1.6$~\cite{Noh2003,Ravasz2003}.

The CDF tails lose their scale-free form when BTC dominance returns. This can be seen in the windows that end in 07/2018 in a form of a sharp jump in $k_{\textrm{max}}^{\textrm{USD}}$, which is the BTC/USD node (bottom panel of Fig.~\ref{fig:rozklwierzch}). The BTC/USD node degree increased by a factor of 5 when 1/7/2018 fell out of the corresponding semi-annual window. On that day BTC's valuation dropped down by 5\%, while ETH's valuation increased by 10\%. Other large-capitalization cryptocurrencies also rallied on that day. The market MST returned to its centralized topology.

\subsection{The largest eigenvalue vs. MST maximum node degree}
\label{compEIGMST}

Interpretations of $\lambda_{\textrm{max}}^{(\textrm{X})}$ and $\gamma^{(\textrm{X})}$ are mutually related. Expressing cryptocurrency values in a base currency X of large capitalization takes some cross-correlation contribution that is associated with the dynamics of X. From the perspective of a dominant cryptocurrency, the market seems to be rather weakly collective and the corresponding network representation is decentralized. On the other hand, if one expresses other cryptocurrencies in a base currency with its own dynamics, the market is collective and the network is strongly centralized.
\begin{figure}[ht!]
\centering
\includegraphics[width=1\textwidth]{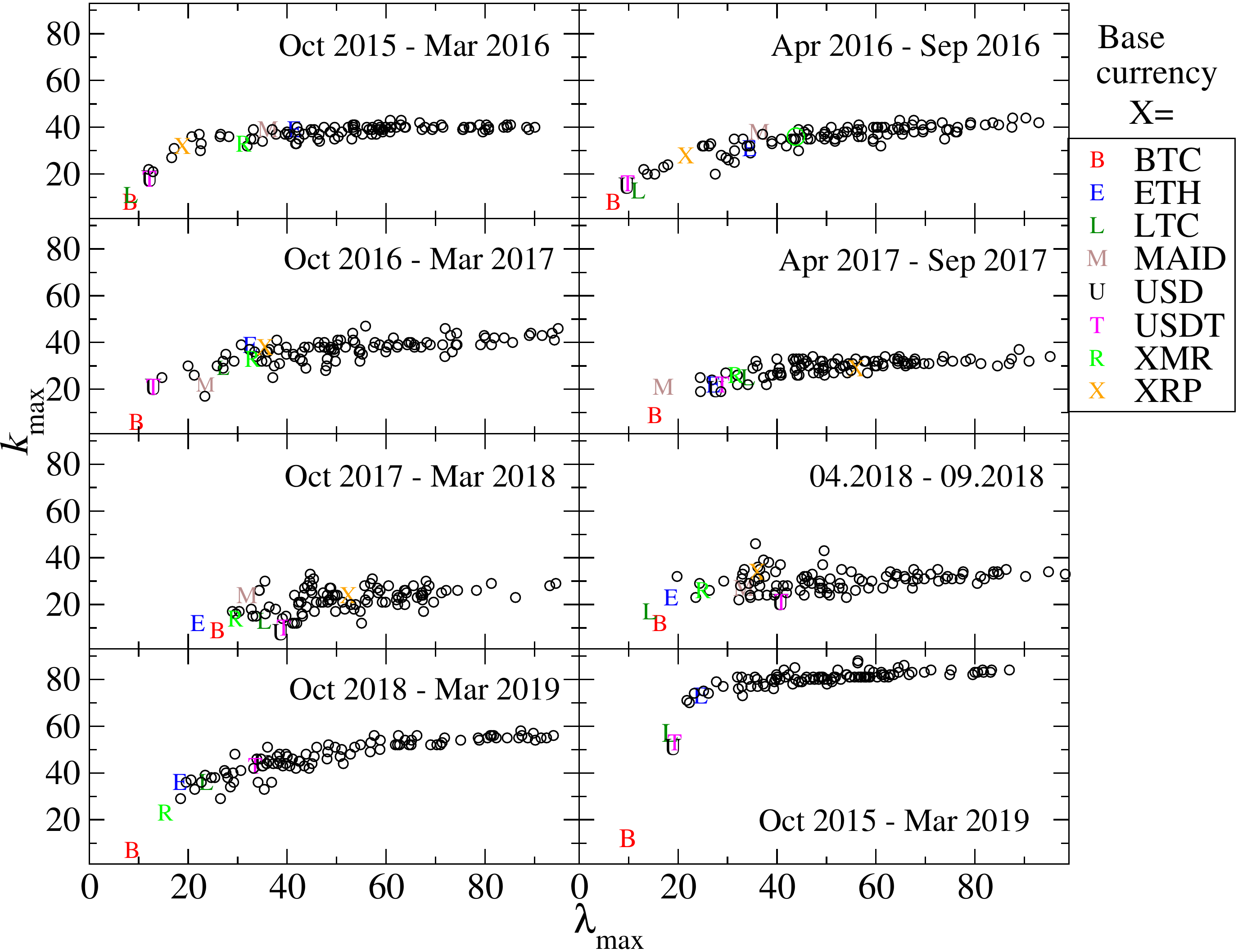} 
\caption{The largest eigenvalue $\lambda_{\textrm{max}}^{(\textrm{X})}$ and the MST maximum node degree $k_{\textrm{max}}^{(\textrm{X})}$) over the whole interval 10/2015--03/2019 (bottom right) and in semi-annual windows. Each circle represents a base currency X, but some characteristic cases are marked with letters.}
\label{fig:EIGkcomp}
\end{figure}

Fig.~\ref{fig:EIGkcomp} shows $\lambda_{\textrm{max}}^{(\textrm{X})}$ vs. $k_{\textrm{max}}^{(\textrm{X})}$ for all the $N$ possible choices of the base currency X calculated in fixed semi-annual windows and over the whole interval 10/2015--3/2019 (bottom left). Each circle or letter denotes different X. (Instead of $\gamma^{(\textrm{X})}$ that describes the CDF tail scaling property, the maximum node degree is used here. This replacement is necessary because for the most choices of X, the scaling quality was not sufficient. It does not alter the interpretation, however.) If a base cryptocurrency is located in the bottom left corner of a plot, its hierarchical position in the market structure is dominant. In all the periods, except for 10/2017--3/2018, BTC holds a dominant position. During 10/2017--2/2018 one could also observe the lowest value of $\lambda_{\textrm{max}}^{(\textrm{X})}$, located below the level of $k_{\textrm{max}}^{(\textrm{X})}=40$. This allowed for the formation of a hierarchical structure with the power-law node degree CDF, which disappeared after the bull market ended in 1/2018 and BTC's dominance returned (Fig.~\ref{fig:EIGkcomp}).

While considering the whole interval, one can observe the highest $k_{\textrm{max}}^{(\textrm{X})} \approx 80$ for a vast majority of the base cryptocurrencies (bottom right panel of Fig.~\ref{fig:EIGkcomp}). This is consistent with the fact that the most nodes are connected to the BTC/USD node and have unit degree, which gives the overall centralized structure of MST. BTC seems to dominate the cryptocurrency market in a way much more significant than USD does~\cite{Drozdz2007curr,Gorski2008,Kwapien2009}, which makes it the most natural base for this market.

\subsection{Cryptocurrency network topology on Binance}
\label{Bi94}

In order to extend sensitivity of the network approach, the $q$MST formalism is a method at hand~\cite{kwapien2017}. It allows one to unveil the network properties on various time scales $s$ and from a  perspective of different fluctuation amplitudes (selected by means of a $q$ parameter). A construction scheme for $q$MST is parallel to the original MST except for a distance $\delta_{ij}(q,s)$ that has to be calculated from the detrended cross-correlation coefficient $\rho(q,s)$ defined by Eq.(\ref{rhoq}) in Sect.~\ref{rho}. It inherits $\rho$'s dependence on the parameters $s$ and $q$:
\begin{equation}
\delta_{ij}^{(\textrm{X})}(q,s) = \sqrt{2 \big( 1 - \rho_{ij}^{(\textrm{X})}(q,s) \big)}, \qquad q > 0.
\label{dijqs} 
\end{equation}
Only positive values of $q$ are considered ($q=1$ and $q=4$), which guarantees that $\delta_{ij}^{(\textrm{X})}(q,s)$ is a metric~\cite{kwapien2017}.

\begin{figure}[ht!]
\centering
\includegraphics[width=1\textwidth]{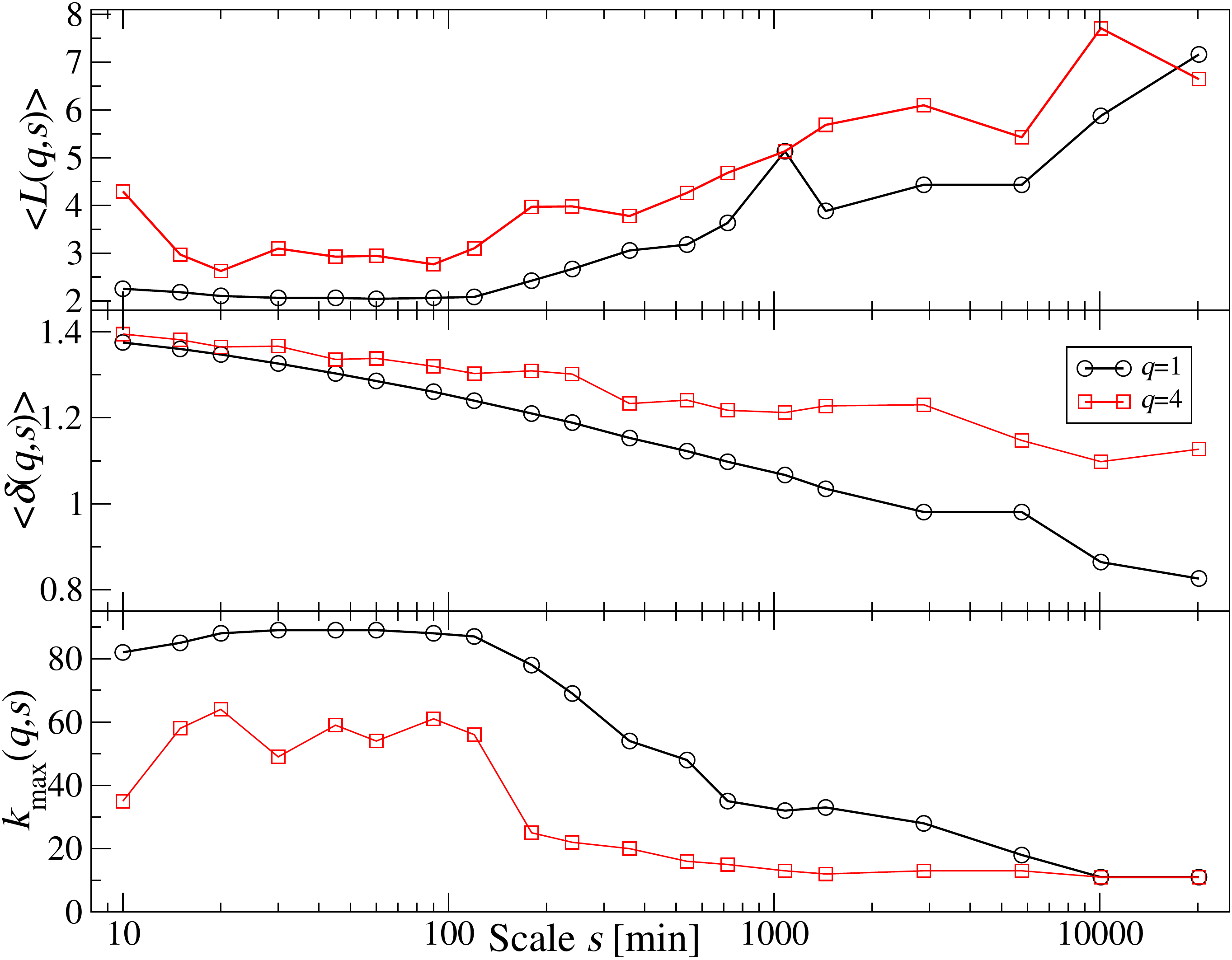} 
\caption{Topological properties of $q$MST with X=BTC: the average shortest binary path length $\langle L(q,s)\rangle$ (top), the mean inter-node metric distance $\langle \delta(q,s) \rangle$ (middle), and the maximum node degree $k_{\textrm{max}}(q,s)$ (bottom) for $q=1$ and $q=4$.}
\label{BiqMST}
\end{figure}
MST are formed from $N=94$ cryptocurrency exchange rate quotes from the Binance platform (a detailed list is given in the Appendix~\ref{listamacBi}). Time series of returns were sampled with 1 min resolution and cover the whole year 2018. Trading on Binance occurs predominantly on the exchange rates where BTC is the base. In 2018 there were 93 such exchange rates. The second most frequently traded cryptocurrency was ETH (90 exchange rates). It was also possible to obtain 30 exchange rates expressed in Binance coin (BNB) and 8 exchange rates expressed in USDT, but here there are only considered those involving BTC. This choice preserves dynamics of the individual cryptocurrencies without being distorted by evolution of the strongest one (see Sect.~\ref{drabinka}). Because of fixing X=BTC, the BTC superscript is dropped in all quantity symbols for clarity.
\begin{figure}
\centering
\includegraphics[width=1\textwidth]{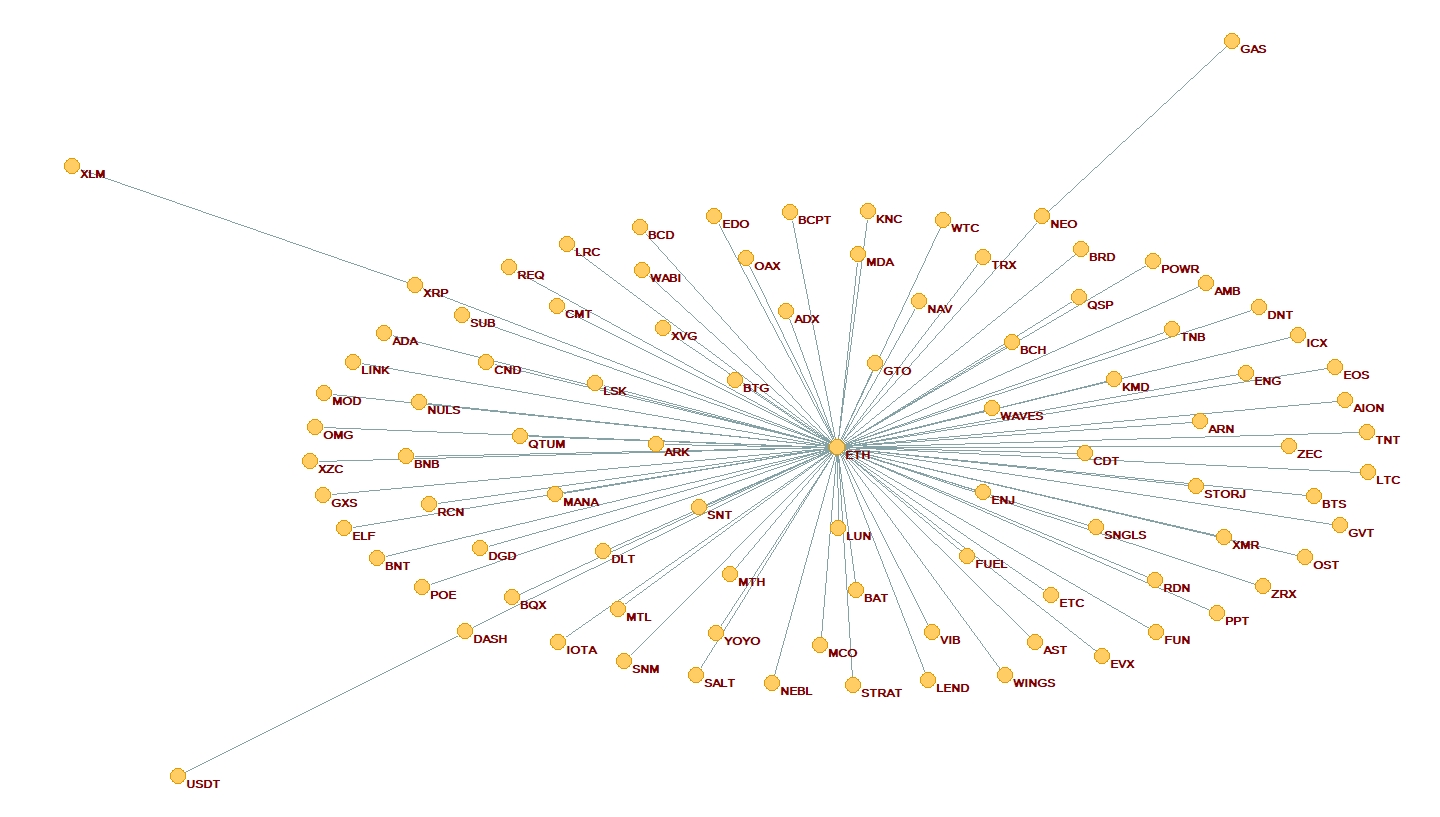} 
\includegraphics[width=1\textwidth]{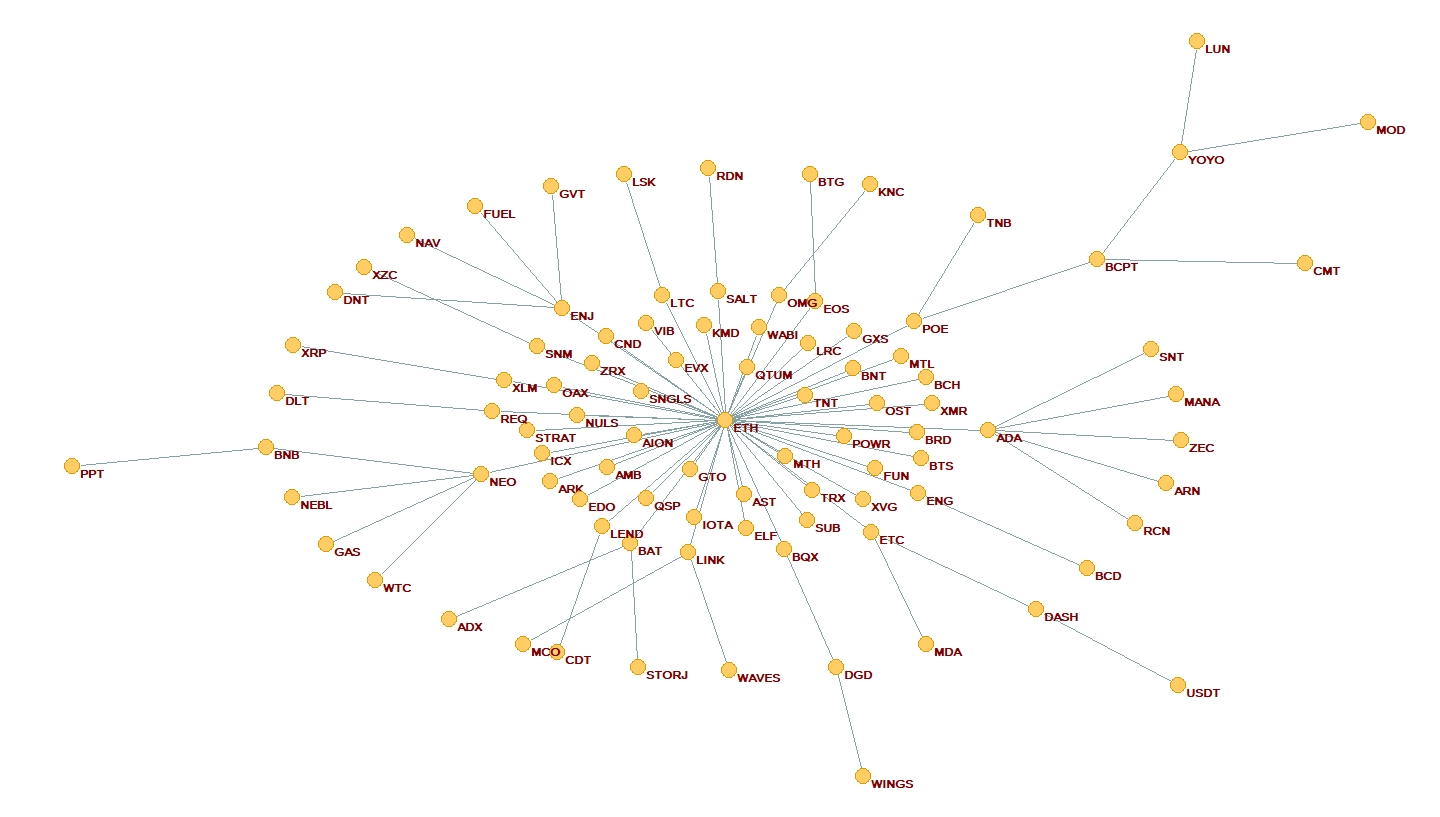} 
\caption{$q$MST with nodes representing cryptocurrencies expressed in BTC for $s=60$ min and $q=1$ (top) and $q=4$ (bottom). The data comes from Binance.}
\label{fig:qMSTs1h}
\end{figure}

\begin{figure}
\centering
\includegraphics[width=1\textwidth]{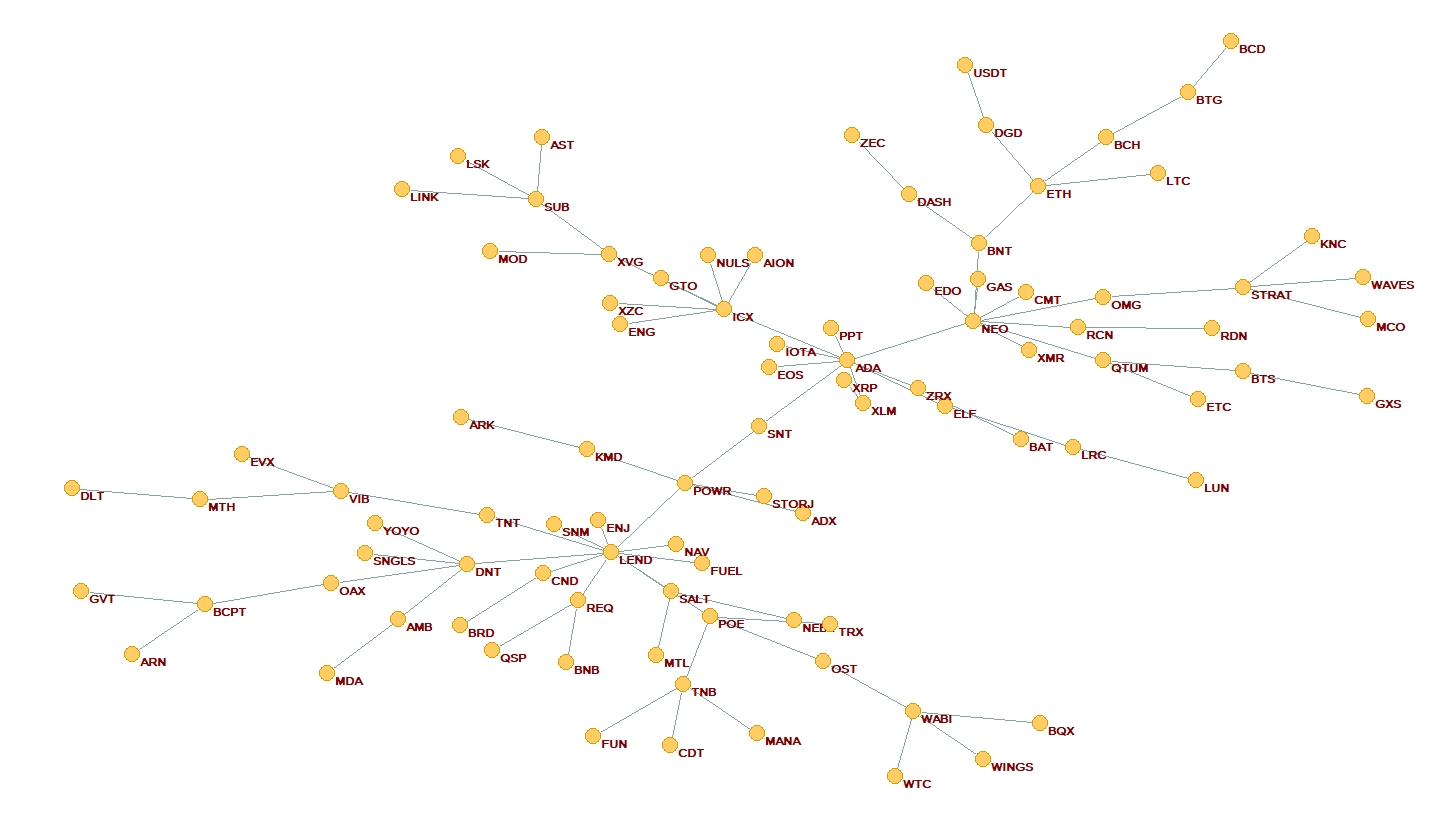} 
\includegraphics[width=1\textwidth]{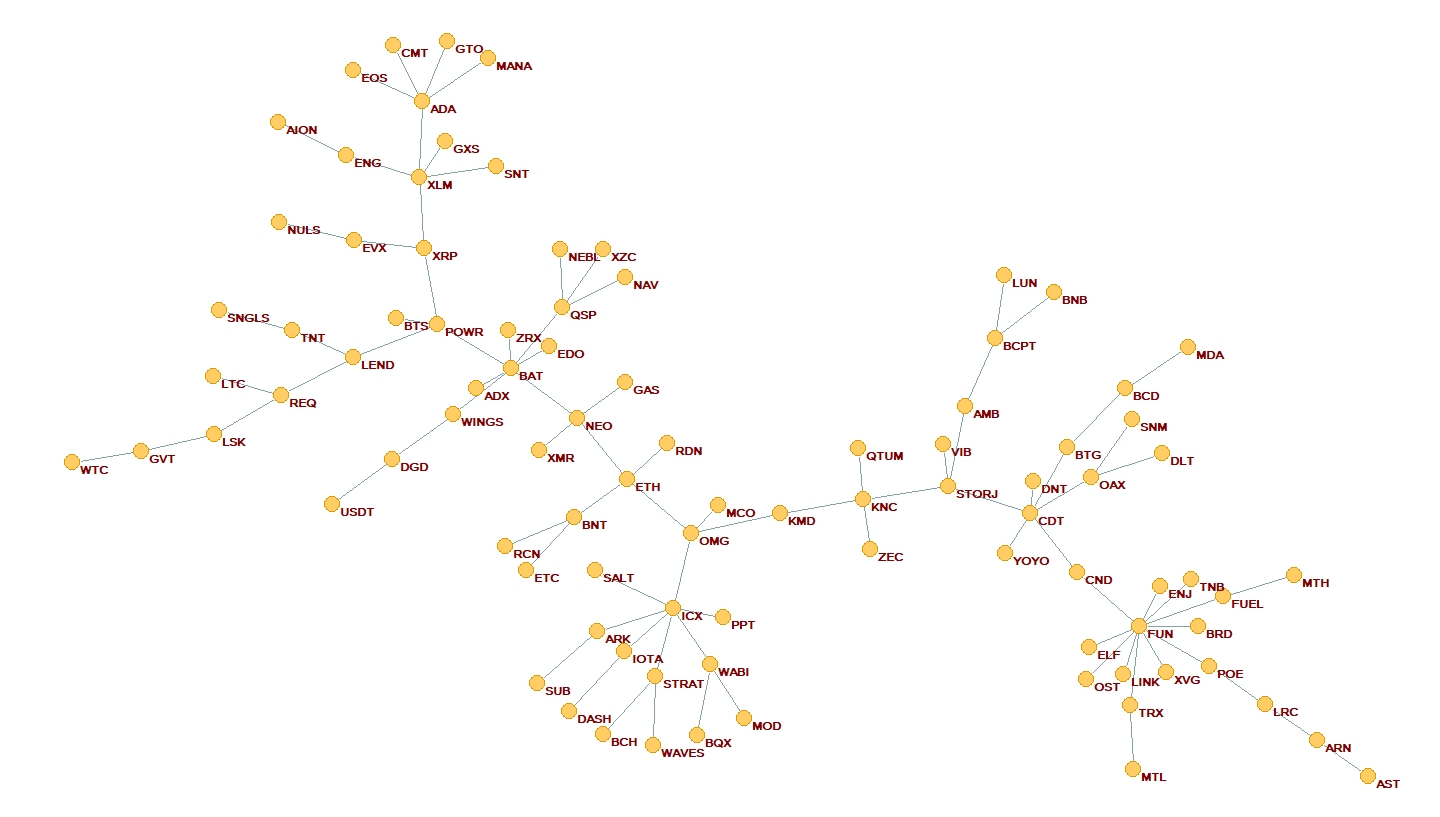} 
\caption{$q$MST with nodes representing cryptocurrencies expressed in BTC for $s=10,080$ min (1 week) and $q=1$ (top) and $q=4$ (bottom). The data comes from Binance.}
\label{fig:qMSTs1w}
\end{figure}

Mean distance $\langle \delta(q,s) \rangle$ decreases with increasing $s$ (middle panel in Fig.~\ref{BiqMST}), which is a natural consequence of the increased cross-correlation strength on longer time scales (Sect.~\ref{sect::CrossCorrelations}). In contrast, it increases with $q$. The average shortest path length $\langle L(q,s)\rangle$ calculated for binary edges (i.e., the $q$MST edges are considered as unweighted here) increases with $s$ and it is larger for $q=4$ than for $q=1$, which indicates that MST becomes less centralized for large returns (top panel in Fig.~\ref{BiqMST}). The MST topology shows a kind of ``phase transition'' for $s \approx 120$ min. Below this scale it is almost perfectly centralized for $q=1$ with $\langle L(q,s)\rangle \approx 2$ and $k_\textrm{max}(q,s) \approx 90$ and significantly centralized for $q=4$ with $\langle L(q,s)\rangle \approx 3$ and $k_\textrm{max}(q,s) \approx 60$. The central node represents always ETH or, more precisely, the exchange rate ETH/BTC (Fig.~\ref{fig:qMSTs1h}), because for short time scales the information flow enters the market principally through the most actively traded cryptocurrency -- ETH. The situation looks different for the time scales longer than 120 min, because for such scales the ETH hub loses centrality and MST becomes decentralized gradually (Fig.~\ref{BiqMST}). For the longest considered time scale of  2 weeks $k_{\textrm{max}}(q,s)$ drops below 20 for both $q=1$ and $q=4$. The corresponding $q$MSTs are distributed (Fig.~\ref{fig:qMSTs1w}).

\begin{figure}[ht!]
\centering
\includegraphics[width=1\textwidth]{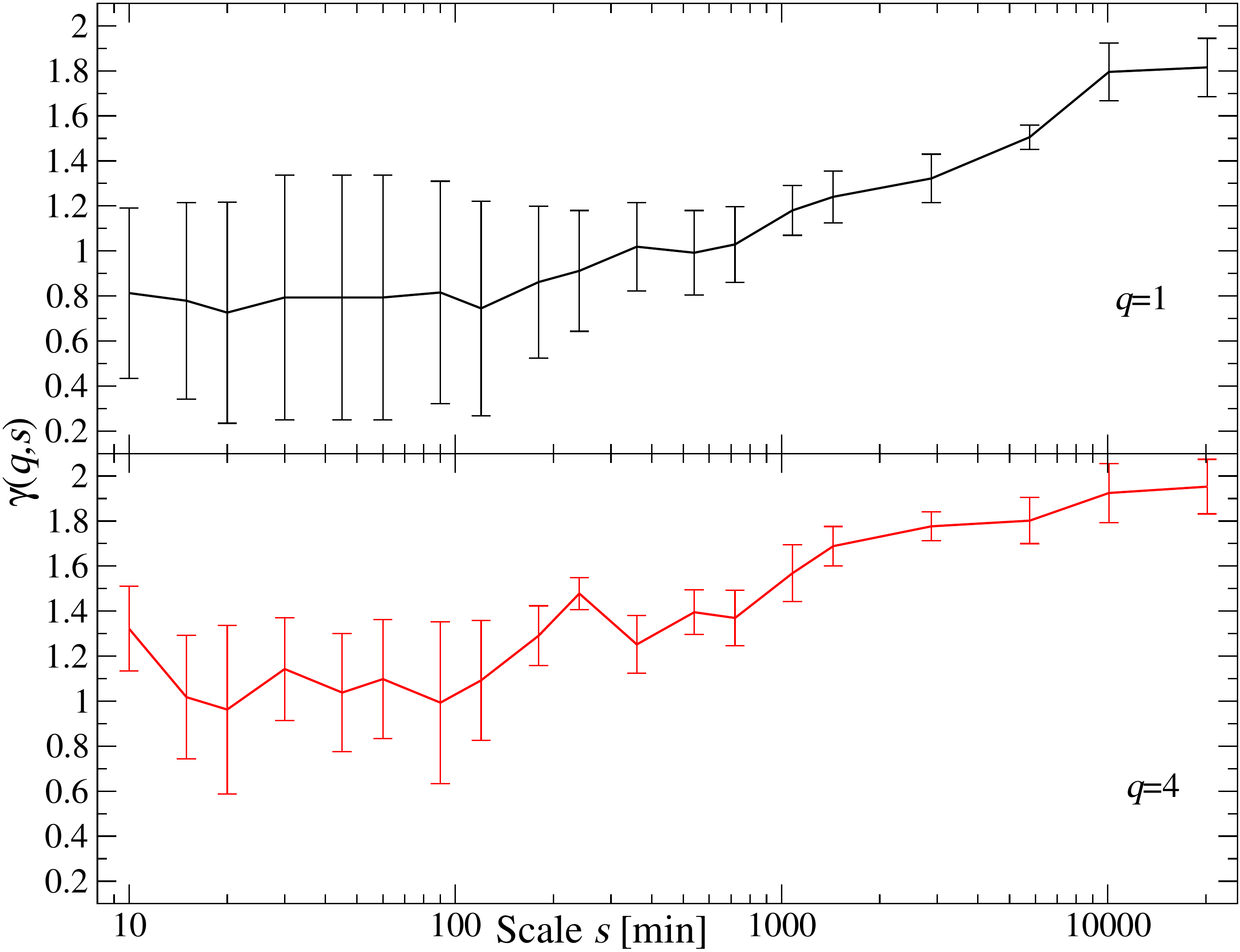} 
\caption{Tail scaling exponent $\gamma(q,s)$ estimated for the node degree CDFs describing topology of $q$MST for different values of $s$ and for $q=1$ (top) and $q=4$ (bottom). Standard error of $\gamma(q,s)$ obtained from a linear regression fit is given by the error bars. $\gamma(q,s)=1.6$ corresponds to the hierarchical network model, while $\gamma(q,s)>1.8$ corresponds to distributed networks.}
\label{gammaBiqMST}
\end{figure}

Despite the fact that $\delta_{ij}(q,s) \approx \sqrt{2}$ for the shortest scales $s=10$ min, the results are statistically significant. This was checked by calculating $\delta_{ij}(q,s)$ and their standard deviation for 100 realizations of surrogate data (randomly shuffled returns of 93 exchange rates). Only in a few cases $\delta_{ij}(q,s)$ exceeded $3\sigma$ threshold.

The above observations are supported by the scaling properties of the $q$MST node degree CDFs. For each $s$ and $q$, the corresponding CDF tails were approximated by a power-law function with the exponent $\gamma(q,s)$ (Eq.(\ref{rozklkrot})). Fig.~\ref{gammaBiqMST} shows values of this exponent together with the standard error of a linear regression fit. Up to $s \approx 120$ min, one does not observe any power-law scaling (large error bars) with a dominant role of the ETH node being responsible for this behaviour (see Fig.~\ref{fig:qMSTs1w}). However, for $s \gtrsim 500$ min the maximum node degree $k_\textrm{max}(q,s)$ decreases and the node degree CDF reveal a scaling tail (a smaller $\gamma(q,s)$ error, Fig.~\ref{gammaBiqMST}). The exponent $\gamma(q,s)$ increases with increasing $s$ reaching values higher than 1.8 for $s=1$ week, which are characteristic for decentralized networks.

The $q$MSTs reach topology of the hierarchical network model~\cite{Noh2003,Ravasz2003} corresponding to $\gamma(q,s) \approx 1.6$ later for $q=1$ ($s \approx 7,000$ min) than for $q=4$ ($s \approx 1,000$ min). This is because for large returns the cryptocurrency exchange rates are globally less correlated than for medium returns, which opens more space for the formation of sub-clusters in $q$MSTs for $q=4$. This result shows that a standard MST construction ($q=1$) offers too little sensitivity and it is thus inferior to the $q$MST approach.

The network approach presented here for X=BTC confirms the results reviewed in earlier sections according to which the cross-correlation between the cryptocurrency exchange rates tend to increase with increasing time scales because of the fact that less frequently traded exchange rate pairs need time to synchronize. With increasing scale, the MST structure becomes more diverse and decentralized. A dominant position of ETH observed for short scales disappears and a hierarchical structure develops. This happens faster for large returns than for medium ones because the cross-correlations are weaker for the former. 

\section{Summary and conclusions}

A striking development of the cryptocurrency market in the last several years -- from a small and peripheral market to a market at the level of a middle-size stock exchange capitalization  -- provides a unique opportunity to observe its quick structural self-organization. Availability of the high frequency data allows one to conduct advanced statistical studies of the processes governing the market evolution from its birth to the present. 

In this review we presented the most important results derived from a few studies carried out based on available high-frequency data from the cryptocurrency exchange platforms. A main stress was put on methods known well from statistical physics like the detrended fluctuation analysis with its  multifractal extensions quantifying nonlinear dependencies in data and the cross-correlation matrix analysis exploiting universal predictions of the random matrix theory and the complex network theory.

A principal conclusion that can be inferred from the presented results is that the cryptocurrency market has gradually been pursuing its way to maturity. It is still not a fully developed market unlike, for example, Forex, since there are still significant differences between both as regards liquidity and the number of transactions, among others. The following characteristics of the market were also discussed in detail: the leptokurtic tails of the return probability distribution functions, the long memory in volatility, the antipersistent dynamics of the returns, the multiscale auto- and cross-correlations between the cryptocurrency exchange rate returns in respect to other cryptocurrencies and the fiat currencies like USD and EUR. Behaviour of the exchange rates on different trading platforms was also a subject of interest.

All these properties, known together as the financial stylized facts, characterize mature financial markets with the stock markets, the commodity markets, and the foreign currency market, among others. In contrast, on developing/emerging markets some or all these properties may not be observed at all. Thus, their existence (or lack thereof) can determine a development stage of a particular market. On the cryptocurrency market one observes a gradual shift from an early, immature stage prior to 2014 towards more developed stages after that year with some parts of the market reaching even a stage of the almost-complete maturity. These parts consist predominantly of the exchange rates involving bitcoin (BTC), ethereum (ETH), a and several other the most liquid cryptocurrencies like ripple (XRP) and litecoin (LTC). That shift was caused by a significant increase of the trading volume and frequency that was seen on the market over the years. This observation points out to these quantities as the principal driving factors of market development.

More specifically, as regards the most mature and long-traded part of the market, i.e., the exchange rates BTC/USD and ETH/USD, the return autocorrelation function was changing its form from short-term memory to long-term one with a power-law decay. In parallel, the return CDF tails lost their initial L\'evy-stable form with the exponent $\gamma \leqslant 2$ form and became inverse cubic power law ($\gamma \approx 3$ or even less). As mentioned above, these shifts took place around 2013-2014 and after that both characteristics do not differ from mature markets. An analogous change of dynamics happened in 2017, when the Hurst exponent rose from $\sim 0.4$ to $\sim 0.5$, i.e., from the antipersistent regime typical for developing markets to the neutral regime proper to mature markets. The same was visible for the multiscale autocorrelations -- the singularity spectra $f(\alpha)$ change shape from a strong left-side asymmetry seen till 2017 to a mature, symmetric form in 2018. As in other markets, multiscaling is observed for large returns, but, contrary to that, the smaller returns are monofractal here unlike the corresponding returns on Forex and other mature markets.

One of the key features of Forex is possibility of expressing any currency in any other currency. On cryptocurrency market such a possibility emerged gradually as new trading platforms became available. Initially, only the exchange rates involving main fiat currencies were traded, but after a few years the cryptocurrencies could also be expressed in each other without participation of the traditional currencies. In this case the most liquid exchange rate -- BTC/ETH -- reveal autocorrelation that resembles the autocorrelation of EUR/USD. However, like in immature markets, $f(\alpha)$ was still left-side asymmetric even in the second half of 2018. Multiscaling and other stylized facts might be viewed as indicators of market complexity. Therefore, a process of market maturing is inevitably associated with complexity increase. A feature that can also be considered in this context is multiscale cross-correlations among the exchange rates. Such cross-correlations were found at the level of medium and large returns and their strength increases with increasing scale (unlike Forex, where such dependence is much weaker). Moreover, for a fixed scale, the detrended cross-correlations were getting stronger over the years, which is especially evident for small scales. This is associated with reduction in the triangular arbitrage opportunities. The less opportunities are detectable, the more liquid and mature the market is, because lack of liquidity slows a rate of information spreading. Since the cryptocurrency market is decentralized and since liquidity varies among the trading platforms, different platforms display different stage of the market maturity -- a feature unique for this market. 

The strength of the multiscale detrended cross-correlations among the exchange rates depends crucially on whether the two exchange rates involve a common base (crypto)currency -- if this is the case, they are substantially stronger. There is a hierarchical structure of cross-correlations in the market: one can identify the exchange rate clusters consisting of stronger-than-average correlated rates -- typically those that remain in a triangle relation with each other. Because of a slower information flow, the clusters do not comprise the exchange rates without triangle relationship, which is a feature distinct from Forex, where there are other factors that correlated currencies, like the geographic proximity.

At present it is still too early to discuss whether the cryptocurrency market will survive for a substantial period of time or in which direction it may evolve except for that it is very likely that its liquidity can improve constantly in the following years. One of the key reasons is that despite its high volatility level at present, this market offers new investment opportunities, thus actually facilitating portfolio diversification. Here an important role is played by the most liquid cryptocurrencies, ETH and BTC, which, if chosen for the exchange rate base, decollectify the market and, in turn, allow one for further portfolio diversification within the cryptocurrency market itself. These cryptocurrencies seem to be also the most mature ones and thus the most credible.

In this context the results that is among the most important is a kind of ``phase transition'' that has been noticed recently in 2020 during the Covid-19 pandemic. The cryptocurrency market evolution used to show no cross-correlation with the evolution of the traditional markets: the prices of BTC or ETH and other financial instruments like the fiat currencies, stock indices, and commodities were uncorrelated on average. However, this suddenly changed during this year's turmoil observed on the markets that was related to the Covid-19 outburst: the major cryptocurrency prices became cross-correlated. First, the positive cross-correlations were seen with the instruments considered safe, like Japanese yen, but later such cross-correlations appeared also with rather the risky instruments like S\&P500, typical currencies, and commodities. Interestingly, this occurred not only during the market panic at the 1st pandemic wave in early Spring 2020, but also during the recovery phase experienced by the traditional markets in late Spring and even during the pandemic slowdown in Summer 2020. It seems that the cryptocurrency market ceased to be viewed as a ``safe haven'' by the investors and become a part of the network constituting the global market. It is an important question whether this coupling is a transient feature only or, maybe, it is another manifestation of the market's maturation and it will continue to be observed in future.

\newpage
\appendix
\renewcommand*{\thesection}{\Alph{section}}
\section{Full names and statistical properties of cryptocurrencies from Binance and Kraken}
\label{BiKrdodatek}

\begin{table}[ht!]
\centering
\caption{List of symbols and full names of cryptocurrencies listed on both Binance and Kraken.}
\begin{tabular}{|c|c|}
\hline
\textbf{Ticker} & \textbf{Full name}           \\ \hline
BAT    & Basic Attention Token \\ \hline
BCH    & Bitcoin Cash          \\ \hline
BTC    & Bitcoin               \\ \hline
BNB    & Binance coin               \\ \hline
DASH   & Dash                  \\ \hline
ETC    & Ethereum Classic      \\ \hline
ETH    & Ethereum              \\ \hline
ICX    & ICON                  \\ \hline
LSK    & Lisk                  \\ \hline
LTC    & Litecoin              \\ \hline
MIOTA  & IOTA               \\ \hline
NEO    & NEO                   \\ \hline
REP    & Augur                 \\ \hline
USDT   & Tether                \\ \hline
XLM    & Stellar               \\ \hline
XMR    & Monero                \\ \hline
XRP    & Ripple                \\ \hline
ZEC    & Zcash                 \\ \hline

\end{tabular}
\label{listanazwBiKr}
\end{table}

\begin{figure}[]
\centering
\includegraphics[width=1\textwidth]{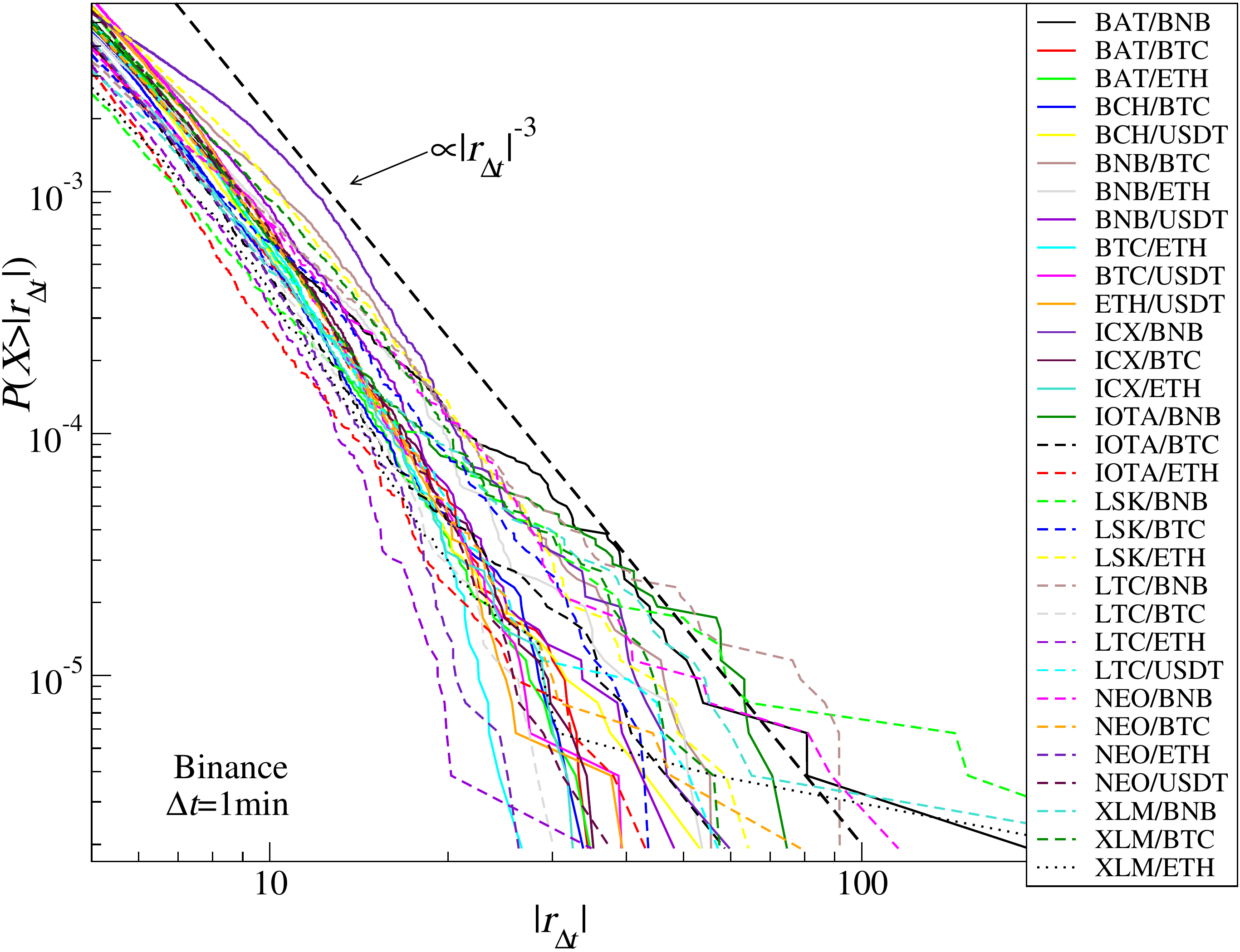}
\caption{Cumulative distributions of absolute normalized  log-returns $r_{\Delta t=1\textrm{min}}$ representing cryptocurrency exchange rates on Binance in 2018.}
\label{fig:Binance_rozklady}
\end{figure}

\begin{table}[]
\centering
\caption{Estimated $\gamma$ exponent, average length of non-trading periods $\langle t_0 \rangle$ (i.e., zero-returns sequences), number of non-trading periods $N_0$, average volume-per-minute in USD and Hurst exponent, for the cryptocurrency pairs listed on Binance in 2018.}
\begin{tabular}{|c|c|c|c|c|c|}
\hline
\textbf{Name}& \boldmath$\gamma$ &\boldmath $\langle N_0 \rangle$ &\boldmath $N_0$ & \boldmath$\langle W \rangle$ &\boldmath$H$    \\ \hline
    BTC/USDT  & 3,45$\pm$0,1      & 1,07   & 8559  & 196331  & 0,47 \\ \hline                    
    ETH/USDT  & 3,39$\pm$0,1      & 1,10   & 16184 & 61430  & 0,48  \\ \hline
    BTC/ETH  & 3,30$\pm$0,1      & 1,09   & 12134 & 60824     & 0,51                     \\ \hline
BNB/BTC  & 2,67$\pm$0,2  & 1,26     & 40098 & 18634 & 0,43 \\ \hline
BCH/BTC  & 3,36$\pm$0,1      & 1,25   & 29973 & 16299  & 0,48                       \\ \hline
BNB/USDT & 3,08$\pm$0,15 & 1,20     & 31320 & 16268 & 0,46\\ \hline
BCH/USDT  & 3,36$\pm$0,1      & 1,40   & 40922 & 15893  & 0,48                       \\ \hline
ICX/BTC  & 3,15$\pm$0,1  & 1,58     & 77707 & 15336& 0,45 \\ \hline
NEO/USDT & 3,29$\pm$0,1  & 1,27     & 42709 & 14830& 0,47  \\ \hline
XLM/BTC  & 2,71$\pm$0,15 & 1,33     & 67637 & 14556 & 0,46 \\ \hline
NEO/BTC  & 3,17$\pm$0,15  & 1,34     & 61330 & 14445 & 0,46 \\ \hline
LTC/BTC  & 3,42$\pm$0,1      & 1,17   & 34360 & 13110  &  0,47                    \\ \hline
LTC/USDT  & 3,23$\pm$0,15     & 1,29   & 45208 & 11793   & 0,47                    \\ \hline
IOTA/BTC & 3,26$\pm$0,15 & 1,30     & 52793 & 7928 & 0,45 \\ \hline
NEO/ETH  & 3,54$\pm$0,1 & 1,94     & 80537 & 3765 & 0,42  \\ \hline
ICX/ETH  & 3,24$\pm$0,15 & 2,20     & 87489 & 3496 & 0,42  \\ \hline
BNB/ETH  & 2,74$\pm$0,2  & 1,47     & 65950 & 2755 & 0,45 \\ \hline
XLM/ETH  & 2,96$\pm$0,2   & 1,95     & 85581 & 2580& 0,43   \\ \hline
LSK/BTC  & 2,72$\pm$0,15  & 1,94     & 92975 & 2489& 0,44  \\ \hline
BAT/BTC  & 3,12$\pm$0,15   & 1,85     & 88702 & 2166 & 0,45  \\ \hline
LTC/ETH  & 3,61$\pm$0,15  & 2,18     & 93044 & 1825& 0,42  \\ \hline
LSK/BNB  & 2,39$\pm$0,35   & 10,27    & 44317 & 1595 & 0,41 \\ \hline
IOTA/ETH & 3,52$\pm$0,15   & 2,29     & 91798 & 1495 & 0,43 \\ \hline
NEO/BNB  & 2,54$\pm$0,2   & 3,66     & 87796 & 1294 & 0,40  \\ \hline
LTC/BNB  & 2,32$\pm$0,25  & 3,86     & 88053 & 830  & 0,41  \\ \hline
BAT/ETH  & 3,48$\pm$0,15 & 2,93     & 95765 & 516  & 0,41 \\ \hline
LSK/ETH  & 2,75$\pm$0,15  & 4,19     & 81720 & 451  & 0,39  \\ \hline
ICX/BNB  & 2,51$\pm$0,2   & 4,80     & 76597 & 260 & 0,36  \\ \hline
IOTA/BNB & 2,88$\pm$0,25  & 4,35     & 81816 & 130 &0,39   \\ \hline
BAT/BNB  & 2,70$\pm$0,3 & 6,84     & 61754 & 72  & 0,36   \\ \hline
XLM/BNB  & 2,59$\pm$0,3 & 3,65     & 90019 & 12  & 0,37   \\ \hline
\end{tabular}

\label{tab:BiApp}
\end{table}

\begin{figure}[]
\centering
\includegraphics[width=1\textwidth]{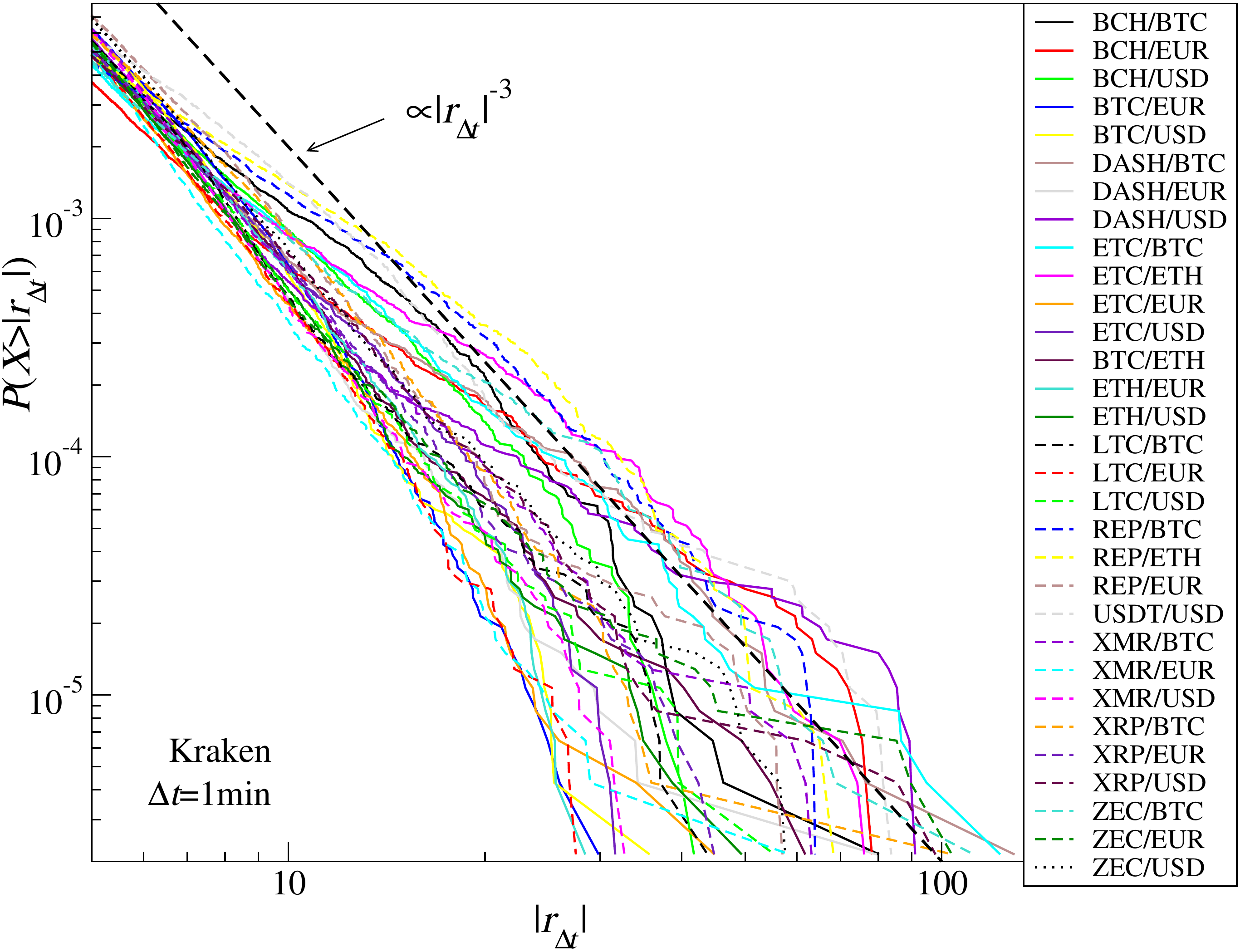}
\caption{Cumulative distributions of absolute normalized  log-returns $r_{\Delta t=1\textrm{min}}$ representing cryptocurrency exchange rates on Kraken in 2018.}
\label{fig:Kraken_rozklady}
\end{figure}

\begin{table}[]
\centering
\caption{Estimated $\gamma$ exponent, average length of non-trading periods $\langle t_0 \rangle$ (i.e., zero-returns sequences), number of non-trading periods $N_0$, average volume-per-minute in USD and Hurst exponent, for the cryptocurrency pairs listed on Binance  in 2018.}
\begin{tabular}{|c|c|c|c|c|c|}
\hline
\textbf{Name}& \boldmath$\gamma$ &\boldmath $\langle N_0 \rangle$ &\boldmath $N_0$ & \boldmath$\langle W \rangle$ &\boldmath$H$    \\ \hline
BTC/EUR  & 3,61$\pm$0,1  & 1,49     & 44310 & 35599 & 0,48\\ \hline
BTC/USD  & 3,63$\pm$0,15 & 1,61 & 55360 & 33998 & 0,48 \\ \hline
ETH/USD  & 3,4$\pm$0,1   & 1,78 & 67807 & 19711 & 0,49 \\ \hline
ETH/EUR  & 3,37$\pm$0,15  & 1,65     & 61097 & 18171 & 0,49 \\ \hline
XRP/EUR & 3,14$\pm$0,15  & 2,28     & 77480 & 7321 & 0,46  \\ \hline
XRP/USD  & 2,86$\pm$0,2  & 2,69     & 80418 & 6259 & 0,46 \\ \hline
BTC/ETH  & 3,18$\pm$0,2  & 2,62 & 82074 & 5250 & 0,48 \\ \hline
XRP/BTC  & 2,99$\pm$0,15 & 3,50 & 72556 & 2763  & 0,45 \\ \hline
BCH/USD  & 2,61$\pm$0,15 & 3,83 & 75427 & 2151 & 0,48  \\ \hline
BCH/EUR  & 2,28$\pm$0,3  & 3,31     & 82295 & 2036  & 0,47 \\ \hline
USDT/USD & 2,37$\pm$0,3  & 8,97     & 43821 & 1747 & 0,31  \\ \hline
LTC/EUR  & 3,61$\pm$0,1 & 3,11     & 84104 & 1732 & 0,48 \\ \hline
LTC/USD  & 3,41$\pm$0,1  & 3,48 & 79983 & 1722& 0,47   \\ \hline
BCH/BTC  & 2,35$\pm$0,15 & 4,39 & 68059 & 1251 & 0,43  \\ \hline
ETC/USD  & 3,16$\pm$0,15  & 4,18     & 74681 & 1114 & 0,47  \\ \hline
ETC/EUR  & 3,53$\pm$0,15 & 4,01     & 75596 & 973 & 0,47   \\ \hline
LTC/BTC  & 3,34$\pm$0,15 & 5,09 & 62675 & 940 & 0,45   \\ \hline
XMR/EUR  & 3,77$\pm$0,15  & 4,16     & 76277 & 865 & 0,49   \\ \hline
XMR/USD  & 3,73$\pm$0,2  & 5,58     & 63385 & 864 & 0,49  \\ \hline
ETC/BTC  & 2,31$\pm$0,2  & 6,34 & 55647 & 709  & 0,45 \\ \hline
XMR/BTC  & 3,21$\pm$0,15 & 6,58 & 54832 & 614  & 0,43  \\ \hline
DASH/EUR & 3,59$\pm$0,15 & 6,47     & 57268 & 526 & 0,48  \\ \hline
DASH/BTC & 2,52$\pm$0,2  & 7,58 & 49134 & 416 & 0,44    \\ \hline
ZEC/USD  & 3,18$\pm$0,2 & 7,45     & 51105 & 354 & 0,48   \\ \hline
DASH/USD & 2,81$\pm$0,3  & 9,36     & 42727 & 394 & 0,47   \\ \hline
REP/EUR  & 3,09$\pm$0,25  & 6,79     & 54740 & 374& 0,45   \\ \hline
ZEC/EUR  & 3,25$\pm$0,25 & 6,20     & 59369 & 345  & 0,47 \\ \hline
ZEC/BTC  & 2,29$\pm$0,25  & 7,51     & 51170 & 345 & 0,42   \\ \hline
ETC/ETH  & 2,18$\pm$0,2  & 9,15 & 41861 & 266 & 0,41  \\ \hline
REP/BTC  & 2,05$\pm$0,25 & 8,43     & 46419 & 247  & 0,41  \\ \hline
REP/ETH  & 2,00$\pm$0,3 & 11,57    & 35701 & 158 & 0,40  \\ \hline
\end{tabular}

\label{tab:KrApp}
\end{table}
\begin{table}[]
\centering
\caption{Average and maximum triangular arbitrage opportunity on Binance in 2018.}
\begin{tabular}{|c|c|c|}
\hline
 \textbf{Triangle}         & \textbf{Mean}    & \textbf{Max}   \\ \hline
BTC-USDT-BNB & 0,00096 & 0,0465 \\ \hline
BTC-USDT-ETH & 0,00059 & 0,0367 \\ \hline
BTC-USDT-LTC & 0,00087 & 0,0931 \\ \hline
BTC-USDT-NEO & 0,00097 & 0,1029  \\ \hline
BTC-USDT-BCH & 0,00129  & 0,0793 \\ \hline
BNB-ETH-USDT & 0,00126  & 0,0582 \\ \hline
LTC-ETH-USDT & 0,00127  & 0,0955 \\ \hline
NEO-ETH-USDT & 0,00131  & 0,0435 \\ \hline
LTC-BNB-USDT & 0,00203  & 0,2517  \\ \hline
NEO-BNB-USDT & 0,00222  & 0,2924  \\ \hline
ETH-BAT-BTC  & 0,00255  & 0,1249  \\ \hline
ETH-BNB-BTC  & 0,00108   & 0,0631 \\ \hline
ETH-BNT-BTC  & 0,00286  & 0,0889 \\ \hline
ETH-ETC-BTC  & 0,00139  & 0,0630 \\ \hline
ETH-ICX-BTC  & 0,00148  & 0,0823 \\ \hline
ETH-IOTA-BTC & 0,00159  & 0,0699  \\ \hline
ETH-LSK-BTC  & 0,00243   & 0,2139   \\ \hline
ETH-LTC-BTC  & 0,00115  & 0,05458 \\ \hline
ETH-NEO-BTC  & 0,00115  & 0,09129 \\ \hline
ETH-XLM-BTC  & 0,00138  & 0,71637  \\ \hline
BTC-BAT-BNB  & 0,00447  & 0,57722  \\ \hline
BTC-ICX-BNB  & 0,00343   & 0,2256  \\ \hline
BTC-IOTA-BNB & 0,00266  & 0,23356  \\ \hline
BTC-LSK-BNB  & 0,00486  & 1,16489   \\ \hline
BTC-LTC-BNB  & 0,00185  & 0,24864  \\ \hline
BTC-NEO-BNB  & 0,00204  & 0,284  \\ \hline
BTC-XLM-BNB  & 0,00228  & 0,7974 \\ \hline
ETH-BAT-BNB  & 0,00483   & 0,5747  \\ \hline
ETH-ICX-BNB  & 0,00355  & 0,2288  \\ \hline
ETH-IOTA-BNB & 0,00292  & 0,24176  \\ \hline
ETH-LSK-BNB  & 0,00519  & 1,16722   \\ \hline
ETH-NEO-BNB  & 0,00227  & 0,28951  \\ \hline
ETH-XLM-BNB  & 0,00258    & 0,79693   \\ \hline
\end{tabular}

\label{tab:BiArbitall}
\end{table}
\begin{table}[]
\centering
\caption{Average and maximum triangular arbitrage opportunity on Kraken in 2018.}
\begin{tabular}{|c|c|c|}
\hline
 \textbf{Triangle}         & \textbf{Mean}    & \textbf{Max}   \\ \hline
BTC-EUR-ETH  & 0,00112 & 0,0776 \\ \hline
BTC-EUR-BCH  & 0,00186 & 0,2003 \\ \hline
BTC-EUR-DASH & 0,00131 & 0,2087 \\ \hline
ETC-EUR-ETH  & 0,00160 & 0,1586 \\ \hline
ETC-EUR-BTC  & 0,00149 & 0,2449 \\ \hline
BTC-EUR-XMR  & 0,00146 & 0,1258 \\ \hline
BTC-EUR-LTC  & 0,00129 & 0,1359 \\ \hline
BTC-EUR-XRP  & 0,00156 & 0,1919 \\ \hline
BTC-EUR-REP  & 0,00176 & 0,2080 \\ \hline
ETH-EUR-REP  & 0,00184 & 0,2920 \\ \hline
BTC-EUR-ZEC  & 0,00139 & 0,2408 \\ \hline

BTC-USD-ETH  & 0,00121 & 0,1147 \\ \hline
BTC-USD-BCH  & 0,00179 & 0,1852 \\ \hline
BTC-USD-DASH & 0,00126 & 0,2179 \\ \hline
ETC-USD-ETH  & 0,00159 & 0,1671 \\ \hline
ETC-USD-BTC  & 0,00145 & 0,2415 \\ \hline
BTC-USD-XMR  & 0,00138 & 0,1187 \\ \hline
BTC-USD-LTC  & 0,00124 & 0,1336 \\ \hline
BTC-USD-XRP  & 0,00155 & 0,2269 \\ \hline
BTC-USD-ZEC  & 0,00133 & 0,2203 \\ \hline

ETC-ETH-BTC  & 0,00119 & 0,2459 \\ \hline
REP-ETH-BTC  & 0,00139 & 0,3511 \\ \hline
\end{tabular}

\label{tab:KrArbitall}
\end{table}

\newpage
\vskip-1cm 
\section{List of 100 cryptocurrency names from CoinMarketCap}
\label{listamac}
\begin{table}[ht!]
\centering
\caption{Full names of the cryptocurrencies from section~\ref{Matrixcorr}.}
\scriptsize
\begin{tabular}{|c|c|c|c|}
\hline
\textbf{Ticker} & \textbf{Full name} & \textbf{Ticker} & \textbf{Full name}    \\ \hline
BTC             & Bitcoin              & GAME            & GameCredits             \\ \hline
ETH             & Ethereum             & GCR             & Global Currency Reserve \\ \hline
XRP             & Ripple               & GRC             & GridCoin                \\ \hline
LTC             & Litecoin             & NLG             & Gulden                  \\ \hline
XLM             & Stellar              & LEO             & LEOcoin                 \\ \hline
USDT            & Tether               & MINT            & MintCoin                \\ \hline
DASH            & Dash                 & NMC             & Namecoin                \\ \hline
XMR             & Monero               & MUE             & MonetaryUnit            \\ \hline
XEM             & NEM                  & XMY             & Myriad                  \\ \hline
BCN             & Bytecoin             & NAV             & NavCoin                 \\ \hline
BTS             & BitShares            & NTRN            & Neutron                 \\ \hline
DGB             & DigiByte             & OK              & OKCash                  \\ \hline
XVG             & Verge                & OMNI            & Omni                    \\ \hline
DOGE            & Dogecoin             & PINK            & Pinkcoin                \\ \hline
SC              & Siacoin              & POT             & PotCoin                 \\ \hline
MAID            & MaidSafeCoin         & PURA            & Pura                    \\ \hline
MONA            & Monacoin             & RBY             & Rubycoin                \\ \hline
EMC             & Emercoin             & SHIFT           & Shift                   \\ \hline
RDD             & ReddCoin             & SIB             & SIBCoin                 \\ \hline
NXT             & Nxt                  & SLR             & SolarCoin               \\ \hline
SYS             & SysCoin              & XST             & Stealth                 \\ \hline
NXS             & Nexus                & VRC             & VeriCoin                \\ \hline
GRS             & Groestlcoin          & XAUR            & Xaurum                  \\ \hline
VTC             & Vertcoin             & NEOS            & NeosCoin                \\ \hline
PPC             & Polishcoin           & IOC             & I/O Coin                \\ \hline
UNO             & Unobtanium           & AUR             & Auroracoin              \\ \hline
XPM             & Primecoin            & CURE            & Curecoin                \\ \hline
XWC             & Whitecoin            & NVC             & Novacoin                \\ \hline
BAY             & BitBay               & USNBT           & NuBits                  \\ \hline
VIA             & Viacoin              & PTC             & Pesetacoin              \\ \hline
AEON            & Aeon                 & GLD             & GoldCoin                \\ \hline
ABY             & ArtByte              & CANN            & CannabisCoin            \\ \hline
BITB            & Bean Cash            & TRC             & Terracoin               \\ \hline
BCY             & Bitcrystals          & QRK             & Quark                   \\ \hline
BSD             & Bitsend              & EFL             & e-Gulden                \\ \hline
BLK             & Blackcoin            & ADC             & AudioCoin               \\ \hline
BURST           & Burst                & CPC             & CPChain                 \\ \hline
BLOCK           & Blocknet             & XMG             & Magi                    \\ \hline
CLAM            & Clams                & HUC             & HunterCoin              \\ \hline
COVAL           & Circuits of Value    & ORB             & Orbitcoin               \\ \hline
CLOAK           & Cloakcoin            & CDN             & Canada eCoin            \\ \hline
XCP             & Counterparty         & DEM             & Deutsche eMark          \\ \hline
DMD             & Diamond              & ZET             & Zetacoin                \\ \hline
XDN             & DigitalNote          & BTA             & Bata                    \\ \hline
EMC2            & Einsteinium          & WDC             & WorldCoin               \\ \hline
EXP             & Expanse              & FJC             & FujiCoin                \\ \hline
FAIR            & Faircoin             & START           & Stratcoin               \\ \hline
FTC             & Feathercoin          & DGC             & Digitalcoin             \\ \hline
FLO             & Flo                  & TEK             & TEKcoin                 \\ \hline
FLDC            & Foldigcoin           & PXI             & Prime-XI                \\ \hline
\end{tabular}
\label{listaKrdaily}

\end{table}
\newpage
\vskip-1cm 
\section{List of 94 cryptocurrency names from Binance}
\label{listamacBi}

\begin{table}[ht!]
\centering
\caption{Full names of the cryptocurrencies from section~\ref{Bi94}.}
\scriptsize
\begin{tabular}{|c|c|c|c|}
\hline
\textbf{Ticker} & \textbf{Full name} & \textbf{Ticker} & \textbf{Full name}    \\ \hline
ADA  & Cardano                    & LTC   & Litecoin             \\ \hline
ADX  & AdEx                       & LUN   & Lunyr                \\ \hline
AION & Aion                       & MANA  & Decentraland         \\ \hline
AMB  & Ambrosus                   & MCO   & MCO                  \\ \hline
ARK  & Ark                        & MDA   & Moeda Loyalty Points \\ \hline
ARN  & Aeron                      & MOD   & Modum                \\ \hline
AST  & AirSwap                    & MTH   & Monetha              \\ \hline
BAT  & Basic Attention Token      & MTL   & Metal                \\ \hline
BCD  & Bitcoin Diamond            & NAV   & NavCoin              \\ \hline
BCH  & Bitcoin Cash               & NEBL  & Neblio               \\ \hline
BCPT & Blockmason Credit Protocol & NEO   & NEO                  \\ \hline
BNB  & Binance coin               & NULS  & NULS                 \\ \hline
BNT  & Bancor                     & OAX   & OAX                  \\ \hline
BQX  & Ethos                      & OMG   & OmiseGO              \\ \hline
BRD  & Bread                      & OST   & OST                  \\ \hline
BTC  & Bitcoin                    & POE   & Po.et                \\ \hline
BTG  & Bitcoin Gold               & POWR  & Power Ledger         \\ \hline
BTS  & BitShares                  & PPT   & Populous             \\ \hline
CDT  & Blox                       & QSP   & Quantstamp           \\ \hline
CMT  & CyberMiles                 & QTUM  & QTUM                 \\ \hline
CND  & Cindicator                 & RCN   & Ripio Credit Network \\ \hline
DASH & Dash                       & RDN   & Raiden Network Token \\ \hline
DGD  & DigixDAO                   & REQ   & Request              \\ \hline
DLT  & Agrello                    & SALT  & SALT                 \\ \hline
DNT  & District0x                 & SNGLS & SingularDTV          \\ \hline
EDO  & Eidoo                      & SNM   & SONM                 \\ \hline
ELF  & Aelf                       & SNT   & Status               \\ \hline
ENG  & Enigma                     & STORJ & Storj                \\ \hline
ENJ  & Enjin Coin                 & STRAT & Stratis              \\ \hline
EOS  & EOS                        & SUB   & Substratum           \\ \hline
ETC  & Ethereum Classic           & TNB   & Time New Bank        \\ \hline
ETH  & Ethereum                   & TNT   & Tierion              \\ \hline
EVX  & Everex                     & TRX   & Tron                 \\ \hline
FUEL & Etherparty                 & USDT  & Tether               \\ \hline
FUN  & FunFair                    & VIB   & Viberate             \\ \hline
GAS  & Gas                        & WABI  & Tael                 \\ \hline
GTO  & Gifto                      & WAVES & Waves                \\ \hline
GVT  & Genesis Vision             & WINGS & Wings                \\ \hline
GXS  & GXChain                    & WTC   & Waltonchain          \\ \hline
ICX  & ICON                       & XLM   & Stellar              \\ \hline
IOTA & MIOTA                      & XMR   & Monero               \\ \hline
KMD  & Komodo                     & XRP   & Ripple               \\ \hline
KNC  & Kyber Network              & XVG   & Verge                \\ \hline
LEND & Aave                       & XZC   & Zcoin                \\ \hline
LINK & Chainlink                  & YOYO  & Yoyow                \\ \hline
LRC  & Loopring                   & ZEC   & Zcash                \\ \hline
LSK  & Lisk                       & ZRX   & 0x                   \\ \hline
\end{tabular}
\label{listanazwBi94}
\end{table}

\newpage
\bibliographystyle{elsarticle-num}

{\footnotesize
\bibliography{PhysRep}
}
\end{document}